\documentclass[final,3p,times]{elsarticle}

\usepackage{eqnarray,amssymb}
\usepackage{lipsum}
\usepackage{graphicx}        
\usepackage{grffile}         
\usepackage{graphics}
\usepackage{epsfig}
\usepackage{verbatim}
\usepackage{dcolumn}
\usepackage{soul}
\usepackage{bm}            
\usepackage{mathtools}
\usepackage{physics}
\usepackage{hyperref}        
\hypersetup{colorlinks,breaklinks,
             linkcolor=black,urlcolor=black,
             anchorcolor=black,citecolor=black}
\usepackage{xcolor}
\usepackage{xspace}
\usepackage{xfrac}			
\usepackage{float}

\newcommand{\EQ}[1]{Eq.~(\ref{eq:#1})}
\newcommand{\EQS}[2]{Eqs.~(\ref{eq:#1}) and (\ref{eq:#2})}

\newcommand{\FIG}[1]{Fig.~\ref{fig:#1}}
\newcommand{\FIGS}[2]{Figs.~\ref{fig:#1} and~\ref{fig:#2}}
\newcommand{\SEC}[1]{Sec.~\ref{sec:#1}}
\newcommand{\SECS}[2]{Secs.~\ref{sec:#1} and ~\ref{sec:#2}}

\def\bal#1\eal{\begin{align}#1\end{align}}

\newcommand*{\tran}{{\mkern-1.5mu\mathsf{T}}}		

\journal{Physics Reports}

\begin{document}

\begin{frontmatter}





\title{Speedups in nonequilibrium thermal relaxation: Mpemba and related effects}


\author[Dresden]{Gianluca Teza}
\affiliation[Dresden]{organization={Max Planck Institute for the Physics of Complex Systems},
            addressline={Nöthnitzer Str. 38}, 
            city={Dresden},
            postcode={01187}, 
            country={Germany}}
\ead{teza@pks.mpg.de}

\author[SFU]{John Bechhoefer}
\affiliation[SFU]{organization={Simon Fraser University, Department of Physics},
            addressline={8888 University Dr.}, 
            city={Burnaby},
            postcode={V5A 1S6}, 
            state={British Columbia},
            country={Canada}}
\ead{johnb@sfu.ca}

\author[Ceuta,Granada,Granada2]{Antonio Lasanta}
\affiliation[Ceuta]{organization={Universidad de Granada, Departamento de Algebra},
            addressline={Cortadura del Valle}, 
            city={Ceuta},
            postcode={51001}, 
            state={Ceuta},
            country={Spain}}

\affiliation[Granada]{organization={Universidad de Granada, Instituto Carlos I de Física Teórica y Computacional},
            addressline={}, 
            city={Granada},
            postcode={18071}, 
            state={Granada},
            country={Spain}}

\affiliation[Granada2]{organization={Universidad de Granada, Nanoparticles Trapping Laboratory},
            addressline={}, 
            city={Granada},
            postcode={18071}, 
            state={Granada},
            country={Spain}}
\ead{alasanta@ugr.es}

\author[WIS]{Oren Raz}
\affiliation[WIS]{organization={Weizmann Institute of Science, Department of Physics of Complex Systems},
            addressline={76100}, 
            city={Rehovot},
            country={Israel}}
\ead{oren.raz@weizmann.ac.il}

\author[UVAphys,UVAmath]{Marija Vucelja}
\affiliation[UVAphys]{organization={University of Virginia, Department of Physics}, 
            addressline={382 McCormick Rd}, 
            city={Charlottesville},
            postcode={22904}, 
            state={VA},
            country={USA}}
\affiliation[UVAmath]{organization={University of Virginia, Department of Mathematics}, 
            addressline={141 Cabel Drive}, 
            city={Charlottesville},
            postcode={22904}, 
            state={VA},
            country={USA}}
\ead{mv8h@virginia.edu}

\begin{abstract}
Most of our intuition about the behavior of physical systems is shaped by observations at or near thermal equilibrium. 
However, even a basic phenomenon such as a thermal quench can lead to states far from any thermal equilibrium, where counterintuitive, anomalous effects can occur.
A prime example of anomalous thermal relaxation is the Mpemba effect, in which a system prepared at a hot temperature cools down to the temperature of the cold environment faster than an identical system prepared at a warm temperature. 
Although originally reported for water more than 2000 years ago by Aristotle, the recent observation of analogous relaxation speedups in a variety of systems has motivated the search for generic explanations from the point of view of  
nonequilibrium statistical mechanics.
Here, we review this and related anomalous relaxation effects, which all share a nonmonotonic dependence of relaxation time versus initial ``distance" from the final state or from the phase transition. The final state can be a thermal equilibrium or a nonequilibrium steady state. 
We first review the early water experiments and classify the zoology of anomalous relaxation phenomena related to the Mpemba effect.
We then introduce general concepts and provide a modern definition of the Mpemba effect, focusing on the theoretical frameworks of stochastic thermodynamics, kinetic theory, Markovian dynamics, and phase transitions.
Finally, we discuss the most recent experimental and numerical developments that followed these theoretical advances. These developments paved the way for the prediction and observation of novel phenomena, such as the inverse Mpemba effect.
The review is self-contained and introduces anomalous relaxation phenomena in single- and many-body systems, both classical and quantum. 
We also discuss the broader relevance of the Mpemba effect, including its relation with equilibrium and dynamical phase transitions and its experimental implications.  
We end with perspectives that connect anomalous speedups to new ideas for designing optimal heating/cooling protocols, heat engines, and efficient samplers.
\end{abstract}





\end{frontmatter}

\tableofcontents


\newpage

\begin{quotation}
\noindent
\textit{This review is dedicated to Erasto Mpemba (1950–2023) and Denis Osborne (1932–2014),\\ whose courage and wisdom made a difference.}
\end{quotation}

\newpage
\section{Introduction}
\label{sec:introduction}
In 1969, Erasto Mpemba and Denis Osborne published a short article titled ``Cool?” in the journal \textit{Physics Education}~\cite{mpemba1969cool}.  In it, they claimed that a container of hot water quenched in a freezer could cool and start to freeze sooner than an otherwise-identical container of cold water.  Although it turns out that similar claims had previously been made, the work of Mpemba and Osborne attracted considerable attention in the popular press for its unintuitive claim and for the inspirational story of its origin (more on this below, Sec.~\ref{sec:historical}), and it soon entered the lore of popular science.  Its allure was perhaps enhanced by the difficulty in reproducing the experiments.  Who would not be fascinated by such a simple, paradoxical, controversial experiment?  Dubbed the \emph{Mpemba effect} to honor the leading role of Erasto Mpemba, it was a challenge for both amateur and professional scientists. 

If the Mpemba effect is real, is it due to some peculiar feature of water, or is some general principle at play?  As journalist Philip Ball memorably stated in 2006, ``it is not clear whether the explanation [of the Mpemba effect] would be trivial or illuminating’’~\cite{ball2006does}. Then, in 2017, two groups independently made a case for a general scenario.  One proposed a ``Markovian scenario where a system can start ``farther” from equilibrium and yet reach equilibrium faster~\cite{lu2017nonequilibrium}.  A second showed numerically and analytically that the effect could occur in models of granular fluids~\cite{lasanta2017hotter}.  Strikingly, both works predicted the possibility of an \emph{inverse} Mpemba effect, where a cold system heats up more quickly than a hot one.  These developments were followed by an influential 2019 paper that predicted the possibility of a \emph{strong} Mpemba effect, where the speedup in relaxation is exponentially faster than the generic relaxation rate to equilibrium~\cite{klich2019mpemba}.  This latter prediction was soon after observed experimentally in a colloidal-particle system~\cite{kumar2020exponentially}.  Around the same time, the Mpemba effect in quantum systems began to be discussed~\cite{nava2019lindblad} and, in 2024, observed~\cite{aharony2024inverse,zhang2024observation,joshi2024observing}.  

These recent theoretical and experimental advances have led to an explosion of interest in such \emph{anomalous} relaxation phenomena.  In this review, we explore briefly the original context of the Mpemba effect, with its fascinating and controversial history, and then focus on the new developments in nonequilibrium statistical mechanics that suggest that a variety of physical systems can display anomalous relaxation phenomena and that there are common mechanisms explaining these anomalies. These nonequilibrium phenomena were studied in three different theoretical frameworks: Markovian dynamics, granular gases, and across phase-transitions.  We discuss work that was published, at least as a preprint, through the end of 2024.

The review is structured as follows:  In Sec.~\ref{sec:historical}, we discuss the history of the Mpemba effect and early attempts to understand it.  The explanations mostly focused on special features of water or on modeling particular experimental set-ups; see Sec.~\ref{sec:water}.  Then, after classifying the ``zoology'' of possible Mpemba effects in Sec.~\ref{sec:zoology}, we present the recent theoretical developments, in the form of three general frameworks that each capture certain aspects of anomalous relaxation:  Section~\ref{sec:markovian} focuses on an abstract and general framework where the relaxation dynamics is assumed to be Markovian. Physically, the Markovian assumption stems from a memoryless thermal bath. In this framework, the anomalous relaxation is a property of the system, including its coupling to the thermal bath, and not of the specific thermal bath. Under this condition, every system can, in principle, be described as Markovian. Although such a formalism usually requires a very detailed description, which is often not practical, it gives a clear physical picture of several anomalous relaxation effects and offers precise criteria for their existence. In this section, we also discuss several variants of the quantum Mpemba effects, since most of them are Markovian. Section~\ref{sec:phase-transitions} reviews the existing theories on anomalous relaxations through first- and second-order phase transitions. In~\SEC{kinetic-framework}, we review the many results established in the kinetic theory framework, which were developed in parallel to the Markovian framework, and give different types of explanations and intuitions to anomalous relaxations. Finally,~\SEC{statistics} broadens the theoretical question and discusses results on the likelihood of the anomalous effect in several ensembles of models. 

The experimental observations of the different types of Mpemba effect in all systems except water are reviewed in~\SEC{experiments}. The experiments include several condensed matter systems apart from water and colloidal systems, along with a few experimental observations of the quantum effects in systems of trapped ions. In addition to experiments, numerical observations of these anomalous relaxation effects played a key role in discerning several types of effects. The numerics are discussed in~\SEC{numerical-observations}. 
In~\SEC{applications}, we discuss several applications of ideas that emerged from the study of the Mpemba effect. 
In~\SEC{other-related}, we review related effects that share some features with the Mpemba effect, including analogous effects in active systems, non-Markovian systems, asymmetry in cooling and heating, the Kovacs effect, and others. Lastly, in~\SEC{perspectives}, we give some perspective on the field.

Before starting a long story, we should briefly mention a couple of related topics that are beyond the scope of our review.  First, there is a long history of studying systems that are glassy and relax anomalously slowly~\cite{debenedetti2001supercooled}.  That is, glassy systems are ones that relax at rates that are much slower than the generic relaxation rates of the system.  For example, one can observe a sharp increase in relaxation time as the temperature is lowered past a ``glass transition.”  By contrast, the Mpemba effect concerns various ways in which relaxation is anomalously \emph{faster}.

Second, there is a vast amount of work on controlled transitions between different states that contrasts with the free relaxation that is our focus.  This includes, for example, recent work on ``shortcuts to adiabaticity” where the adiabatic transformation of a system from one ground state to another (including from a hot state to a cold one) is accelerated by adding ``counterdiabatic” forcing~\cite{gueryodelin2019shortcuts}.  We return briefly to this topic in Sec.~\ref{sec:controlled}.

\subsection{Historical background: from Aristotle to the present}
\label{sec:historical}
Scattered throughout the historical record are accounts of a curious phenomenon, that hot water may sometimes freeze more quickly than warm or cool water. The first mention seems to be from Aristotle, who in about 350 BCE wrote in his treatise \textit{Meteorologica} that~\cite{webster1923aristotle} 
\begin{quotation}
    The fact that the water has previously been warmed contributes to its freezing quickly: for so it cools sooner. Hence many people, when they want to cool hot water quickly, begin by putting it in the sun.
\end{quotation}

Further claims along these lines were made by Roger Bacon in his 1267 \textit{Opus Majus}~\cite{bacon1900opus}, by Giovanni Marliani~\cite{clagett1941giovanni} around 1461, by Francis Bacon in his 1620 work \textit{Novum Organum}~\cite{bacon1878novum}, and by Ren\'e Descartes in his 1637 essay \textit{Meteorology} published in conjunction with \textit{Discourses on the Method}~\cite{descartes1637discours}.  Descartes asserts that water which has heated over a fire for a long time will freeze more quickly than water not heated and describes briefly observations made on flasks with long thin vertical necks that allow changes in volume to be easily assessed.  What is particularly interesting is that Descartes returns to the subject in a letter to Mersenne dated 1 March 1638.  Mersenne had apparently expressed skepticism about Descartes's claims.  Descartes responds~\cite{descartes1638mersenne},
\begin{quotation}
	I am surprised also that you should speak of noting down the points in my book which you regard as having been falsified by experience, for I venture to assure you that there are no such false points in it, because I made all the observations for myself, and especially the one you mention concerning warm water freezing faster than cold water. In the book I wrote, not ``hot and cold,'' but ``water which has been heated over a fire for a long time freezes more quickly than other water.'' In order to perform this experiment properly, the water must be allowed to cool down after it has boiled, until it has attained the same degree of coldness as water from a spring. After taking the temperature with a thermometer, you should take some water from a spring and pour equal quantities of the two sorts of water into identical vessels. But there are some experiments which only a few people are capable of performing, and when they are performed badly, the result is often the very opposite of what it ought to be.
\end{quotation}
Thus, Descartes makes clear that he has performed experiments himself.  Beyond the specifics of these experiments and the discussion about whether the water has first cooled somewhat or not,
the exchange is notable (1) as a very early example of controversy about claims regarding the cooling and freezing of water and (2) for recognizing the sensitivity of experimental observations to small procedural details.   As we shall see, both the controversy and the difficulties in obtaining reproducible results for experiments on water have persisted to this day.

Although Marliani, Francis Bacon, and Descartes all seem to have performed at least some experiments, they describe little if any of their methods.  The first experiment that records experimental procedures in some detail dates to 1775.  That year, the journal \textit{Philosophical Transactions} published two letters.  In one, Robert Barker gave an anecdotal report on ``The Process of making Ice in the East Indies" that describes the process of ice-making in the region and notes that ``boiling the water is esteemed a necessary preparative to the method of congelation"~\cite{barker1775the}.   Barker's report seems to have inspired experimental investigations by Joseph Black, a Professor of Chemistry at Edinburgh, who wrote on ``The Supposed Effect of boiling upon Water, in disposing it to freeze more readily, ascertained by Experiments"~\cite{black1775the}.  Black compares freezing of water boiled in a tea kettle for four hours and then placed in a tea cup outside on a day when the temperature was 29~$^\circ$F.  He compared the freezing to a similar amount of water in the same kind of tea cup that was not boiled and found that ice appeared first in the boiled water.  He speculated that the degassing that occurs upon boiling makes it absorb air again as it cools, perhaps increasing the likelihood of ice nucleation.  In other words, it might affect the limit of supercooling.  

Both experiments by Black and Descartes compare samples of water that have identical initial conditions, with one sample boiled and cooled and the other not.  This setup differs from the common scenario for the Mpemba effect, which compares samples of water that start at two different temperatures.  Nonetheless, these early observations demonstrate a persistent ``folklore" of anomalous behavior connected with the freezing of water.  A more detailed discussion of this ``pre-history" of the Mpemba effect (although not including the papers by Barker and Black~\cite{kumar1980Mpemba}) has been given by Jeng~\cite{jeng2006mpemba}.

Interestingly, making ice cubes faster was also a topic of conversation at a dinner party in Los Alamos during the Manhattan Project. General Leslie R. Groves, in his book \emph{Now It Can be Told: The Story Of The Manhattan Project}~\cite{groves1962now}, recalls that on social occasions, Oppenheimer, Tyler, Parsons, and Ashbridge tried to bring scientists and the armed services closer together. During one such evening, the crowd entertained their thoughts on a daily column, syndicated in several Eastern newspapers, which proposed that
\begin{quotation}
...if one wished to expedite the freezing of ice cubes in a refrigerator he might do so by filling the ice trays with boiling hot water.
\end{quotation}
General Groves recalls that opinions of the scientists about the effect ranged from ``ridiculous'' to ``quite possibly true.'' Some of the guests reportedly hurried home to carry out experiments in their refrigerators that night.  

\subsection{The cooling and freezing of water}
\label{sec:water}
If early accounts of anomalies in the cooling and freezing of water amount to a combination of folklore and rudimentary experiments spread diffusely over the centuries, the modern interest in such phenomena dates more precisely to 1969, when Erasto Mpemba and Denis Osborne published the paper ``Cool?"~\cite{mpemba1969cool}.  Mpemba was a Tanzanian schoolboy (in Form 3, equivalent to 9th grade) in 1963, who wanted to make ice cream at school.  The procedure was to boil a flavored mixture, let it cool down to room temperature, and then put it in a freezer.  Worried that another boy would fill up the freezer before his recipe had cooled, Mpemba put the hot mixture directly in the freezer.  Unexpectedly, his hot mixture froze before the other boy's.  With admirable tenacity in the face of skeptical teachers and classmates, Mpemba sought to understand why his mixture had frozen first.  Later, in high school, he convinced a visiting scientist, Denis Osborne, to investigate further. 

As Osborne recounts in his introduction to the paper,
\begin{quote}
    The headmaster of Mkwawa High School invited me to speak to the students on \textit{Physics and National Development} .... One student raised a laugh from his colleagues with a question I remember as: ``If you take two beakers with equal volumes of water, one at 35~$^\circ$C and the other at 100~$^\circ$C, and put them into a refrigerator, the one that started at 100~$^\circ$C freezes first. Why?"  
    
    ... I promised I would put the claim to the test of experiment ....
\end{quote}

\begin{figure}
    \centering
    \includegraphics[width=0.45\columnwidth]{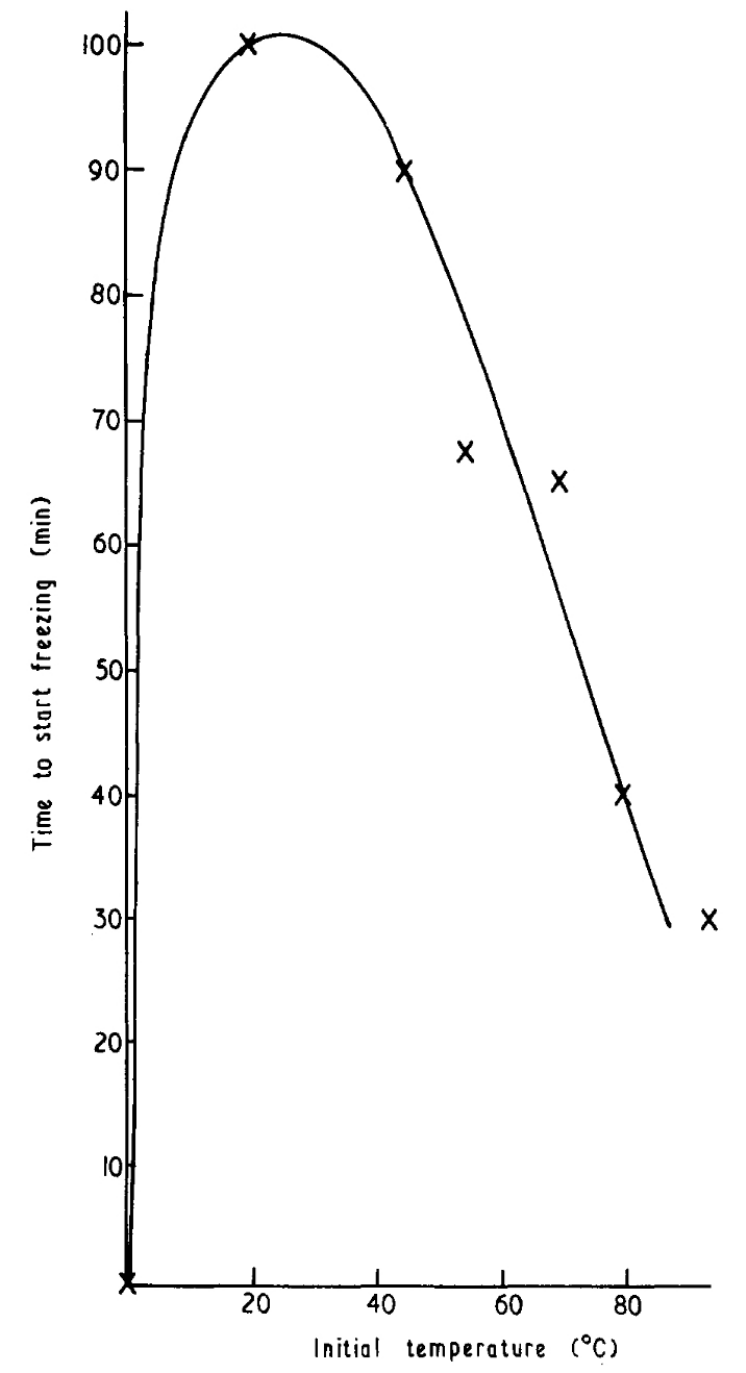}
    \caption{Time to start freezing as a function of initial sample temperature. 
    \textit{Source:} Reprinted with permission from~\cite{mpemba1969cool}.}
    \label{fig:mpemba69}
\end{figure}

The main result of Osborne's study,~\FIG{mpemba69}, claims a dramatic decrease, for increasing initial temperature above about 20 $^\circ$C, in the time to cool and start to freeze for samples of water. The experiments placed 70 cm$^3$ of water in 100 cm$^3$ Pyrex beakers in a domestic freezer chest. The water had been boiled to eliminate dissolved gases (perhaps unknowingly conforming to the procedure adopted by Descartes, Black, and others). Insulation on the bottom and side ensured that most heat was lost from the top. Although one might imagine that a hot sample could lead to evaporation and a reduced time to freeze the remaining mass, that effect was deemed too small to account for the results. Rather, the authors speculated that the hot fluid might lead to more convection that would bring hot water to the surface and increase the rate of heat loss to the atmosphere above.

The study received wide attention in popular science magazines and then in the press more generally, and the anomalous freezing results became known as the Mpemba effect.  As one indication of the experiment's popularity and Mpemba's story, the Royal Society of Chemistry sponsored a competition in 2012 to mark the 50th anniversary of the paper by asking for the ``best explanation'' of the effect.  The organizers received 22,000 submissions, and the competition was featured in a BBC Newsnight television program.  Results were announced in January 2013 at a symposium featuring oral histories by Mpemba and Osborne on their work in the 1960s~\cite{rsc2013oral-history,burridge2016questioning}.

In the years following the work of Mpemba and Osborne, numerous studies attempted to reproduce and explain the ``Mpemba effect."  The work ranged from efforts by professional scientists to efforts by students at different levels.   We can broadly classify them according to six overall categories: environmental coupling, solutes, supercooling, convection, hydrogen bonding, and null results. We examine each in turn:

\begin{enumerate}

\item \textit{Environmental coupling}. The most straightforward explanations of the Mpemba effect involve evaporation and frost:

\begin{enumerate}
\item \textit{Evaporation}: In an article that appeared the same month in 1969 as the work of Mpemba and Osborne (and thus was done independently), G. S. Kell in Canada reported on ``The freezing of hot and cold water"~\cite{kell1969the}.  His study was inspired by Canadian folklore that a wooden pail of hot water set out in freezing weather will begin to freeze sooner than a similar pail of cold water.  Kell supposed that evaporation was responsible and, indeed, measured a 6\% percent mass loss in an experiment with water at 80 $^\circ$C put into a wide-mouthed, open Dewar flask and set outside at a temperature of $-6.5~^\circ$C.  The results were consistent with a simple lumped-element model that also predicted a shorter time to freeze for the hot-water sample.

\hspace{3 mm} In 2010, Vynnycky and Mitchell refined the lumped-element model of Kell and confirmed that evaporation alone is sufficient to account for a Mpemba effect~\cite{vynnycky2010evaporative}.  The calculations were later extended to model one-dimensional (1D) ~\cite{vynnycky2012axisymmetric} and two-dimensional (2D)~\cite{mirabedin2017numerical} spatial profiles. These numerical studies predicted an even stronger Mpemba effect; however, the experiments conducted in Ref.~\cite{vynnycky2012axisymmetric} were less conclusive:  pairs of individual runs were consistent with a Mpemba effect while other, nominally identical experiments failed to show the effect.

\item \textit{Frost}: Another situation in which the physics of the Mpemba effect seems clear was considered by Brownridge~\cite{brownridge2011when}. He placed two identical copper cups on a bed of frost.  One cup contained hot water and heated the frost, melting it and thereby forming a layer of water underneath the cup. There was good thermal contact with the freezer bottom, and solidification was rapid. The other cup contained cold water and did not melt the frost.  Because frost is a mixture of ice and water and thus a relatively good insulator, heat conduction out of the sample was reduced:  the cold water took longer to freeze. A control experiment where the cups were placed on a piece of insulating plastic did not show the Mpemba effect. Also, a similar experiment using glass rather than copper cups showed no Mpemba effect with or without frost~\cite{deeson1971cooler}, showing that this mechanism depends on having a high enough thermal conductance between the water and refrigerator bottom.

\end{enumerate}

Both the evaporation and frost scenarios give intuitive situations where the Mpemba effect is easy to understand, and both share the feature that the mechanism leading to the Mpemba effect may be easily eliminated. To prevent evaporation, one can work with a sealed container or even one with a narrow opening.  To prevent the effects of frost, containers can be placed on a piece of plastic or other insulating material. Putting the necessary controls in place, Mpemba and Osborne~\cite{mpemba1969cool}, as well as Brownridge~\cite{brownridge2011when}, claimed to see the Mpemba effect even in experiments with negligible evaporative losses and no frost. Thus, there are other mechanisms leading to the Mpemba effect.

\item \textit{Solutes}.
The equilibrium concentration of solutes can vary strongly with temperature. A number of experiments have claimed a role for dissolved carbon dioxide, which is less soluble in hot water than in cold. For example, Freeman concluded that dissolved carbon dioxide causes the Mpemba effect~\cite{freeman1979cooler}, in direct contradiction to similar experiments by Deeson~\cite{deeson1971cooler}.   The importance of dissolved air was also affirmed by Wojciechowski et al.~\cite{wojciechowski1988freezing}.  Nonetheless, starting with the paper of Mpemba and Osborne~\cite{mpemba1969cool}, the Mpemba effect was also reported in works that degassed water by boiling it shortly before the experiment~\cite{auerbach1995supercooling}. 

\hspace{3 mm} A related but more complicated solute effect was proposed by Katz~\cite{katz2009hot}.  He considered the effect of dissolved mineral salts, such as calcium bicarbonate, which are responsible for the hardness of the water.  His mechanism draws on two observations: (1) Freezing concentrates these impurities  (``zone refining'') and depresses the solid-liquid equilibrium temperature. (2) Heating water removes CO$_2$, turning bicarbonates into carbonates, precipitating out much of the dissolved calcium, and reducing the freezing point depression.  The result is that hot water would have a lower impurity concentration and thus might reach the solid-liquid equilibrium more quickly than cold water with a higher impurity concentration and greater freezing-point depression. This scenario does not seem to have been investigated experimentally so far.

\hspace{3 mm} A recent proposal by W. B. Zimmerman argues that the Mpemba effect results from microbubbles (with diameters of order 100 microns) due to dissolved oxygen in water~\cite{zimmerman2021towards,zimmerman2022in,zimmerman2022mediating}.  Such microbubbles are small enough that their temperature and internal humidity (water vapor pressure) equilibrate in O($10^{-3}$) s and thus are in local equilibrium.  A bubble at high temperature will have a high vapor pressure, which implies a transient evaporation from the surrounding water that cools the surrounding water.  When the bubble moves to a cooler region, the reduced internal vapor pressure implies a transient heating due to the latent heat of condensation.  In a convection cycle, fluid is periodically heated at the bottom and cooled at the top of a container.  The accompanying evaporation and condensation then enhances heat transport beyond that due to the convection alone.  Zimmerman argues that the effect, which is proportional to the fraction $\phi$ of dissolved gas, can be significant even with $\phi$ of order 1\%.  Zimmerman claims these ideas are consistent with experiments, where, for example, the original experiments of Mpemba and Osborne are done in a way that might lead to a high fraction of dissolved gas (and therefore microbubbles), whereas other negative experiments such as those done by Burridge and Linden~\cite{burridge2016questioning} were done on pure water that had been degassed.  Unfortunately, with no direct measurements of dissolved gases in either of these experiments (or others), the arguments made require complex indirect inferences.  It would be interesting to have a direct experimental test of this scenario.

\item \textit{Supercooling}.  In order to form ice, samples of water must typically be supercooled below the water-ice coexistence temperature (0 $^\circ$C for pure water) so that small crystals of ice can nucleate and grow.  In practice, nucleation is heterogeneous, meaning that it tends to occur on surfaces or dust particles (``motes")~\cite{lifshitz1981physical}.  Mechanisms involving supercooling focus on changes to the amount of supercooling needed to nucleate ice.

\hspace{3 mm} The first investigation to look extensively at the role of supercooling in the Mpemba effect was by Auerbach~\cite{auerbach1995supercooling}. He found that initially hot water could be supercooled less than initially cold water.  However, as pointed out by Jeng~\cite{jeng2006mpemba}, earlier work by Dorsey~\cite{dorsey1948the} reached the opposite conclusion.  Esposito et al.~\cite{esposito2008mpemba} emphasized that the important role of stochasticity for nucleation implies that the Mpemba effect should be considered statistically:  statements apply to ensemble averages of freezing scenarios, not individual trials. A more careful and extensive set of experiments was reported by Brownridge~\cite{brownridge2011when}, who looked at ``several thousand" trials, in contrast to Auerbach and Esposito et al., who each considered approximately a hundred trials. Consistent with earlier experiments, he found a high degree of variability in the time to supercool and start freezing.  Indeed, to have consistent results, he had to preselect the vials used for the samples.  Different vials have different sample-wall roughnesses and different amounts of dust and, thus, tend to have different supercooling thresholds for the observation of ice. Brownridge found two vials that had a 5.5 $^\circ$C difference between nucleation thresholds. If hot water was put into the vial with a small-supercooling vial, it consistently (28 times out of 28) froze sooner than cold water put into the vial with large-supercooling threshold.  

\hspace{3 mm} More recently, Burridge and Hallstadius demonstrated a similar scenario~\cite{burridge2020observing}:  They prepared two containers that were nominally identical. One had a deliberately roughened interior, whereas the other had a smooth interior (but roughened exterior to preserve mass and thermal properties).  They consistently saw a Mpemba effect when hot water was inserted into rough containers and compared to cold water in smooth containers.  Thus, the experiments of Brownridge and of Burridge and Hallstadius both require two correlated conditions (container and initial temperature) rather than a single condition involving the initial temperature.

\item \textit{Convection}.  Mpemba and Osborne suggested already in their 1969 paper that thermal convection may contribute to the more rapid cooling of a hot sample by helping to maintain a ``hot top'' that accelerates the cooling of an initially hot sample~\cite{mpemba1969cool}. Deeson also concluded that convection is relevant~\cite{deeson1971cooler}. In subsequent work, Maciejewski~\cite{maciejewski1996evidence}, Vynnycky and Kimura~\cite{vynnycky2015convection}, and Mirabedin and Farhadi~\cite{mirabedin2017numerical} all confirmed that convection does have a major influence. Indeed, in convective heat transport, heat is carried by fluid motion more quickly than by diffusion and thus tends to be the dominant factor in fixing the temperature field of water inside the container.  Thus, it may well be a part of scenarios leading to the Mpemba effect without necessarily being a cause of it. Of these works, only Maciejewski~\cite{maciejewski1996evidence} suggests a causal role for convection. His explanation, which is only sketched, depends on the density maximum of water at 4 $^\circ$C. This maximum would occur at different heights in the container for different initial conditions and thus lead to spatially differing profiles.  However, the more detailed numerical studies by Vynnycky and Kimura ~\cite{vynnycky2015convection} and by Mirabedin and Farhadi~\cite{mirabedin2017numerical} did not find evidence for Maciejewski's scenario.

\item \textit{Hydrogen bonding}. A very different mechanism, involving hydrogen bonding, has been proposed by Zhang et al.~\cite{zhang2014hydrogen,sun2023the}. That scenario depends on the relaxation of the O:H--O bond in water. The bond releases energy at lower temperatures at a rate that depends on the initial temperature.  

\hspace{3 mm} Two other studies involving molecular dynamics simulations also point to the potential importance of hydrogen bonds in water.  Jin and Goddard~\cite{jin2015mechanisms} conduct molecular dynamics simulations of water in systems ranging from a few hundred up to 1000 water monomers and followed the dynamics over times as long as 500 ns.  They observed that initially warm water had a higher population of water hexagon ice nuclei than did initially cooler water.  The hexagons could serve as nucleation sites.   Likewise, Tao et al.~\cite{tao2017different} suggest hydrogen-bonded structures again increase the number of initiation sites in initially hot water.

\item \textit{Null results}. Finally, there are papers that search for but do \textit{not} observe the Mpemba effect.  Soon after Mpemba and Osborne's original work, Ahtee tried and failed to reproduce their results under conditions as close as possible to those of Mpemba and Osborne~\cite{ahtee1969investigation}. However, Ahtee did not indicate the number of trials used. As we have seen, experiments show a large variability~\cite{brownridge2011when}, and a null result might simply indicate that the number of trials was too small. In 2016, Burridge and Linden made a careful, systematic study of cooling and found no evidence for any Mpemba effect~\cite{burridge2016questioning} and questioned whether the effect itself is real. That work was criticized for neglecting the role of freezing~\cite{katz2017reply}. A follow-up work by Burridge and Hallstadius that included freezing, as noted above, did show evidence for a kind of Mpemba effect, but only when samples with rough walls were used for the initially hot sample~\cite{burridge2020observing}. Finally, another follow-up study, by Tang et al.~\cite{ tang2022direct}, claims evidence for a Mpemba effect in cooling under conditions where Burridge and Linden did not see the effect.
\end{enumerate}

What lessons can one draw from all these efforts from the past sixty years regarding the Mpemba effect in water? The first is that, despite doubts raised about particular experiments and modeling, there is strong evidence for at least some kinds of Mpemba effects, notably those linked to evaporation and frost.  While these scenarios are often deliberately excluded as ``pitfalls"~\cite{elton2021pathological}, they do meet the criterion that the time to cool and start freezing in an otherwise-identical apparatus is shorter on average for hot water rather than cold.   Other plausible mechanisms such as solutes and supercooling are less supported by experimental evidence but seem likely to be relevant under at least some conditions.   

A second lesson implied by these proven or plausible mechanisms is that no single mechanism ``explains'' the Mpemba effect; rather, there are several, perhaps many ways for it to arise, and scenarios may involve several coupled physical factors.

A third lesson is that experiments on freezing water are difficult, for two overall reasons: First,  the many mechanisms that affect cooling make sorting out the different roles problematic. The nucleation of ice crystals is extraordinarily sensitive to small details of sample preparation (roughness of the sample walls, the presence of dust or precipitates). Second, the relatively large scale of samples used in these experiments ($\sim 10$ cm) implies that each trial of cooling and then freezing takes a long time ($\sim 10^3$ to $10^4$ s), limiting the number of trials used.  In order to make progress, new experimental techniques may help.  In this regard, a recent study on cooling curves of small droplets based on IR video recordings is promising~\cite{ortega2024thermographic}.  

One might consider reducing the scale in order to speed up the cooling dynamics. However, the various mechanisms implicated in some scenarios of the Mpemba effect show different scalings with size. For samples of height $\ell$, thermal convection is quantified by the Rayleigh number, which scales as $\ell^3$, whereas diffusion times scale as $\ell^2$. Experiments with smaller volumes of water will tend to strongly suppress or even eliminate convection. At the same time, reducing the volume will also reduce the number of heterogeneous nucleation sites, increasing the supercooling threshold for observing spontaneous nucleation.  (Indeed, studying small droplets is the classic method for reaching large supercoolings~\cite{turnbull1956phase}.) Again, simple scale changes can alter the dominating physics.

Also, one might consider using numerical experiments as a way to explore the relative roles of different types of physical mechanisms. Mirabedin and Farhadi~\cite{mirabedin2017numerical} did just that, exploring what happens when evaporation is or is not present during cooling. Unfortunately, their work studied only the cooling process and did not attempt to model ice nucleation in the supercooled regime. The sensitivity of nucleation to rare events (nucleation sites) makes such modeling problematic, yet experiments do indicate that freezing is likely an essential element of the Mpemba effect in water~\cite{burridge2016questioning}.  

Finally, the hydrogen-bonding scenario and, to some extent, the convection scenario (insofar as the density maximum at 4 $^\circ$C plays a significant role) depend on physics that is special to water~\cite{ball1999h2O,sun2016the}. The search for explanations that depend on such special features of one material lost some motivation once it was realized that the Mpemba effect is not unique to water and can occur in other experimental systems, as discussed in~\SEC{experiments}.

\subsection{The zoology of Mpemba-like effects}
\label{sec:zoology}
\label{Sec:Zoology}
The terms ``Mpemba effect" and ``Mpemba-like effect" are often used to describe various phenomena generated by different physical mechanisms. Even in the case of water, where the term ``Mpemba effect" was originally coined as the observation that under some conditions, hot water freezes before cold water, there are different mechanisms that lead to the same observation. For example, both evaporation~\cite{kell1969the} and dissolved gases and solids~\cite{freeman1979cooler} can explain, under some conditions, the same experimental observation---``initially hot water freezes before colder water when both are put in the freezer''. Nevertheless, ``hot water freezes faster because a larger fraction of the water evaporates" and ``hot water freezes faster because there are less dissolved gases in it'' are two different physical effects. Moreover, the term ``Mpemba effect," which originally was used to describe the freezing of hot water before cold water, is now used to describe many additional effects, ranging from quantum systems to abstract models of magnetism. We believe that it is useful to categorize the different effects discussed in the literature and propose a terminology to differentiate between the various classes. 

At this stage, the mechanisms behind several experimental observations are not fully understood. Therefore, it is most useful to define the terminology based on the applied conditions and observed nonmonotonic relaxation, and not by the underlying physical mechanisms. We suggest categorizing the different effects based on the following properties:
\begin{enumerate}
    \item We distinguish between nonmonotonic relaxation effects in cooling processes, to which we refer as ``\emph{direct Mpemba effects}," nonmonotonic relaxation in 
    heating processes, which are ``\emph{inverse Mpemba effects}" and effects where the quenched parameter is not the temperature. We refer to the latter as  ``\emph{Mpemba-like effects}.''
    \item The final state of the system after the end of the relaxation process can be equilibrium, namely thermal distributions (as was the case in most experiments done in water, magnetic alloys, and other systems), nonequilibrium steady states (as in driven granular gases), or neither (as in spin glass systems). We refer to these effects as ``\emph{equilibrium Mpemba}," ``\emph{steady state Mpemba}" and ``\emph{glassy Mpemba}," respectively.
    
    \item In various scenarios, different types of initial conditions can be discussed. In systems that relax to a final equilibrium state corresponding to a final temperature $T_f$, it is natural to consider the initial conditions as equilibrium distributions corresponding to different temperatures. Therefore, when the system relaxes to an equilibrium distribution, it is assumed that the initial distribution is also an equilibrium distribution. However, in several systems as driven granular gases the final state is not an equilibrium, but rather a nonequilibrium steady state. In this case, there are two different types of initial conditions: (i) A steady state corresponding to a different value of the ``temperature'' parameter in the system, while keeping all other parameters fixed. This is the analog of the equilibrium case, where the different initial conditions are identical in all aspects except for their initial temperature. (ii) A generic nonequilibrium distribution. In these cases, one has to be careful in determining which initial condition is ``closer'' to the final state of the system. 
    
    \item In all related phenomena, there is some notion of nonmonotonic relaxation time, but the way this time is defined and measured is drastically different in the various suggested setups: in some cases, it is the time to the phase transition that dictates if the effect exists or not \cite{holtzman2022landau,zhang2022theoretical}; in others, it is the time for the system to reach the new equilibrium / nonequilibrium steady state; and in yet other cases, it is the crossing in the relaxation curves of some other property that signals the distance from the final state, e.g., energy in spin glass~\cite{baity2019mpemba}. Although it is plausible that there are some connections among these cases---for example, a cup of water that freezes first also reaches the freezer's temperature faster---we distinguish among these various definitions. When the relaxation time is measured to a first or second-order phase transition, we refer to the effect as ``\emph{Mpemba effect through a first / second-order phase transition}". When the nonmonotonic relaxation is measured in the long time limit, namely in the final stage of the relaxation, it is referred to as ``\emph{final relaxation Mpemba effect}."  A very strict definition was suggested in \cite{van2024thermomajorization}, where a system is defined to have a ``\emph{thermomajorization Mpemba effect}" if the crossing through relaxation happens for every monotone, namely for any quantity that is assured to decay.
    
    \item In all related phenomena, there is some notion of nonmonotonic relaxation time, but the way this time is defined and measured is drastically different in the various suggested setups: in some cases, it is the time to the phase transition that dictates if the effect exists or not \cite{holtzman2022landau,zhang2022theoretical}; in others, it is the time for the system to reach the new equilibrium / nonequilibrium steady state; and in yet other cases, it is the crossing in the relaxation curves of some other property that signals the distance from the final state, e.g., energy in spin glass~\cite{baity2019mpemba}. Although it is plausible that there are some connections among these cases---for example, a cup of water that freezes first also reaches the freezer's temperature faster---we distinguish among these various definitions. When the relaxation time is measured to a first or second-order phase transition, we refer to the effect as ``\emph{Mpemba effect through a first / second-order phase transition}." When the nonmonotonic relaxation is measured in the long time limit, namely in the final stage of the relaxation, it is referred to as ``\emph{final relaxation Mpemba effect}." A very strict definition was suggested in \cite{van2024thermomajorization}, where a system is defined to have a ``\emph{thermomajorization Mpemba effect}" if the crossing through relaxation happens for every monotone, namely for any quantity that always decays
    monotonically with the relaxation dynamics. Lastly, when the effect is declared based on the crossing of some observed quantity as effective temperature or average energy, we refer to it as a ``\emph{crossing Mpemba effect}" for the relevant property, e.g., in the spin glass, there is an ``\emph{Energy crossing Mpemba effect}." Note that, in principle, a system may have one type of Mpemba effect without the other, e.g., a final relaxation Mpemba effect without the existence of the effect through a phase transition that can occur in the system. The exact relations between these, e.g., whether crossing the phase transition first implies faster relaxation to the steady state, are yet unknown. 
   
    \item Following \cite{klich2019mpemba}, we distinguish between the common case where the relaxation timescale is independent of the initial condition but the specific coefficient changes, the ``\emph{weak Mpemba effect}," and cases where for the specific initial condition the relaxation timescale itself has a jump in its value, which is referred to as the ``\emph{strong Mpemba effect}." We note that although this distinction was so far made only for final relaxation \cite{klich2019mpemba} and distance crossing \cite{biswas2020mpemba} Mpemba effects, it is very plausible that a similar distinction exists also in phase-transition Mpemba effects.
    
    \item Recently, several types of nonmonotonic relaxation effects were measured in quantum systems. At least in some of these cases~\cite{aharony2024inverse,nava2024mpemba,chatterjee2023quantum}, the quantum nature of the system is crucial for the existence of the effect. When this is the case, we refer to it as a ``\emph{Quantum Mpemba effect}."
    
    \item Relaxation phenomena can occur due to interaction of the system with a thermal environment with which it exchanges heat (and, possibly, particles) or in a closed system of many interacting particles, where the relaxation is towards an equilibrium distribution for the single particle distribution. When a nonmonotonic relaxation happens in a system that is coupled to a thermal environment, we refer to the effect as a ``\emph{thermal Mpemba effect},'' and if the system is a closed system, we refer to the effect as a ``\emph{non-thermal Mpemba effect}."  
\end{enumerate}

The above different ways of categorizing experimental, analytical and numerical observations of the effect should be combined. For example, the first experimental observation of an inverse Mpemba effect in Ref. \cite{kumar2022inverse} should be classified as a ``final relaxation inverse equilibrium thermal classical Mpemba effect,'' and, in fact, both the strong and the weak versions of the effects were measured in this case. Driven granular gases have both direct and inverse steady-state distance crossing classical thermal Mpemba effects, where the crossing is either in the effective temperature of the system or some other parameter of the velocity distribution. However, it is not known if a weak only version exists for the macroscopic mean-field antiferromagnetic Ising model \cite{klich2019mpemba} or for a large 2D Ising model \cite{gal2020precooling}, where we only know how to find the strong effect. Similarly, it is not known if the direct first-order phase transition equilibrium effect in clathrate hydrates is a weak or a strong effect.    

The above zoological classification is very useful when comparing different effects but can be cumbersome when discussing a single effect. Therefore, whenever the context is clear, we refer to the effect without the full classification, e.g., ``direct strong Mpemba effect," and so on.

\subsection{Notation}
We use arrows to denote classical vectors, $\vec{v}$, Dirac notation with ``bra'', $\bra{v}$,  and ``ket'', $\ket{v}$, to denote wavefunctions, bold for matrices, $\textbf{A}$, and add a ``hat'', $\hat{\textbf{L}}$, or introduce script notation, $\mathcal{L}$, for operators. The partition function, $\mathcal{Z}$, and the Hamiltonian, $\mathcal{H}$, are denoted by script notation. 
Typically, the probability that a system is in a state $i$ is denoted by $p_i$. The probability density is also denoted with small $p$ or $f$ (a more customary notation in kinetic theory, used in \SEC{kinetic-framework}). 
From the context, we make it clear whether $p$ denotes the probability itself or the probability density.
The Boltzmann distribution corresponding to a temperature $T$ is denoted by $\pi(T)$ or $\vec{\pi}(T)$.
The inverse temperature is denoted by $\beta \equiv 1/(k_BT)$, where $k_B$ is the Boltzmann constant. The Planck constant, $\hbar$, is set to unity, $\hbar = 1$. 
The $i$-th Ising spin is denoted by $\sigma_i$. In the granular section, $\vec \sigma$ is the standard notation for a vector connecting two centers of granular particles. Unless otherwise noted, the mass of a colloidal particle, $m$, and the damping coefficient $\gamma$ are set to unity. The reaction rate in Markovian jumps, $\Gamma$, is set to unity.

\section{Markovian framework}
\label{sec:markovian}
From a modern perspective, the observation that hot water can freeze sooner than cold water, when all other conditions are identical, is consistent with similar experimental and numerical observations in other systems; see Sec.~\ref{sec:experiments} for discussion of the relevant experiments and Sec.~\ref{sec:numerical-observations} for numerical observations. A natural approach is to address such nonmonotonic relaxation dynamics from a more general, abstract point of view.
The framework of Markovian dynamics \cite{zwanzig2001nonequilibrium} represents the ideal nonequilibrium abstract setting to uncover the basic mechanisms that enable the emergence of these phenomena. The Markovian approach to nonequilibrium statistical physics has successfully described a wide variety of complex phenomena far from equilibrium, from the micro to the macro scale, with discrete or continuous degrees of freedom, for single-body and many-body systems.

Markovianity expresses the physical assumption of a ``lack of memory" in the thermal environment and corresponds to a comprehensive, exhaustive picture of the system that includes direct access to all degrees of freedom. Because states have no memory,  observations cannot be induced by some ``hidden" mechanism that we are not able to detect; rather, they stem from the intrinsic properties of the system and its dynamics. As a consequence, the Markovian framework is not only the ideal setting to explain the observed phenomenology of anomalous relaxation phenomena, but it also provides solid ground on which to explore and uncover novel anomalous relaxation phenomena. For similar reasons, Markovian dynamics are widely used to describe nonequilibrium relaxation dynamics---from the Kramers-Fokker-Planck dynamics of a Brownian particle to the Monte Carlo dynamics (Glauber, Metropolis, or Kawasaki) of a classical spin system to the Lindbladian dynamics of an open quantum system.  The solid foundation provided by the Markovian approach, however, comes with a price: exact results for macroscopically large systems are known for only a few systems, such as single-body, 1D, or mean-field models. 

In this section, we discuss a general framework for anomalous relaxation in both classical and quantum setups. First, we introduce a simple classical setup to describe thermal quenches: a system prepared in equilibrium at a given temperature $T_0$ is coupled to a thermal bath that has a different temperature $T_b$. This coupling initiates a relaxation process towards a new equilibrium at temperature $T_b$ that is driven by exchanges of heat with the thermal bath. We introduce a Markovian characterization of the time evolution after a thermal quench.  The description is based on a master equation for systems with discrete degrees of freedom or on a Fokker-Planck equation for systems with continuous degrees of freedom. We then give criteria for the existence of a Mpemba effect. In contrast to the case of water, these criteria are based not on the phase transition time but rather on the properties of the last stage of relaxation. Nevertheless, as we discuss in Sec.~\ref{sec:phase-transitions}, the Markovian framework gives insight into the traditional definition of the Mpemba effect, too.  Several examples highlighting the different phenomenology encountered in Markovian systems are discussed.
Finally, we provide an overview of analogous effects observed in quantum systems, highlighting the strong analogies in the characterization that can be issued with Markovian open quantum systems, as well as parallels with non-thermal effects in closed quantum systems. Not all the quantum effects discussed in this section are Markovian, but since most are, we discuss them all in this section.

\subsection{Classical systems}
\label{sec:classical-markovian}
Markovian dynamics can describe both discrete degrees of freedom (DoF) (e.g., in spin systems) or continuous DoF (e.g., in a Brownian particle). In what follows, we discuss both cases.  Each concept is first defined for continuous degrees of freedom and then for the discrete case.

We consider a system with $n$ continuous DoF, which we identify with a vector $\vec{x}=\{x^1,\dots,x^n\}$  in phase space (or configuration space) $\Omega$. Each DoF represents an internal continuous parameter of the system, such as the coordinates and momenta of a classical particle. A point $\vec x \in\Omega$ is also referred to as a \textit{microscopic configuration}, or a \textit{microstate} of the system.  The Hamiltonian of the system, $\mathcal{H}(\vec{x})$, explicitly depends on the DoF and determines the energy of the system at each microscopic configuration. For a system with discrete DoF, e.g., a system of spins, the configuration space, the set of all possible states of the system, is also discrete. We can then denote each microstate, namely each possible configuration of the system, as $x_i$, with $i = 1, 2, \ldots, N$. The energy of the system in microstate $x_i$ is $E_i$.

If a system with a continuous DoF is isolated, it follows a deterministic Hamiltonian dynamics that conserves energy and dictates $\vec x(t)$. However, when the system is put in contact with a bath at temperature $T_b$ such that the system and the bath can exchange energy, the dynamics of the system are stochastic rather than deterministic. The system is no longer described by a specific microstate $\vec x(t)$ but rather by a probability distribution, $p(\vec{x},t)$, the probability to find the system at microstate $\vec x$ at time $t$.  For the discrete case, $p_i(t)$ denotes the probability to be in microstate $x_i$ at time $t$, and we denote by $\vec p(t) = (p_1(t),...,p_N(t))^\tran$. Here, $N$ specifies the total number of microstates. Normalization of the probability distribution implies that $\int \dd \vec x \, p(\vec x,t)=1$ or $\sum_{i=1}^N p_i = 1$, respectively.

A system that is in contact with a bath at temperature $T_b$ relaxes, after long-enough time, to a thermal equilibrium with the thermal bath, regardless of its initial condition. The distribution corresponding to the thermal equilibrium is referred to as the \textit{Boltzmann distribution}.  For continuous DoF, it is given by
\begin{align}
    \pi(\vec{x};T_b)\propto e^{-\frac{1}{k_B T_b}\mathcal{H}(\vec{x})}.
\end{align}
The normalization of the equilibrium distribution, known as the \textit{partition function}, is fixed by integrating over all the microscopic configurations as 
\begin{align}
    \mathcal{Z}(T_b)=\int_{\Omega}\dd\vec{x}\, e^{-\frac{1}{k_B T_b}\mathcal{H}(\vec{x})}.
\end{align}
For discrete DoF, the Boltzmann probability is given by
\begin{align}
    \pi_i(T_b)\propto e^{-\frac{E_i}{k_B T_b}},
\end{align}
with \emph{partition sum}
\begin{align}
    \mathcal{Z}(T_b)=\sum_{i} e^{-\frac{E_i}{k_B T_b}}.
\end{align}
This characterization of the system demonstrates how temperature affects the statistics of the configurations for a system at equilibrium. At low temperatures, we expect to find the system in a configuration that minimizes the energy. As we increase the temperature, the probability distribution progressively flattens, ultimately becoming uniform for all configurations at infinite temperature.  

\begin{figure}
    \centering
    \includegraphics[width=0.4\columnwidth]{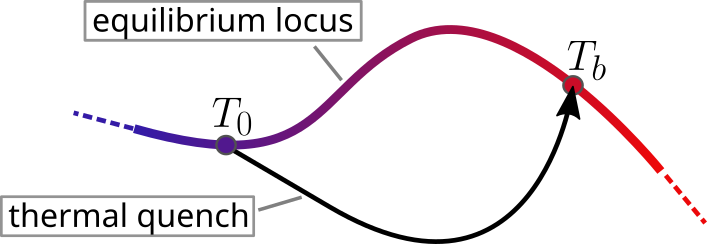}
    \caption{Sketch of an equilibrium locus in a high-dimensional distribution space. Each point along the blue-to-red line represents a Boltzmann equilibrium $\pi(T)$ at a different bath temperature $T\in[T_0,T_b]$.  A quasistatic protocol brings the system from an equilibrium configuration $T_0$ to an equilibrium temperature $T_b$ via a path closely following the equilibrium locus.
    A thermal quench generally leads the evolution through a path (black line), along which the system explores configurations that do not correspond to any Boltzmann equilibrium.    
    }
    \label{fig:manifold_locus}
\end{figure}

For both discrete and continuous DoF, it is instructive to view the probability distribution over microstates, $p(\vec x,t)$ or $p_i(t)$, as a point in an abstract \emph{distribution space}.
The set of points in distribution space that correspond to all equilibrium distributions at different temperatures defines the \textit{equilibrium locus}.
Although this is a line characterized by a single parameter---the temperature---it lives in the distribution space, which, even for systems having few DoF, has a dimension so high that it is impractical to visualize directly. An abstract picture is shown in Fig. \ref{fig:manifold_locus}.

\subsubsection{Markovian dynamics after a thermal quench}
\label{sec:markovian-quench}
In a Mpemba scenario, a system that is prepared at an equilibrium distribution of some initial temperature $T_0$ is coupled to a thermal bath at a different temperature, $T_b$. Its distribution therefore starts at $\pi(\vec x, T_0)$, or $\vec\pi(T_0)$, and relaxes towards $\pi(\vec x, T_b)$, or $\vec\pi(T_b)$. Our goal is to characterize the evolution of the probability distribution, $p(\vec x, t)$ or $\vec p(t)$, following this thermal quench.

An ideal thermal bath consists of a reservoir that is so much larger than the system that it is effectively unaltered by interactions with the system. In this case, the heat exchange between the ideal bath and the system does not change any of the bath properties: the bath is memoryless and characterized by only its temperature, $T_b$. This assumption is necessary to formulate a general theory that explains the core mechanisms at the basis of the Mpemba effect, as it allows us to exclude artifacts stemming from complex model-specific mechanisms that could be ascribed, e.g., to memory effects of the system or to a nonideal thermal bath.
Under this assumption, for a system with a continuous DoF, the dynamics of the probability distribution evolves as
\begin{align}
\label{eq:markovian_operator}
    \frac{\partial p(\vec{x},t)}{\partial t}= \mathcal{L}_{T_b} p(\vec{x},t),
\end{align}
where $\mathcal{L}_{T_b}$ is a Markovian operator regulating the evolution of the system. For a system with discrete DoF, the corresponding evolution is given by
\begin{align}
\label{eq:master_equation}
  \frac{\partial p_i(t)}{\partial t}=\sum_{j=1}^N W_{i\!j}(T_b)p_j(t) .
\end{align}
The matrix $\mathbf{W}(T_b)$ denotes the \textit{transition-rate matrix}. Its off-diagonal elements $W_{i\!j}(T_b)$ represent the transition rate from state $j$ to state $i$, while the diagonal terms $W_{ii}$ are negative and quantify the escape rate from the $i$-th state. Setting $W_{ii}=-\sum_j W_{\!ji}$ ensures probability conservation as states evolve. 

The exact forms of $\mathbf{W}$ or $\mathcal{L}$ are determined by the system, the bath temperature $T_b$, and the coupling with the thermal reservoir.
For an ergodic transition-rate matrix $\mathbf{W}$, convergence to the Boltzmann equilibrium $\vec \pi(T_b)$ is ensured by the \textit{detailed balance} (DB) condition, which imposes the rate  $W_{i\!j}$  of transitions between two generic microstates $x_i$ and $x_j$ and the reverse, $W_{ji}$. Transitions satisfying DB obey the relation
\begin{align}
\label{eq:detailed_balance}
  \frac{W_{i\!j}}{W_{ji}} = e^{-\frac{E_i-E_j}{k_B T_{b}}} .
\end{align}
This condition is not restrictive and allows one to define a dynamical model that captures most of the relevant features of an experimental setup. Note that DB is obeyed with respect to the target distribution. For Markovian operators, the detailed balance condition is far more complicated; see Ref.~\cite{horowitz2009exact} for a discussion.

\paragraph{Thermal quench, modern definition of the Mpemba effect} 

The equilibrium locus is often  referred to as the \emph{quasistatic locus}: indeed, a quasistatic protocol forces the system to evolve from one equilibrium (at some temperature $T_0$) to another equilibrium (at temperature $T_b$) along the quasistatic locus of points, as depicted in Fig.~\ref{fig:manifold_locus}. More specifically, the probability distribution at any instant of time has relaxed to a Boltzmann distribution, implying that one can define a time-dependent temperature function $T(t)$ such that $p(\vec{x},t)\equiv \pi(\vec{x},T(t))$.  The initial and final distributions are given by the equilibrium distributions at $T_0$ and $T_b$, respectively.  Quasistatic protocols do not lead to a Mpemba effect, since different relaxation trajectories cannot cross each other.

In a thermal quench, by contrast, the system is prepared at some initial distribution $p_0\equiv p(t=0)$ and then suddenly coupled to a thermal bath that leads it to explore configurations that generally do \emph{not} lie along the equilibrium locus. In the long-time limit, we expect the system to approach the  Boltzmann distribution at the bath temperature, namely $\lim_{t\to\infty}p(\vec{x},t)=\pi(\vec{x},T_b)$; compare the black curve in \FIG{manifold_locus}
with the blue-to-red curve of the equilibrium locus.

In the context of Markovian relaxation, we associate a Mpemba effect with cases where preparing a system in equilibrium at a hot temperature $T_h$ speeds up the cooling process towards a bath temperature $T_b$ in comparison to an analogous quench initiated from a 
colder temperature $T_b$ such that $T_h>T_c>T_b$.
We refer to the temporal evolution of a system initially prepared in equilibrium at $T_h$ and $T_c$ as $p_h(\vec{x},t)$ and $p_c(\vec{x},t)$, respectively.
In other words, we set $p_{\{h,c\}}(\vec{x},t=0)=\pi\left(\vec{x},T_{\{h,c\}}\right)$ and follow the out-of-equilibrium dynamics of these distributions towards $\pi(\vec{x},T_b)$. 
If  $p_h(\vec x,t)$ converges ``more quickly" to $\pi(\vec{x};T_b)$ than $p_c(\vec x,t)$, then a Mpemba effect exists for this system.

In order to establish which of the two distributions converges more quickly, we introduce a distance measure between $p(\vec x,t)$ and $\pi(\vec{x};T_b)$, that we denote by ${D}[p_{\{h,c\}}(\vec{x},t),\pi(\vec{x};T_b)]$, to properly quantify the distance of a system from the target equilibrium. The exact conditions for this distance function are discussed in Sec. \ref{sec:distance-function}, but for the following discussion both the Kullback-Leibler and the $L_1$ norm can serve as an example. 

\begin{figure}
    \centering
    \includegraphics[width=0.8\columnwidth]{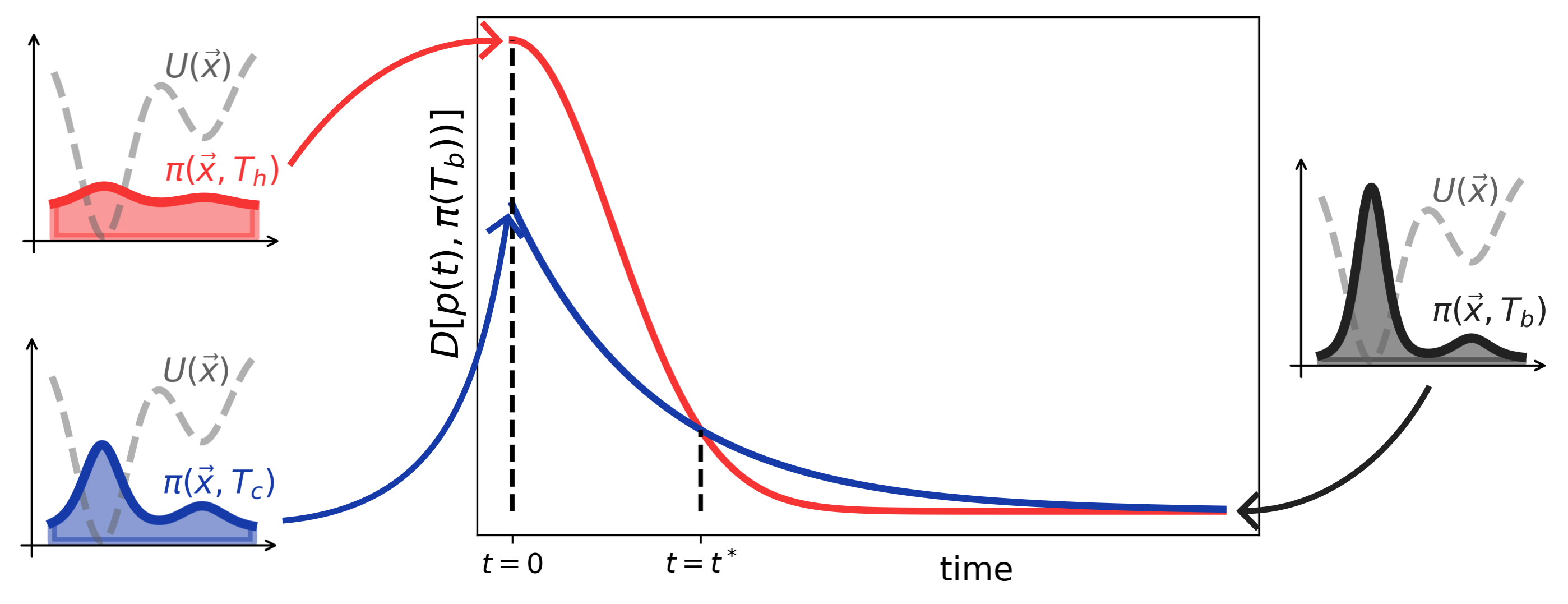}
    \caption{
    Sketch of the evolution of a distance measure $D$ in a thermal quench towards a bath temperature $T_b$.
    The red line describes the evolution of a system prepared at an initial Boltzmann equilibrium $\pi(\vec{x},T_h)$ at a hot temperature $T_h$, while the blue line results from a quench from a Boltzmann equilibrium $\pi(\vec{x},T_c)$ at a cold temperature $T_c$.
    The Mpemba effect is indicated by the crossing of the distance measures at time $t^*$.
    The small insets illustrate the different shapes of Boltzmann equilibria in a system described by the dashed-line potential, $U(\vec{x})$.
    }
    \label{fig:distance_measure}
\end{figure}

At some time $t^*$, we should see the two distances become equal,
${D}[p_h(\vec{x},t^*),\pi(\vec{x},T_b)]={D}[p_c(\vec{x},t^*),\pi(\vec{x},T_b)]$. At times greater than $t^*$, the initially hotter system is closer to equilibrium than the initially colder one: 
\begin{align} \label{eq:MpembaDef}
    {D}[p_h(\vec{x},t),\pi(\vec{x};T_b)]&<{D}[p_c(\vec{x},t),\pi(\vec{x};T_b)]\,,
\end{align}
indicating that the relaxation of the initially hotter system has overtaken that of the initially colder one 
(see Fig.~\ref{fig:distance_measure}). 

A few comments on this definition are in order:
\begin{enumerate}
    \item The definition suggests that the effect strongly depends on the specific dynamics, as well as on the distance function. As we discuss in Sec. \ref{sec:dynamics}, the details of the dynamics do play an important role; however, the exact form of the distance function used might not matter at all (``thermomajorization Mpemba effect"), or it might matter only for atypical distance functions. See the discussion in Sec. \ref{sec:distance-function}.
    
    \item The definition is based on the distance from the bath temperature equilibrium distribution, toward which all initial conditions eventually relax. It is a \textit{final relaxation} Mpemba effect, according to the terminology introduced in Sec.~\ref{sec:zoology}.
    
    \item Equation~\eqref{eq:MpembaDef} holds for every $t>t^*$. A ``transient Mpemba effect" that imposes Eq.~\eqref{eq:MpembaDef}  only for some finite time interval would strongly depend on the exact distance function used. 
    
    \item The definition is easily generalized to the inverse Mpemba effect discussed in Sec.~\ref{sec:inverse}.
\end{enumerate}

\paragraph{Characterization of the Mpemba effect via spectral decomposition}
The above definition for the Mpemba effect implies a need to scan all possible equilibrium initial conditions, calculate (or measure) all the corresponding relaxation trajectories, and compare their distances as a function of time to observe crossing and deduce the presence of the effect. Although such a procedure is possible, there is a much simpler way to test for a Mpemba effect, as we next discuss. For simplicity, we consider systems with discrete DoF; generalizing to continuous DoF is straightforward for most cases of interest.  

If $\mathbf{W}$ satisfies the DB conditions, its eigenvalues are real. However, in general, $\mathbf{W}$ is not a symmetric matrix, meaning that its left eigenvectors $\vec u_i$ differ from its right eigenvectors $\vec v_i$. Moreover, both sets of eigenvectors are not orthogonal. The left and right eigenvectors obey
\begin{align}
    \textbf{W}\vec v_i = \lambda_i\vec{v}_i , \\
    \vec u_i^\tran \textbf{W} = \lambda_i \vec u_i^\tran ,
\end{align}
respectively.  Assuming ergodicity (which, for rate matrices, means irreducibility of $W$), the Perron-Frobenius theorem ensures convergence to a unique steady state \cite{schnakenberg1976network}. This implies that the dominant eigenvalue, the one with the largest real part, is identically zero and nondegenerate. It is convenient to label the eigenvalues in the order $0=\lambda_1>\lambda_2\geq\dots\geq\lambda_N$. The right eigenvector $\vec{v}_1$ represents the steady state; the DB condition then implies that $\vec{v}_1=\vec{\pi}(T_{b})$.

The master equation can be formally integrated as
\begin{align}
    \vec{p}(t)=e^{\mathbf{W}(T_{b})t} \vec{p}_0, 
\end{align}
with initial condition $\vec{p}(t=0)=\vec{p}_0$.
In principle, the initial condition $\vec{p}_0$ could be arbitrary; however, in the context of the Mpemba effect, we restrict our analysis to initial conditions corresponding to a Boltzmann equilibrium $\vec{\pi}(T_0)$ at some temperature $T_0$. Using the right eigenvectors of $\mathbf{W}$, we can decompose the solution of the master equation as
\begin{align}
\label{eq:m}
	\vec{p}(t,T_0,T_{b})=\vec{\pi}(T_{b})+\sum_{i=2}^{N} a_i(T_0,T_{b}) \, e^{\lambda_i(T_{b})t} \vec{v}_{i}(T_{b}),
\end{align}
where we write explicitly the dependence of the eigenvalues and eigenvectors on the bath temperature and introduce the projection coefficient
\begin{align}
\label{eq:ai-coef}
 a_i(T_0,T_b)=\frac{\vec{u}_{i}(T_{b}) \cdot \vec{\pi}(T_0)}{\vec{u}_{i}(T_{b}) \cdot \vec{v}_{i}(T_{b})},
\end{align}
representing the projection of the initial condition on the $i$-th left eigenvector. We underline that the sign of $\vec u_i$ is arbitrary, but it determines the sign of $a_i$. Note that the projection coefficient $a_i$ is in terms of the \emph{left} eigenvectors $\vec{u}_i$, whereas $\vec{p}$ is expanded over its \emph{right} eigenvectors $\vec{v}_i$.  If \textbf{W} is symmetric, then $\vec{u}_i = \vec{v}_i$; otherwise, they are different. In general, the eigenvectors are biorthogonal and can be normalized so that $\vec u_i\cdot \vec{v}_i = \delta_{ij}$. The eigenvector decomposition highlights how all the amplitudes of the non-dominant eigenvectors decay exponentially in time, with a rate proportional to the (negative) eigenvalues $\lambda_i$.

Assume for simplicity that $\lambda_2$ is strictly larger than $\lambda_3$ or, equivalently, that $|\lambda_2| < |\lambda_3|$.  Then, after a long-enough time, the time-dependent probability is  approximately
\begin{align}
    \vec{p}(t,T_0,T_b)\approx \vec{\pi}(T_b)+a_2(T_0,T_b) e^{\lambda_2(T_b)t}\vec{v}_2(T_b),
\label{eq:longtime_a2}
\end{align}
Thus, we see that $a_2$ is the only term retaining information on the initial condition in the long-time limit, implying that it can determine the existence of the Mpemba effect for a given pair of initial temperatures $T_h$ and $T_c$.
Choosing as a distance measure the $L_1$ norm gives, for initial temperature $T_0$, 
\begin{align}
    {D}[\vec{p}(t,T_0,T_b),\vec{\pi}(T_b)]\approx \left|a_2(T_0,T_b)\right| \, e^{\lambda_2(T_b)t}\sum_{k}\left|\left(\vec v_2\right)_k(T_b)\right|.
\end{align}
It is clear from the above equation how a nonmonotonic dependence of $a_2$ on the initial temperature provides a sufficient condition to determine the existence of the Mpemba effect.
Indeed, if 
\begin{align}
|a_2(T_c,T_b)|>|a_2(T_h,T_b)|,
\end{align}
then, at long times, ${D}[\vec{p}(t,T_h,T_b),\vec{\pi}(T_b)]<{D}[\vec{p}(t,T_c,T_b),\vec{\pi}(T_b)]$, implying that there exists some finite time $t^*>0$ after which the system that started from $T_h$ is closer to the target equilibrium at $T_b$.

For a Markovian operator with a discrete set of eigenvalues, the same analysis holds, essentially. To state more exactly the connection between the nonmonotonicity of $|a_2(T_0,T_b)|$ as a function of $T_0$ and the existence of a Mpemba effect, we need to discuss the distance function in more detail. Before doing that, we demonstrate the overall idea using two simple examples of the effect in discrete and continuous DoF.

\subsubsection{Example: Finite number of microscopic configurations}
\label{Sec:Example-3-states}
\begin{figure}
    \centering
    \includegraphics[width=0.7\columnwidth]{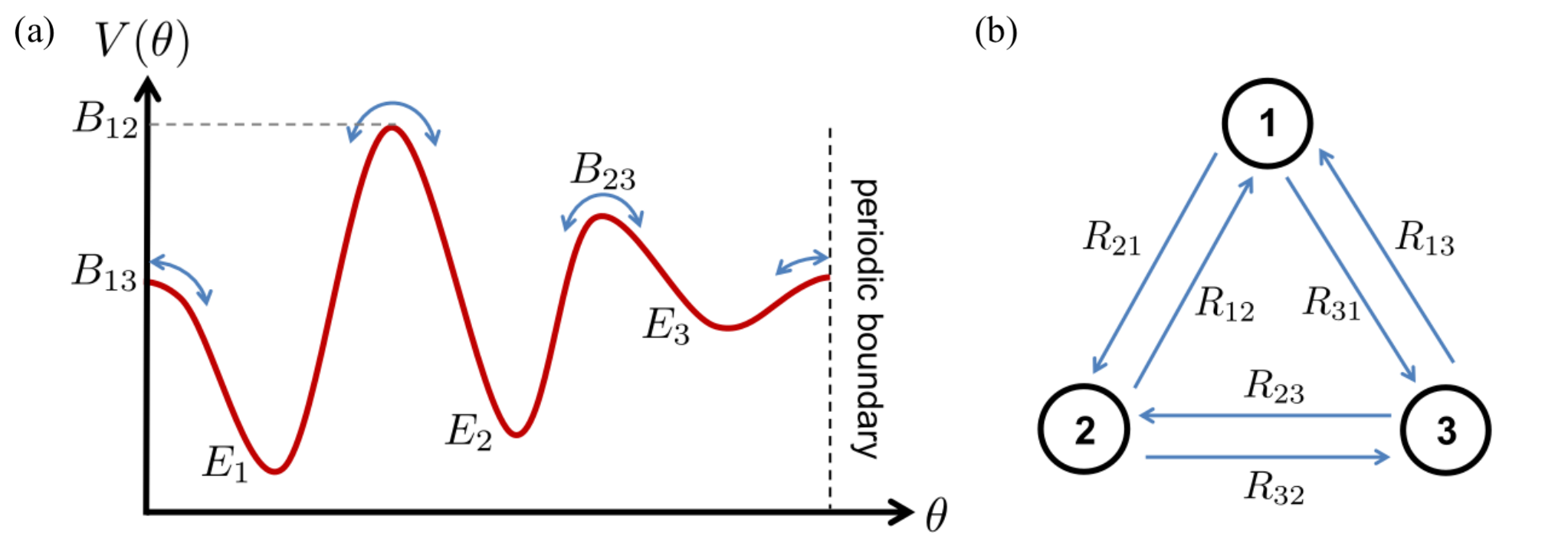}
    \caption{(a) Sketch of an energy landscape with three metastable states.
    Each state's energy is denoted by $E_i$, and the barrier between two states is denoted by $B_{i\!j}$.
    (b) An effective description of the energy landscape in (a) as a three-state system, where $R_{i\!j}$ denotes the transition rate from $j$ to $i$ and is given by Eq.~\ref{eq:arrhenius}. \textit{Source:} Reprinted with permission from~\cite{lu2017nonequilibrium}.}
\label{fig:3_state_system}
\end{figure}

\begin{figure}
    \centering
    \includegraphics[width=0.825\columnwidth]{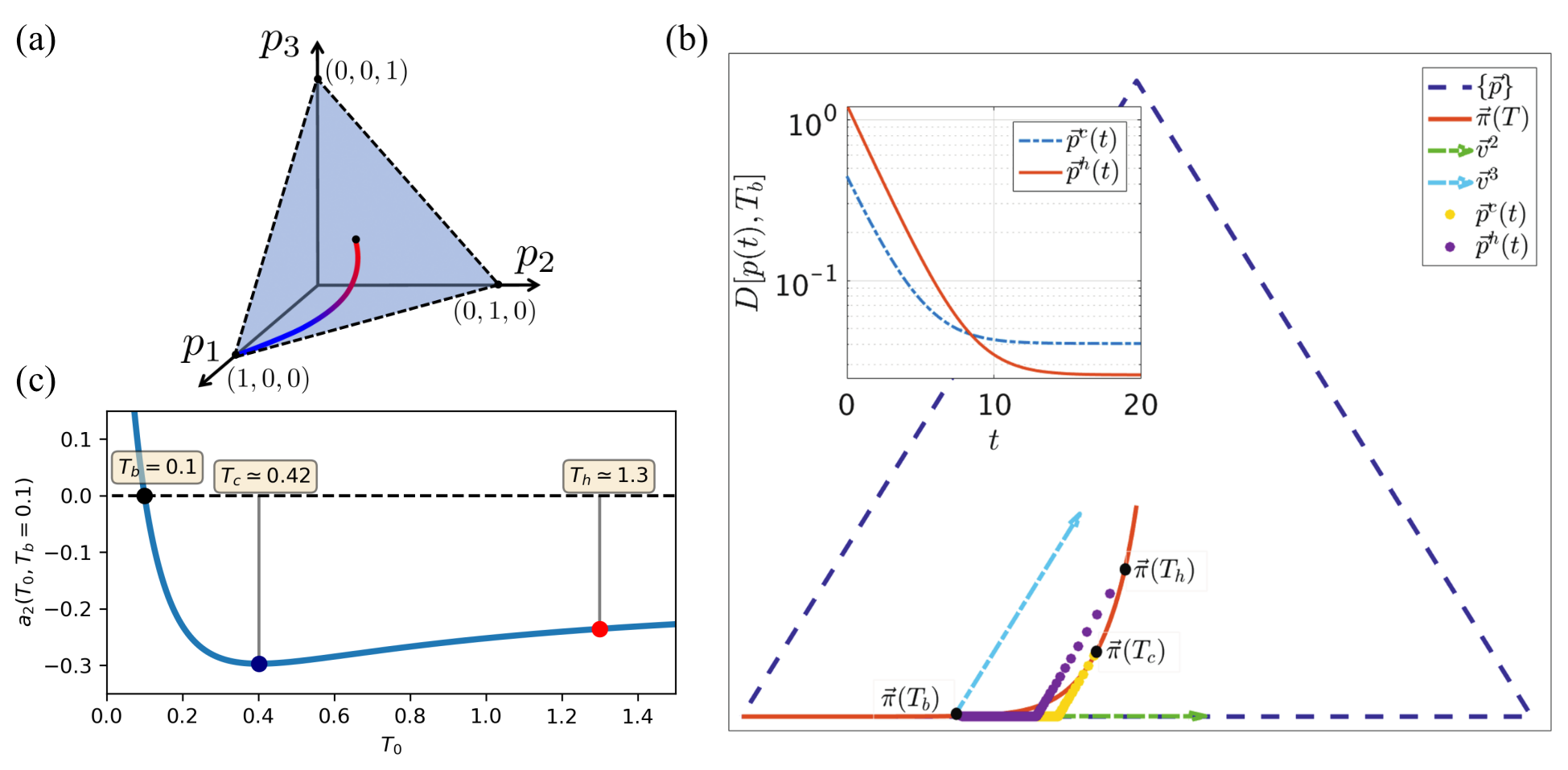}
    \caption{(a) The probability distribution among the three states can be described by the vector $\vec{p} = (p_1 , p_2 , p_3 )$, and all possible values of $\vec{p}$ form a simplex in $(p_1 , p_2 , p_3)$ (shaded triangle).
    The curved line is the quasistatic locus, namely the set of Boltzmann distributions $\vec{\pi}(T)$ corresponding to different temperatures from $0$ (blue end) to $\infty$ (red end).
    (b) The projection coefficient $a_2(T_0,T_b)$ at $T_b=0.1$ exhibits a nonmonotonic dependence on $T_0$, with $\partial_{T_0}a_2(T_0,T_b)=0$ at $T_0\simeq 0.42$, which is a sufficient condition to prove the existence of a Mpemba effect in this system.
    (c) The set of all normalized probability distributions forms a simplex, the dashed blue triangle.
    The red solid curve is the quasistatic locus $\vec{\pi}(T)$.
    The dashed arrows are the two right eigenvectors of the rate matrix, $\mathbf{W}$, $\vec{v} _2$, and $\vec{v} _3$, associated with slow (green) and fast (blue) relaxation modes.
    The dotted lines represent the relaxation process, $\vec{p} ^{\, c} (t)$ starting at $T_c =0.42$ (orange) and $\vec{p} ^{\, h} (t)$ starting at $T_h = 1.3$ (purple). In this example with $T_b=0.1$, the ratio $\frac{a_2(T_0,T_b) \lambda_2(T_b)}{a_3(T_0,T_b) \lambda_3(T_b)}\lesssim 10^{-3}$ for every $T_0>0$, implying that, in the first stages of the relaxation, the out-of-equilibrium relaxation is along $\vec{v}_3$. 
    The inset shows that the distances ${D} [\vec{p} ^{\, h}(t),\vec{\pi}(T_b) ]$ and ${D} [\vec{p} ^{\, c} (t), \vec{\pi}(T_b) ]$ both decrease with time.  The initially hot system starts at a larger distance, but after some time, its distance from equilibrium is smaller than that of the initially cold system. \textit{Source:} Panels (a) and (c) are reprinted with permission from~\cite{lu2017nonequilibrium}.}
    \label{fig:3_state_simplex} 
\end{figure}

Let us first demonstrate the effect in the simplest possible system, a three-state system~\cite{lu2017nonequilibrium}, as depicted in Fig.~\ref{fig:3_state_system}. The main advantages of the three-state system are that the distribution space, the equilibrium locus, and the relaxation trajectories in the distribution space can all be plotted; see, e.g., Fig. \ref{fig:3_state_simplex}.

To define a three-state example, we first need to specify the rate matrix describing transitions between states. A common temperature dependence of the rate matrix  elements that satisfy the detailed balance condition~\EQ{detailed_balance} is the Arrhenius form  \cite{mandal2011proof}, 
\begin{align}
\label{eq:arrhenius}
    W_{i\!j}(T_b)=\Gamma\, e^{-\frac{B_{i\!j}-E_j}{k_B T_b}},
\end{align}
where $\Gamma$ is an amplitude that sets the timescale of the system and $B_{i\!j}=B_{ji}$ represents a potential barrier that separates the two states (hence $B_{i\!j}\geq E_i$ and $B_{i\!j}\geq E_j$).
Such a form is convenient, as it emerges naturally when spatially coarse graining a continuous-space model; see Fig.~\ref{fig:3_state_system}.

To observe a Mpemba effect, consider a specific example where the energies of the three states are given by $E_1=0$, $E_2=0.1$, and $E_3=0.7$, and the barriers between them are chosen to be $B_{12}=1.5$, $B_{13}=0.8$, and $B_{23}=1.2$.
The distribution space of this system is identified by the set of all probability vectors $\vec{p}=(p_1 , p_2 , p_3 )$ with components satisfying $0\leq p_i \leq 1$ and $p_1+p_2+p_3=1$.
Thus, the distribution space is a two-dimensional simplex (triangle) whose vertices are at $(1,0,0)$, $(0,1,0)$ and $(0,0,1)$; see Fig.\ref{fig:3_state_simplex}a. The quasistatic locus $\vec{\pi}(T)$ of this system is shown as a gradient-color curve in distribution space in Fig. \ref{fig:3_state_simplex}a. This parametric curve represents all the Boltzmann equilibrium configurations at temperatures $T\in[0,\infty]$. 
At zero temperature, the system is in its lowest-energy configuration (here, $E_1$), which corresponds to $\vec{\pi}(T=0)=(1,0,0)$.  At infinite temperature, the Boltzmann distribution is uniform over the states, with $\vec{\pi}(T=\infty)=(\sfrac{1}{3},\sfrac{1}{3},\sfrac{1}{3})$.

A temperature quench causes the probability vector to evolve from an initial equilibrium distribution $\vec \pi(T_0)$ at temperature $T_0$ to a final equilibrium at temperature $T_b$. This relaxation trajectory can also be represented by a curve in the distribution simplex shown in Fig.~\ref{fig:3_state_simplex}c.
The eigenvalue problem of the $3\times 3$ rate matrix $\mathbf{W}$ can be reduced to an exactly solvable second-order polynomial equation, which provides an analytical expression for the second and third eigenvalues and eigenvectors.
The exact evolution of the system is given by
\begin{align}
    \vec{p}(t,T_0,T_b)= \vec{\pi}(T_{b})+a_2(T_0,T_{b}) e^{\lambda_2(T_{b})t}\vec{v}_{2}(T_{b})
    +a_3(T_0,T_{b}) e^{\lambda_3(T_{b})t}\vec{v}_{3}(T_{b}).
\end{align}
As long as the coefficients of $\vec v_2$ and $\vec v_3$ are comparable in size (i.e., $t\sim [\lambda_3-\lambda_2]^{-1} \log (a_2/a_3)$), the direction of the relaxation curve is a linear combination of the second and third eigenvectors.
However, after a long time, the system approaches the equilibrium locus at the target bath temperature equilibrium $\vec{\pi}(T_b)$ along a direction parallel to $\vec{v}_2$.

To prove the existence of the Mpemba effect in this setup, we choose a bath temperature $T_b=0.1$ and analyze the $a_2(T_0,T_b=0.1)$ coefficient during cooling quenches. In particular, if its dependence on the initial temperature $T_0>T_b$ is nonmonotonic, then $a_2$ reaches an extremum at some temperature, which we set to be our cold temperature; i.e., $\partial_{T_0}a_2|_{T_0\equiv T_c}=0$.
This is the case for our system (Fig. \ref{fig:3_state_simplex}b) with $T_c\simeq0.42$ maximizing locally $|a_2|$.
Thus, we can find a hot temperature $T_h>T_c$ such that $|a_2(T_0=T_h)|<|a_2(T_0=T_c)|$.
This is verified in the example shown in Fig.~\ref{fig:3_state_simplex} for $T_h=1.3$, which proves that the Mpemba effect exists in that system.
In~\FIG{3_state_simplex}c, we show the trajectories of the cooling quench starting from this set of hot and cold temperatures, highlighting how the last stage of the evolution occurs along a direction parallel to $\vec{v}_{2}$.
Additionally, we evaluate the ``entropic distance measure''
\begin{align}
    {D}[\vec{p}(t,T_0,T_b),\vec{\pi}(T_b)]=\sum_{i=1}^M \left( \frac{E_i}{k_B T_b} \right)\left[p_i(t;T_0,T_b)-\pi_i(T_b)\right]+p_i(t,T_0,T_b) \ln p_i(t,T_0,T_b) -\pi_i(T_b) \ln \pi_i(T_b),
\end{align}
for the two initial conditions at $T_h$ and $T_c$.
In the inset, a plot of the distances from equilibrium of the two evolutions cross at $t^*\approx 9$, highlights how the initially hotter system has become closer to the target Boltzmann equilibrium at $T_b$.  More formally, ${D}[\vec{p}(t,T_h,T_b),\vec{\pi}(T_b)]>{D}[\vec{p}(t,T_c,T_b),\vec{\pi}(T_b)]$ for $t>t^*$, proving the existence of the Mpemba effect.

\paragraph{No Mpemba effect in a two-state system}
One might think a two-state system would be an even simpler setting to demonstrate the Mpemba effect, but there can be no Mpemba effect in a Markovian system with only two states. To understand why, recall that the distribution space is one dimensional, since the probability vector can be written as $\vec p = (p,1-p)$. The rate matrix can be written as
\begin{align}
\textbf{W} = 
    \left( \begin{array}{cc}
       -a & b \\
       a & -b
    \end{array} \right).
\end{align}   
The evolution of the probabilities follows a simple, exponential relaxation in one variable:
\begin{align}
    \dot p = b - (a+b)p, 
\end{align}
with a solution given by
\begin{align}
    p(t) = \frac{1}{a/b+1} + \left(p(0) - \frac{1}{a/b+1}\right)e^{-(a+b)t}, 
\end{align}
and no crossing between trajectories happens regardless of the ratio $a/b$ or specific initial conditions.  (In~\SEC{boundary_coupling}, we will see that this is also the case for certain systems weakly coupled to the thermal bath~\cite{teza2023relaxation}.)
The Mpemba effect might nevertheless exist in systems with only two states but having memory effects or for dynamics with stochastic resetting; see~\SEC{stochastic-resetting}.

\subsubsection{Example: Brownian particle in a potential}
\label{Sec:Brownian}
The discrete-state example outlined above enabled us to plot the distribution space, with its equilibrium locus and also the relaxation trajectories. However, very little physical insight and intuition was gained from it. Our next example demonstrates the same effect in a system with a continuous DoF, a Brownian particle in a potential. In this case, the distribution space and trajectories within this space cannot be directly plotted. However, it does give heuristic insight into the Mpemba effect, as it allows us to link the relaxation process with the system's energy landscape.

\begin{figure}
    \centering
    \includegraphics[width=0.8\columnwidth]{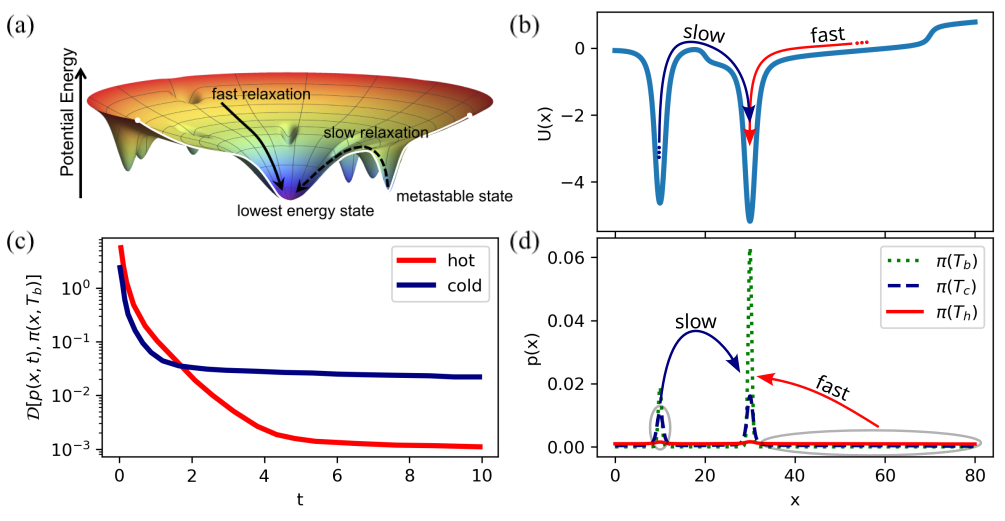}
    \caption{
    (a) Sketch of an energy landscape in a high-dimensional configuration space as a 2D energy funnel, cut open for illustrative purposes.
    The solid arrow represents a fast relaxation, and the dashed arrow a slow one.
    (b) Example 1D energy landscape to demonstrate the Mpemba effect in the corresponding Fokker–Planck dynamics. The left well coarse-grains to a metastable state, whereas the right well coarse-grains to the lowest-energy state.
    The width of the basin of attraction of the deeper well is significantly greater than that of the shallower well.
    (c) Decay of the distance measure $D[p(x,t),\pi(x,T_b)]$, as in \EQ{distance_measure_fokker_planck}, from the target Boltzmann equilibrium (here $T_b=0.45$).
    A system prepared at an initially hotter temperature ($T_h=10$, red line) relaxes faster than one prepared at a colder temperature ($T_c=1.38$, blue line). (d) The Boltzmann distributions at different temperatures.
    After the quench, both initial distributions relax toward the final equilibrium distribution (dotted green line).
    The initially colder (dashed blue line) system is more populated in the lowest well than the initially hot system (solid red line); however, after a short relaxation, the hot system ends up with a higher population in the lowest well, on account of the fast relaxation from initial states within its basin. This implies that, statistically, the relaxation from an initially hot system is faster than that from an initially colder system, and the Mpemba effect occurs. \textit{Source:} Panel (a) is reprinted with permission from~\cite{lu2017nonequilibrium}.}
    \label{fig:brownian_system}
\end{figure}
In the specific example, we consider the position coordinate $x$ of a 1D Brownian particle placed in contact with a thermal bath at temperature $T_{b}$ and subject to a potential $U(x)$. The state $x(t)$ is described by the overdamped Langevin equation,
\begin{align}
   \dot{x}=-\frac{1}{ \gamma}\partial_x U(x)+\sqrt{2D}\,\xi(t),
\end{align}
where $\gamma$ is the drag coefficient, $D$ the diffusion constant, and $\xi(t)$ is white Gaussian noise obeying $\left< \xi(t) \right> = 0$ and $\left< \xi(t) \xi(t') \right> = \delta(t-t')$. The diffusion constant, $D(T_{b})=k_B T_{b}/\gamma$, is related to the bath temperature through the Einstein-Smoluchowski relation. 

The associated Fokker-Planck equation describes the time evolution of the probability density $p(x,t)$ to find the particle at a position $x$ at time $t$. Note that $p$ is a density in $x$ but not in $t$, which is simply a parameter. The time evolution obeys
\begin{align}
    \partial_t p(x,t)=-\frac{1}{ \gamma} \partial_x[(\partial_xU(x))p(x,t)] + D(T_b)\partial_x^2 p(x,t) = \mathcal{L}_{T_b} p(x,t), 
\label{eq:Fokker-Planck}
\end{align}
where $\mathcal{L}_{T_b}$ is the Fokker-Planck operator. In the long-time limit, the system relaxes toward the Boltzmann distribution,
\begin{align}
    \pi(x,T_b)=\frac{e^{-U(x)/k_B T_b}}{\mathcal{Z}(T_b)},
\end{align}
where the normalization constant is given by the equilibrium partition function $\mathcal{Z}(T_b)=\int _{\Omega} \dd x \, e^{-U(x)/k_B T_b}$.

Figure~\ref{fig:brownian_system}b illustrates how a Mpemba effect can arise, using a double-well potential $U(x)$ in the interval $x\in[0,80]$.
The potential has a local minimum at $x\approx 10$ and an absolute minimum at $x\approx 30$.

The eigenfunctions of $\mathcal{L}_{T_b}$, solving  $\mathcal{L}_{T_b}v_i(x,T_b)=\lambda_i(T_b) v_i(x,T_b)$ with $0=\lambda_1\geq\lambda_2\geq\lambda_3\geq\dots$, form a complete basis. As in the discrete case above, the system is initially prepared at a Boltzmann equilibrium $\pi(x;T_0)$ at temperature $T_0$. The time evolution can be decomposed as
\begin{align}
    p(x,t)=\pi(x,T_{b})+\sum_{i>1}a_i(T_0,T_{b})e^{\lambda_i(T_{b})t}v_i(x,T_{b})\,,
\label{eq:eigenfunc_continuous}
\end{align}
where $a_i$ is a coefficient retaining information about the initial condition and hence depends explicitly on $T_0$.
We set as a target bath temperature for the cooling process $T_{b}=0.45$.  At this temperature, the distribution is mostly peaked around the second (deeper) well; see the green line in Fig.~\ref{fig:brownian_system}d. In the long-time limit, $a_2$ controls the relaxation; hence, we can study its dependence on the initial temperature $T_0$ and search for a nonmonotonic dependence on it.  For the specific potential chosen here with $T_b=0.45$, the projection coefficient $|a_2(T_0,T_b)|$ has a local maximum as a function of $T_0$ at $1.38$. We choose the cold temperature to be $T_c = 1.38$, as this choice maximizes the difference between $a_2(T_c,T_b)$ and $a_2(T_h,T_b)$ coefficients. (This idea is later illustrated in the discussion of the strong Mpemba effect; see Fig.~\ref{fig:strong_sketch.png}a.)

The nonmonotonicity of $a_2(T_0,T_b)$ as a function of $T_0$ for fixed $T_b$ is the hallmark of the Mpemba effect; indeed, preparing the system at any temperature $T_h>T_c$  implies $|a_2(T_h,T_{b})|<|a_2(T_c,T_{b})|$, leading to a quicker relaxation towards the equilibrium steady state at $T_{b}$.
As an example, we choose $T_h=10$ to see (Fig. \ref{fig:brownian_system}c) how the relaxation process towards equilibrium is faster when starting from such a distribution.

As in the discrete case, we can analyze the distance measure between probability density functions. Instead of the $L_1$ norm used there, we choose here to use the Kullback-Leibler divergence, 
\begin{align} \label{eq:distance_measure_fokker_planck}
    D_{\mathrm{KL}}[p(x,t),\pi(x,T_b)]&\equiv \int_\Omega \dd x \, p (x,t) \ln \left(\frac{p(x,t)}{\pi(x,T_b)}\right) 
   \\
   & = 
    \int_\Omega \dd x  \left[ \left( \frac{U(x)}{k_B T_b} \right) \, \left( p(x,t)-\pi(x,T_b) \right) +p(x,t) \ln p(x,t) - \pi(x,T_b) \ln \pi(x,T_b)\right].
\end{align}
Evaluating $D_{\mathrm{KL}}$ with respect to the two initial conditions at $T_{c}$ and $T_{h}$, we observe the expected crossing at $t\approx 2$, demonstrating the Mpemba effect; see~\FIG{brownian_system}c.

In this specific choice of $U(x)$, there is a simple physical interpretation for the Mpemba effect: the potential has two wells separated by a relatively high barrier, and the deep well has a wider basin of attraction. At the lowest temperature, $T_b=0.45$ and the equilibrium distribution is tightly concentrated at the bottom of the two wells. The ratio between the accumulated probability in the two wells is dictated by the energy difference between the energy minima of the wells, relative to $T_b$; see green line in Fig.~\ref{fig:brownian_system}d. For the cold temperature $T_c$, the distribution is still peaked at the two minima, but the ratio between the probabilities within each well is different. Lastly, in the high temperature (red line in Fig. \ref{fig:brownian_system}d), the probability is almost uniformly distributed across all space. The slowest relaxation mode in this case has a clear physical interpretation: it corresponds to diffusive motion from one well to the other. However, as the equilibrium distributions of both $T_b$ and $T_h$ have the same probability to be within each well, the relaxation takes place without the slowest mode. In contrast, the two wells are not balanced in terms of accumulated probabilities when initiating the system at $T_c$. Thus, in this case, a significant portion of the dynamics occurs in the slowest relaxation mode. Therefore, the relaxation from $T_h$ is faster than from $T_c$, as can be seen in Fig.~\ref{fig:brownian_system}c.

While this specific choice of $U(x)$ allows a clear understanding of the geometric mechanism behind the Mpemba effect, it is important to stress that the effect can be observed in less heavily engineered potentials. A simpler double-well potential can also exhibit the Mpemba effect~\cite{teza2023relaxation}, as seen in an experiment \cite{kumar2020exponentially,kumar2022inverse}.  The Mpemba effect is also predicted in an even simpler asymmetric, single-well potential in a box~\cite{biswas2023mpemba_a}.

\subsubsection{The distance from equilibrium}
\label{sec:distance-function}
The definition of the Mpemba effect given above is based on a distance function $D$ between two distributions.  Some natural choices for $D$ include the Kullback-Leibler divergence (or relative entropy), which for systems with discrete DoF has the form $D_{\mathrm{KL}}[\vec p,\vec q] = \sum _i p_i \ln (p_i/q_i)$, and the $L_1$ distance, which for systems with discrete DoF has the form $L_1(\vec{p},\vec{q}) = \|\vec p - \vec q\|_1=\sum_i|p_i - q_i|$. These two are probably the easiest to motivate physically; nevertheless, other choices can be relevant. In~\cite{lu2017nonequilibrium}, it was suggested that $D$ can be any distance function that has the following three properties:
\begin{itemize}
\item[(i)] $D[p,p]=0$, and $D[\pi(\vec{x},T),\pi(\vec{x},T_b)]$ is a monotonically increasing function of $|T-T_b|$.  That is, the greater the temperature difference, the greater the distance to equilibrium. Otherwise, it would be possible to have a direct or an inverse Mpemba effect already at $t=0$, which would not be a statement about the relaxation process but rather a quirk of the distance function. 

\item[(ii)] $\partial_t D[p(t),\pi] \le 0$ for any $\vec p(t)$, namely, the distance function is a monotonically non-increasing function of time, for all distributions. If such a distance function has an H-theorem (as do, e.g., the KL-divergence and the $L_1$ norm), it is a \emph{monotone}. If this is the case, the distance from equilibrium does not increase spontaneously with time.  A nonmonotonicity in time  would imply that a system that started relaxing first can nevertheless be farther from equilibrium, which 
is not a reasonable way to measure how far the system is from its final stage. 

\item[(iii)] $D[\vec p(t),\vec\pi(T_b)]$ must be a continuous convex function of $\vec p(t)$. Convexity ensures that the functional derivative of $D[\vec p(t),\vec\pi(T_b)]$  with respect to $\vec p(t)$ is bounded in the limit $\vec p(t)\to \vec\pi(T_b)$. 
\end{itemize}

The three conditions on the distance function discussed above are, in fact, not enough to avoid several possible anomalies. This was noticed already in \cite{klich2019mpemba}, where an additional demand was added somewhat ``heuristically": that the exponential ratio between the time-dependent coefficients of $\vec{v}_{2}$ and $\vec{v}_{3}$ not be compensated for by a distance function with an exponential weight in some directions relative to others. In other words, the metric should not grow exponentially faster in one direction than another, so that it compensates for the temporal exponential relaxation in different directions. This point was cleared up only in~\cite{van2024thermomajorization}, where the exact conditions for the existence of the Mpemba effect using any monotone as a distance function were found, using recent results on thermomajorization \cite{horodecki2013fundamental}. In particular, a sufficient condition for having a Mpemba effect with any monotone as a distance function is
\begin{align}
\label{eq:thermomajorization-Mpemba}
    \left\|\frac{a_2(T_h)}{a_2(T_w)}\vec v_2 
        - z\vec\pi(T_b) \right\|_1 
    \leq  \left\|\vec v_2 - z\vec\pi(T_b) \right\|_1 
    \leq  \left\|\frac{a_2(T_w)}{a_2(T_h)}\vec v_2 - z\vec\pi(T_b) \right\|_1, 
\end{align}
where $z\in\{(\vec v_2)_i/\vec \pi_i, \; 1\leq i \leq N\}$ is one of the element-by-element ratios of the vectors $\vec v_2$ and $\vec \pi$.  The inequality must hold for all $N$ values. It was also shown that the condition we used so far---nonmonotonicity of $|a_2(T,T_b)|$ of the initial temperature for $T$, assuming $\lambda_2>\lambda_3$---is not enough: 
a monotone for which no crossing of the distance function occurs might still exist, even in the $t\to\infty$ limit. 

It is important to note that the ``thermomajorization Mpemba effect," where the crossing between the distance function through relaxation must occur for any monotone, is a strong condition to impose. In most cases, a crossing of Kullback-Leibler divergences suffices to identify a Mpemba effect. The nonmonotonicity of $|a_2(T)|$ is then a sufficient condition for the existence of the effect: in the long-time limit, we can expand
\begin{align}
\label{eq:DKL-large-time-limit}
    \lim_{t\to\infty}D_{\rm KL}\left[\vec p(t),\vec \pi(T_b)\right]&=\lim_{t\to\infty} D_{\rm KL}\left[\vec \pi(T_b) + a_2e^{\lambda_2 t} \vec v_2 + a_3e^{\lambda_3 t} \vec v_3,\vec \pi (T_b)\right]\nonumber\\
    &= \lim_{t\to\infty}\sum _i \left(a_2 e^{\lambda_2t}\left(\vec v _2\right)_i + a_3e^{\lambda_3t}\left(\vec v _3\right)_i\right) \,,
\end{align}
which implies that when $\lambda_2 > \lambda_3$, the distance at large enough time is dominated by $a_2$, and nonmonotonicity in $a_2$ implies nonmonotonicity in the distance function. 

The above properties, (i)--(iii) and sufficient conditions~\EQS{thermomajorization-Mpemba}{DKL-large-time-limit}, readily generalize to the continuous case by replacing the sum over states with an integral. Also, note that it is not necessary to demand that $D[\vec{p},\vec{\pi}]$ be symmetric, with respect to $\vec p$ and $\vec \pi$, for it to be a good quantifier of distance from the equilibrium $\vec \pi$ during a relaxation process, $\lim_{t\to\infty}\vec p(t) = \vec \pi$. Indeed, $D_{\mathrm{KL}}[\vec p, \vec \pi]$ is not symmetric with respect to $\vec p$ and $\vec \pi$; it is technically a ``divergence'' but not a metric. 

The mean energy in the system has sometimes been used as a proxy for the distance from equilibrium \cite{baity2019mpemba,vadakkayil2021should,ohga2024microscopic}. This quantity is often easier to measure than quantities such as the KL divergence or $L_1$ norm in large systems, as the latter explicitly depends on the full distribution $\vec p(t)$ of system states. This is especially true in numerical simulations of large systems, where the mean energy can be measured quite accurately but the distribution space is too large to estimate $\vec p(t)$ and its distance from equilibrium.

\begin{figure}
    \centering
    \includegraphics[width=6.0in]{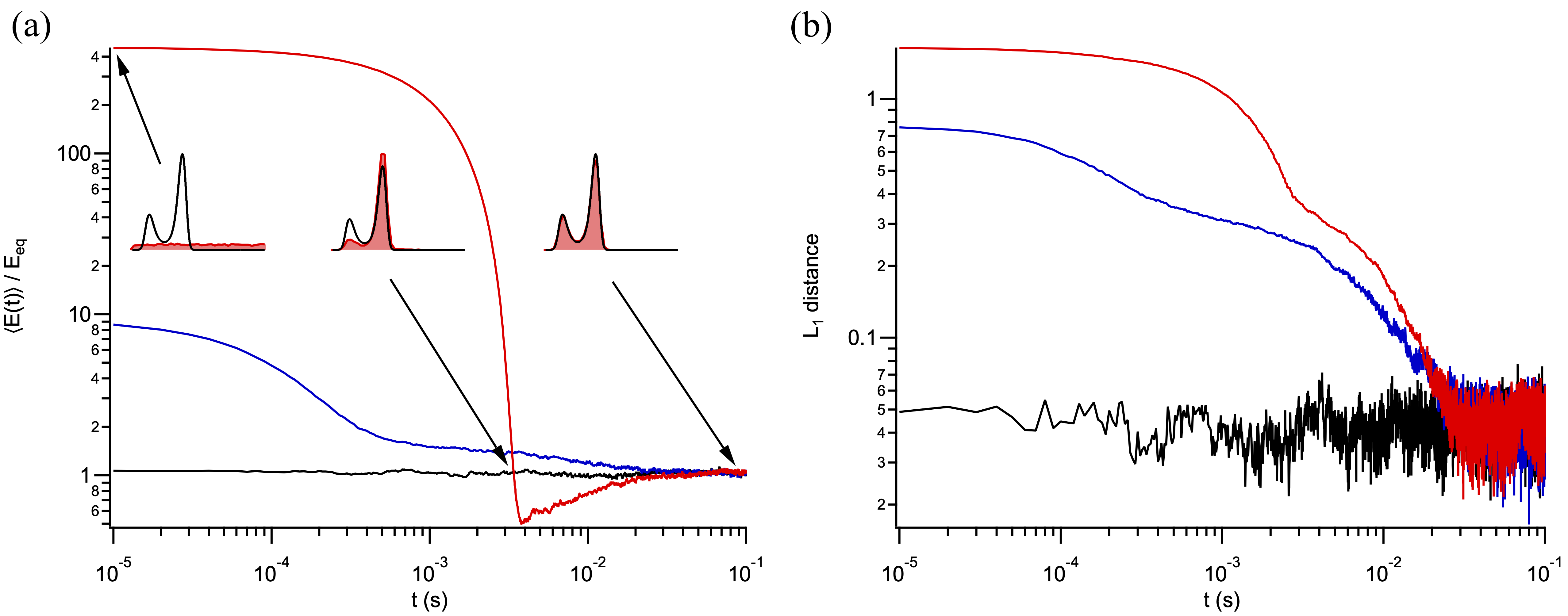}
    \caption{Average energy is not always a reliable indicator of the Mpemba effect. (a)  Numerical simulation of the average energy evolution relative to the equilibrium value for an ensemble of single-particle systems relaxing to equilibrium from hot (red curve) and cold (blue curve) initial temperatures, along with a control (black curve) that is initially in equilibrium with the bath.  The initial ``hot" system has a nearly uniform distribution (left inset, where the shaded pink distribution is nearly flat), and differs significantly from the distribution in equilibrium with the bath (right insert).  The average energy of the middle, ``cold" temperature initial state (blue curve)  equals that of the hot initial temperature (red curve) at $t \approx 0.04$ s. (b) $L_1$ distance for the same numerical simulations show no crossing of hot (red) and cold (blue) curves and thus no Mpemba effect. Numerical simulations courtesy of Siddharth Sane.}
    \label{fig:average-energy}
\end{figure}

To illustrate the issues with measuring the distance to equilibrium via the average energy, \FIG{average-energy}a shows a numerical simulation of the average energy of an ensemble of particles relaxing in a tilted double-well potential that corresponds to the experiments in Ref.~\cite{kumar2020exponentially}; see~\SEC{colloids}.  To understand the striking nonmonotonic relaxation, we note that the relaxation in a double-well potential proceeds in two stages (see also~\SEC{metastability}, below).  In the first stage, particles typically relax to the nearest local potential well.  In the second stage, the relative probabilities of being in each state (well) relax to their equilibrium values~\cite{chetrite2021metastable}.  In the particular case shown in~\FIG{average-energy}, more particles relax to the lower-energy well, implying an average energy that is below the equilibrium value.  The energy increases to that value in the second stage of relaxation.  The fact that two states (here, at $t \approx 0.04$ s and at long times, as illustrated by the center and right inset distributions) have the same average energy violates one of the basic axioms for a metric, that $D[p,\pi] > 0$ for all $p \neq \pi$.  The nonmonotonicity of relaxation implies that two relaxation curves could cross without necessarily implying a Mpemba effect.  Indeed, \FIG{average-energy}b shows the ``standard" L$_1$ distance plot for the same simulation and confirms that there is no Mpemba effect for this case (no crossing of hot and cold curves), even though the average energy curves do cross.

Depending on the specifics of the system at hand, including its dynamics, alternative analyses can act effectively as a proxy to determine the existence of a crossing of the distance measures. For example, determining the system's slowest relaxation mode allows for avoiding a characterization of the evolution of the whole distribution. Alternatively, a geometric analysis of the trajectories of relevant observables can help for extremely large systems whose size makes it effectively impossible to reliably determine $D$ (e.g., spin systems). In~\SEC{kinetic-framework}, we discuss the analogous issue of defining a suitable distance function in the granular-gas framework.

\subsubsection{The inverse Mpemba effect}
\label{sec:inverse}
The characterization of the ME outlined above in the stochastic thermodynamics framework can, in principle, be applied to test for an analogous effect when rapidly heating---i.e., in a ``heating thermal quench.'' The terminology \textit{inverse Mpemba effect} refers to the possibility of observing the ME in a heating quench protocol, by contrast to the previously discussed \textit{direct Mpemba effect}. The possibility of observing an inverse effect was first proposed in \cite{lu2017nonequilibrium} and later observed in an experimental colloidal particle setup~\cite{kumar2022inverse}; see Sec.~\ref{sec:colloids}.
\begin{figure}
    \centering
    \includegraphics[width=\columnwidth]{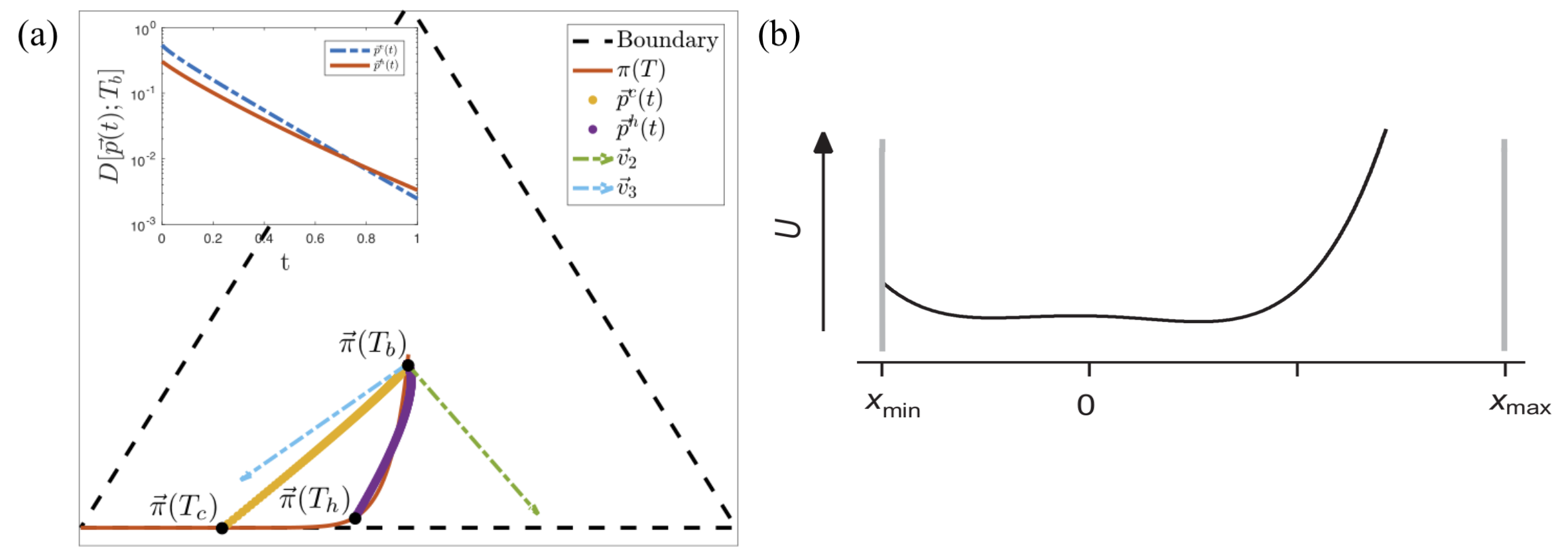} \caption{Inverse Markovian Mpemba effect in a three-state system. (a) Two identical systems, initiated at a cold temperature $T_c$ and a hot temperature $T_h$, are coupled to a hot bath with temperature $T_b > T_h > T_c$. The initially cold system has a very small component along the slow relaxation mode $\vec{v} _2$ (green dashed arrow). In contrast, the initially hot system has a much larger coefficient along the slow direction (even though it is at a smaller distance from the equilibrium). Hence, it decays more slowly than the initially cold system. Inset: Distance from equilibrium for both systems, as a function of time after quench.  Initially, the cold system decays more quickly, and the inverse Markovian Mpemba effect occurs.  Both distances ${D} [\vec{p}_h(t),\vec{\pi}(T_b)]$ and ${D} [\vec{p}_c(t),\vec{\pi}(T_b)]$ decrease with time. The initially colder system (blue dashed line) decays faster than the initially hotter one (red line), and the inverse Markovian Mpemba effect occurs. (b) Schematic of the asymmetric energy landscape $U(x)$ used to explore the inverse Mpemba effect, set asymmetrically within a one-dimensional box $[x_{\rm min}, x_{\rm max} ]$ with potential walls with large finite slope at the domain boundaries. \textit{Source:} Reprinted with permission from~\cite{lu2017nonequilibrium} and~\cite{kumar2020exponentially} respectively.}
    \label{fig:inverse}
\end{figure}

Let us begin with an example of a three-state system. We set the energy levels of the three states to $E_1=0$, $E_2=0.1$ and $E_3=1$, while the heights of the barriers separating the wells are $B_{12}=2$, $B_{13}=1.01$ and $B_{23}=10$. The dynamics of the system is regulated by a rate matrix $\mathbf{W}$, whose components have an Arrhenius form, as outlined in~\EQ{arrhenius}. In contrast to the direct case, the inverse effect involves a heating protocol, meaning that the bath temperature is higher than both initial conditions. Setting, for example, the bath temperature at $T_b=10$, we need to find a pair of temperatures $T_c<T_h<T_b=10$ 
for which one can
observe a finite-time crossing of a distance measure between the probabilities $\vec{p}_{h,c}(t)$
and the target Boltzmann distribution $\vec{\pi}(T_b)$.

Analyzing the spectrum of the evolution operator $\mathbf{W}(T_b=10)$, one finds that the second eigenvalue $\lambda_2$ is strictly greater than the third one $\lambda_3$, so that a sufficient condition for the existence of the inverse effect is to find $T_c<T_h<T_b=10$ such that $|a_2(T_h,T_b)|>|a_2(T_c,T_b)|$.  This condition ensures the existence of a crossing time $t^*$ such that, for $t>t^*$,
\begin{align}
    {D} [\vec{p}_h(t),\vec{\pi}(T_b) ]>{D} [\vec{p}_{c}(t),\vec{\pi}(T_b) ]\,,
\end{align}
along the relaxation process. Choosing, e.g., $T_c=0.3$ and $T_h=0.78$ provides the conditions for the projection on $\vec{v}^{\,(2)}$ (green arrow in Fig. \ref{fig:inverse}a) of the Boltzmann equilibrium of $\vec{\pi}(T_h)$ to be much larger than that of $\vec{\pi}(T_h)$. This causes $\vec{p}_{h}(t)$ to lag behind $\vec{p}_{\rm c}(t)$ in the long-time limit. We find a crossing of the distance function at $t^*\approx 0.77$, as shown in the inset to~\FIG{inverse}a.

The inverse effect can also be found in systems with continuous degrees of freedom, such as the paradigmatic model of a Brownian particle diffusing in a 1D potential $U(x)$. This case was also experimentally observed in~\cite{kumar2020exponentially}, for an overdamped Brownian colloidal particle moving in the asymmetric double-well potential depicted in~\FIG{inverse}b. Such a scenario provides an intuition for the mechanism that regulates the dynamics: having a metastable state that is progressively more populated by an equilibrium probability density with increasing temperature can hinder the redistribution of the probability towards a uniform distribution (expected at very high temperatures) when there is a pronounced asymmetry in the landscape; see also~\FIG{brownian_system}. This example also shows that the inverse Mpemba effect can be observed in a setup in which metastability does not play a dominant role, as the dynamics of the system could be modeled accurately as subject to a quench to an infinite-temperature bath: here, the potential defines the low-temperature probability densities but does not play any role after the quench.

\subsubsection{The strong Mpemba effect}
\label{sec:strong}
The standard definition of the Mpemba effect involves a relaxation towards an equilibrium at some temperature $T_b$. We already distinguished between direct and inverse effects, depending on whether the quenching protocol involves cooling or heating.  But within this classification, there is another important distinction to be made, depending on \textit{how quickly} the system approaches the target equilibrium.  Below, we shall show that not only can a hot system relax more quickly than an initially colder one, but it can actually show an \emph{exponential} speedup in the relaxation through a jump in the relaxation rate.  This qualitatively new phenomenon is referred to as the \emph{strong Mpemba effect}. By contrast, the \emph{weak Mpemba effect} describes a relaxation that, although faster, nevertheless has the same exponential rate, $e^{\lambda_2 t}$.  (This is the type of Mpemba effect that we have discussed so far.)
 
\paragraph{Characterization of the strong Mpemba effect}
\begin{figure}
    \centering
    \includegraphics[width=0.8\columnwidth]{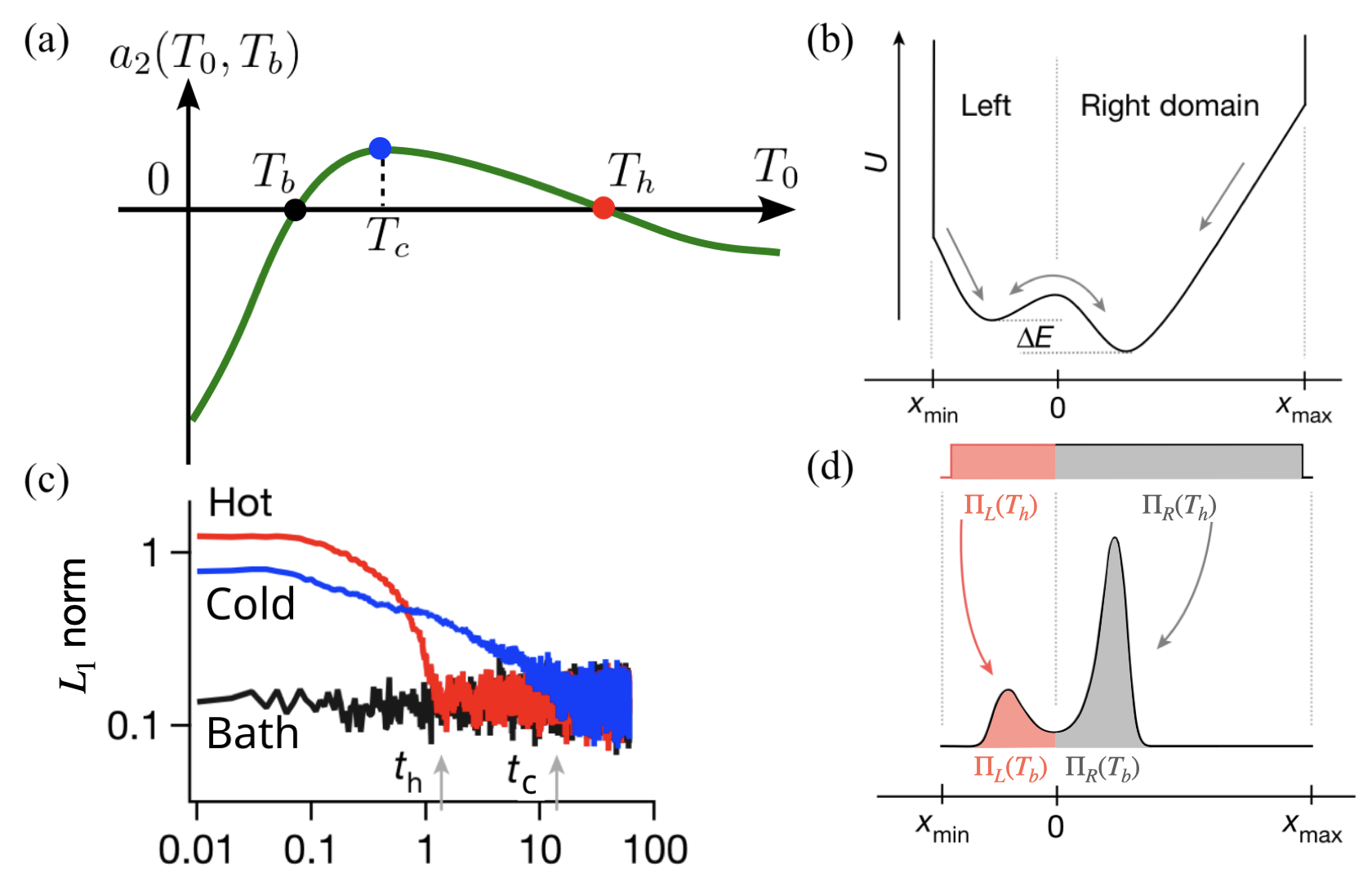}
    \caption{(a) Strong direct Mpemba effect, illustrated with a sketch highlighting the existence of a $T_0\neq T_b$ such that $a_2(T_0,T_b)=0$, which allows for exponentially faster cooling.
    (b) 1D potential for a Brownian colloidal particle that allows the observation of a manifestation of the strong Mpemba effect (see \cite{kumar2020exponentially}), in which the distance measure ${D}$ exhibits an exponentially faster relaxation, as shown in panel (c).
    (d) The Boltzmann distribution in the well at infinite and bath temperature $T_b$, where the red (gray) area highlights how the total probability is partitioned between the two wells for both temperatures. The Mpemba effect is the strongest when the two red (gray) areas match for the initial (here sketched at $T_h \to \infty$) and the final (bath) temperature $T_b$: $\Pi_L(T_h)\approx \Pi_L(T_b)$. \textit{Source:} Panels (b-d) are reprinted with permission from~\cite{kumar2020exponentially}.}
    \label{fig:strong_sketch.png}
\end{figure}

Recall that the evolution operator $\mathcal{L}_{T_b}$ describes relaxation towards a Boltzmann equilibrium at $T_b$.  In terms of its real eigenvalues $\lambda_i(T_b)$ and right eigenfunctions $v_i(\vec{x},T_b)$, we can express the evolution of the probability density $p$ of a system prepared at initial temperature $T_0$ throughout the relaxation process as
\begin{align}\label{eq:evolution_p}
    p(\vec{x},t)=\pi(\vec{x},T_b)+a_2(T_0,T_b)e^{\lambda_2(T_b)t}v_2(\vec{x},T_b)
    +a_3(T_0,T_b)e^{\lambda_3(T_b)t}v_3(\vec{x},T_b)+\dots,
\end{align}
where we ordered the eigenvalues according to their value, such that $0=\lambda_1>\lambda_2\geq\lambda_3\geq\dots$.
Here, $a_i$ are coefficients that retain information on the initial condition and, therefore, depend explicitly on $T_0$.  They are given by
\begin{align}
    \label{eq:ai-coefs-cont}
    a_i(T_0,T_b)=\frac{\int_\Omega \dd \vec{x}\ u_i(\vec{x},T_b) \pi(\vec{x},T_0)}
    {\int_\Omega \dd\vec{x}\ u_i(\vec{x},T_b) v_i(\vec{x},T_b)},
\end{align}
where $u_i(\vec{x},T_b)$ represent the left eigenfunctions of $\mathcal{L}_{T_b}$.
Now, if there exists some temperature $T_0\neq T_b$ such that
\begin{align}
    a_2(T_0,T_b)=0 ,
\end{align}
then, since $\lambda_2>\lambda_3$, the relaxation process of~\EQ{evolution_p} reduces to
\begin{align}\label{eq:evolution_p_no_a2}
    p(\vec{x},t)=\pi(\vec{x},T_b)  +a_3(T_0,T_b)e^{\lambda_3(T_b)t}v_3(\vec{x},T_b)+\dots,
\end{align}
implying an \textit{exponentially} faster convergence towards $\pi(\vec{x},T_b)$, with a corresponding rate given by $\lambda_3$ rather than by $\lambda_2$.
The $a_2$ temperature dependence is exemplified in a sketch in Fig. \ref{fig:strong_sketch.png}a, which illustrates the strong, direct Mpemba effect scenario. In this example, there exists some $T_b<T_c<T_h$ such that $a_2(T_h,T_b)\equiv 0$.  {Note that $a_2(T_0=T_b,T_b)=0$, as well, since a system initially at equilibrium with the bath at temperature $T_b$ obeys $p(\vec{x},t)=\pi(\vec{x},T_b)$ exactly, with all $a_{i > 1} = 0$.}
An analogous picture applies for a strong inverse effect for a heating protocol.
The existence of a strong Mpemba effect implies the existence of the weak effect too, as $a_2(T_0,T_b)$ is a continuous function of $T_0$.  
{Thus, if $a_2$ has two isolated zeroes in the strong-Mpemba-effect scenario, it must be a nonmonotonic function of initial temperature, implying the existence of the weak Mpemba effect in between the two zeros of $a_2$.}

Figure~\ref{fig:strong_sketch.png} shows an experimental observation of the strong direct Mpemba effect in a colloidal setup, where a Brownian particle is diffusing in the 1D potential depicted in~\FIG{strong_sketch.png}b. The $L_1$ distance measure for this system is observed to relax from an initial hot state (represented by the red line) and a cold state (represented by the blue line) towards an even colder bath temperature (indicated by the black line). Both the distance and time axes are shown on a logarithmic scale, which helps visualize the decay curve. The initially hot system exhibits a decay curve with a different slope compared to the relaxation from the cold initial condition, highlighting the exponentially faster cooling phenomenon~\cite{kumar2020exponentially}. Panel (d) provides a sketch of the experimental findings~\cite{kumar2020exponentially}, which are theoretically confirmed in Ref.~\cite{walker2022mpemba}. It shows that the Mpemba effect is strongest when the equilibrium probability of the particle being in the left well at the bath temperature, $ \Pi_{L}(T_b)$, closely matches the probability of the particle being in the left well at the initial hot temperature, $\Pi_L(T_h)$. This condition enables the system to attain global equilibrium by locally adjusting the probability distribution within each well rather than depending on transitions over the barrier. It is important to note that transitions between the wells are infrequent because of the small diffusion coefficient and the large, wide barrier. 

\paragraph{Parity index and geometry of the strong effect}

\begin{figure}
    \centering
    \includegraphics[width=0.45\columnwidth]{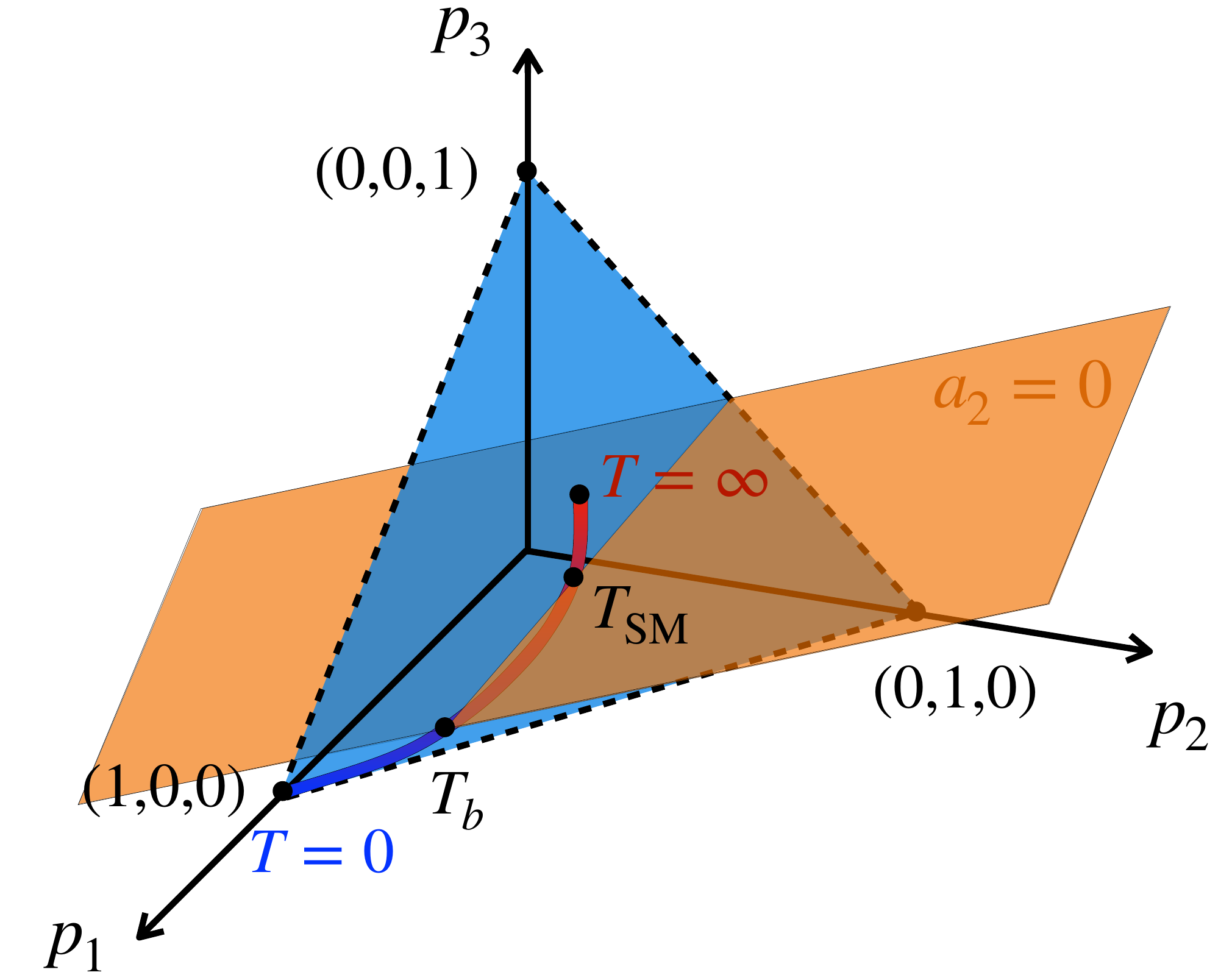}
    \caption{Geometry of the strong Mpemba effect in a three-state system.
    The probability simplex is illustrated by the blue triangle.
    The set of all equilibrium distributions forms the gradient-colored curve. 
    The blue and red points on the curve correspond to $T=0$ and $T=\infty$.
    The Mpemba index is nonzero if the equilibrium curve crosses the $a_2=0$ plane (orange) at the bath temperature $T_b$ and at some other temperature, here labeled as $T_{\rm SM}$.
    }
    \label{fig:strong_effect_geometry}
\end{figure}

The strong Mpemba effect is associated with the existence of temperatures $T\neq T_b$ for which the coefficient $a_2(T_0,T_b)$ becomes identically zero.
Since $a_2$ is a continuous function of the initial temperature, $T_0$, the total number of zeros can provide additional sufficient criteria to characterize the Mpemba effect.
We introduce, therefore, the following Mpemba indices
\begin{align} \nonumber
    &\mathcal{I}_M^{\textrm{dir}}\equiv \textrm{\# of zeros of } a_2(T_0,T_b)\textrm{ in } T_b<T_0<\infty \\  
    &\mathcal{I}_M^{\textrm{inv}}\equiv \textrm{\# of zeros of } a_2(T_0,T_b) \textrm{ in } 0<T_0<T_b
\end{align}
and the total index as
\begin{align}
    \mathcal{I}_M=\mathcal{I}_M^{\textrm{dir}}+\mathcal{I}_M^{\textrm{inv}} .
\end{align}

As a simple example of these ideas, we return to the discrete three-state system introduced in Fig. \ref{fig:3_state_system}, which can be considered an effective representation of an overdamped particle in a three-well potential.
The system can be described by a normalized probability vector $\vec{p}=(p_1,p_2,p_3)$, with $\sum_i p_i=1$ and $p_i\geq0$.  States lie on the probability simplex highlighted with a blue triangle in Fig.~\ref{fig:strong_effect_geometry}, where the equilibrium locus, representing all the Boltzmann equilibrium vectors $\vec{\pi}(T)$ for $T\in[0,\infty]$ is highlighted as the blue to red curve lying on the probability simplex.
Each component is related to the energies of the states by a Boltzmann factor $\pi_i (T)\propto e^{-E_i/k_B T}$.
The locus is representative (without loss of generality) of a system with the lowest energy configuration in State 1. For $T=0$, it approaches the probability vector $(1,0,0)$ (blue side of the curve), while in the limit of infinite temperature the locus approaches the equipartition probability vector $(\sfrac{1}{3},\sfrac{1}{3},\sfrac{1}{3})$ lying at the center of the simplex (red side of the curve).
For a given bath temperature $T_b$ (black point marked by $T_b$ on the simplex), the set of all points $\vec{x}$ such that $\vec{x}\cdot \vec{u}_2(T_b)=0$ is highlighted with an orange surface.  Its intersection with the simplex provides all the probability vectors $\vec{p}$ such that $\vec{p}\cdot \vec{u}_2(T_b)=0$. While the equilibrium locus intersects the $a_2=0$ plane at $\vec{\pi}(T_b)$ by definition, the set of $\vec{p}$ resulting from the intersection with the simplex generally represents an out-of-equilibrium system.
However, in the specific example highlighted in \FIG{strong_effect_geometry}, this line intersects the equilibrium locus at some temperature $\vec{\pi}(T_M)$ with $T_M\neq T_b$ (black dot marked by $T_{\rm{SM}}$) since the two extremes of the equilibrium locus (at $T=0$ and $T=\infty$) lie on the same side of the $a_2=0$ hyperplane: this provides a case for the strong Mpemba effect.

Because the hyperplane $a_2=0$ has codimension one, it separates the probability simplex into two disjoint sets. Therefore, assuming that the equilibrium locus is a continuous function of the temperature, namely that there are no phase transitions in the system, the parity of the number of times the equilibrium locus crosses $a_2=0$ depends solely on the position of its boundaries at $T=0$ and $T=\infty$. Given a certain $T_b$, a sufficient condition for the strong Mpemba effect is that  $\textrm{sign} [a_2(T_0=0,T_b)]=\textrm{sign} [a_2(T_0=+\infty,T_b)]$. With this motivation, we can define the parity of $\mathcal{I}_M$ in terms of the Heaviside step function $\theta$ as
\begin{align}
    \mathcal{P}(\mathcal{I}_M)=\theta\left( a_2(T_0=0,T_b) a_2(T_0=\infty,T_b) \right) ,
\end{align}
where we note that the crossing at $T_0=T_b$ does not contribute to the Mpemba index so that an even number of crossings corresponds to an odd parity of $\mathcal{I}_M$. A distinction between direct and inverse effects can be made if one knows the derivative $\partial_{T_0}a(T_0=T_b,T_b)$, which determines the change of sign of $a_2$ at the bath temperature.
We can use it to determine the parity of $\mathcal{I}_M^{\textrm{inv}}$ as
\begin{align}
\label{eq:Parity-inv}
    \mathcal{P}\left(\mathcal{I}_M^{\textrm{inv}}\right)=
    \theta \left(\left. \partial_{T_0}a(T_0,T_b)\right|_{T_0 = T_b} a_2(T_0=0,T_b)\right) .
\end{align}
Similarly, we can evaluate the parity of $\mathcal{I}_M^{\textrm{dir}}$ as
\begin{align}
\label{eq:Parity-dir}
    \mathcal{P}\left(\mathcal{I}_M^{\textrm{dir}}\right)=
    \theta \left( -\left.\partial_{T_0}a(T_0,T_b)\right.|_{T_0 = T_b} a_2(T_0=\infty,T_b)\right) .
\end{align}
Mathematically, a situation in which $a_2(T_0,T_b)=0$ without a sign change (i.e., $\partial_{T_0}a_2(T_0,T_b)=0$) can occur.
{This scenario differs from the parity characterization outlined above but requires fine-tuning of the system parameters: a small perturbation would generally destroy the needed degeneracy.}
Nevertheless, while in these cases, $\mathcal{P}(\mathcal{I}_M)$ does not correspond to the exact parity, it still serves as a lower bound to the number of crossings.
We note that a discontinuity in $a_2(T_0,T_b)$ with respect to a change in the bath temperature can appear 
{if the second and third eigenvalues become equal at some $T_b$ (i.e., if $\lambda_2=\lambda_3$).}
This case is discussed in Sec.~\ref{sec:eigenvalue-crossing} and, in fact, generates the transition into the strong inverse effect in the Ising antiferromagnet at low values of the magnetic field \cite{klich2019mpemba,teza2023relaxation}.

In Sec.~\ref{SubSec:NumericalMonteCarlo}, it is shown that the strong effect turned out to be extremely useful in identifying the Mpemba effect in numerical simulations of large systems, where the number of microstates is so large that a reasonable sampling of the evolving distribution is impossible. In such cases, Monte-Carlo simulations can be used to identify the strong Mpemba (direct and inverse) effects, but no similar method is known to identify the weak effects.  

Finally, a recent example where $a_2=0$ at two different temperatures greater than the bath $T_b$ (odd parity index) was given in Ref.~\cite{malhotra2024double}.  The authors describe the behavior as a ``double Mpemba effect" because the cooling time shows two local minima (where there is a strong Mpemba effect) as the temperature is increased from $T_b$.  The theory setup corresponds closely to the experimental setup reproduced in Fig.~\ref{fig:strong_sketch.png} from Ref.~\cite{kumar2020exponentially}.

\subsubsection{Mpemba effect in boundary and weak coupling setups} \label{sec:boundary_coupling}
\noindent The theoretical models exhibiting the Mpemba effect discussed so far assumed that all relevant degrees of freedom (DoF) are directly coupled to the thermal bath.
However, in most experiments demonstrating the Mpemba effect---the colloidal system \cite{kumar2020exponentially,kumar2022inverse} being the only exception as it has a single DoF---the system is coupled to the heat bath only through its boundaries, so that the heat is exchanged through a limited portion of the system (e.g., its surface). One might then worry that instances of the Mpemba effect studied so far  \emph{require} that all the DoF be fully coupled to the thermal bath and that the models of the Mpemba effect discussed above may not be relevant to the observations of anomalous relaxation in other systems, which are spatially extended.
Intuitively, one would expect the effect to be more visible as the coupling with the thermal bath strengthens, while a weak coupling would make it harder to observe, if not impossible.
A Markovian framework for a thermal quench through boundary coupling was developed in Ref. \cite{teza2023relaxation} and used to study the influence of the coupling strength on the presence of a Mpemba effect.

Let us consider a system that is coupled to a thermal bath only through its boundaries. In other words, heat can be transferred between the system and the bath only through those DoF sitting on the boundaries. All other transitions between microstates are ``bulk transitions," and no energy is exchanged with the bath in those transitions. From conservation of total energy, they can happen only between states with the same energy.
These bulk transitions serve as a \emph{self-ergodizing} (SE) mechanism, whereas the \emph{boundary transitions} (BT) couple the system to the bath and enable transitions between different energy shells.
This structure can be modeled by
\begin{align}\label{eq:full_rate_matrix}
    \mathbf{W}\left(\Gamma^{\rm SE},\Gamma^{\rm BC}\right)=\Gamma^{\rm SE}\mathbf{W}^{\rm SE}+\Gamma^{\rm BC}\mathbf{W}^{\rm BC}.
\end{align}
Here $\mathbf{W}^{\rm SE}$ and $\mathbf{W}^{\rm BC}$ are normalized rate matrices corresponding to the self-ergodizing and boundary coupling transitions, respectively, and $\Gamma^{\{\rm SE,\rm BC\}}$ are coupling constants modulating the rates amplitude. The ratio  $C=\Gamma^{\rm BC}/\Gamma^{\rm SE}$ determines the coupling strength:
in the limit $C\ll 1$, boundary flips---in which energy can be exchanged with the bath---occur rarely compared to ergodization flips, and the system ergodizes quickly after each energy exchange with the bath. In the $C\gg1$ limit, the boundaries exchange heat much faster than the ergodization rate. Thus, the boundaries are in thermal equilibrium with the bath, and the diffusion of energy into the system sets the timescale for the relaxation.
The specific normalization chosen for $\mathbf{W}^{\{\rm BC,\rm SE\}}$ changes the value of $C$, but not its limiting cases.
As an example, one can choose to normalize with respect to the maximum rate so that $\max\left(W_{i\!j}^{\rm SE}\right)=\max\left(W_{i\!j}^{\rm BC}\right)\equiv 1$.

By construction, $\mathbf{W}^{\rm SE}$ is a rate matrix that contains only transitions between states that have the same energy.
Generically, it is a reducible matrix with a zero-eigenvalue degeneracy equal to the number of different energy macrostates.
As for $\mathbf{W}^{\rm BC}$, it is generically a sparse matrix, as most transitions involve more than the boundary spins.
The interaction of the system with a thermal bath at temperature $T_b$ satisfies detailed balance; therefore, $W^{\rm BC}_{ji} = W^{\rm BC}_{i\!j}e^{-(E_j-E_i)/k_B T_b}$, where the $E_{\{i,j\}}$ are the energies corresponding to the two states.

A naive perturbation scheme with $C\ll1$ would not prove useful: for $C=0$, the matrix $\mathbf{W}$ is reducible, and its zero eigenvalue is highly degenerate, meaning that one cannot apply the standard analysis.
Instead, it is instructive in this case to aggregate all the microstates that share the same energy into a single \emph{macrostate} and construct the effective dynamics by summing all the microscopic transitions between them \cite{teza2020exact}. 
The dynamics is then dictated only by the boundary flips, and the diffusion within each energy shell is assumed to happen instantaneously. 
Similarly, for $C\gg 1$, microstates can be aggregated into macrostates by combining all the microstates connected by boundary flips. Mathematically, the two aggregation procedures can be done by arranging the states such that $\mathbf{W}^{\rm BC}$ or $\mathbf{W}^{\rm SE}$ is block diagonal where each block corresponds to transitions within a macrostate and coarsening over these blocks.

\paragraph{Example: Ising antiferromagnet with boundary coupling}

\begin{figure}
    \centering
    \includegraphics[width=0.8\columnwidth]{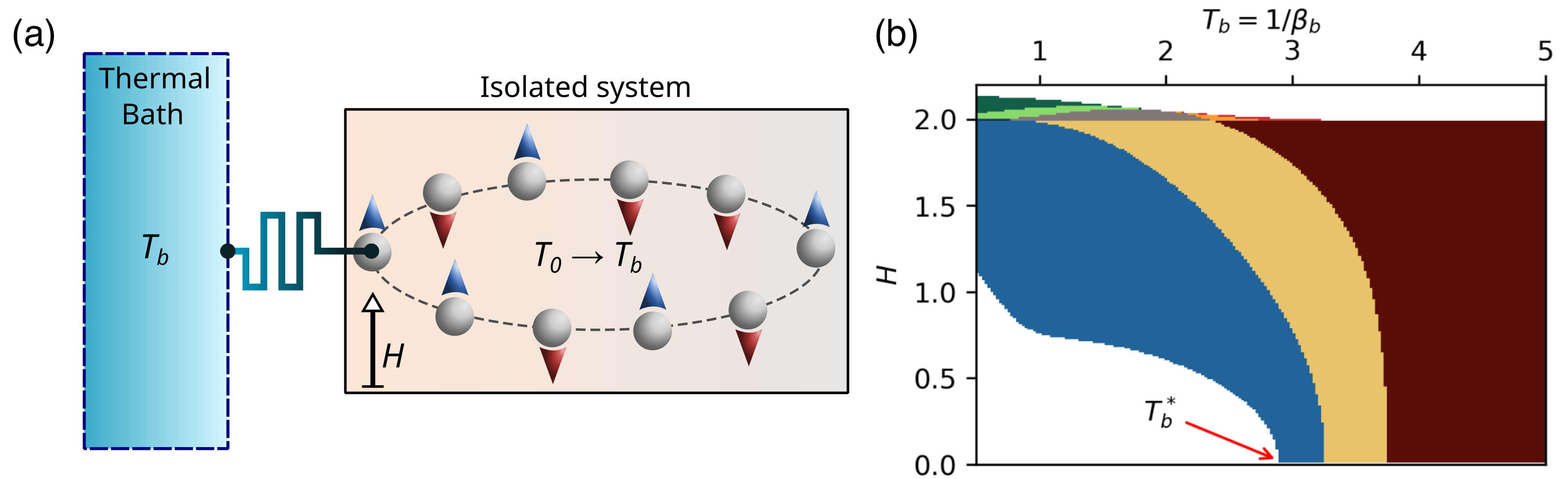}    
    \caption{(a) Antiferromagnet Ising chain with a single spin coupled to the thermal bath.
    Heat is exchanged with the bath only through the single spin connected to it.
    (b) Mpemba phase diagram of a partially coupled Ising antiferromagnet (with $N=10$ and coupling strength $C=1$).
    The white area corresponds to a system not exhibiting any Mpemba effect, while different colors are associated with a different flavor of the effect; see Fig.~\ref{fig:ising_1DAF_phase_diag} for the specific details of each Mpemba effect.
    The horizontal line at $H=2$ is related to the phase transition of the system at zero temperature. 
    }
    \label{fig:ising_1dAF_boundary_sketch}
\end{figure}
As sketched in \FIG{ising_1dAF_boundary_sketch}a, we illustrate the above construction using a one-dimensional ring of $N$ Ising spins with nearest-neighbor antiferromagnetic interactions.
Each spin $\{\sigma_s\}_{s=1\dots N}$ in the chain can either be in an up or down ($\pm1$) state, giving a total of $M=2^N$ different microstates, identified by the $N$-dimensional vector $\vec{\sigma}$. The energy of a microstate $\vec{\sigma}$ is provided by the Hamiltonian function
\begin{equation}\label{eq:ham_AF}
    \mathcal{H}(\vec{\sigma})= - J\sum_{ s } \sigma_{s} \sigma_{s+1} -H\sum_{s=1}^{N} \sigma_{s},
\end{equation}
where $J<0$ is the coupling constant, $H$ is an external magnetic field and $\sigma_{N+1}\equiv \sigma_1$. For simplicity, we set $J=-1$.

To model boundary coupling in this system, we couple a specific spin (say $\sigma_1$) to the bath. A general microstate $\vec\sigma$ is then connected through thermal flips only with a single state $\vec\sigma'$ in which the first spin is flipped, $\sigma_{1}\to -\sigma_{1}$, while the remaining spins are unaltered. For two general microstates $\vec \sigma^i$ and $\vec \sigma^j$, the transition is therefore
\begin{align}
\label{eq:W_BC}
    W^{\rm BC}_{i\!j}= \frac{\delta_{\sigma_{1}^i,-\sigma_{1}^j}\prod_{s>1}\delta_{\sigma_{s}^{i},\sigma_{s}^j}}{1+e^{\beta_b \left(\mathcal{H}\left(\vec{\sigma}^i\right)-\mathcal{H}\left(\vec{\sigma}^j\right)\right)}},
\end{align}
where $\delta_{i\!j}$ is the Kronecker delta, $\sigma_s^i$ is the $s$ spin in the microstate $\vec{\sigma}^i$, and we used standard Glauber dynamics as the transition weight \cite{glauber1963time,felderhof1971spin}, ensuring that the equilibrium distribution is the Boltzmann distribution.

\begin{figure}
    \centering
    \includegraphics[width=0.9\columnwidth]{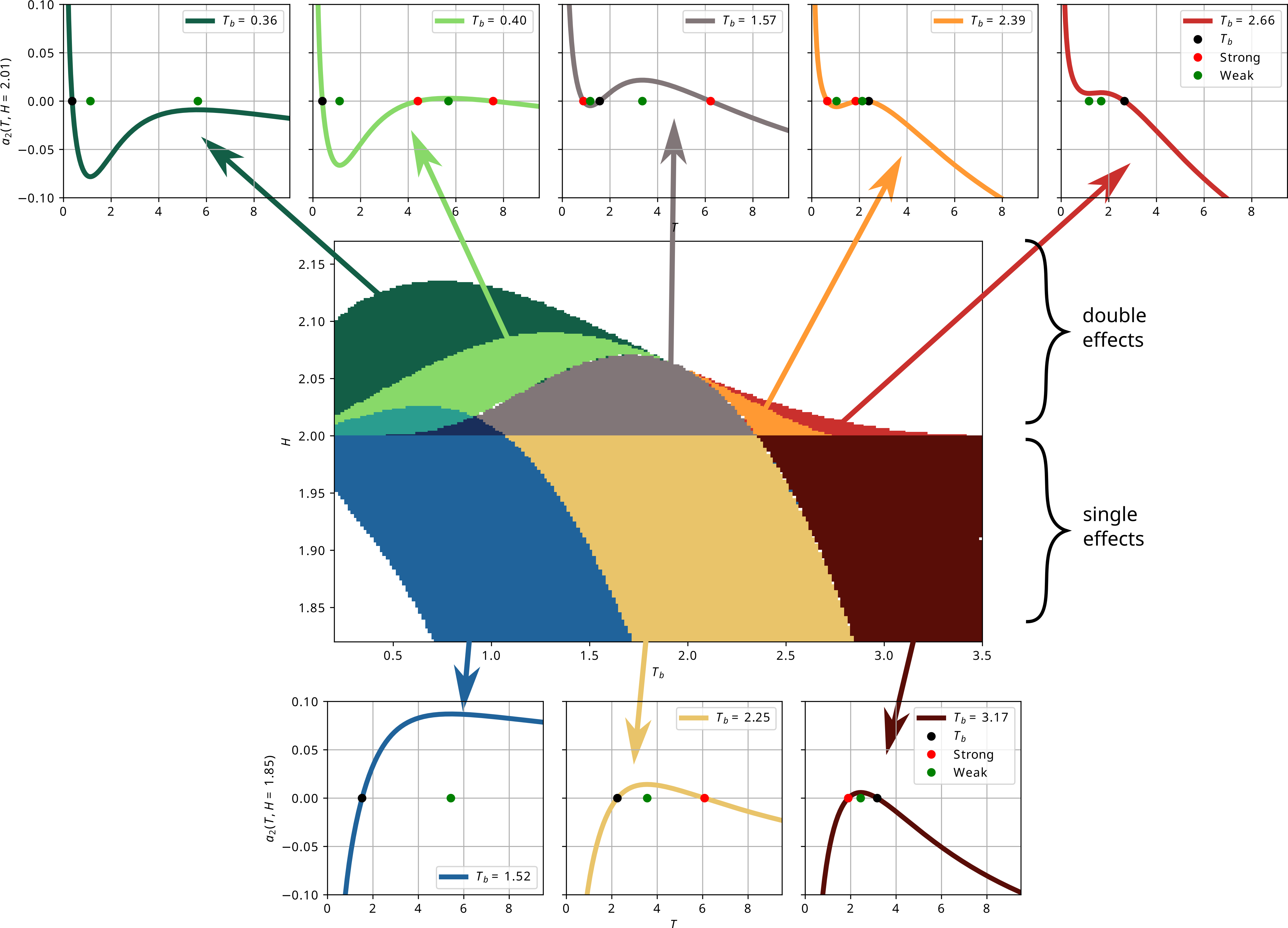}
    \caption{Different flavors of Mpemba effect in an antiferromagnetic spin chain subject to boundary coupling.
    This plot offers a zoom of the phase diagram shown in Fig.~\ref{fig:ising_1dAF_boundary_sketch}b to better highlight where the different effects occur.
    The horizontal line at $H=2$ is associated with the zero-temperature phase transition of the 1D Ising ferromagnet.
    Below this line, there are only single Mpemba effects: a weak direct (blue), a strong direct (yellow) and a strong inverse (burgundy). Above, double effects lead to a much richer variety, as highlighted by all the insets, which provide exemplars of the different cases of $a_2(T_0,T_b)$ curves for each colored area.
    Fixing the external magnetic field $H=2$ highlights how $a_2$ changes shape smoothly with respect to the bath temperature $T_b$, yielding different varieties of the Mpemba effect.
    Here, black markers highlight the bath temperature $T_b$ (where $a_2\equiv 0$ is implied); red markers indicate additional zeros of $a_2$, implying a strong direct effect if $T_0>T_b$ or a strong inverse effect if $T_0<T_b$; and green markers denote extrema points of $a_2$, implying the possibility to observe a weak effect. 
    }
    \label{fig:ising_1DAF_phase_diag}
\end{figure}

To model bulk transitions, we use rates that decay exponentially as $2^{-d_{i\!j}}$, where $d_{i\!j}=\sum_s \delta_{\sigma_{s}^i,-\sigma_{s}^j}$ is the Hamming distance \cite{hamming1950error} that counts the number of spins that must be flipped between the two configurations, as suggested by decimation-like procedures operated on Markov jump processes \cite{teza2020exact}. We therefore formalize bulk transitions between two states $i\neq j$ as
\begin{align}
\label{eq:W_ST}
    W^{\rm SE}_{i\!j}= \delta_{\mathcal{H}\left(\vec{\sigma}^{\,i}\right),\mathcal{H}\left(\vec{\sigma}^{j}\right)}2^{-d_{i\!j}}.
\end{align}
The full transition matrix for the model is finally built as a linear combination of the two rate matrices, as in~\EQ{full_rate_matrix}. With this construction, let us consider the persistence of the Mpemba effect in a boundary-coupling setup. In~\FIG{ising_1dAF_boundary_sketch}b, we plot the Mpemba phase diagram for $N=10$ (implying $M=1024$ microstates) as a function of bath temperature $\beta_b^{-1}$ and external magnetic field $H$.
The resemblance with that of the model studied in~\cite{klich2019mpemba} is striking, especially considering the profound dissimilarities between the two models: the all-to-all connectivity of a mean-field description makes it relevant for high-dimensional setups, while the model presented here addresses a one-dimensional spin chain.
Additionally, in this model, we imposed a partial coupling with the thermal bath, while in the standard mean-field Ising antiferromagnet, every spin exchanges heat with the thermal bath.
The similarity is not limited only to the examples where the Mpemba effect can be observed but extends to the many specific flavors of the Mpemba effect; see \FIG{ising_1DAF_phase_diag} for a detailed analysis.
In general, we observe both direct/inverse, weak/strong and single/multiple effects, which are encoded by the different colors in the figure: here, boundary coupling merely shifts phase-diagram borders, and all flavors of the Mpemba effect can be observed.
This hints that the antiferromagnetic character lies at the origin of the Mpemba effect, as it is the only feature shared among these very different models.
The horizontal line at $H=2$ is determined by the phase transition that characterizes the system at $T_b=0$: for $0<H\leq2$ (alternating spins ground state), only single Mpemba effects can be observed, while for $H>2$ (aligned spins ground state) double effects occur.
For $0<H\leq2$, a strong inverse Mpemba effect effect exists for $T_b \ge T_b^*=1/\beta^*$ (highlighted with a red arrow).
The reason that no Mpemba effect exists for $T < T_b^*$, for small external magnetic fields, traces back to a phase transition of the relaxation dynamics~\cite{teza2023eigenvalue}; see the discussion of applications in Sec.~\ref{sec:applications}.

Additional analysis that demonstrates the Mpemba effect in this setup can be obtained by projecting the relaxation paths in the (high-dimensional) probability space onto two order parameters, the average $\langle m \rangle$ and staggered magnetization $\langle m_s \rangle$ \cite{gal2020precooling}.
This projection method, explained in Sec. \ref{SubSec:NumericalMonteCarlo}, allows Monte-Carlo simulation in large systems where a direct calculation of $a_2$ is impractical.  We can nonetheless compare the relaxation trajectory with the equilibrium distributions at different temperatures; see Sec.~\ref{SubSec:NumericalMonteCarlo} for an example in a 2D Ising antiferromagnet.

\begin{figure}
    \centering
    \includegraphics[width=0.80\linewidth]{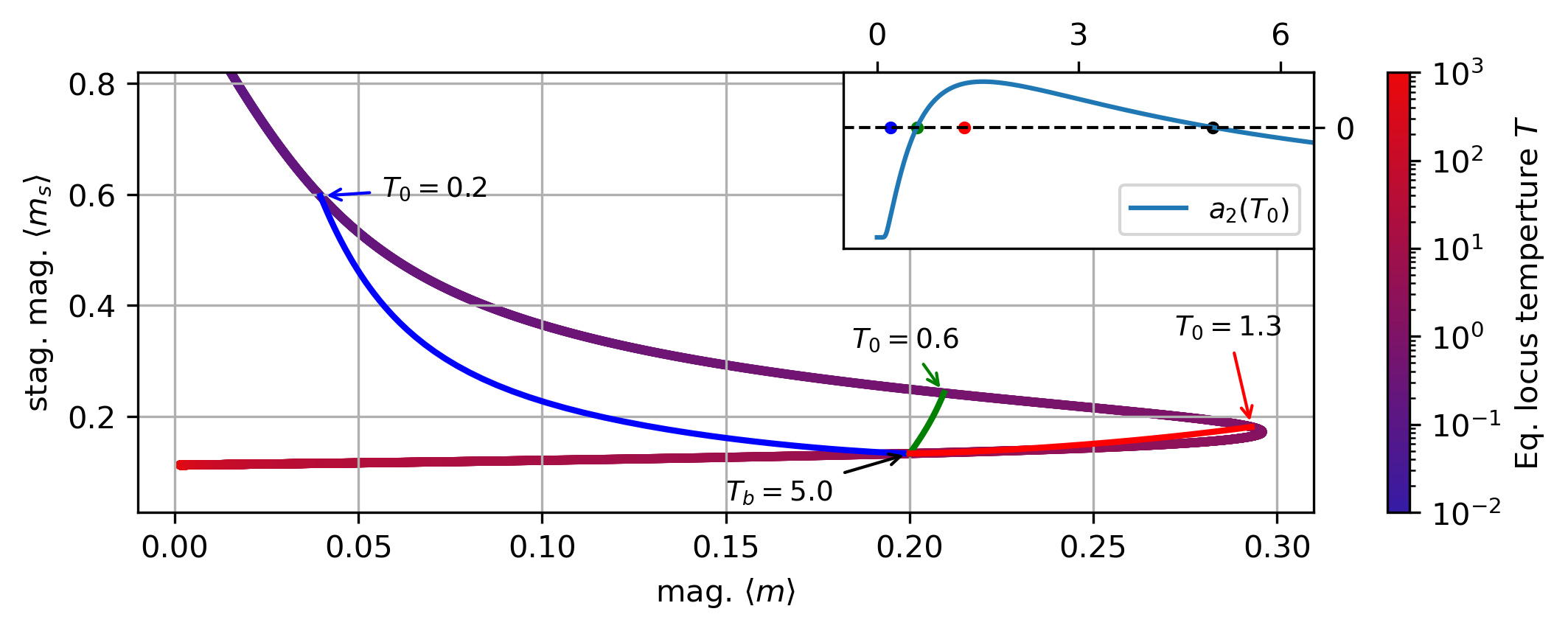}
    \caption{Relaxation paths for a weakly coupled system ($N=50$) on the $\left< m \right>$-$\left< m_s \right>$ plane ($H=\sqrt{2}$). Initial temperatures are chosen before (blue), at (green) and after (red) the zero of $a_2(T_0)$ as shown in the inset.
    Red and green paths approach the equilibrium locus along the same vector but from opposite directions, indicating the existence of a Mpemba effect. \textit{Source:} Adapted with permission from~\cite{teza2023relaxation}. 
    }
    \label{fig:relax_path}
\end{figure}

Figure~\ref{fig:relax_path} illustrates different relaxation paths in a weakly coupled chain of $N=50$ spins in comparison to the equilibrium line.
Initial temperatures are chosen before, at, and after the zero of $a_2$ at $T_0\sim0.6<T_b$. The hot (red) and cold (blue) relaxation paths approach the bath equilibrium at $T_b=5$  along the same direction, which is the projection of $\vec{v}_2$ into the observable space, but with opposite direction. This {corresponds to} a sign-change in $a_2$, which indeed vanishes at the blue trajectory implying nonmonotonicity of $a_2(T_0)$ (inset of Fig. \ref{fig:relax_path}) and consequently a Mpemba effect in the system.

\paragraph{Example: Colloidal setup with boundary coupling}
Even if all the relevant degrees of freedom are coupled to the thermal bath, as in the example of a Brownian particle diffusing in a confining potential discussed in Sec \ref{Sec:Brownian}, boundary coupling still might be an issue if the fluid in which the particle diffuses is heated from the boundary.
When the system is quenched to some temperature, the fluid does not change its local temperature instantaneously and uniformly.
Rather, its boundaries are coupled to a thermal bath, and the temperature profile changes according to some internal dynamics.
If this dynamics is much faster than the particle diffusion, the liquid reaches its uniform temperature before the distribution of the relevant DoF (i.e., the position) changes in any way.
This case coincides with the common assumption of instantaneous uniform quench in the temperature.
In the opposite limit, the equilibration temperature profile is much slower than the diffusion, and the position of the particle follows the steady-state distribution associated with the instantaneous temperature profile. 
In this limit, anomalous relaxations can exist only if they already exist in the local bath temperature profile, which is assumed not to be the case. This implies that a  ``critical minimal coupling'' is needed to observe anomalous relaxation.

\begin{figure}
    \centering
    \includegraphics[width=0.65\linewidth]{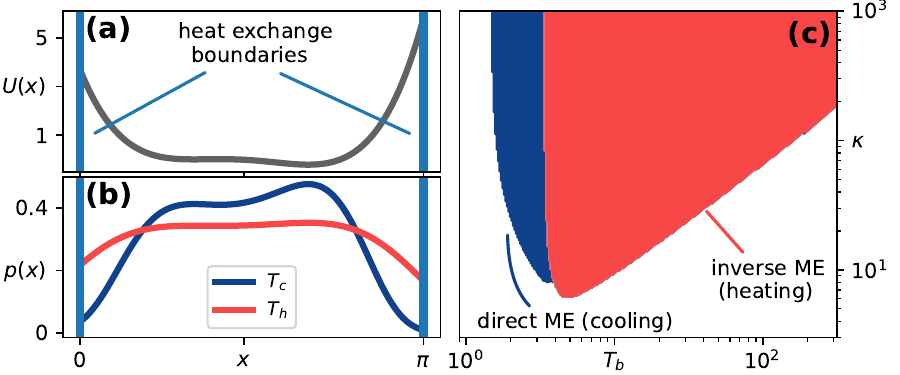}
    \caption{(a) Potential $U(x)$ of the colloidal system used in \cite{kumar2022inverse}. (b) Examples of Boltzmann distributions at cold and hot bath temperatures for which the direct (cooling) and inverse (heating) Mpemba effect exist. (c) Mpemba phase diagram as a function of the bath temperature $T_b$ and thermal diffusivity $\kappa$. \textit{Source:} Reprinted with permission from~\cite{teza2023relaxation}.
    }
    \label{fig:colloidal_boundary}
\end{figure}

To illustrate these ideas, we return to the Brownian particle in a potential used to demonstrate experimentally the inverse \cite{kumar2022inverse} and strong \cite{kumar2020exponentially} Mpemba effects, illustrated in Fig. \ref{fig:colloidal_boundary}a.
Let us consider a simplified 1D model of the setup in which the system is coupled to the environment only at its two boundary points.
We assume that the temperature profile of the fluid follows the heat equation $\partial_t T(x,t) = \kappa \partial_x^2 T(x,t)$, with thermal diffusivity $\kappa$. The initial temperature profile is spatially uniform, $T(x,t=0) = T_{0}$.

The associated Fokker-Planck equation  $\partial_t p(x,t)  = \mathcal{L}(t) p(x,t)$ has a time-dependent operator,
\begin{align} 
\label{eq:FokkerPlanck}
    \partial_t p(x,t) = \frac{1}{\gamma}\partial_x \Big[ \big(\partial_x U(x)\big) p(x,t) \Big] + \frac{k_B}{\gamma} \partial_x^2 \Big( T(x,t) p(x,t) \Big) ,
\end{align}
where $U(x)$ is the potential and $\gamma$ is the drag coefficient of the Brownian particle.
Although the Fokker-Plank operator is time-dependent, the long-time limit $T(x)\to T_b$ implies $\mathcal{L}(t)\to\mathcal{L}_{T_b}$ and ensures convergence to the Boltzmann equilibrium $\pi(x,T_b) \propto e^{-U(x)/k_B T_b}$. The eigenfunctions of $\mathcal{L}_{T_b}$, solving $\mathcal{L}_{T_b}v_i(x,T_b) = \lambda_i v_i(x,T_b)$ with $0=\lambda_1>\lambda_2\geq\lambda_3\geq\dots,$ form a complete basis; as a result,
\begin{align}
\label{eq:FP_sol}
    p(x,t)=\pi(x,T_b) +\sum_{i>1} a_i(T_0,T_b,t)e^{\lambda_i(T_b) t} v_i(x,T_b) \,,
\end{align}
where the $a_i(t)$ coefficients that dictate the current probability density also retain information about the initial conditions of the system.

In the limit of an instantaneous quench, $a_i$ are time-independent, and $a_2$ encodes the existence of the Mpemba effect \cite{lu2017nonequilibrium}: a nonmonotonic dependence in $T_0$ implies the existence of a relaxation shortcut when quenching the system to $T_b$, which can be exponentially faster if $a_2=0$ for some initial condition (a \emph{strong} Mpemba effect).
If the timescale of the quench is comparable to that of the diffusing particle, one cannot rely on this straightforward characterization, as $a_2$ explicitly depends on time.
In the last stages of relaxation, one can nevertheless approximate the difference from equilibrium $\Delta p(x,t) = p(x,t) - \pi(x,T_b) \simeq a_2(t)v_2(x)$ for some $a_2(t)$ decaying exponentially fast. Identifying a sign change in $a_2(t)$ is, therefore, enough to ensure the existence of a strong Mpemba effect.
Formally, this can be done through the Mpemba parity index (see \cite{klich2019mpemba} and Sec.~\ref{sec:strong}),
\begin{align}
    I^{\pm}(t,T_b)=\text{sgn}\int \dd x\ \Delta p^{\pm}(x,t) \Delta p^{\pm\delta T}(x,t),
\end{align}
where the differences $\Delta p^{\pm}$ refer to quenches to $T_b$ starting from initial temperatures $T_0=\{+\infty,0\}$, while $\Delta p^{\pm\delta T}$ to quenches starting from $T_0=T_b\pm\delta T$, for some $\delta T>0$.
A negative sign of $I^{+}$ ($I^{-}$) in the long-time limit implies that for some $T>T_b$ (or $T<T_b$) the coefficient $a_2\equiv 0$, ensuring the existence of a strong direct (or inverse) Mpemba effect.
For the given potential $U(x)$ shown in \FIG{colloidal_boundary}a, and for each set of $T_b$ and $\kappa$, it is possible to evaluate numerically $I^{\pm}$, as demonstrated in \FIG{colloidal_boundary}(c). As expected, for a given value of $T_b$, there exists a critical $\kappa$, below which the Mpemba effect cannot be observed.

\subsubsection{Metastability}
\label{sec:metastability}
Thermal relaxation in a system with metastability occurs in distinct stages. At low temperatures, a physical system is likely to remain ``trapped" in a metastable state for extended periods. A common example of a system with metastability is a particle diffusing over a potential energy landscape with deep wells. In this scenario, thermal relaxation proceeds in two stages. The initial stage is \emph{fast}, during which the particle falls into a local energy well. This is followed by a \emph{slow} equilibration process between the wells. A key characteristic of metastability is \emph{timescale separation}, which is indicated by a gap in the eigenvalue spectrum between $\lambda_2$ and the higher eigenvalues. The relaxation time, which is proportional to $1/\lambda_2$, signifies the escape time from metastable states. In the simplest case of a single energy well with a barrier that the particle must diffuse over, the Kramers' escape time is given by $1/\lambda_2$. Kramers was the first to analyze the problem of escape from a potential well~\cite{kramers1940brownian}. This problem is tractable for small diffusion coefficients and high barriers, and it has been extensively explored in the literature (see e.g.~\cite{chandrasekhar1943stochastic, wang1945theory, van1992stochastic, landauer1961frequency, langer1969statistical, risken1996fokker}).

\begin{figure}
    \centering
    \includegraphics[width=0.7\linewidth]{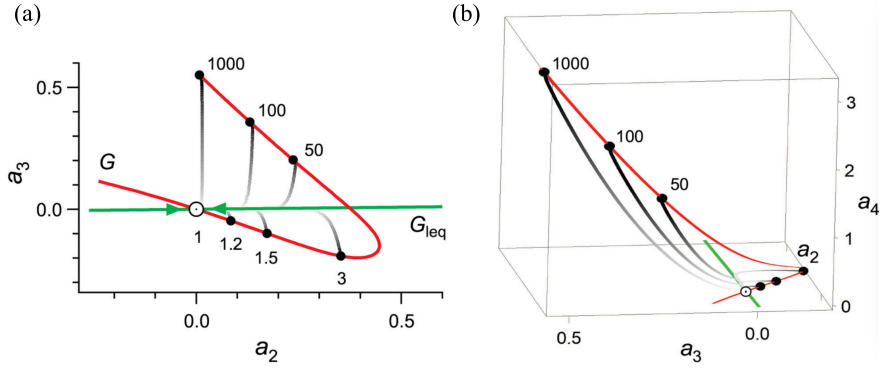}
    \caption{Probability density dynamics in the space of projection coefficients, for the double-well potential described in~\cite{kumar2020exponentially, chetrite2021metastable}. (a) Two-dimensional projection of the dynamics onto coefficients $a_2$ and $a_3$. Red line, labeled $G$, denotes the locus of \textit{global} equilibrium probability densities (equilibria); green line, labeled $G_{\rm leq}$, denotes the locus of \textit{local} equilibria. Gray lines denote individual trajectories starting from initial temperatures $T_0 = \{1000, 100, \ldots \}$, labeled by black markers, to the global equilibrium at $T_b = 1$ (large hollow marker with central dot). The time evolution of the trajectories is indicated by a color gradient from dark to light gray. Time along the local equilibrium locus (green line) is proportional to the relaxation time, as the initial stage is fast. The nonmonotonicity of initial temperatures with respect to length along $G_{\rm leq}$ indicates a Mpemba effect. (b) Three-dimensional projection onto the space of $a_2$-$a_3$-$a_4$ coefficients shows that the red and green curves intersect only at $T_0=T_b$. Both panels are based on data from experiments~\cite{chen2023memory}. \textit{Source:} Reprinted with permission from~\cite{chetrite2021metastable}.}
    \label{fig:fig-metastable-dynamics-Chetrite-Kumar-Bechhoefer.png}
\end{figure}

The Mpemba effect instances related to the overdamped diffusion of a particle in a double-well potential, characterized by deep and well-separated minima, were experimentally observed by Kumar and Bechhofer in their study \cite{kumar2020exponentially}. This phenomenon was further analyzed in collaboration with Ch\' etrite in \cite{chetrite2021metastable}. In the latter reference, the authors presented the dynamics of the probability density in the space of projection coefficients $\{a_i\}$, illustrating the two stages of relaxation. Figure~\ref{fig:fig-metastable-dynamics-Chetrite-Kumar-Bechhoefer.png} shows the projections of this dynamics in both the $a_2a_3$-plane and the $a_2a_3a_4$-space. The gray lines represent probability density trajectories that begin at the locus of global equilibria, denoted as $G$ (the red curve). These trajectories quickly descend to the locus of local equilibria, labeled $G_{\rm leq}$ (the green curve), and then slowly relax toward the global equilibrium at $T_b = 1$, indicated by the white hollow marker. The trajectories were constructed from experimental data, as outlined in detail in \cite{kumar2020exponentially}.

In the scenario considered, the initial stage of evolution occurs rapidly; it thus does not significantly contribute to the overall relaxation time. Consequently, the length of a trajectory on the local equilibrium locus (the green line) is directly proportional to the time required to reach global equilibrium. Therefore, the nonmonotonic behavior of the trajectory length along the local equilibrium locus concerning the initial temperature suggests the presence of a Mpemba effect. As shown in \FIG{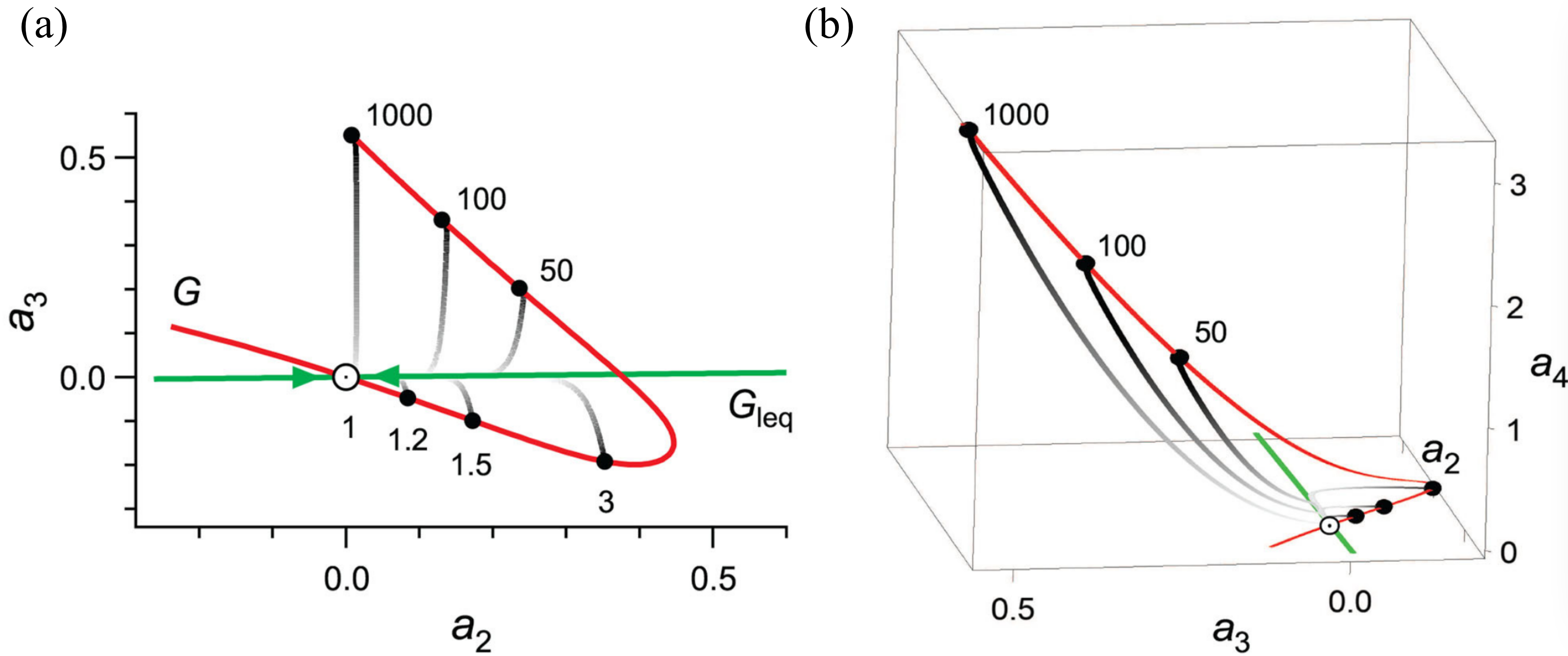}a, when moving away from $T_b$ toward higher temperatures, the length along $G_{\rm leq}$ initially increases and then decreases. Notably, for an initial temperature of $T = 1000$, the rapid portion of the trajectory approaches very close to the intersection of the local equilibrium manifold $G_{\rm leq}$ and the global equilibrium manifold $G$ (where $a_2=a_3=a_4=0$), which indicates that the system is close to a strong Mpemba effect. On the figure, a trajectory corresponding to the strong Mpemba effect would have no projection onto the local equilibrium locus, $G_{\rm leq}$.  

Ch\' etrite, Bechhoefer, and Kumar also provide a clear physical interpretation of the Mpemba effect in the context of a two-stage relaxation~\cite{chetrite2021metastable}. They demonstrate that the Mpemba effect in this system is linked to a nonmonotonic relationship between the initial temperature and the maximum amount of work that can be extracted from local equilibrium. According to the second law of thermodynamics for a system in contact with a single thermal bath at temperature $T_b$, the relationship between work and free energy states can be expressed as
\begin{align}
\label{eq:IIlaw-work}
    W \leq -\Delta F_{{\rm neq}}(T_b),
\end{align}
where $W$ represents the amount of work extracted from a nonequilibrium isothermal protocol, and $\Delta F_{{\rm neq}}$ denotes the difference in the nonequilibrium free energies between the final and initial states of the protocol~\cite{gavrilov2017direct}. In analogy with the equilibrium free energy, $F[\pi(T_b)] = - T_b \ln \left(Z(T_b)\right)$, the nonequilibrium free energy, $F_{{\rm neq}}[p,T_b]$, is also defined to have the \emph{Gibbs-Shannon form}
\begin{align}
    F_{{\rm neq}}[p,T_b] = E(p) - T_b S[p], 
\end{align}
with average energy $\langle E\rangle _p$ and \emph{Shannon entropy} $S(p)$ defined as 
\begin{align}
    \langle E \rangle _p = \int \dd{x} p(x,t) U(x)\quad \text{and} \quad 
    S[p] = -\int \dd{x} p(x,t)\ln p(x,t) = - \langle \ln p \rangle_ p.
\end{align}
The maximum amount of work that can be extracted from a nonequilibrium isothermal protocol is the equality case of~\EQ{IIlaw-work}, which the authors refer to as the \emph{extractable work}, 
\begin{align}
\label{eq:Wextractable}
    W_{\rm ex} \equiv - \Delta F_{{\rm neq}}(T_b). 
\end{align}
At the end of the first stage, the particle has quickly descended into a well. The \emph{local equilibrium probability density}, $\pi_{{\rm leq}}(x,T,T_b)$, locally resembles the (global) equilibrium density $\pi(x,{T_b})$, but with altered statistical weights in the left and right wells,  
\begin{align}
    \pi_{\rm leq}(x,T,T_b) = 
    \begin{cases} \Pi_L(T) \displaystyle\frac{\pi(x,T_b)}{\Pi_L(T_b)}, & x \in D_L 
    \\
    \\
    \Pi_R(T) \displaystyle \frac{\pi(x,T_b)}{\Pi_R(T_b)}, & x \in D_R 
    \end{cases}
\end{align}
where $D_{L}$, $D_{R}$ are the left and right well domains and $\Pi_{L}(T) \equiv \int _{D_L} \dd x \, \pi (x,T)$ is the probability to be in the left well, while $\Pi _R = 1 - \Pi _L$ is the probability to be in the right well. The \emph{extractable work} from the time that the local equilibrium is established (end of the initial stage) to the final relaxation to global equilibrium is, from~\EQ{Wextractable},
\begin{align}
    W_{\rm ex} = F[\pi(T_b)]- F_{{\rm neq},T_b} [\pi_{\rm leq}(T,T_b)] = T_b D_\mathrm{KL} [\pi_{\rm leq}(T,T_b),\pi(T_b)].
\end{align}
Thus, the extractable work between the local equilibrium state and the global equilibrium is proportional to the Kullback-Leibler divergence.  Recalling the definition of the equilibrium Mpemba effect as a nonmonotonicity of a distance function with respect to the initial temperature, we see that the metastable Mpemba effect occurs when the extractable work from the local equilibrium state is a nonmonotonic function of the initial temperature $T$. 

The experimental setup described by Kumar and Bechhoefer~\cite{kumar2020exponentially} involved a colloidal particle in water, held in place by optical tweezers. The optical tweezers incorporated a virtual feedback loop, allowing the particle to be positioned within a double-well potential that had two well-separated minima; see Sec.~\ref{sec:colloids}. The authors were the first to experimentally demonstrate the strong Mpemba effect, which had been theoretically predicted in~\cite{klich2019mpemba} (Sec.~\ref{sec:strong}). They also provided valuable insights into the conditions under which this effect is apparent in their systems. Given that the two minima of the potential are deep and well-separated, with a barrier in between, the authors argued that it is ``faster” to rearrange the statistical weight locally within each well separately rather than to overcome the barrier. Thus, Kumar and Bechhoefer proposed that the initial temperature leading to the strong Mpemba effect, denoted as $T_{\rm SM}$, should be the temperature at which the initial probability of the particle being in one well is equal to the final probability of being in the same well (equilibrium probability at the bath temperature $T_b$). This relation can be expressed mathematically as
\begin{align}
\label{eq:SMcond}
    \Pi_L(T_{\rm SM}) \approx \Pi _L (T_b).   
\end{align}
This situation is also illustrated in~\FIG{strong_sketch.png}d. Additionally, it is noted that after the first stage, the system is in global equilibrium, since $\pi_{\rm leq}(T_{\rm SM},T_b) = \pi(T_b)$. Consequently, the extractable work in going from the local equilibrium at $T_{\rm SM}$ to the global equilibrium is zero. Importantly, the double-well potential used in their result has a finite domain, defined by $x\in[x_\mathrm{min},x_\mathrm{max}]$. The asymmetry of the domain, attributed to the left and right wells, was controlled by the parameter $\alpha = |x_\mathrm{max}/x_\mathrm{min}|$. For specific asymmetric values of $\alpha$, they observed both weak and strong cases of the Mpemba effect.

A different approach for the same problem was taken by Walker and Vucelja~\cite{walker2022mpemba}. They showed that, in the limit of low diffusion and high energy barriers, \EQ{SMcond} holds. By solving the Kramers escape problem perturbatively~ \cite{kramers1940brownian, risken1996fokker, zwanzig2001nonequilibrium}, they found an approximate expression for the second eigenfunction, $\vec v_2(x,T_b)$. Using this eigenfunction, they derive the conditions for both weak and strong Mpemba effects. Interestingly, their corrections include the mean first passage time (MFPT) required to escape a potential well. The condition for the strong Mpemba effect is given by
\begin{align}
\label{eq:SMEcond}
    &0=
    \left(\frac{\Pi _L (T)}{\Pi _L(T_b)} - \frac{\Pi _R (T)}{\Pi _R(T_b)}\right)
+\lambda_2 \left( A_L(T) \frac{\Pi _L (T)}{\Pi _L(T_b)} -  A_R(T)\frac{\Pi _R (T)}{\Pi _R(T_b)}\right), 
\end{align}
where $A_L$ and $A_R$ depend on the mean first passage times from their respective wells. They are defined as 
\begin{align}
    A_L \equiv \tau_L(x_{\rm min}) - \int _{D_L} \dd x \, \tau_L(x)\frac{e^{-U(x)/k_B T_b}}{\Pi_L(T)} \quad \text{and} \quad
    A_R \equiv \tau_R(x_{\rm max}) - \int _{D_R} \dd x \, \tau_R(x)\frac{e^{-U(x)/k_B T_b}}{\Pi_R(T)}.
\end{align}
Here, $A_L$ represents the difference between the mean first passage time from the left well at the left edge of the domain $x_{\rm min}$, $\tau_L(x_{\rm min})$, and the equilibrium average mean first passage time over the left well. Similarly, $A_R$ is defined using $x_{\rm max}$ as the right edge of the domain and $\tau_R(x)$, which is the mean first passage time from $x\in D_R$ to the left well.

As the switching rate $\lambda_2$ increases, the diffusion over the barrier also increases. In addition to the geometry of the potential, the probability current densities begin to play an important role, hence making the dependence on the mean first passage time significant. Further details and the condition for the weak Mpemba effect, in terms of the mean first passage time, can be found in Ref.~\cite{walker2022mpemba}.

The shape of the potential plays an important role in determining whether there is a Mpemba effect. In this discussion, we focused on cases with metastability in the system, but other aspects of the potential's geometry are also significant. For example, in ~\cite{lu2017nonequilibrium} the authors propose a one-dimensional potential that exhibits the Mpemba effect. This potential features a metastable state and a ground state with a wide and shallow basin of attraction;  see~\FIG{brownian_system}. The energy landscape has both a ``fast" and a ``slow" direction for relaxation. The relaxation from the metastable state is ``sluggish" because of the barrier. Starting at a high temperature is advantageous because most of the statistical weight resides in the broad parts of the potential, allowing for a ``fast" relaxation to the ground state.

The presence of metastability significantly affects the type and occurrence of the Mpemba effect, as illustrated in both the spin-glass example in Section~\ref{Sec:NumericalSpinSystems} and the Random Energy Model discussed in Section~\ref{sec:statistics}. However, metastability is not a prerequisite for observing the Mpemba effect. Walker and Vucelja in~\cite{walker2021anomalous} examined an overdamped particle diffusing within a piecewise-continuous potential characterized by energetically flat regions. In these flat regions, the particle can diffuse freely but has a high entropy, making it time-consuming to ``escape" from them. This piecewise-continuous potential can be coarse-grained into three states, where the energy levels correspond to different free-energy levels, leading to a form of "hidden" metastability. A generalization of the above results to piecewise linear potential was done by Biswas, Rajesh, and Pal in~\cite{biswas2023mpemba} as well as in \cite{biswas2025mpemba}. They investigated analytically a piecewise-linear double potential well \cite{biswas2023mpemba} (as well as the single well, with and without activity, in \cite{biswas2025mpemba}) and concluded that neither metastability nor the asymmetry of the potential is necessary or sufficient for the Mpemba effect to occur. A similar conclusion appeared in Kumar’s thesis~\cite{kumar2022anomalous}, where it was numerically demonstrated that the Mpemba effect exists in a single-well potential without metastability. 

It would be interesting to explore whether a similar scenario could be applied to a generalized ``lazy" random walk, where a particle can remain in the same state for a while instead of immediately transitioning to a neighboring state. The Mpemba effect has been observed in scenarios both with and without metastable states. Notably, adding a metastable state by adjusting the lengths of the flanks of the piecewise-continuous potential did not result in the emergence of the Mpemba effect~\cite{walker2021anomalous}. 

It is important to note that all the cases described here focus on finite-size spatial domains. It would be insightful to repeat some of these cases using natural boundary conditions, specifically with an infinite spatial domain where potentials with finite slopes confine the particles. Additionally, it would be interesting to relax the metastability requirements and investigate the nature of the Mpemba effect in this context. Further extensions could also include diffusion on landscapes of higher dimensions, or several interacting particles.

\subsubsection{The role of dynamics} 
\label{sec:dynamics}
In some systems, e.g., the 1D antiferromagnetic system discussed above, the existence of some type of Mpemba effect seems natural from the shape of the equilibrium line; see, for example,~\FIG{relax_path}. It is important to stress, however, that this is \emph{not} the case: the existence or non-existence of these effects is a consequence of the dynamics and cannot be deduced from the equilibrium properties of the system alone. This is because the effect is based on the projection of the equilibrium distribution on $\vec v_2$, which is a property of the dynamics, and is not an equilibrium property. In contrast to the equilibrium properties, the exact dynamics are non-universal and affected by many specific issues, such as the precise type of coupling between the system and the bath. 

One might imagine that if the equilibrium line is sufficiently complex, the equilibrium projections on $\vec{v}_2$ could be nonmonotonic, regardless of the direction of $\vec{v}_2$. However, this is never the case: for any system that obeys detailed balance, there is at least one dynamical law in which no Mpemba effect occurs. In fact, a fully connected graph with Metropolis dynamics does not exhibit any type of Mpemba effect, regardless of the equilibrium distribution~\cite{klich2018solution}. Interestingly, in some cases, removing a single link, which corresponds to introducing an infinite barrier, in a fully connected graph with Metropolis dynamics can be enough to induce the Mpemba effect~\cite{klich2018solution}. Similarly, in \cite{bera2023effect}, it was shown that in chemical reactions with identical initial conditions and the same equilibrium target distribution, the existence of the Mpemba effect depends on the precise details of the dynamics. Altering the dynamics, in some cases, one can make the effect appear or disappear. Such implications of dynamics in the context of the rates of chemical reactions are discussed in \SEC{rate-of-chemical-reactions}. Additionally, applications related to optimal transport and engine efficiency can be found in~\SEC{optimal-transport-heat-engine-efficiency}.

\subsection{Quantum systems}
\label{sec:quantum-framework}
Until recently, the Mpemba effect was studied exclusively in the context of classical systems.  Indeed, a 2018 review of ``Quantum quench dynamics’’~\cite{mitra18quantum} has no mention of a Mpemba or even a Mpemba-related effect. In the same year, though, a phenomena reminiscent of the Mpemba effect was observed in a numerical simulation of a dilute atomic gas in an optical resonator~\cite{keller2018quenches}.  In that case, there was a faster buildup of the amplitudes of the order-parameter correlation function, for low-frequency modes. Then, in the last five years, the situation changed dramatically, and this review includes {over fifty} references that relate to the Mpemba effect or analogs seen in quantum systems. A very recent review gives a short overview of these results~\cite{ares2025quantum}. 

It is largely believed that none of the quantum effects are at the origin of the classical effects. Instead, these are quantum effects that share properties that are similar to classical effects. Much as in classical systems, several different quantum effects have been considered, and some were observed experimentally. Below, we give a brief conceptual overview of this fast-developing field.  We discuss experiments from 2024 in~\SEC{experiments}. 

\subsubsection{Theoretical framework: Markovian open quantum systems} 
\label{sec:QuantumMarkov}

The simplest type of quantum Mpemba effect is a quantum analog of the Markovian dynamics discussed above. The density matrix $\rho(t)$ plays a role similar to the probability vector $\vec p(t)$. Its diagonal elements encode $\vec p(t)$, with $\rho_{ii}(t) = p_i(t)$.  The density matrix evolves according to both its internal Hamiltonian and its coupling to the environment, as described by the Gorini–Kossakowski–Sudarshan–Lindblad (GKSL) quantum master equation; see, e.g.,~\cite{chruscinski2017brief}, 
\begin{align}
\label{eq:LindbladianDyna}
\dot{\rho}(t)&=\mathcal{L}[\rho(t)],
\\
\mathcal{L}[\rho(t)]&=-i[H,\rho(t)]+\sum_{\mu=1}^{N_J}\left(L_\mu \rho(t) L^\dagger_\mu-\frac{1}{2}\left\{L^\dagger_\mu L_\mu,\rho(t)\right\}\right)\, .
\label{eq:Lind}
\end{align}
Here, $\mathcal{L}$ is the Lindblad operator, $H=H^\dagger$ is the system Hamiltonian, and the $N_J$ jump operators $L_\mu$ describe the dissipative effects resulting from an environment. Note that here we put $\hbar = 1$. The Lindbladian $\mathcal{L}$ preserves the trace, $\Tr\left(\mathcal{L}[\rho(t)]\right)=0$, and hermiticity, $(\mathcal{L}[\rho(t)])^\dagger=\mathcal{L}[\rho^\dagger(t)]$, $\forall \rho(t)$, and generates completely positive (physical) dynamics of the quantum state $\rho(t)$. The formal solution to the quantum master equation is then given by $\rho(t)=e^{t\mathcal{L}}[\rho(0)]$.

Assuming $\mathcal{L}$ to be diagonalizable, one can find the right eigenmatrices, $\rho^r_k$, such that 
\begin{align}
\mathcal{L}[\rho^r_k]=\lambda_k\, \rho^r_k\, .
\label{eq:right-diag}
\end{align}
The complex numbers $\lambda_k$ are the eigenvalues of the Lindblad map. Note that, due to the Hermiticity-preservation of $\mathcal{L}$, if $\lambda_k$ is a complex eigenvalue, then $\lambda_k^*$ must also be an eigenvalue. For the same reason, one can also show that if $\lambda_k$ is real, then $\rho_k^r$ can be chosen as Hermitian. Associated with the map defined in~\EQ{Lind}, there is a dual map, also called the adjoint Lindblad map, which implements the evolution of observables:
\begin{align}
\mathcal{L}^{\dagger}[O]=i[H,O]+\sum_{\mu=1}^{N_J} \left(L^\dagger_\mu O L_\mu-\frac{1}{2}\left\{O,L^\dagger_\mu L_\mu\right\}\right).
\end{align}
This dual map, $\mathcal{L}^{\dagger}$, is diagonalized by the left eigenmatrices $\rho^\ell_k$, 
\begin{align}
\mathcal{L}^{\dagger}[\rho^\ell_k]=\lambda_k\, \rho^\ell_k\, .
\label{left-diag}
\end{align}
The matrices $\rho^\ell_k$ are in principle different from the matrices $\rho_k^r$ in \EQ{right-diag}. However, $\rho_k^\ell$ and $\rho_k^r$ still form a basis for the space of matrices and can always be defined with the property $\Tr\left(\rho^\ell_k \rho^r_h\right)=\delta_{kh}$. 

Since the dynamics generated by $\mathcal{L}$ is completely positive, all the eigenvalues of the Lindblad map have a non-positive real part, ${\rm Re}\left(\lambda_k\right)\le0$. Furthermore, trace preservation enforces that at least one eigenvalue is zero, $\lambda_1=0$. If such an eigenvalue is non-degenerate, which here we assume to be the case, the (asymptotic) stationary state of the open quantum system
\begin{align}
    \rho_{\rm ss}=\lim_{t\to\infty}\rho(t)
\label{rho_ss}
\end{align}
is unique and given by the right eigenmatrix $\rho^r_1$. Since the left eigenmatrix associated with $\lambda_1$ is the identity, $\rho^\ell_1={\bf 1}$, one has $\Tr \left(\rho^r_1\right)=1$. Finally, the matrix $\rho^r_1$ is guaranteed to be positive due to the complete positivity of $e^{t\mathcal{L}}$.

The spectral decomposition of $\mathcal{L}$ allows us to write the dynamics of any initial density matrix as 
\begin{align}
    e^{t\mathcal{L}}\left[\rho_0\right]=\rho^r_1+\sum_{k=2}^{d^2}e^{t\lambda_k}{\rm Tr}\left(\rho^\ell_k\, \rho_0\right)\rho^r_k\, ,
\label{eq:dyn}
\end{align}
where $d$ is the dimension of the Hilbert space of the system. This decomposition shows that the matrices $\rho^r_k$ are nothing but the excitation modes of the system, each one characterized by its decay rate $|{\rm Re}\,\lambda_k|$. For long times, the relevant terms are those related to the $\lambda_k$ with the smallest real part in modulus. We order the eigenvalues $\lambda_k$ so that ${\rm Re} \,\lambda_2\ge {\rm Re} \,\lambda_3\ge \dots \ge {\rm Re}\,\lambda_m$ and assume that the eigenvalue $\lambda_2$ is real and unique. In this case, the slowest timescale for relaxation is given by the inverse of the spectral gap 

\begin{align}
    \tau=\frac{1}{|\lambda_2|}\, , 
\label{time-scale}
\end{align}
and $\rho_2^r$ 
is, in fact, the slowest decaying excitation mode of the Markovian open quantum dynamics. 

Overall, this framework is an exact analog of the ``classical final relaxation equilibrium (direct or inverse) Mpemba effect." To the best of our knowledge, the first manuscript to address the quantum version of the Mpemba effect is \cite{nava2019lindblad}, where an exactly solvable fully connected quantum Ising model with $n$-spin exchange with Lindbladian dynamics was considered. Although the framework used in this work is similar to the Lindbladian framework discussed above, an analysis of the effect in terms of the slowest relaxation mode was not given. Instead, it was numerically observed that a (direct) Mpemba effect exist in this model. In \cite{chatterjee2023quantum}, a quantum dot system coupled to two thermal baths was theoretically analyzed. A farther analysis of the Mpemba effect in quantum dots that includes electron-electron interactions appears in \cite{graf2024role}. A similar framework for the evolution of the density matrix describing the quantum dot was used. Interestingly, it was found that in this specific system $a_2=0$ for all initial temperatures because of a symmetry of the problem; hence, the effect is encoded in $a_3$ rather than $a_2$, and the relevant timescale is $\tau = |\lambda_3|^{-1}$, except at the initial condition where $a_3=0$ (a strong effect initial condition), where the system relaxes with a characteristic timescale of $\tau=|\lambda_4|^{-1}$. The case of vanishing gap between $\lambda_2$ and $\lambda_3$ in the same setup was studied in \cite{zhou2023accelerating}, where it was also suggested to accelerate the relaxation by reducing $\lambda_2$, through Floquet engineering. A different system that was analyzed with a similar framework and was implemented experimentally in a trapped ions setup was discussed in \cite{aharony2024inverse}. In this system, only a strong inverse Mpemba effect was observed, but not a direct effect. The nonequilibrium-steady-state analog of the effect was studied in \cite{wang2024mpemba,nava2024mpemba}. In \cite{chatterjee2024multiple,kheirandish2024mpemba}, a two-level driven dissipative system subject to an oscillatory electric field and dissipative coupling with the environment \cite{chatterjee2024multiple} and using the trace distance \cite{kheirandish2024mpemba} were studied. This system has a transition between real to complex-valued eigenvalues, which turned out to play a role in anomalous relaxation, as it leads to multiple strong Mpemba effects. A perturbative approach to the problem was suggested in \cite{ivander2023hyperacceleration}. A different approach, investigating the dynamics of Lindbladians with thermal steady states, was considered in~\cite{moroder2024thermodynamics}. The effect was also addressed using information geometry approach in \cite{bettmann2024information}. In~\cite{wang2024mpembaa}, the effect was discussed in the famous Sachdev-Ye-Kitaev (SYK) model, and some connections to quantum chaos were made. A non-Markovian framework was studied in \cite{strachan2024non}, where systems with a finite, non-negligible memory time due to the interactions between the (finite) bath and the system were considered. An analog effect for imaginary time,  which is a theoretical tool to evaluate ground-state properties
in simulations, was considered in~\cite{chang2024imaginary}. Additional example of a Mpemba effect in a non-Markovian open quantum system was suggested in \cite{wang2024going}. A similar effect was also discussed in the context of Mosaic models coupled to bosonic environment~\cite{dong2411quantum}. 
A sufficient condition for the existence of the effect was developed in \cite{furtado2024strong}. They introduced a ``Mpemba parameter'', which is analogous to $a_2(T_i)$ discussed in the classical case, to quantify how strong the Mpemba effect can be. The case where the initial condition is not a thermal state, but rather some nonequilibrium state, was studied in \cite{longhi2024mpemba} for the quantum harmonic oscillator coupled to a finite temperature bath.  It was demonstrated that any initial distribution of populations with a mean excitation number exactly equal to the mean number of quanta in the thermal equilibrium state displays an accelerated relaxation to equilibrium, namely it corresponds to a strong Mpemba effect. In \cite{qian2024intrinsic}, the issue of how to measure the distance from equilibrium in the quantum setup (in analogy with the discussion in Sec. \ref{sec:distance-function}) was considered. 
An interesting suggestion to apply the same formalism in photonic setup was given in \cite{longhi2024photonic,longhi2024bosonic}. In the first work \cite{longhi2024photonic}, light diffusion in finite-sized photonic lattices under incoherent (dephasing) dynamics is theoretically considered. It is shown that certain highly localized initial light distributions can diffuse faster than initial, broadly delocalized distributions.  This is analogous to the classical Mpemba effect in diffusing systems studied, e.g., in \cite{kumar2020exponentially}. In the second work \cite{longhi2024bosonic},  a leaky optical resonator or waveguide was studied, and it was shown that the Mpemba effect  can be observed when comparing the decay dynamics of coherent states (classical light) with certain non-classical states. Lastly, in \cite{medina2024anomalous}, a signature of the Mpemba effect in the context of quantum batteries was identified, where situations of batteries in higher charge states can discharge faster than less charged state was studied.

\subsubsection{Unitary transformation of initial state}
\label{sec:QuantumUnitary}

A related approach that was first suggested in \cite{carollo2021exponentially}, further developed in \cite{kochsiek2022accelerating} and experimentally demonstrated on trapped ions in \cite{zhang2024observation}, is to accelerate the relaxation towards the equilibrium by a unitary transformation that sets $a_2=0$. The idea of first establishing $a_2=0$ and only then quenching the system to a thermal bath at $T_{\rm final}$ was discussed also in the classical setup in \cite{gal2020precooling}; however, in that work, setting $a_2=0$ was done by coupling to a different bath (pre-heating or pre-cooling), whereas in the quantum scenario a natural approach is to use a unitary transformation instead. In principle, this technique is an application of Mpemba effect that belongs to Sec.~\ref{sec:applications} along with its classical analog. But since it played an important role in the study of quantum Mpemba effects, we review it here.  

A generic initial state has an overlap with all decaying modes of a Lindblad dynamics. In particular, there is an overlap with the slowest decaying mode. As such, the approach to the stationary state will take place in a time which is of the order of the relaxation time $\tau$ defined in \eqref{time-scale}. However, from Eq.~\eqref{eq:dyn}, it is clear that this becomes completely irrelevant for the dynamics if $\Tr\left(\ell_2 \rho_0\right)=0$, in analogy with the strong Mpemba effect. In such a case,  the state would relax at a faster rate with  $1/|{\rm Re}(\lambda_3)|$, which implies an exponential speedup of the convergence to stationarity. Such an acceleration may always be achieved when starting from an initial pure state, $\rho_0=\ket{\psi}\bra{\psi}$, by performing a unitary rotation to the state before the actual time-evolution takes place. 

Given an initial pure state $\rho_{0}=\ket{\psi}\bra{\psi}$, there always exists a unitary transformation $U$, which depends on the state, such that
\begin{align}
    \Tr\left(\ell_2 \, U\rho_0 U^\dagger \right)=0\, ,
\label{goal}
\end{align}
if the slowest decaying mode is unique. This can be shown as follows:  First, we notice that the matrix $\ell_2$ must be Hermitian since we have assumed that $\lambda_2$ is real and nondegenerate. As such, we can write it in its spectral form 
\begin{align}
    \ell_2=\sum_{k=1}^d\alpha_k\ket{\varphi_k}\bra{\varphi_k}\, ,
\end{align}
where $\bra{\varphi_k}\ket{\varphi_h}=\delta_{kh}$. We then note that, since ${\rm Tr}\left(\ell_2\, r_1\right)=0$ and $r_1$ is positive, the set of eigenvalues $\alpha_k$ either consists of both positive and negative real numbers or must contain at least one zero. Introducing an auxiliary orthonormal basis $\{\ket{\psi_k}\}_{k=1}^d$ for which $\ket{\psi}=\ket{\psi_1}$ (i.e., the initial state is a basis state) and using the spectral decomposition we find for the left-hand side of Eq.~\eqref{goal}:
\begin{align}
    \Tr\left(\ell_2 \, U\rho_0 U^\dagger \right)=\sum_{k=1}^d\alpha_k\bra{\psi_1}U^\dagger\ket{\varphi_k}\bra{\varphi_k}U\ket{\psi_1}\, .
\end{align}
To simplify the construction of the unitary transformation, we divide it into two parts, $U=U_2\, U_1$. The first unitary is chosen such that it maps the auxiliary basis $\ket{\psi_k}$ onto the basis $\ket{\varphi_k}$, which is simply achieved by  $U_1=\sum_k\ket{\varphi_k}\bra{\psi_k}$, yielding
\begin{align}
    \Tr\left(\ell_2 \, U\rho_0 U^\dagger \right)=\sum_k\alpha_k\bra{\varphi_1}U_2^\dagger\ket{\varphi_k}\bra{\varphi_k}U_2\ket{\varphi_1}\, .
\end{align}
In the next step, we construct $U_2$ such that the right-hand side of this expression becomes zero. Recalling that $\alpha_k$ are real numbers, two cases need to be considered: If one of the $\alpha_k$ is zero, it is sufficient that $U_2$ performs a permutation of the basis $\{\ket{\varphi_k}\}$, mapping $\ket{\varphi_1}$ onto the eigenstate $\ket{\varphi_h}$, for which $\alpha_h=0$. 

In the non-trivial case, in which $\ell_2$ does not have a zero eigenvalue, we can make a construction based on the following observation: the eigenvalue $\alpha_1$ is a real number and can be either positive or negative. Since $\ell_2$ cannot be a positive (or negative) eigenmatrix, there must be an eigenvalue $\alpha_n$ such that ${\rm sign}(\alpha_n)=-{\rm sign}(\alpha_1)$. We then construct the Hermitean operator $F=\ket{\varphi_1}\bra{\varphi_n}+\ket{\varphi_n}\bra{\varphi_1}$, 
which we use to define the unitary
\begin{align}
    U(s)\equiv e^{-is\, F}={\bf 1}+\left(\cos(s)-1\right) F^2 -i\sin(s)F, 
\end{align}
where $F^2=\ket{\varphi_1}\bra{\varphi_1}+\ket{\varphi_n}\bra{\varphi_n}$. Using this unitary operator, we find that 
\begin{align}
\begin{split}
    \Tr\left(\ell_2 \, U(s)U_1\rho_0 U_1^\dagger U^\dagger(s) \right)=\alpha_1\cos^2(s)+\alpha_n\sin^2(s)\, .
\end{split}
\label{fin-part}
\end{align}
The above quantity has the same sign as $\alpha_1$ for $s=0$ but has the same sign as $\alpha_n$ for $s=\pi/2$. In particular, it vanishes for $\bar{s}=\arctan \left( \sqrt{\left|\alpha_1/\alpha_n\right| } \right)$, so that if we take the unitary transformation $U=U(\bar{s})U_1$, Eq.~\eqref{goal} is satisfied. This implies that the initial state is rotated into one that is orthogonal to the slowest decaying mode and will thus relax, in general, on a timescale  $1/|{\rm Re}(\lambda_3)|$. The approach to stationarity has thus been exponentially accelerated by a factor $|{\rm Re}(\lambda_3)|-|{\rm Re}(\lambda_2)|$.

Another way to obtain a quantum Markovian Mpemba effect is using a global unitary transformation when the previous transformation is not applicable. A very common case is when the eigenvalues of the lowest excited modes of the master operator form a complex-conjugate pair. In this case, the
evolving system state oscillates as it decays to steady state.

\subsubsection{Entanglement asymmetry} \label{sec:EntanglementAsymmetry}
A different type of effect that is also commonly referred to as the ``quantum Mpemba effect" was recently suggested in~\cite{ares2023entanglement}, further studied in~\cite{rylands2024microscopic,murciano2024entanglement,caceffo2024entangled,liu2024symmetry,turkeshi2024quantum,chalas2024multiple,ares2024quantum,foligno2024non} and experimentally demonstrated in a trapped ions setup~\cite{joshi2024observing}. In the terminology defined in Sec.~\ref{Sec:Zoology}, this effect can be categorized as a ``non-thermal quantum Mpemba effect." In this case, the dynamics of an isolated system rather than a system coupled to a bath is considered. In other words, the system is not coupled to any environment and follows Hamiltonian dynamics. To nevertheless observe some relaxation, one divides the Hamiltonian system into two spatially separated parts.  One part can be viewed as ``the system," whereas the other part serves as a ``local bath"; however, in contrast to the usual case, the dynamics of this ``bath'' are also taken into account. Interestingly, this is closely related to the framework for classical macroscopic system used in~\cite{yang2020non} to study the Mpemba effect through a first-order phase transition, which we discussed in~\SEC{FirstOrderPT}. 

The following scenario is considered in this effect: The system, which includes both the ``system" and the ``bath," is prepared in some initial condition, usually a pure state. The full system then evolves under a quantum Hamiltonian dynamics that has some global symmetry. For example, the particle number symmetry, i.e. $U(1)$, in~\cite{ares2023entanglement,joshi2024observing,rylands2024dynamical,yamashika2024quenching,yamashika2024entanglement}, $S\!U(2)$ and $Z_2$ in~\cite{ferro2024non,liu2024symmetry} and ``space-time symmetry" in~\cite{klobas2024translation}. The initial condition of the system is chosen such that the symmetry is not respected; namely, it is not an eigenstate of the system, which can be static or periodically driven Hamiltonian~\cite{banerjee2024entanglement}. This asymmetry can be quantified by an ``asymmetry parameter," the \emph{entanglement asymmetry}, which is the relative entropy between the density matrix projected onto the conserved quantity subspace and the density matrix traced over the bath. During the dynamics, this asymmetry decays because of the global symmetry of the Hamiltonian, and thus the symmetry of the system is also restored. When the entanglement asymmetry decays faster for the initial condition with the larger asymmetry in the initial condition, there is a ``quantum entanglement asymmetry Mpemba effect."  The terminology is chosen in analogy to the classical Mpemba effect. This scenario is demonstrated in the experimental measurements shown in Fig.~\ref{fig:quantum_asymmetry} and discussed in~\SEC{quantum-experiment}.

In this framework, two initial conditions with different entanglement asymmetry values are compared. However, as pointed out in~\cite{liu2024symmetry}, the specific value of the entanglement asymmetry does not uniquely dictate the initial condition, and different initial conditions with the same initial entanglement asymmetry might relax in a different way. Therefore, one has to be careful in choosing the initial conditions; otherwise, the effect can be trivial, e.g., if the fastest relaxation among the high-valued entanglement asymmetry is chosen as the initial state and is compared to the slowest relaxation among the low entanglement asymmetry states. In a similar study in the context of many-body localization~\cite{liu2408quantum}, it was shown that the Mpemba effect is universal for any initial tilted product state, in contrast to the cases in the chaotic systems where the presence of the quantum Mpemba effect relies on the choice of initial states.

\subsection{Discussion}
\label{sec:discussion-markovian}
The Markovian framework provides a clear picture for some types of anomalous relaxations: the long-time characteristics of the system are controlled by the slowest dynamical mode, with the relevant information encapsulated in $a_2$, $\vec v_2$, and $\lambda_2$. These properties are also responsible for several other interesting phenomena, as will be discussed in Secs.~\ref{sec:statistics}, \ref{sec:applications}, and \ref{sec:other-related}.

The main assumptions of the Markovian framework are (i) Markovianity, (ii) instantaneous temperature quench, and (iii) linearity. Since we have already elaborated on the Markovian assumption in Sec.~\ref{sec:introduction}, we discuss the other two assumptions here. 
\begin{enumerate}
    \item \emph{Instantaneous Temperature Quench:} In addition to the ``perfect bath'' assumption that implies Markovianity, our discussion assumed that the environment temperature is instantaneously changed from $T_i$ to $T_f$. This is a reasonable assumption in many cases, for both small and thermodynamically large systems, e.g., when suddenly placing a cup of water into a freezer; it is also a good description for the experimental system discussed in \cite{kumar2020exponentially,kumar2022inverse,aharony2024inverse,hu2018conformation}. However, this assumption is not always valid; in several experiments, the temperature changes from $T_i$ to $T_f$ at a (constant or varying) rate (e.g., 5 K/min in the differential scanning calorimetry used in \cite{ahn2016experimental}). When the temperature is not instantaneously quenched but rather changes gradually, the system can still be analyzed using a Markovian dynamics; however, in this case, the Markovian operator (Fokker-Planck operator or rate matrix) is not constant but rather changes with time. This 
    complicates considerably the analysis, since the eigenvectors of the Markovian operator change, too. It is therefore possible that the coefficient along $\vec v_2$ corresponding to the final temperature operator is zero in the initial state, but evolution with a different operator that corresponds to a different temperature increases the corresponding coefficient. However, it is still possible to find the strong version of the Mpemba effects, using essentially the same idea presented in the example of colloidal setup with boundary coupling discussed in Sec.~\ref{sec:boundary_coupling}; see Fig.~\ref{fig:colloidal_boundary}. In this example, the operator is time dependent because the temperature profile of the fluid varies with time.   
    \item \emph{Linearity.} Assuming linear dynamics for the probability distribution is quite natural because Liouville's equation for the probability distribution in Hamiltonian dynamics is linear. Commonly, the dynamics of a system are modeled as Hamiltonian dynamics plus dissipation and random forcing; both forces are generated as a result of the coupling of the system to the thermal environment. Assuming memory-less random force as well as dissipation, which is linear in the momentum, results in linear dynamics for the probability distribution. 
    
    Since the analysis presented in this section was heavily based on the eigenvectors and eigenvalues of the linear dynamics, one might worry that the linear assumption is crucial for the analysis. However, this is not the case: even if the time evolution of the probability is nonlinear, we nevertheless assume that in the long-time limit the probability relaxes to a unique steady state, the equilibrium distribution. From a dynamical-system point of view, the equilibrium distribution can thus be considered as the unique stable fixed point of the dynamics. It is, therefore, possible to linearize the dynamics in the long-time limit and identify the slowest-dynamics manifold from which all initial conditions relax towards the equilibrium point, except for the initial condition that relaxes from the codimension-one ``fast manifold" that has no coefficient along the slowest dynamical variable. Therefore, it is possible to identify strong direct and inverse Mpemba effects, even in this case. An example of nonlinear dynamics of the probability distribution was given in \cite{klich2018solution} for the mean-field anti-ferromagnetic with Glauber dynamics.  
\end{enumerate}

\section{Phase transitions}
\label{sec:phase-transitions}
Historically, the Mpemba effect was first discussed in the context of observations and experiments involving freezing water, and its initial definitions were often phrased in terms such as ``the time to begin freezing.’’  
However, the exact definition of the phase transition time in a first-order phase transition, where the two phases coexist for some nonzero fraction of the time, is tricky. Indeed, different definitions were used in different experiments to measure the ``freezing time,'' one being the moment at which the temperature at a specific point in the system (e.g., the point where the thermometer is placed) reaches the critical temperature~\cite{burridge2016questioning} and another being the moment at which the latent heat begins to be released~\cite{jeng2006mpemba,ahn2016experimental}. Under some experimental conditions, the differences in freezing time seem large enough that the exact definition does not play a role, and the effect exists regardless of the exact definition. However, for the case of water, the Mpemba effect is observed reliably only when freezing occurs and when its timing is a part of the definition~\cite{burridge2016questioning,katz2017reply}.

From a theoretical perspective, defining the effect based on the phase transition time raises several difficulties, and indeed, the majority of works on the Mpemba effect used frameworks that are independent of a phase transition in the system. Some of the challenges raised by incorporating phase transitions into the definition of the Mpemba effect are as follows: (i)  First-order phase transitions are often dominated by rare events (especially in an inhomogeneous nucleation process). Consequently, there is a wide temporal distribution for the phase-transition time. Estimating the phase-transition-time distribution when the process is dominated by a rare event is difficult theoretically, numerically, and even experimentally. (ii) To observe the phase transition time, one needs to follow the dynamics of a large system that is far from the linear-response regime after a thermal quench. There are very few cases for which analytical results for the dynamic are known, and these are often mean-field results. (iii) Even when it is possible to follow the dynamics of the system analytically or numerically, identifying the moment at which the phase transition happens is not easy. A possible way to bypass these issues is to use mean-field models. Indeed, all the works on Mpemba effects through a phase transition to date have used some kind of mean-field approximation and are thus quite far from being directly applicable to any of the experimental results. 

In what follows, we discuss separately anomalous relaxations through first- and second-order phase transitions. A key feature in first-order phase transitions is the existence of a metastable state. The transition from the metastable state to the global minimum of the free energy---the equilibrium state---is assumed to be the slowest process in the relaxation, and by analyzing this transition, it is possible to identify mechanisms for a Mpemba effect. The physical picture differs in second-order phase transitions, where there are no metastable states. A major obstacle, in this case, is to define a useful ``phase transition time''; however, once such a time is determined, identifying a Mpemba effect is straightforward. 

\subsection{Mpemba effect through a first-order phase transition}
\label{sec:FirstOrderPT}

\begin{figure}[htb]
    \centering
    \includegraphics[width=0.7\columnwidth]{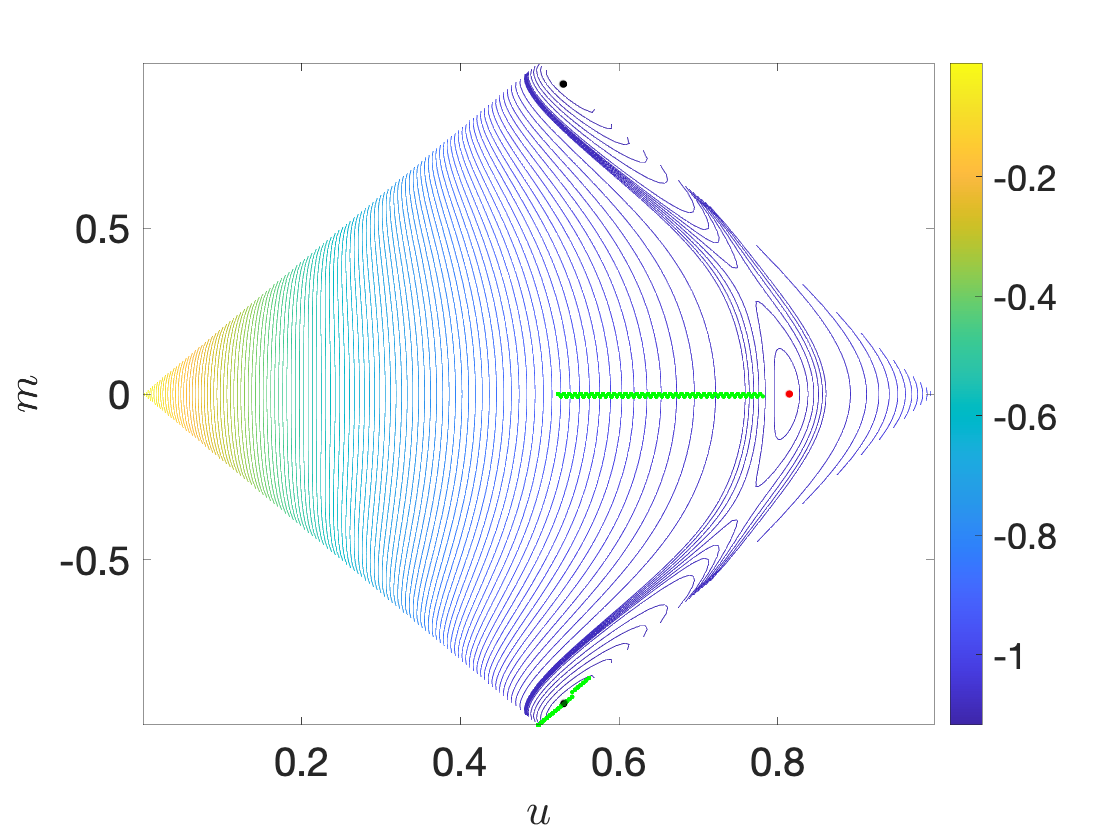}
    \caption{Level sets of the free energy $F(u,m)$ for the mean-field Ising model described by \EQ{H_FirstOrder1}, as a function of staggered magnetization per spin $u$ and magnetization per spin  $m$, at $K=-0.95$, $J=-1$, $I=1$ and $\beta = 1.5385$. The red marker indicates a local minimum of $F(u,m)$, which is a metastable state. The global minima are denoted by the two black markers. The two green lines represent the equilibrium locus of the system for these values of $K$, $J$, and $I$. As the system has a first-order phase transition, there is an abrupt jump from $m=0$ to $m\neq0$, which splits the equilibrium locus to two disjoint lines. A system initiated at the hot phase with $m=0$ first relaxes quickly towards the metastable state (the red marker) and then more slowly to equilibrium (the black markers).}
    \label{fig:FirstOrderFreeEnergy1}
\end{figure}

Two different mechanisms have been proposed to explain the Mpemba effect through a first-order phase transition.
The first, suggested in \cite{yang2020non} and developed in \cite{yang2022mpemba}, applies to any system with a metastable state associated with the first-order phase transition and coupled to a finite bath. Recall from Sec.~\ref{sec:metastability} that a system with a metastable state relaxes following two timescales towards the final, stable equilibrium state: A fast relaxation (i) of the system from its initial state to the metastable state is followed by a slow transition (ii) from the metastable state to the stable, equilibrium state. 
The first stage, whether starting from a hot or cold state, is typically much faster than the second stage.  But---and this is the key point---if the thermal bath is finite, then the energy transferred from system to bath during the first stage will increase the bath temperature. The initially hot system, which transfers \emph{more} heat into the thermal bath during its relaxation to the metastable state, is thus coupled to an effectively \emph{higher} temperature thermal bath and hence reaches sooner the stable equilibrium state.

The mechanism described above is generic. So far, it has been demonstrated on a relatively simple model, the mean-field Ising model with staggered magnetic field; see \FIG{FirstOrderFreeEnergy1}. The Hamiltonian of the system is given by
\begin{align}
\label{eq:H_FirstOrder1}
    \mathcal{H} = -\frac{K}{2}\left(\sum_{i=1}^N (-1)^i \sigma_i\right) -\frac{J}{2N}\left(\sum_{i=1}^N \sigma_i\right)^2 - \frac{I}{4N^3}\left(\sum_{i=1}^N \sigma_i \right)^4 ,
\end{align}
where $\sigma_i\in\{-1,+1\}$ is a spin degree of freedom at site $i$, $K$ the staggered magnetic field magnitude, $J$ the two-spin coupling constant, and $I$ a four-spin coupling constant. Interestingly, this specific model is very similar to the antiferromagnet system studied in~\cite{klich2019mpemba}, which has a rich phase diagram of final relaxation Mpemba effects. Being a mean-field model, a microstate of the system can be uniquely characterized by two numbers, for example the staggered magnetization per spin $u = N^{-1}\sum_{i=1}^N (-1)^i \sigma_i$ and the total magnetization, $m = N^{-1}\sum_{i=1}^N \sigma_i$. The free energy landscape of this model with $K=-0.95$, $J=-1$, $I=1$ at $\beta = 1.5385$ is shown in \FIG{FirstOrderFreeEnergy1}. It has a local minimum (the red marker at $m=0$, $u=0.815$)  that is a metastable state, whereas the global minima are at low/high values of $m$ (the black markers at top and bottom of~\FIG{FirstOrderFreeEnergy1}). During the first stage of the relaxation, the system stays in the metastable state for a long time until a rare fluctuation takes it to the final equilibrium state. The mean time for the system to go through the phase transition can be simply modeled as an Arrhenius law, $\tau\sim\exp(N\Delta E)$ \cite{yang2020non}, but a more sophisticated calculation \cite{yang2022mpemba} showed that $\tau\sim \exp(\Delta S_\textrm{tot})$, where $\Delta S_\textrm{tot}$ is the total entropy change (for system and bath) between the metastable state and the global minimum.  

\begin{figure}
    \centering
    \includegraphics[width=0.6\columnwidth]{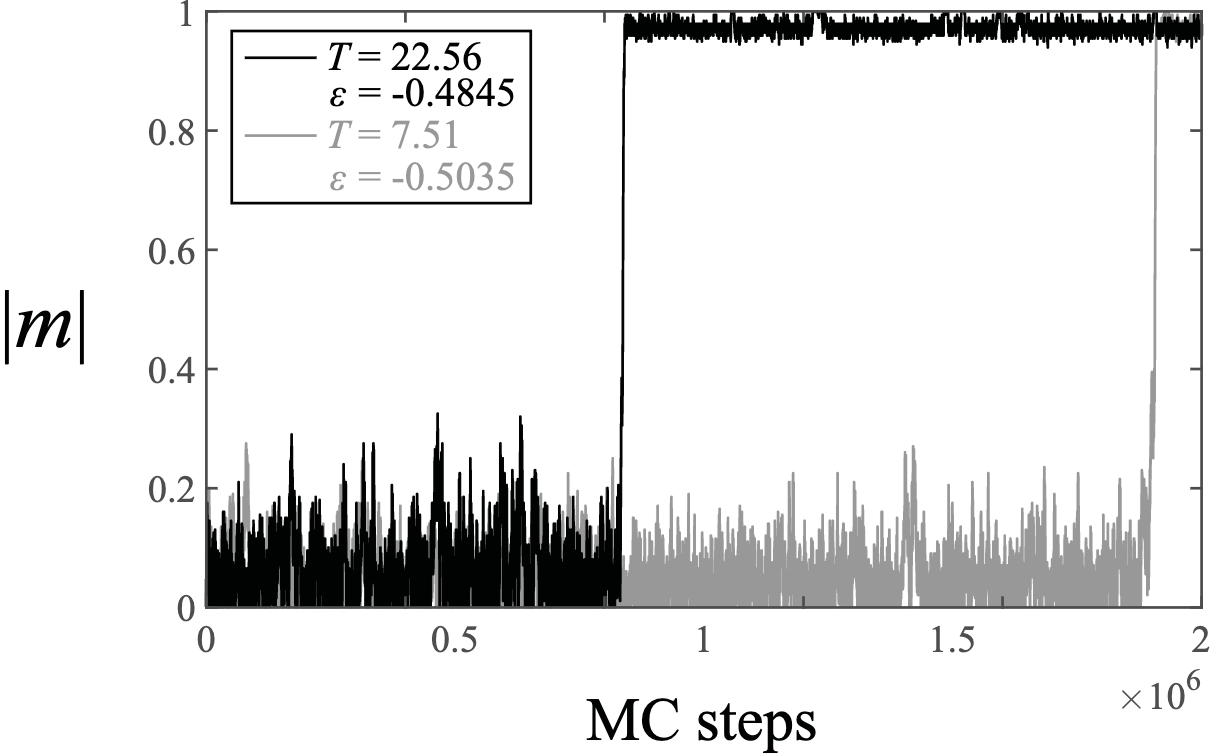}
    \caption{Creutz microcanonical Monte Carlo simulations of the system described by \EQ{H_FirstOrder1}. The black and gray graphs show the mean magnetization per spin, in absolute value, as a function of the Monte Carlo steps for two initial conditions sampled from different temperatures. The phase transition is declared by the abrupt jump from fluctuations around $|m|=0$ to $|m|\sim 1$. \textit{Source:} Reprinted with permission from~\cite{yang2020non}.}
    \label{fig:MCFirstOrder}
\end{figure}

The effect in this model was demonstrated via a microcanonical Monte Carlo simulation, based on the Creutz algorithm, for a system of $N=400$ spins in \cite{yang2020non} and $N=50$ spins in \cite{yang2022mpemba}.  The spins were coupled to a finite bath, with $N_\textrm{bath} = 20N$ in \cite{yang2020non} and $50N$ in \cite{yang2022mpemba}. The system was initiated at the equilibrium distributions for various initial temperatures, and the bath was sampled from a temperature far below the phase transition temperature. A phase transition was identified by the ``jump" in mean magnetization from the local, metastable minimum at zero magnetization to one of the two global minima at non-zero magnetization; see \FIG{MCFirstOrder}. Indeed, the initially hot system is observed to go through the phase transition before the initially cold system.  

This mechanism for a Mpemba effect through a first-order phase transition is generic: first-order phase transitions are usually associated with a metastable state, and finite baths are common. Clearly, however, this approach is not relevant for an inverse Mpemba effect, where the signs are all reversed.  That is, the colder initial state transfers more heat \emph{out} of the thermal bath, reducing its temperature and thereby slowing the second stage.
We note that an inverse Mpemba effect has never been measured through a first-order phase transition.

There are also a few difficulties with this proposed mechanism: (1) The phase transition time, as well as its difference between the cold and hot initial conditions, diverges exponentially in the thermodynamic limit, but this limit is essential for the existence of a phase transition. (2) The existence of the effect strongly depends on the finite size of the bath, even though some effect should exist even when the thermal bath is large~\cite{yang2022mpemba}. (3) This mechanism is useful for mean-field models, where the phase transition is not dominated by nucleation (homogeneous or inhomogeneous) processes, by the coupling of the system to the bath, and other factors that are known to be important in practice. In finite-dimension systems, the relaxation from the metastable state to the final equilibrium might be affected by additional factors.

A different mechanism for a Mpemba effect through a first-order phase transition was suggested in \cite{zhang2022theoretical}, where it was also demonstrated on a carefully constructed \emph{Blume-Emery-Griffiths} mean-field model. This mechanism is relevant when, after the quench, the free energy landscape of the system has more than one minimum: a local minimum corresponding to metastable states and a global minimum corresponding to the equilibrium state. If the hot initial condition happens to be in the basin of attraction of the equilibrium minimum, and if, in addition, the colder temperature initial condition flows to the metastable state, then the hot initial condition takes less time to get to the cold equilibrium, since it does not have to escape the local metastable state. A similar idea was, in fact, suggested in the context of water \cite{jeng2006mpemba}, where it was conjectured that the effect has to do with supercooling of the initially cold water. Indeed, supercooled water is in a metastable state, and if it so happens that the initially cold water is in the basin of attraction of this metastable state, whereas the hot water is in the basin of attraction of ice, then this generates a Mpemba effect. 

\begin{figure}
    \centering
    \includegraphics[width=0.9\columnwidth]{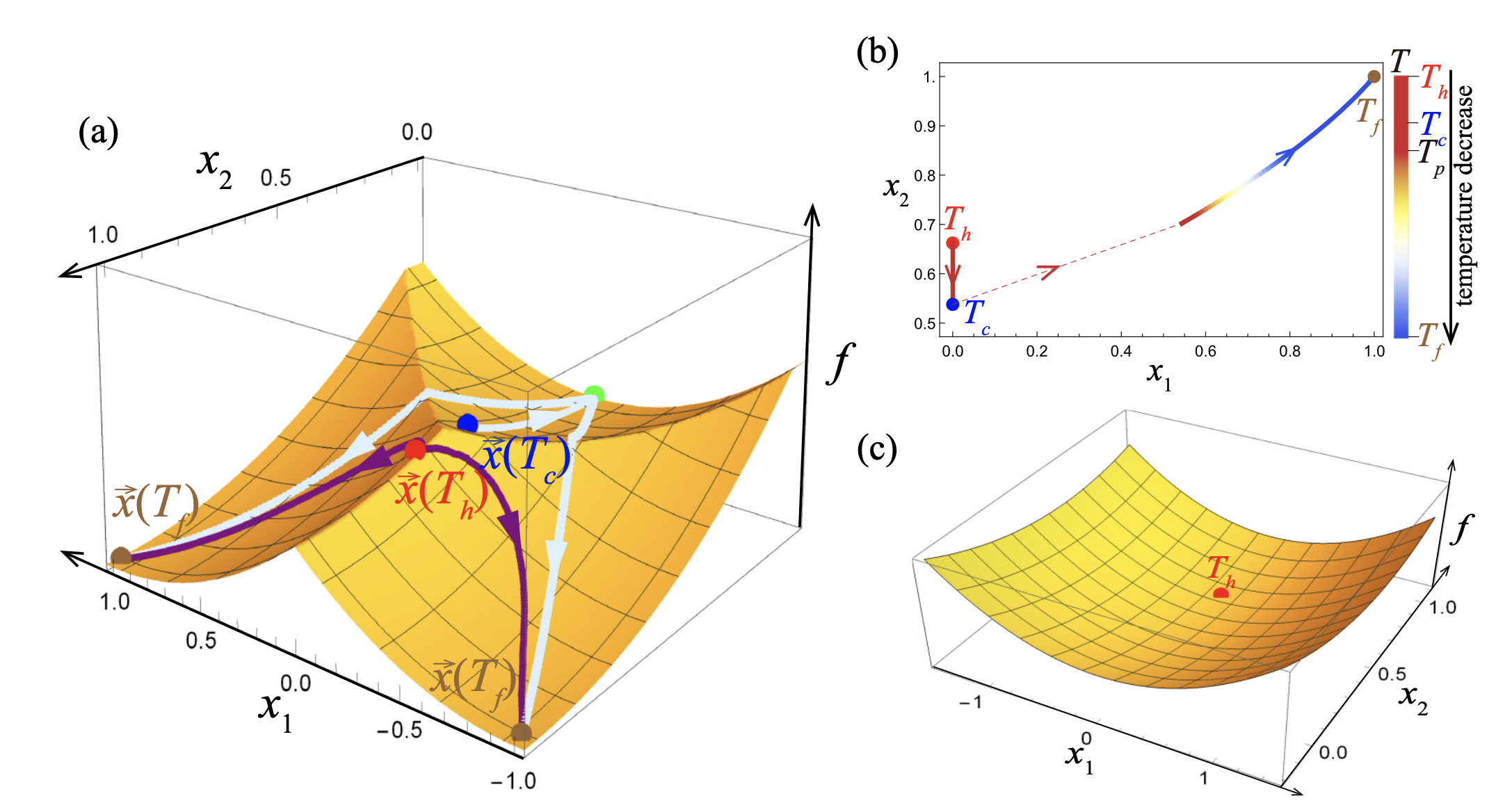}
    \caption{The free energy, $F(x_1, x_2)$, landscape of the Blume-Emery-Griffiths mean-field model. (a) $F(x_1,x_2)$ for $T=T_\mathrm{final}$. This landscape has three minima: two symmetric global minima $\vec x(T_f)$ (brown points) corresponding to the equilibrium state at $T_\mathrm{final}$, and one local minimum (green point). Each minimum has a basin of attraction, and the three basins are separated by ridges. The initial condition at the hot temperature $\vec x(T_h)$ (red point) is on the ridge between the basins of the two equilibrium points, whereas the cold initial condition, $\vec x(T_c)$ (blue point) is in the basin of attraction of the metastable local minimum. Therefore, it will take the cold initial condition much longer time to relax towards the equilibrium state of the model. Panel (b): The equilibrium line of the model at the relevant parameters in the $(x_1,x_2)$ plane. Panel (c): Free energy $F(x_1,x_2)$ at temperature $T_h$. \textit{Source:} Reprinted with permission from~\cite{zhang2022theoretical}.}
    \label{fig:FirstOrder_secondModel}
\end{figure}

A carefully constructed mean field  Blume-Emery-Griffiths model was used to demonstrate this effect. Because the model is mean field, the state of the system in the thermodynamic limit is fully characterized by two macroscopic parameters ($x_1$ and $x_2$ in \FIG{FirstOrder_secondModel}).  The parameters of the system were thus chosen such that the hot initial condition is on the ridge between the basin of attractions corresponding to the two minima, while the medium temperature initial condition is in the basin of attraction of the metastable state; see \FIG{FirstOrder_secondModel}a. Thus, the time to escape the unstable initial condition of the hot system grows with the logarithm of the system size, whereas the time to exist in the metastable state in the medium temperature initial condition grows exponentially with system size. 

This mechanism is generic and might exist in any system that has the relevant structure of metastable state and initial equilibrium distributions that flow into the equilibrium or metastable state basins of attraction. It bypasses the need to carefully define the phase transition time in a way that does not diverge in the thermodynamic limit since, regardless of how it is defined, it is clear that initiating the system in the basin of attraction of a metastable state for relatively cold initial conditions and of the equilibrium state for hot initial conditions can generate the effect. On the other hand, it is also clear that this effect has drastically different characteristics from the effects observed experimentally: for this mechanism to exist, there should be a clear range of parameters at which the system relaxes to a metastable state, and a range of parameters from which it never does. It 
seems likely that this mechanism exists in some experimental systems; however, it is challenging to analyze it theoretically or even numerically beyond a mean-field model, namely in a model whose state during relaxation depends on many macroscopic order parameters. In addition, we lack a simple physical picture to connect the basin of attractions of the metastable and equilibrium states to the hot/cold temperature initial conditions. Such a connection between the equilibrium line and the basin of attraction at low temperatures, which is a property of the dynamics and not just the equilibrium state of the model, might also shed light on how generic is the effect: Is it expected in many models, or does it require some careful tuning or special mechanism?     

\subsection{Relaxation through a second-order phase transition}
As far as we know, a Mpemba effect through a second-order phase transition has never been experimentally observed, despite a theoretical 
argument for its existence~\cite{holtzman2022landau}. Identifying the phase transition time is a main challenge in addressing a Mpemba effect through a second-order phase transition. The two mechanisms suggested for a first-order phase transition bypassed the need to exactly define the phase transition time to observe a Mpemba effect: The ``finite bath" mechanism suggested in \cite{yang2020non} assumes that the hot and cold temperature reach essentially the same metastable state and that it is a property of the thermal bath (its temperature) that differentiates the evolution from the metastable state to the final equilibrium state. This makes it 
unnecessary to exactly define the phase transition time, which indeed diverges in the thermodynamic limit, because it is clear that a rare fluctuation that takes the system from the metastable state to the final equilibrium state is more probable---and therefore, its average time is smaller---when the system is coupled to higher temperature bath. In the ``different basin of attraction" mechanism suggested in \cite{zhang2022theoretical}, the exact definition of the phase transition time is again not essential, since it is clear that the hot initial condition relaxes directly to the final equilibrium, whereas the cold initial condition relaxes first to a metastable state and thus takes longer (in the thermodynamic limit, an infinite time) to reach the equilibrium state, no matter how the phase transition time is defined.   

\begin{figure}
    \centering
    \includegraphics[width=0.7\columnwidth]{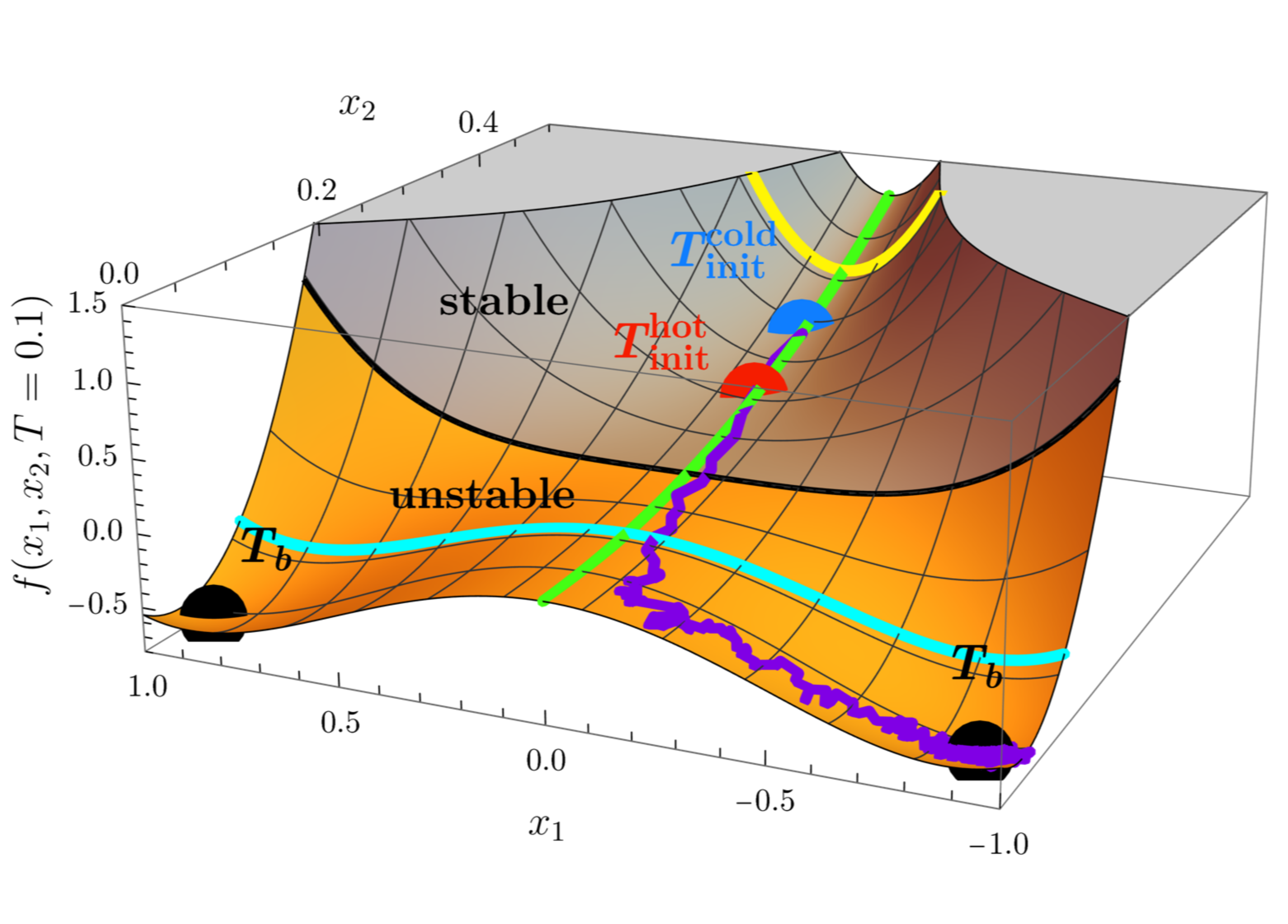}
    \caption{A system with two order parameters can have a phase transition at a finite time. The phase of the system is defined based on the value of the order parameter $x_1$ and its stability: $x_1=0$ is the disordered phase, and $x_1\neq0$ is the ordered phase. However, the stability of $x_1=0$ can change as a function of the other order parameter, $x_2$, as shown in the figure: in the gray area $x_1=0$ is still stable, and in the yellow area it is not. Therefore, the phase transition happens when the system crosses the black line. A representative trajectory is demonstrated by the pink line. If the two initial conditions are arranged as the blue (cold) and red (hot) points in the figure, then there will be a Mpemba effect since the initially hot system crosses the phase transition line first. \textit{Source:} Reprinted with permission from~\cite{holtzman2022landau}.}
    \label{fig:SecondOrderPT_1}
\end{figure}

In contrast to the two mechanisms for a Mpemba effect through a first-order phase transition discussed above, the definition of the phase-transition time plays a key role in the Mpemba effect through a second-order phase transition. These transitions are usually not associated with a metastable state, and thus, the transition is not expected to be dominated by fluctuations, as in a first-order phase transition. However, defining the exact time at which the phase transition happens is somewhat subtle since many of the equilibrium characteristics of an equilibrium second-order phase transition, e.g., divergence of the correlation length or the percolation transition, take infinite time in the thermodynamic limit. The definition of the phase-transition time used in \cite{holtzman2022landau} is based on the following picture of a second-order phase transition: the system has an order parameter whose physical value minimizes the (Landau) free energy. At the phase transition temperature, the minimum of the free energy splits into two minima, with a maximum in between. The maximum corresponds to the instability of the system in the disordered state. The phase transition time is defined as the time at which the disordered phase becomes unstable. For systems with a single order parameter, the phase transition time must then be zero. However, as demonstrated in \FIG{SecondOrderPT_1}, for systems with more than a single order parameter, this definition can provide a non-zero time for the phase transition, which does not diverge in the thermodynamic limit. It can thus be used to compare the phase transition time for different initial conditions and, therefore, to define and detect a Mpemba effect. 

Let us demonstrate the phase transition time on a simple system, the mean field antiferromagnetic Ising model.  In this model, each spin lives on a node. Half of the nodes belong to sublattice A, and the other half to sublattice B. The two sublattices represent the two sublattices of an antiferromagnetic system. Each spin in sublattice A is coupled antiferromagnetically to all spins in sublattice B and vice versa, but there is no coupling between the spins in the same sublattice. In addition, there is an external magnetic field coupled to all the spins in the system. We denote the number of spins up (down) in sublattice A as $N^A_\uparrow$ ($N^A_\downarrow$) and for sublattice B as $N^B_\uparrow$ ($N^B_\downarrow$). The total magnetization is denoted as $m = (N^A_\uparrow + N^B_\uparrow - N^A_\downarrow - N^B_\downarrow)/N$ and the staggered magnetization, which is the magnetization difference between the two sublattices, as $m_s = (N^A_\uparrow - N^B_\uparrow - N^A_\downarrow + N^B_\downarrow)/N$. With these, the Hamiltonian of the system is given by 
 \begin{align}
     \mathcal{H}(m,m_s) = \frac{1}{N}\left(-J(m_s^2-m^2) - hm\right) ,
 \end{align}
 where $J<0$ is the (antiferromagnetic) coupling constant, and $h$ is the external magnetic field. The single spin Glauber dynamics leads, in the thermodynamic limit, to equations of motion for $s$ and $m$ \cite{klich2019mpemba},
 \begin{align}
\label{eq:sdot}
\dot{m_s} &= \frac{1}{4} \left[ \tanh \left( \frac{h-m+m_s}{T_{b}} \right) - \tanh \left(\frac{h-m-m_s}{T_{b}} \right) \right] -\frac{m_s}{2},\\
  \label{eq:mdot}
\dot{m} &= \frac{1}{4} \left[ \tanh \left( \frac{h-m+m_s}{T_{b}} \right) + \tanh \left(\frac{h-m-m_s}{T_{b}} \right) \right] -\frac{m}{2} ,
\end{align}
where $T_b$ is the temperature of the thermal bath to which the system is coupled. 

\begin{figure}
    \centering
    \includegraphics[width=0.8\columnwidth]{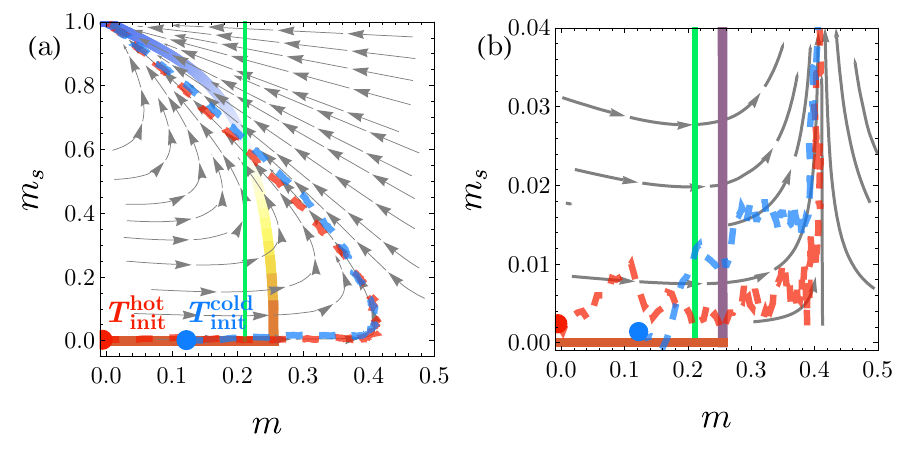}
    \caption{Mpemba effect in a mean-field antiferromagnetic Ising model. (a) The arrows represent the dynamics of the staggered magnetization and magnetization, from~ \EQS{sdot}{mdot}, on the $(m_s,m)$ plane. The green line is the value of the magnetization, $m$, at which the stability of the $m_s=0$ changes: $m_s=0$ is stable for smaller values of $m$ and unstable for higher values of $m$. Therefore, the phase transition time is the time $t^*$ at which (on average) the system crosses the green line.
    Two realizations (with noise) are plotted. The red line is initiated from the hot initial condition and the blue is initiated from the cold. (b) Zoom in around $m_s=0$. \textit{Source:} Adapted with permission from~\cite{holtzman2022landau}.}
    \label{fig:MeanFieldAF_PT}
\end{figure}

In this system, the phase of the system is dictated by the value of the staggered magnetization $m_s$: the system is in the ordered, antiferromagnetic state when $m_s\neq 0$ and in the disordered, paramagnetic phase when $m_s=0$. If the system is initiated at the disordered phase, then $m_s=0$ is a solution of \EQ{sdot} regardless of the value of $m$. However, for some values of $m$, the solution, $m_s=0$ is stable, and for others, it is unstable.
We can then use the change in stability of the $m_s=0$ state as a criterion to define the phase transition time.  Specifically, the transition occurs at a time $t^*$ when the evolution $m(t)$ causes an initially stable $m_s=0$ state to become unstable. Figure~\ref{fig:MeanFieldAF_PT} illustrates the scenario: At $t=t^*$, the system crosses the phase transition line (the green line in the figure), and the $m_s=0$ solution changes its stability.

The mean-field, antiferromagnetic Ising model described above demonstrates that a second-order phase transition defined through the instability of the order parameter is possible at finite, non-zero time. Once the phase transition time has been defined, it is easy to identify direct and inverse Mpemba effects. However, a concrete microscopic model with dynamics that demonstrates any of these effects is not yet known. Moreover, it is not even obvious if this definition is useful beyond mean-field models. A direct Mpemba effect of this type was nevertheless demonstrated through a phenomenological Landau free energy function, shown in \FIG{SecondOrderPT_1}, where the temperature dependence was constructed such that a direct Mpemba effect through a second-order phase transition exists.    

\subsection{Discussion}
Although some progress has been made in understanding Mpemba effects through phase transitions in the last few years, and even though many of the experimental observations of Mpemba effects are, in fact, through a phase transition, many questions remain open: 
(i) Are these effects related to any of the other Mpemba effects, e.g., final relaxation effects? (ii) Is there a theory for such effects beyond the mean-field approximation? (iii) Are these theoretical suggestions relevant to any of the corresponding experimental observations? (iv) Is there a simple, physical picture for systems that have such effects?  

\section{Kinetic theory: granular and molecular gases}
\label{sec:kinetic-framework}
So far, we have discussed two frameworks that are relevant to Mpemba effects: the Markovian framework and Mpemba effects through phase transitions. Here we present a third framework, driven granular and molecular gases, which has been used to develop Mpemba-related concepts in parallel to the Markovian framework. Because the Markovian framework discussed in Sec.~\ref{sec:markovian} requires knowledge of all details of the microstates and the transition rates connecting them, it is useful primarily for small or few-body systems. The thermodynamic limit can be taken in only a very few cases, such as 1D or mean-field systems. The phase-transition analysis in Sec.~\ref{sec:phase-transitions}, on the other hand, relies on the dynamics of a macroscopic system away from equilibrium and thus has been applied so far only to mean-field models. The driven-granular-gas framework offers a different yet complementary approach that allows analytic study of many-body systems in the thermodynamic limit. 
An important difference from the Markovian systems of~\SEC{markovian} and the phase transitions of~\SEC{phase-transitions} is that granular gases are inherently out of equilibrium and need the injection of energy to sustain a fluidized state, implying that stationary states are nonequilibrium steady states.

Anomalous relaxation effects can be expected to be found in granular gases, for several reasons~\cite{patron2023non}. Such effects commonly appear in complex systems that consist of many (identical or different) structural units, for which a continuum description is appropriate in some limits. Within this kind of description, it is often assumed that the instantaneous value of the complete set of macroscopic variables (pressure, temperature, density, etc.) fully characterizes the system's temporal evolution. This is indeed the case in simple systems such as an ideal gas, but in more complicated systems, relaxation processes often require additional variables to be tracked. 
It is precisely in these nonequilibrium scenarios that the equilibrium macroscopic description fails, and that failure is responsible for anomalous relaxation effects.

The kinetic theory of gases offers a natural framework for studying these phenomena. This well-developed and established theory, in turn, offers insights into the emergence of anomalous relaxation effects. In the illustrative example of a granular gas of rough spheres \cite{torrente2019large,megias2022mpemba} (see below), the kinetic temperature is characterized by the average of translational and rotational kinetic energies of the spheres.
However, when an external parameter of such a system is quenched to a different value, the relaxation times of these velocities can be drastically different, and the energy transfer between the angular and linear velocity due to collisions is highly nonlinear.
This nonlinearity can generate anomalous relaxation phenomena. 

Two different mechanisms for the emergence of anomalous effects in granular gases have been studied.
The first leverages the coupling between the non-Gaussian part of the velocity distribution and the kinetic temperature \cite{lasanta2017hotter,santos2020mpemba,patron2023non,megias2022thermal,takada2021mpemba,mompo2021memory}. This mechanism works even with smooth spheres, where the angular velocities do not play any role \cite{lasanta2017hotter}. The second mechanism exploits the energy exchange between linear and angular components of the kinetic energy~\cite{gijon2019paths,torrente2019large,megias2022mpemba}. Prototypical examples include several models of granular spheres, hard and viscoelastic. The latter include Maxwell models with and without external driving.  Another example is rough, hard spheres with translational driving. Others include disks and also rotational driving. Finally, there are molecular gases, with ingredients such as drag force, viscosity, and even mixtures of particles. In the context of anomalous relaxations, the main difference between granular and molecular fluids is that, in the former, the Mpemba effect can emerge naturally, as granular fluids are intrinsically out of equilibrium. Namely, even when granular gases evolve freely, the Mpemba effect can appear.  The dissipative interactions between material particles need an injection of energy to sustain a fluidized state. Note that even in a homogeneous cooling state of granular fluids \cite{brey1996homogeneous}, the Mpemba can be observed. In contrast, in molecular gases, the interactions and collisions are elastic (conserve energy), and the system must be driven out of equilibrium by introducing additional ingredients, namely, external nonlinear drag \cite{santos2020mpemba}, another species \cite{gomez2021mpemba}, or immersing it in a surrounding fluid \cite{santos2020mpemba,megias2022thermal}. 

In what follows, we introduce the kinetic theory of granular gases and provide a few examples of how this framework leads to anomalous relaxation phenomena. We also present examples of molecular gases where analogous effects can be reproduced. 

\subsection{The Boltzmann-Fokker-Planck equation and hydrodynamics} 
\label{Theokinetic}

The kinetic theory of gases is a classical approach to modeling dilute gases \cite{brush1965kinetic}. In this theory, the system is composed of an ensemble of particles interacting through binary collisions obeying the molecular-chaos hypothesis. The Boltzmann-Fokker-Planck equation for the homogeneous states of an ensemble of hard spheres is
\begin{align}
    \partial_t f(\vec{v} , \vec{w};t) -\nabla _{\vec{v}} \cdot\left( \frac{\vec{F}(\vec{v})}{m}+\frac{\chi_0^2}{2}\nabla _{\vec{v}}\right) f(\vec{v} , \vec{w};t)={\mathcal{J}[\vec{v}, \vec{w}|f(t)]},
\label{GBE}
\end{align}
where
\begin{itemize}
    \item $f(\vec{v}, \vec{w};t)$ is the one-body velocity distribution function, with $\vec{v}$ and $ \vec{w}$ the translational and angular velocities, respectively;
    \item $\mathcal{J}[\vec{v}, \vec{w}|f(t)]$ is the binary collision operator (or kernel) \cite{brush1965kinetic,cercignani1988boltzmann,puglisi2014transport}. It characterizes the properties of the interaction in the specific model, such as the smoothness of the spheres, which ranges from smooth to rough, and the elasticity of their collisions, ranging from elastic to inelastic; 
    \item $\vec{F}(\vec{v})$ is a velocity-dependent force and $m$ the mass of particles;
    \item the term  $\frac{\chi_0^2}{2}\nabla^2_{\vec v}$ represents the influence of a stochastic homogeneous volume force, which corresponds to white noise.
\end{itemize}  
The stochastic force results from an external stochastic driving that compensates for the energy that is lost in each collision and thereby avoids relaxation towards zero temperature. 
It represents the influence of a stochastic, white noise  $\vec{F}^{\mathrm{wn}}$ that acts on each particle, and it satisfies $\langle \vec{ F}_{i}^{\mathrm{wn}} (t) \rangle = \vec{0}$ and $\langle F_{i,q} ^{\mathrm{wn}} (t) F_{j,s} ^{\mathrm{wn}} (t') \rangle = m^2 \chi_0^2\delta_{i\!j} \delta_{q,s}\delta(t-t')$, where indices $i,j$ refer to particle numbers and the indices $q,s\in\{x,y,z\}$ to the spatial coordinates.
Note that, for simplicity, we assume a stochastic force but no stochastic torque; thus, there is no diffusion term for the angular velocity.
Additional rotational driving can be considered \cite{megias2022mpemba}, but translational driving illustrates better the situation encountered in most experimental systems and physical scenarios, in which the stochastic force tries to mimic the injection of energy produced in granular gases immersed in thermal fluids or a fluidization produced by vibrations. 

Equation~\eqref{GBE} describes the evolution of the single-particle velocity distribution for a collection of identical spheres at low particle density.
The density is assumed constant, so that the collision rate is independent of the particle position.
Additionally, Eq.~\eqref{GBE} assumes the molecular chaos hypothesis and that all collisions are binary and instantaneous.
An important observation is that while this equation shares many features with the Fokker-Planck Eq.~\eqref{eq:FokkerPlanck}, there is a fundamental difference: Eq.~\eqref{GBE} describes the evolution in time of a \textit{velocity} distribution $f(\vec v, \vec w;t)$, regulating the probability of finding a particle at time $t$ with velocities $\vec v$ and $\vec w$, whereas Eq.~\eqref{eq:FokkerPlanck} is an evolution equation for the probability to find the system in some specific microscopic configuration $\vec x$.
An additional key difference between the two descriptions lies in the physical mechanisms that can enable the emergence of anomalous relaxation effects.
On the one hand, in the Markovian framework, the potential $U(\vec x)$ in Eq.~\eqref{eq:FokkerPlanck} can be arbitrarily complicated (Fig.~\ref{fig:brownian_system}a), and the roughness of the potential is ultimately responsible for much of the anomalous effects that one observes in these setups.
On the other hand, in Eq.~\eqref{GBE}, anomalous effects can be caused by a nonlinear, drag-velocity-dependent force  $\vec F(\vec v)$ in molecular gases. They can also be caused by a shear force and/or the inelasticity of collisions in granular gases (expressed in $\mathcal{J}[\vec{v}, \vec{w}|f(t)]$) that produces a non-Gaussian distribution function that couples the kinetic temperature with higher-order moments. 

To keep the derivation simple, we analyze a relatively easy case, where there is no deterministic force, and the energy lost in each collision is compensated only by the stochastic homogeneous volume force. This simplification gives a Boltzmann-Fokker-Planck equation of the form 
\begin{align}
    \left( \partial_t - \frac{\chi_0^2}{2} \nabla _{\vec{v}}^2\right) f(\vec{v} , \vec{w};t) = \mathcal{J}[\vec{v}, \vec{w}|f(t)].
\label{BE}
\end{align}
Using the above equation for the evolution of the single-particle velocity probability distribution, we can calculate the average over particles for any single-particle function  $A(\vec{v},\vec{w})$ of the particle velocity $\vec v$ and angular velocity $\vec{w}$,  at time $t$,
\begin{align}
    \langle A(t)\rangle=n^{-1}\int \dd \vec{v}\int \dd \vec{w}\,
    A( \vec{v}, \vec{w})f(\vec{v}, \vec{w};t),
\end{align}
where the number density $n$ is given by
\begin{align}
    n=\int \dd \vec{v}\int \dd \vec{w}\, f( \vec{v}, \vec{w};t),
\end{align}
and $f(\vec{v}, \vec{w};t)$ follows the evolution of Eq. (\ref{BE}) with suitable initial conditions.
The basic physical properties of interest in the context of Mpemba and related phenomena are the translational, $T_t$, and rotational, $T_r$, granular temperatures, defined as
\begin{align}
\label{eq:Tt,Tr}
    &T_t={\frac{m}{3}}\left\langle v^2\right\rangle,\quad T_r={\frac{I}{3}}\left\langle
    w^2\right\rangle, 
\end{align}
where $I$ is the moment of inertia of the gas spheres. It is convenient and useful for the analysis that follows to also define the temperature ratio $\theta$ and total granular temperature $T$, which is the mean total kinetic energy per particle,
\begin{align}
\label{eq:T,theta}
    &\theta\equiv \frac{T_r}{T_t}, \quad T \equiv \frac{T_t+T_r}{2}= T_t \frac{1+\theta}{2}.  
\end{align}

\paragraph{Rescaling to dimensionless variables}
It is often convenient to change to dimensionless variables, scaled by the (time-dependent) temperatures, as this procedure removes most of the time dependencies. To this end, we first define the instantaneous collision frequency
\begin{align}
   \nu(t)=4n\sigma^2 \sqrt{\pi T_t(t)/m},
\end{align}
where $\sigma$ is the diameter of a gas sphere. Using this frequency, we rescale time as
\begin{align}
  \tau=\frac{1}{2}\int_0^t \dd t'\,\nu(t'),
\end{align}
which roughly accounts for the accumulated number of collisions per particle. In the case of thermostatted systems, time can be rescaled with the stationary temperature $T_s$,where $T_s$ corresponds to the temperature the system reaches when dissipation and injection of energy are balanced; see Sec.~\ref{SGHS} and \ref{Sec:BreakingEP}. Dimensionless velocities then follow
\begin{align}
  \vec{c}(t)\equiv\frac{\vec{v}}{\sqrt{2 T_t(t)/m}}, \quad
  \vec{j}(t)\equiv\frac{{\vec{w}}}{\sqrt{2 T_r(t)/I}},
\label{cw}
\end{align}
and the corresponding reduced  velocity distribution function is
\begin{align}
\phi(\vec{c},\vec{j};\tau)\equiv \frac{1}{n}\left[\frac{4 T_t(t)
    T_r(t)}{m I}\right]^{3/2}f(\vec{v}, {\vec{w}};t).
\end{align}
Finally, we write the dimensionless collision kernel
\begin{align}
  \label{Cop}
\mathcal{J}[\vec{c},
\vec{j}|\phi(\tau)]=\frac{2}{n\nu(t)}\left[\frac{4 T_t(t) T_r(t)}{m I}\right]^{3/2} \mathcal{J}[\vec{v}, {\vec{w}}|f(t)].
\end{align}

Note that rescaling variables does not affect the phenomenology of the particular model of interest and, therefore, the existence of the Mpemba effect in molecular and granular systems.  

\paragraph{Expansion Around a Maxwellian Distribution}
Our goal is to use dimensionless variables and write evolution equations for the various temperatures defined above, to characterize crossings and hence anomalous relaxation effects.
To achieve this goal, we formally expand the exact solution to the Boltzmann-Fokker-Planck equation Eq.~\eqref{BE} around a Maxwellian distribution, which here is a Gaussian distribution with variances $T_t$ and $T_r$ for the linear and angular velocities, respectively~\cite{vega2015steady}.
Explicitly, we write
\begin{align}
    \phi_M(c,j) =n \pi^{-3}e^{-c^2-j^2}.
\end{align}
Note that with our dimensionless velocities $\vec c$ and $\vec v$, the variances are normalized by the temperatures. We then expand the dimensionless velocity distribution as 
\begin{align}
\label{ap:phi_sonine}
    \phi(\vec{c},\vec{j};\tau) = \phi_M(c,j)\sum_{i=0}^\infty \sum_{k=0}^\infty \sum_{\ell=0}^\infty a_{ik}^{(\ell)}(\tau)\Psi_{ik}^{(\ell)}\left( \vec{c},\vec{j} \right),
\end{align}
where $\Psi_{jk}^{(\ell)}(\vec{c},\vec{j})$ are products of Laguerre and Legendre polynomials \cite{resibois1977classical}.
The coefficients are given by 
\begin{align}
a_{ik}^{(\ell)}(\tau) = \frac{1}{N_{ik}^{(\ell)}}\left\langle \Psi_{ik}^{(\ell)}\left(\vec{c},\vec{j}\, \right)\right\rangle,
\end{align}
where  $N_{ik}^{(\ell)}$ is a normalization constant, and the averaging is done with the time-dependent distribution $\phi(\vec c, \vec j; \tau)$. It is useful to express $a_{ik}^{(\ell)}$ through the velocity moments as
\begin{align}
M_{pq}^{(r)}(\tau) \equiv  \langle c^p j^q (\vec{c} \cdot \vec{j})^r \rangle = \int \dd \vec{c} \int \dd \vec{j}\  c^p j^q (\vec{c} \cdot \vec{j})^r \phi(\vec{c},\vec{j};\tau). 
\end{align}
Lastly, we introduce the reduced collisional moments, which are velocity moments calculated using the collision kernel rather than the distribution itself, as 
\begin{align}
  \label{mupq}
\mu_{pq}^{(l)}(\tau)\equiv-\int \dd\vec{c}\int \dd\vec{j}\,
c^pj^q
(\vec{c}\cdot\vec{j})^l
{\mathcal{J}[\vec{c},\vec{j}|\phi(\tau)]}.
\end{align}
With the above definitions and some algebra, we write evolution equations for the temperatures defined in~\EQ{Tt,Tr},
\begin{align}
\label{evol_gamma}
    \frac{\partial \ln\theta(\tau)}{\partial \tau}&=\frac{2}{3}\left[\mu_{20}^{(0)}(\tau)-\mu_{02}^{(0)}(\tau)-\gamma(\tau)\right],
\\
\label{evol_gamma2}
    \frac{\partial\ln\gamma(\tau)}{\partial \tau}&=\mu_{20}^{(0)}(\tau)-\gamma(\tau),
\end{align}
where 
\begin{align}
\label{eq:gamma_granular}
  \gamma \equiv \left(\frac{T_{\mathrm{noise}}}{T}\frac{1+\theta}{2}\right)^{3/2} \quad \text{and} \quad
  T_{\mathrm{noise}}\equiv m\left(3\chi_0^2/4\sqrt{\pi}n \sigma^2\right)^{2/3}
\end{align}
are a dimensionless measure of noise intensity and the temperature associated to a given noise intensity respectively.

Given the temporal evolution of the temperature, we can define various anomalous relaxation phenomena: If, in two relaxation processes with different initial conditions but identical steady state, the corresponding curves of $T(\tau)$ cross, then there is a direct/inverse Mpemba effect. The specific type of the effect depends on the relaxation process -- heating or cooling. The Mpemba effect has been studied using numerical, analytical techniques and two simulation methods detailed in the numerical observation; see  Sec.~\ref{Sec:NumericalKinetic}. 

Note that the evolution of the temperature ratio $\partial_\tau \theta(\tau)$ cannot be expressed by a closed equation, since $\theta(\tau)$ is also a function of the reduced velocity distribution $\phi(\vec c,\vec j;\tau)$ through the collisional moments $\mu_{20}^{(0)}$ and $\mu_{02}^{(0)}$ defined in Eq.~\eqref{mupq}. This explains why two trajectories of $\theta(\tau)$ can cross. This also implies that the above evolution equations are not closed: in principle, one has to know $\phi(\vec c,\vec j;\tau)$ in order to use them. In order to close Eqs.~(\ref{evol_gamma},\ref{evol_gamma2}), we use the \textit{first Sonine approximation} \cite{vega2015steady}, which is the first nontrivial truncation of the aforementioned exact infinite expansion. To this end, we need to incorporate the evolution equations for $\mu^{(0)}_{02}$ and $\mu^{(0)}_{20}$ with Eqs.~\eqref{evol_gamma} and \eqref{evol_gamma2}. 

To evaluate the evolution of $\mu_{02}^{(0)}$ and $\mu_{20}^{(0)}$, we go back to the Boltzmann-Fokker-Planck equation, Eq.~\eqref{BE}. Multiplying both sides of the equation by $c^p j^q (\vec{c} \cdot \vec{j})^r$ and integrating over $\vec{c}$ and $\vec{j}$, we obtain the equation for the moment hierarchy,
\begin{align}
    \frac{\partial \ln M_{pq}^{(r)}(\tau)}{\partial \tau}- \frac{\gamma(\tau)}{6} \frac{p (p+1+2r) M_{p-2,q}^{(r)} (\tau)+r(r-1)M_{p,q+2}^{(r-2)}}{M_{pq}^{(r)}} =\frac{p+r}{3} \left[\mu_{20}^{(0)}(\tau)-\gamma(\tau) \right]+ \frac{q+r}{3}\mu_{02}^{(0)} - \frac{\mu_{pq}^{(r)}}{M_{pq}^{(r)}} .
\label{MoBE}
\end{align}
In the above equation, the evolution of $M_{pq}^{(r)}$ is a function of $M_{pq}^{(r)}$ and $\mu_{pq}^{(r)}$, but also of $M_{p-2,q}^{(r)}$ and $M_{p,q+2}^{(r-2)}$. This hierarchical set of equations is truncated at fourth order. By normalization, $a_{00}^{(0)}=1$, $a_{10}^{(0)}=a_{01}^{(0)}=0$, and the lowest nontrivial coefficients are  those associated with moments of  degree four, namely
\begin{align}
\label{cumulant}
    a_{20}^{(0)}=\frac{4}{15}\langle c^4\rangle-1,
\quad
a_{02}^{(0)}=\frac{4}{15}\langle j^4\rangle-1,
\end{align}
\begin{align}
\label{a20}
    a_{11}^{(0)}=\frac{4}{9}\langle c^2j^2\rangle-1,\quad
a_{00}^{(1)}=\frac{8}{15}\left[\langle
  (\vec{c}\cdot\vec{j})^2\rangle-\frac{1}{3}\langle
  c^2j^2\rangle\right],
\end{align}
which we call the fourth-order cumulants.

\begin{figure}
    \centering
    \includegraphics[width=0.7\columnwidth]{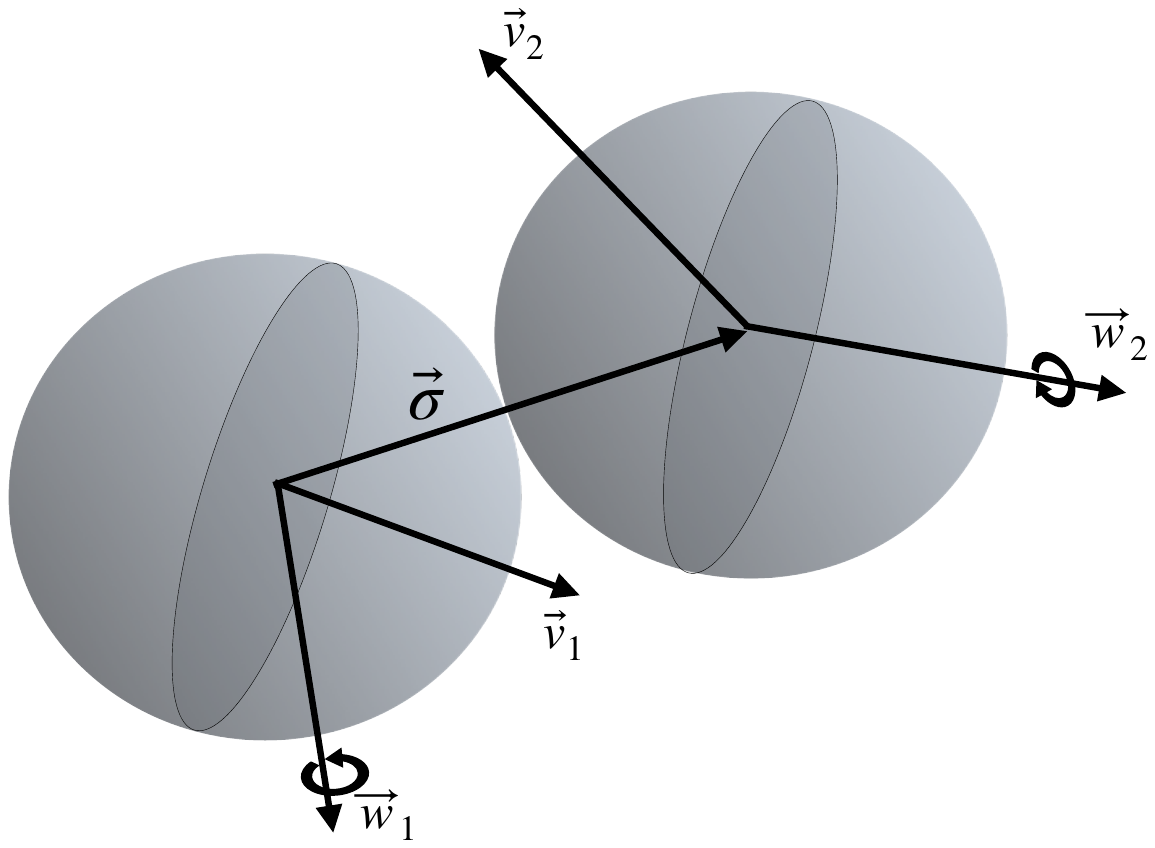}
    \caption{Sketch of a collision between two inelastic rough hard spheres. }
    \label{fig:coll_sketch}
\end{figure}

\paragraph{The collision kernel}
Our discussion so far did not consider any specific property of the collision itself.  These properties were all encapsulated into the collision kernel $\mathcal{J}[\vec{v}, \vec{w}|f]$, or equivalently the dimensionless kernel $\mathcal{J}[\vec{c}, \vec{j}|\phi]$. Therefore, we next consider the collision integral for inelastic rough granular gas particles, which expresses the effect of collisions on the time evolution of the one-particle distribution function. A general collision model for rough granular spheres with constant coefficients is characterized by normal and tangential restitution coefficients $\alpha$ and $\beta$, respectively. This is a good approximation for many fluids at low particle density. Figure~\ref{fig:coll_sketch} sketches the geometry of a collision: two particles enter the collision with initial velocities $\vec v_1$, $\vec v_2$ and angular velocities $\vec w_1$, $\vec w_2$. The vector $\vec\sigma$ that connects their centers has magnitude $\sigma$ equal to the diameter of one sphere. The collision rules (conservation of momentum and reduction of energy) with the linear and angular coefficients of restitution imply that
\begin{align}    
    \vec{\sigma}\cdot\vec{u}'=-\alpha\,{\vec{\sigma}}\cdot\vec{u},
\qquad \vec{\sigma}\times\vec{u}'=-\beta \,
    \vec{\sigma}\times\vec{u},
\label{collrule}
\end{align}
where primes denote post-collision values and $\vec{u}=\vec{v}_1-\vec{v}_2-{\frac{1}{2}}{ \vec{\sigma}}\times({ \vec{w}}_1+{ \vec{w}}_2)$
is the relative velocity of the spheres at their contact point during a binary collision. The coefficient of normal restitution $\alpha$ takes values between $0$ (completely inelastic collision) and $1$ (completely elastic collision), while the coefficient of tangential restitution $\beta$ takes values between $-1$ (completely smooth collision, unchanged angular velocities) and $+1$ (completely rough
collision).

The explicit expression of the collision terms in this case is given by
\begin{align}
\label{coll_ter_expli}
    \mathcal{J}[\vec{v}_1, \vec{w}_1|f(t)]=\sigma^2 \!\!
 \int_{\vec{\sigma} \cdot \vec{u}>0}\! \dd\vec{\sigma}\ \dd\vec{w}_2 \  \dd\vec{v}_2 \ (\vec{\sigma} \cdot \vec{u}) \left(\frac{f(\vec{v}^*_1, \vec{w}^*_1,t) f(\vec{v}^*_2, \vec{w}^*_2,t)}{\alpha^2 \beta^2} -f(\vec{v}_1, \vec{w}_1,t) f(\vec{v}_2, \vec{w}_2,t) \right),
\end{align}
where the asterisks ($\vec{v}^*, \vec{w}^*$)  indicate precollisional velocities, namely, the velocities that after collision give as results ($\vec{v}, \vec{w}$). Note that in the case of molecular gases, the collision kernel is the same, but $\alpha$ must be replaced by $1$ and $\beta$ by $\pm 1$, which corresponds to the case of elastic collisions.

Using the setup defined in the above section, we next consider several specific cases, which shed some light on the specific mechanism that enables anomalous relaxation phenomena in granular gases.

\subsection{Example: Granular smooth hard spheres}
\label{SGHS}
The simplest case of a dilute granular gas is a system composed of $N$ smooth, hard spheres of mass $m$ and diameter $\sigma$. We can analyze it within the framework presented above theoretically and numerically by means of molecular dynamics (MD) and direct simulation Monte Carlo (DSMC) methods (see Sec. ~\ref{Sec:NumericalKinetic}).
Here, ``hard'' means that the particles can interact only through collisions; and when they collide, momentum is conserved, while part of the kinetic energy is lost, as quantified by the \textit{restitution coefficient} $0<\alpha<1$.
By ``smooth," we mean that there is no friction between the spheres during the collision.  The angular velocity of each particle is then conserved and does not affect the linear velocities.
This feature is accounted by the \textit{angular restitution} coefficient $\beta=-1$.
In this simple scenario, where the angular velocities play no role, it is less convenient to use the dimensionless parameters introduced in the previous section.
We thus introduce a single relevant granular temperature $T$, which corresponds to $T_t$ in the notation of Sec. \ref{Theokinetic}.
In addition, moments that in general require three indices here need only a single index.  Thus, $a_{i,j}^{(k)}$ is written as $a_i$, where $i$ refers to the order of the linear velocity. Nevertheless, the results outlined in the previous section still apply to the smooth, hard-sphere scenario.

The collision rule for this model is that the component of the relative velocity along the line joining the centers of the two colliding particles is reversed and shrunk by a factor $\alpha$, the coefficient of normal restitution.
In other words, the collision rule of Eq.~\eqref{collrule} needs just the first condition, 
\begin{align}
    \vec{\sigma}\cdot\vec{u}'=-\alpha\,\vec{\sigma}\cdot\vec{u}.
\end{align}
As outlined in the previous section, a stationary, finite-temperature state is maintained because particles are subject to an independent white-noise force with variance $m^{2}\chi_0^{2}$.  (We restrict ourselves to homogeneous situations.)
Therefore, the velocity distribution function reduces to $f(\vec{v},t)$ and obeys a Boltzmann-Fokker-Planck kinetic equation
\cite{van1998velocity,brey1998hydrodynamics}.

The time evolution of the granular temperature $T$ is coupled to that of the excess kurtosis, also referred to as the second Sonine coefficient.
Specifically, $a_2$ which is the first Sonine correction in Eq.~\eqref{ap:phi_sonine} to the Maxwellian distribution, simplifies dramatically to  $a_{2}=\frac{3}{5} \langle v^{4} \rangle / \langle v^{2} \rangle^{2}-1$.
From the kinetic equation for the moments of velocity distribution function, Eq.~\eqref{MoBE}, one finds
\begin{subequations}
\label{evol-eqs}
\bal
\dv{T}{t}=& -\frac{2\kappa}{3}\left(\mu_{2}T^{3/2}-\chi\right),\\
\dv{\ln(1+a_{2})}{t}=& \frac{4\kappa}{3T} \left(\mu_{2}T^{3/2}-\chi-\frac{\frac{1}{5}\mu_{4}T^{3/2}-\chi}{1+a_{2}}\right),
\eal
\end{subequations}
where $\kappa=2ng(n)\sigma^{2}\sqrt{\pi/m}$, with $g(n)$ being the pair correlation function at contact \cite{carnahan1969equation},  $\chi=\frac{3}{2}m\chi_0^{2}/\nu$, and $\mu_{2}$ and $\mu_{4}$ are the dimensionless collisional moments defined in Eq.~\eqref{mupq}, with $q=l=0$.
As in the general case discussed in the previous section, Eqs.~\eqref{evol-eqs} are exact but not closed.
To close them, we consider the first Sonine approximation for $\mu_{2}$ and $\mu_{4}$, i.e.,
$\mu_{n}\simeq \,^0\mu_{n}+ \,^{1}\mu_{n} a_{2}$ (here, the left upper index represents the order of the perturbation), with 
$\,^0 \mu_2 =1-\alpha^2$, $\,^1 \mu_2=\frac{3}{16} \,^0 \mu_2$,
 $\,^0 \mu_4=\left(\frac{9}{2}+\alpha^2\right) \,^0 \mu_2$, and $\,^1 \mu_4=  (1+\alpha)\left[2+\frac{3}{32}(69+10\alpha^2)(1-\alpha)\right]$. 

In the long-time limit, when the average energy loss in collisions is balanced by the average energy input from the stochastic force, the granular gas reaches a steady state.
The stationary values of the temperature and the excess kurtosis can be obtained by the time-independent solution of Eqs.~\eqref{evol-eqs}, with the corresponding results from the first Sonine approximation and neglecting the nonlinear terms.
These steady state values are given by $T_s = \left(\chi/\mu_{2}^{s}\right)^{2/3}$ and $a_2^s=16(1-\alpha)(1-2\alpha^2)/[241-177\alpha+30\alpha^{2}(1-\alpha)]$, while the corresponding steady state collision moments are given by $\mu_{n}^{s}=\,^0 \mu_{n}+ \, ^1 \mu_{n} a_{2}^{s}$.
At this level of approximation, one can close Eqs.~\eqref{evol-eqs} and linearize them around the stationary state.
Defining $\tilde T(\tau) \equiv T(\tau)/T_s$ with $\tau=\kappa \sqrt{T_s} t$ and the linearized variables $\delta \tilde T \equiv \tilde T -1$ and $\delta a_2 \equiv a_2- a_2 ^s$, we express the linearized Eqs.~\eqref{evol-eqs} as
\begin{align}
\label{evol-eqs-linear}
   \dv{ }{\tau}
\begin{pmatrix} \delta \tilde T\\ \delta
     a_{2}
\end{pmatrix}=-
\begin{pmatrix}
    \mu^{st}_2 & \frac{2}{3} \,^1 \mu_2 \\
    -2 \mu^{st}_2 a_2^{st} & \frac{4}{15} [\,^1 \mu_4 -5 \,^1\mu_2 (1+a_2^{st}) ]
\end{pmatrix}
\cdot
\begin{pmatrix} 
    \delta \tilde T\\  
    \delta a_{2}
\end{pmatrix},
\end{align}

\begin{figure}
    \centering
    \includegraphics[width=0.8\columnwidth]{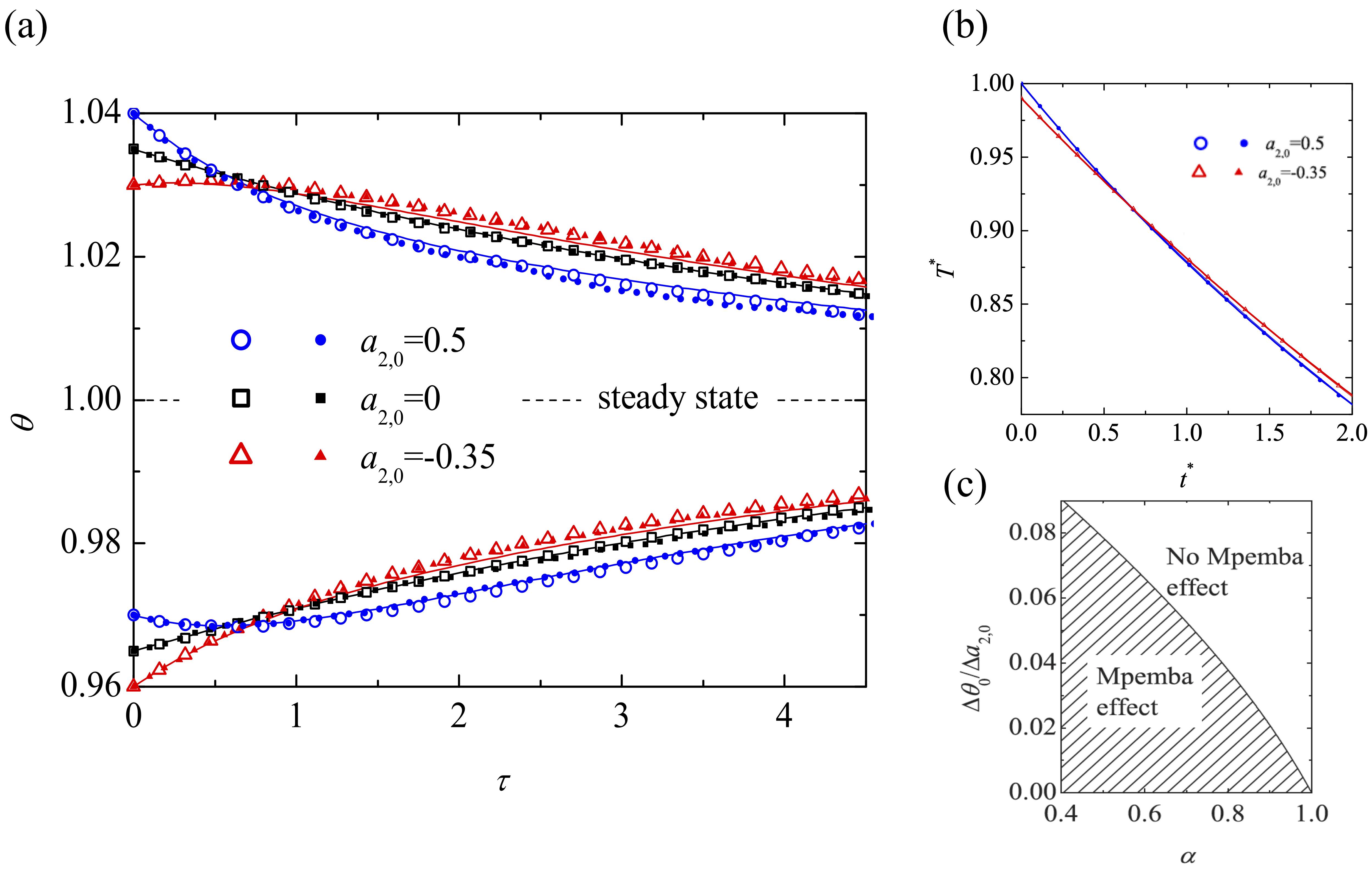}
    \caption{(a) Example of Mpemba effect and its inverse in a granular gas of smooth, hard spheres and homogeneous cooling state  (b) Results from two simulation methods, DSMC (open symbols) and MD (filled symbols) data show excellent agreement with the theoretical prediction (lines); see (\ref{Sec:NumericalKinetic}). (c) Phase diagram of  $\alpha$  and initial conditions. \textit{Source:} Reprinted with permission from~\cite{lasanta2017hotter}.
    }
    \label{fig:mpembasmooth}
\end{figure}
The coupled equations for the temperature $\tilde{T}$ and $a_2$ can produce several types of nonmonotonic relaxation phenomena, including the direct, the inverse Mpemba effects and others \cite{lasanta2017hotter,prados2010the,santos2020mpemba,patron2023non,mompo2021memory}.
By choosing the appropriate initial conditions for $a_2$ (initial state kurtosis), one can speed up or slow down the transient relaxation of a granular gas temperature for a given value of $T$.
This is demonstrated in Fig.~\ref{fig:mpembasmooth}c, which shows the occurrence of the Mpemba effect as a function of the inelasticity parameter $\alpha$ and on the initial kurtosis and temperature.
Note that, in this example, the effect is small in the range of possible choices of initial temperatures. Also, the initial temperatures in Fig.~\ref{fig:mpembasmooth}a have a maximum difference of about $0.01 \% $.
Of particular interest is Fig.~\ref{fig:mpembasmooth} b), where a non-thermostated granular gas is considered. This ``homogeneous cooling state'' relaxes indefinitely towards the zero-temperature state, with $\chi_0=0$ continues to cool.
In the homogeneous cooling state, the Mpemba effect is even smaller than in the heated case, but nevertheless, it exists; see  Fig. \ref{fig:mpembasmooth} b).
Note that, in contrast to other cases, the homogeneous cooling state has no natural scale for temperature. In this system, one thus uses an arbitrary reference temperature $T_\mathrm{ref}$.  The rescaled variables are $T^*=T/T_\mathrm{ref}$ and $t^*=\kappa \sqrt{T_\mathrm{ref}} \, t$.

Equation~\eqref{evol-eqs-linear} shares strong similarities with the Markov framework discussed in Sec.~\ref{sec:markovian}, and one can indeed identify the existence of anomalous relaxations by analyzing the eigenvalues and eigenvectors of the corresponding matrix.
Nevertheless, there are profound differences between the nature of the Mpemba effect in granular gases and in a general Markovian system.
One such difference is that temperature in Eq.~\eqref{evol-eqs-linear} is both one of the evolving quantities and also determines the existence of anomalous effect. By contrast, in the Markovian framework, the existence of the effect is encoded in the projection of the initial condition onto the second-slowest left eigenvector of the rate matrix (see, e.g., Eq.~\eqref{eq:eigenfunc_continuous}).
An additional distinction is that in granular gases the initial conditions are set by choosing both $T_0$ and $a_{2,0}$, so that both evolve in time towards a steady state: This follows from the fact that in granular gases the experimentally controlled parameters are $\alpha$ and $\chi$, not $T$ and $a_2$. For smooth granular gases, it is therefore natural to ask whether one can tune the external parameters ($\chi$ and $\alpha$) to quench the temperature while leaving $a_2$ unchanged, and nevertheless to observe anomalous relaxation.
Figure~\ref{fig:mpembasmooth} c) directly answers this question: it shows that the quotient $\Delta \theta_0 / \Delta a_{2,0}$ must remain inside the dashed area in order to observe the effect in the system discussed above.
Therefore, in the specific model discussed here, when $a_{2,0}$ is kept fixed and only the steady-state temperature varies, the effect cannot appear (this can be seen in Fig. \ref{fig:mpembasmooth} a) for $\Delta a_{2,0}=0$).
Note that the same happens in the viscoelastic model Sec.~\ref{subsub:othergranular}; see also Fig. \ref{fig:mpembavisco} b). This, however, is not a general property of granular gases. As discussed below, it can also happen in other models.

\subsection{Example: Granular breaking equipartition}\label{Sec:BreakingEP}

In the smooth-sphere granular gases discussed above (Sec.~\ref{SGHS}), the kurtosis is typically small, a feature that facilitates the theoretical characterization of the effect.
Indeed, it allows perturbative approximations that lead to a quantitative estimate of the initial kurtosis required to reproduce a Mpemba effect.
However, a small initial kurtosis limits the range of the parameters of the effect itself: The two initial temperatures must be relatively close in order to observe the effect; see  Fig.~\ref{fig:mpembasmooth}b. 

In a granular gas of \emph{rough} spheres, the difference in initial temperatures that allows us to observe the Mpemba effect can be much larger~\cite{torrente2019large}. 
As we discuss below, the mechanism that makes it larger is the coupling between the translational and rotational temperatures, which are of the same order of magnitude.
This is related to the fact that granular gases do not reach an equilibrium state but rather a nonequilibrium steady state.
Rough granular gases then do not follow the equilibrium equipartition theorem, and the energy in each degree of freedom depends on the values of the restitution coefficients, $\alpha$ and $\beta$.
In addition, the basic features of the effect can be understood within the Maxwellian (Gaussian) approximation without resorting to higher-order cumulants.
A similar reasoning was considered in  \cite{megias2022mpemba} for rough disks.

The simplest description of the rough-sphere granular gas is provided by the Maxwellian approximation, which assumes a bivariate Gaussian distribution for the linear and angular velocities,
\begin{align}
f(\vec{v}, {\vec{w}};t)=\frac{1}{n}\left[\frac{4 T_t(t)
    T_r(t)}{m I}\right]^{3/2}\exp\left[- \frac{m \vec{v}^2
    }{2T_t(t)}-\frac{I \vec{w}^2
    }{2T_r(t)} \right]. 
\end{align}
This description is obtained by substituting a Maxwellian velocity distribution into the collision integrals in Eq. (\ref{mupq}). Equivalently, one may require that all the nontrivial cumulants vanish in this approach, which yields
\begin{align} 
\label{mu2002M}
    \mu_{20,M}&=1-\alpha^2+\frac{\iota(1+\beta)}{(1+\iota)^2}\left[2+\iota(1-\beta)-\theta(1+\beta)\right],
\\
    \mu_{02,M}&=\frac{\iota(1+\beta)}{(1+\iota)^2}\left[2+\kappa^{-1}(1-\beta)-\theta^{-1}(1+\beta)\right], 
\end{align}
where $M$ indicates the Maxwellian approximation and $\iota\equiv 4I/m\sigma^2$ is the dimensionless moment of inertia.
Substituting Eq. (\ref{mu2002M}) into the evolution equations (\ref{evol_gamma}) and  (\ref{evol_gamma2}) provides us with the Maxwellian approximation. 

As explained more in detail in Sec.~\ref{Theokinetic}, in the long-time limit, the granular gas reaches a steady state that balances the action of the stochastic force against energy loss in collisions.
For non-smooth granular gases, in which angular velocity plays an important role, this steady state is characterized by the temperature ratio $\theta^{\rm st}$ and total granular temperature $T^{\text{st}}$ in the Maxwellian approximation; see Eq.~\eqref{eq:T,theta}.
In terms of the coefficients of restitution and the stochastic forcing intensity, the temperature ratio and the total granular temperature are, see \cite{vega2015steady},  
\begin{subequations}
\label{eq:theta-gamma-st}
\begin{align}
    \theta^{\text{st}}=&\frac{1+\beta}{2+\iota^{-1}(1-\beta)}, \quad T^{\rm st}=\frac{1+\theta^{\text{st}}}{2}\left(\frac{3m^{3/2}\chi_{0}^{2}}{4\sqrt{\pi}n\sigma^{2}\gamma^{\text{st}}}\right)^{2/3}, \\
    \gamma^{\text{st}}\equiv& 1-\alpha^{2}+\frac{2(1-\beta^{2})}{2+\iota^{-1}(1-\beta)}.
\end{align}
\end{subequations}
where $\gamma^{\text{st}}$ is the dimensionless measure of noise intensity at stationarity; see Eq.~\eqref{eq:gamma_granular}.
Note that $\theta^{\text{st}}\leq 1$ is independent of $\alpha$ in the Maxwellian approximation. Higher-order approximations introduce a weak dependence on $\alpha$; see~\cite{torrente2019large}.

As outlined in the general framework of kinetic theory, Sec.~\ref{Theokinetic}, it is useful to introduce dimensionless variables for temperature and time.
Thus, we define $T^{*}\equiv T/T^{\text{st}}$, so that $\tau$ is now defined as $\tau\equiv 2n\sigma^{2}\sqrt{\pi T^{\text{st}}/m}t$.
In the Maxwellian approximation, $T^{*}$ and $\theta$ evolve according to 

\begin{subequations}\label{Phi-Psi}
\begin{align}
\partial_{\tau}\ln T^*=&\Phi(T^*,\theta), \\
\label{Psi}
\partial_{\tau}\ln\theta=&-(1+\theta)\left(\frac{2}{3}\left(\frac{\gamma^{\text{st}}}{T^*(1+\theta^{\text{st}})}-
\sqrt{\frac{T^* ( 1+\theta^{\text{st}})}{\left( 1+\theta \right)^3}}\gamma^{\text{st}}+ K\sqrt{{\frac{T^*(1+\theta^{\text{st}})}{\left(1+\theta\right)^3}}}\left(1-\frac{\theta}{\theta^{\text{st}}}\right)(1-\theta^{\text{st}})\right)-
              K\sqrt{{\frac{T^*(1+\theta^{\text{st}})}{\left(1+\theta\right)^3}}}\left(1-\frac{\theta}{\theta^{\text{st}}}\right)\right),
\end{align}
\end{subequations}
where we introduced the constant $K\equiv \iota (1+\beta)^{2}/(1+\iota)^{2}$ and the auxiliary function
\begin{align}
\label{Phi}
    \Phi(T^*,\theta) = 
    \frac{2}{3}\left( \frac{\gamma^{\text{st}}}{T^*(1+\theta^{\text{st}})}-
    \sqrt{\frac{T^* ( 1+\theta^{\text{st}})}{\left( 1+\theta \right)^3}}\gamma^{\text{st}} + K\sqrt{{\frac{T^*(1+\theta^{\text{st}})}{\left( 1+\theta \right)^3}}} \left( 1-\frac{\theta}{\theta^{\text{st}}} \right)(1-\theta^{\text{st}}) \right), 
\end{align}
accounting for the rate of change of  (the logarithm of) the relative temperature. We underline here that the time evolution of $T^*$ is governed by a function that depends on both $T^{*}$ and $\theta$. As described below, this is a prerequisite for the emergence of a Mpemba effect; otherwise, no crossing between the temperature relaxation curves is possible.

\begin{figure}
    \centering
    \includegraphics[width=0.8\columnwidth]{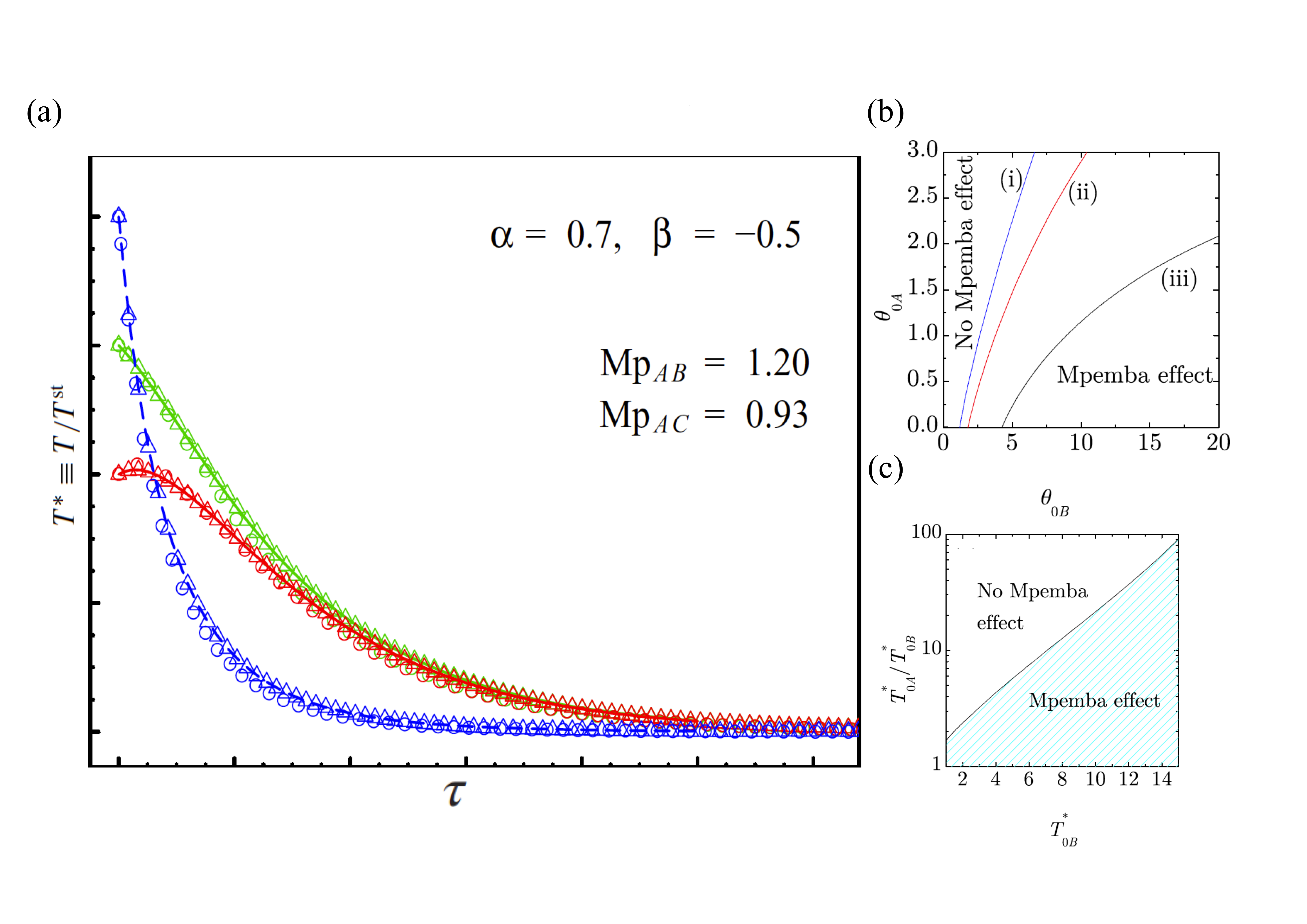}
    \caption{Example of large Mpemba effect in a granular gas of rough hard spheres, uniformly heated. (a) Time occurrence for two given values of $\beta$ and  $\alpha$. Results are given by two simulation methods, DSMC (open symbols) and MD (filled symbols) data show an excellent agreement with the theoretical prediction (lines) see (\ref{Sec:NumericalKinetic}). (b) Phase diagram for the appearance of the effect for three different choices of initial conditions $(T^*_{0A},T^*_{0B})$ ($(i)$: $(5,4)$, $(ii)$: $(4,3)$, $(iii)$: $(5,3)$), for each case below the lines the effect happens. (c) Shows the plane ($T^*_{0B}$,$T^*_{0A}/T^*_{0B}$). Below the plotted line the Mpemba effect may appear, provided that the kinetic energy of the initially hotter (colder) sample is sufficiently concentrated in the translational (rotational) degrees of freedom. \textit{Source:} Reprinted with permission from~\cite{torrente2019large}.
} 
    \label{fig:mpembarough}
\end{figure}

To understand the mechanisms enabling a Mpemba effect, we consider two initial conditions $(T_{0A}^*,\theta_{0A})$ and $(T_{0B}^*,\theta_{0B})$, with $T_{0A}^*>T_{0B}^*>1$, for the same granular gas, i.e., with the same values of the coefficients of restitution $\alpha$, $\beta$ and the same stochastic force $\chi$.
Let us denote by $T_A^*(\tau)$ and $T_B^*(\tau)$ the associated time evolution of the temperature to the steady state. A Mpemba effect then exists in the system if there is a crossing time $\tau_\times$ such that $T_A^*(\tau)<T_B^*(\tau)$ for $\tau>\tau_\times$.
Note that, in this framework, we do not demand that $\theta_{0A}=\theta_{0B}$, which again differs from the Markovian framework.
In this system, it is not possible to have a crossing for $\theta_{0A}=\theta_{0B}$; see Fig. \ref{fig:mpembarough}, right part of panel b.

Let us discuss the possibility of observing a Mpemba effect at short times, where the system retains a ``memory'' of its initial conditions.
For such times, the system cools exponentially, with a characteristic rate roughly equal to the initial value of $-\Phi$.
Thus, a necessary condition for the  Mpemba effect at short times is
\begin{align}
\label{eq:phiA-phiB}
  \Phi(T_{0A}^*,\theta_{0A})<  \Phi(T_{0B}^*,\theta_{0B}).
\end{align}
Consider now two initial conditions, $A \equiv (T^*_{0A},\theta_{0A})$  and $B \equiv (T^*_{0B},\theta_{0B})$  with $T^*_{0A} > T^*_{0B}$, corresponding to different initial conditions.
Since $\Phi(T^*,\theta)$ is monotonic in $\theta$, taking $\theta_{0A} < \theta_{0B}$ fulfills Eq.~\eqref{eq:phiA-phiB}. Physically, the initially hotter system has more of its kinetic energy in the form of translational velocity. In Fig.~\ref{fig:mpembarough}, we analyze the emergence of the Mpemba effect for several pairs of initial temperatures $T^*_{0A}$ and $T^*_{0B}$.
For each such pair, we plot the lines in the ($\theta_{0A}$, $\theta_{0B}$) plane delimiting the regions with and without the Mpemba effect. We can measure the magnitude of the Mpemba effect by defining a \emph{Mpemba parameter} Mp$_{AB}$  as the extremum of the difference of dimensionless temperatures $T^*_B(\tau)-T^*_A(\tau)$ for $\tau > \tau_\times$.

The most favorable conditions for the emergence of the Mpemba effect is when the kinetic energy is concentrated in (i) the translational velocity for the higher temperature $T^*_{0A}$, i.e., $\theta_{0A} = 0$, and (ii) the rotational velocity for the lower temperature $T^*_{0B}$, i.e., $\theta_{0B} \rightarrow \infty$. Figure \ref{fig:mpembarough} shows the time evolution of the granular temperature, for $\alpha=0.7$  and $\beta = -0.5$.
The Mpemba effect is clearly observed as the initially hottest sample (a) is the fastest one to reach the steady state. 

A similar mechanism was studied numerically in \cite{gijon2019paths}, using several models for the interactions between water molecules (SPC, TIP3P, and TIP4P rigid models, as well as a flexible version of the TIP3P) and is discussed in Sec.~\ref{Sec:NumericalWater}. 

\subsection{Example: Other granular models exhibiting anomalous relaxations} \label{subsub:othergranular}
\begin{figure}
    \centering
    \includegraphics[width=0.85\columnwidth]{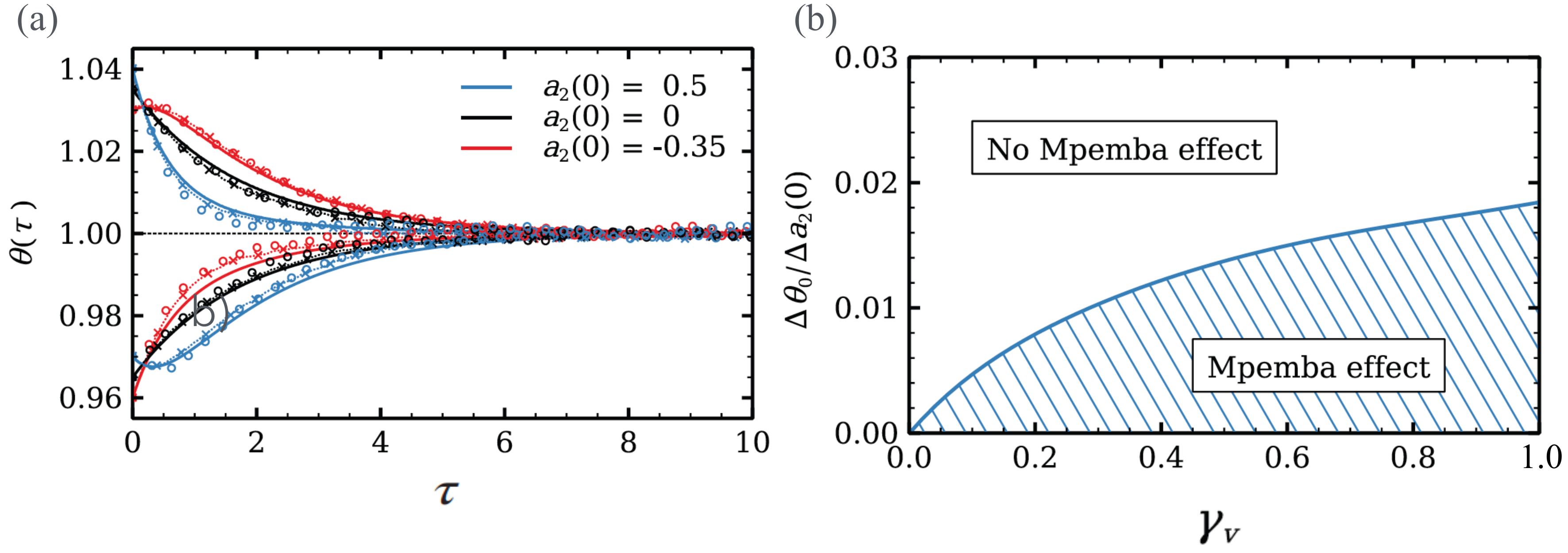}
    \caption{(a) Example of Mpemba and its inverse in a granular gas of viscoelastic spheres.  Note that the difference in initial temperatures is slightly larger than in the hard sphere case. Results are given by two simulation methods, DSMC (dotted lines with crosses) and MD (open symbols). The data show an excellent agreement with the theoretical predictions (lines); see (\ref{Sec:NumericalKinetic}). (b) Phase diagram as a function of  $\gamma_v$  and initial conditions. \textit{Source:} Adapted with permission from~\cite{mompo2021memory}. }
    \label{fig:mpembavisco}
\end{figure}

\noindent \textit{Viscoelastic granular particles}.
A more realistic model for granular gases is the viscoelastic granular particles model, for which the Mpemba effect was considered in \cite{mompo2021memory}.
As in the smooth-hard spheres case (Sec. \ref{SGHS}), the spheres are smooth and angular velocities play no role.
The main difference between the two models is the restitution coefficient, $\alpha$. For smooth, hard spheres, it is assumed to be constant, whereas for viscoelastic granular particles, it depends on velocity.
The velocity dependence is modeled by $\alpha= \alpha(v) \backsimeq 1 - \gamma_v \vert \vec{v}_{1,2} \cdot \vec{\sigma}_{1,2} \vert ^{1/5} + (3/5) \gamma_v^2 \vert \vec{v}_{1,2} \cdot \vec{\sigma}_{1,2} \vert ^{3/5} $ \cite{brilliantov1996model}. Note that, in this case, the difference in initial temperatures that allows us to observe the Mpemba effect can be slightly larger than for smooth, hard spheres; see Fig. \ref{fig:mpembavisco}a. 
Also, the third cumulant  ($a_3=\frac{4}{5}\langle v^{4}\rangle/\langle v^{2}\rangle^{2}-\frac{8}{105}\langle v^{6}\rangle/\langle v^{2}\rangle^{3}-2$) is significant and must be taken into account.
Interestingly, the Kovacs effect, which we discuss in Sec.~\ref{sec:kovacs}, was also observed in this specific model.

\noindent \textit{Maxwell model}. 
A different simplification that allows to study smooth hard spheres systems is the Maxwell model. In this approximation, the collision frequency is not a function of the particle velocity, but rather is a fixed parameter of the system.  It affects significantly the collision kernel.
A series of manuscripts considered the Mpemba and related effects in this and related models, \cite{biswas2022mpemba_a,biswas2022mpemba_b,biswas2023mpemba_distance}, as well as driven monodispersed and bidispersed granular gases \cite{biswas2020mpemba}.
Interestingly, the Mpemba effect exists in monodispersed gas  only if both temperature and additional parameters are quenched; however, in bidispersed granular systems, the Mpemba effect exists even if only the temperature is quenched, as in the Markovian framework.
Moreover, the latter case contains a strong version of the Mpemba effect: for specially chosen initial temperatures, the asymptotic relaxation has a different exponential rate~\cite{biswas2020mpemba}.

\begin{figure}
    \centering
    \includegraphics[width=0.6\columnwidth]{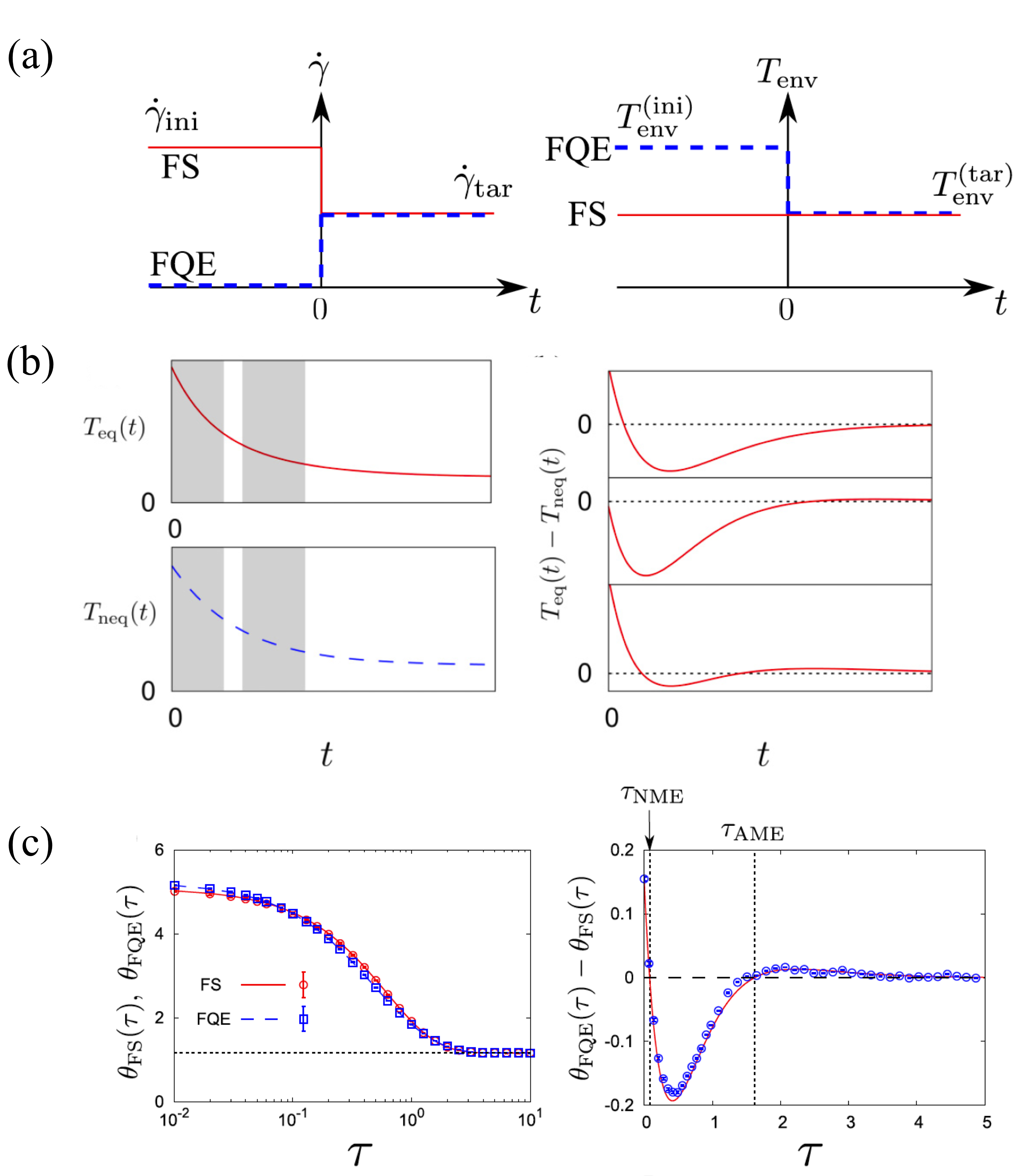}
    \caption{(a) Sketch of the two ME protocols used in an inertial suspension. The protocol for FQE (from a quasi-equilibrium initial state, dashed lines) and FS (from a sheared steady initial state, solid line). For the FQE, the shear rate and the environmental temperature are changed at $t = 0$, (left) from $0$ to $\dot{\gamma}_{tar}$, (right) from $T_{ini}$ environmental to $T_{tar}$ environmental respectively. For FS, (left) the shear rate is changed at t = 0 from $ \dot{\gamma}_{ini}$ to $\dot{\gamma}_{tar}$ while (right) the environmental temperature is kept at $T_{tar}$ environmental. (b) Left: Diagram of the evolutions of the temperatures from the equilibrium (top) and from the nonequilibrium (bottom). Right: illustrates the temperature differences during the (top) normal Mpemba effect (NME), (middle) anomalous Mpemba effect (AME), and (bottom) NME + AME. (c) Numerical simulations show the two types of effects, where $\theta_i=T/T_{\rm env}$ is the dimensionless temperature and $i=$ FS,FQE indicates the protocol. The left panel shows the $\theta_i$ evolution, and the right panel shows the difference $\theta_{\rm FQE}-\theta_{\rm FS}$, where anomalous and normal Mpemba effects can be observed. \textit{Source:} Adapted with permission from~\cite{takada2021mpemba}.
    }
    \label{fig:mpembainertial}
\end{figure}
\noindent \textit{Inertial suspension of granular particles}.
Another granular system where the normal and what the authors denote as ``anomalous Mpemba effect" (as well as its inverse) were found is an inertial suspension of granular particles under shear \cite{takada2021mpemba}.
The system that was considered is a suspension in a shear flow, namely in the presence of a fluid flow with velocity field given by $\vec{u}=(\dot{\gamma} y, \vec{u}_{\perp}=0)$, and subject to a uniform kinetic temperature $T$. Each suspension particle is therefore described by the Langevin equation
\begin{align}
     \dv{\vec{p}_i}{t}= -\zeta \vec{p}_i+ \vec{F}_i^{imp}+m \vec{\xi}_i,
\end{align}
where $\vec p_i = m(\vec v_i - \dot\gamma y_i \vec e_x )$ with $\vec e_x$ is a unit vector in the $x$ direction, $\vec{F}_i^{imp}$ expresses the impulsive force coming from collisions between grains, and $\vec{\xi}_i(t)=\xi_{i,q} \vec{e}_q$ is delta correlated noise from the surrounding thermal fluid: $\langle \vec{\xi}_i \rangle =0$ and $\langle \xi_{i,q}(t) \xi_{j,s}(t')  \rangle = 2 \zeta T_{\rm env} \delta_{i\!j} \delta_{qs} \delta (t,-t')$, where indices $i,j$ refer to particle numbers and $q,s={x,y,z}$. 
This model, therefore, has two external parameters, the environmental temperature $T_{\rm env}$, and the shear rate $\dot \gamma$. These quantities together determine the (nonequilibrium) granular temperature. Note that the Enskog equation \cite{cercignani1988boltzmann} is used here at finite density to describe the system and the influence of the packing fraction on the ME is also analyzed. Within this framework, the authors considered ``normal" and ``anomalous" Mpemba effects. For simplicity, we use these terms in this specific subsection.

The ``normal" Mpemba effect stems from a result which can be shown from the above equation: when a high-enough temperature equilibrium initial condition (here equilibrium is the steady state corresponding to no shearing, $\dot\gamma=0$) is compared to a nonequilibrium initial condition (the steady state of a system with $\dot\gamma\neq 0$) with the same initial temperature, then the nonequilibrium initial condition relaxes faster, namely $\dot T_{\rm eq}(t=0)<\dot T_{\rm neq}(t=0)$. This implies that for small-enough difference in initial temperature, the nonequilibrium initial condition can cool faster and overcome the equilibrium initial condition. Note that this effect concerns the initial stage of the relaxation, in contrast to the Markovian effect that concerns the long-time limit.

In the ``anomalous" Mpemba effect, the temperature crossing does not happen at the early time, but rather at a later time as a consequence of nonlinear temperature dynamics; see Fig.~\ref{fig:mpembainertial}. Note that, in this framework, the initial conditions not only have different temperatures but also one has an equilibrium and the other a nonequilibrium distribution.

\subsection{Example: Molecular gases}

In models of molecular gases, the particle dynamics are described by the Enskog-Fokker-Planck equation \cite{cercignani1988boltzmann}. 
The main difference between this case and previously discussed granular gases is that the collisions in molecular gases are \textit{elastic}. Observing Mpemba and related effects then requires additional ingredients. 

To explain these models, consider a molecular gas immersed in a background fluid. The motion of the Brownian particles can be described by Eq.~\eqref{GBE}, with $\mathcal{J}[\vec{v}, \vec{w}|f(t)]$ a Boltzmann-Enskog (elastic) collision term.
The fluid now produces both thermal fluctuations and a nonlinear viscous force  \cite{santos2020mpemba,megias2022thermal} $\vec{F}(\vec{v})=-\vec{v}\zeta (v)$.
The simplest nonlinear model can be obtained by assuming a quadratic dependence of the drag force on the velocity,
\begin{equation}
\label{non_linear}
  \zeta (v)=\zeta_0 \left(1 + \gamma \frac{m v^2}{k_B T_s}  \right),
\end{equation}
where $\gamma > 0$ is a dimensionless parameter that sets the size of the nonlinear force, and $T_s$ is the equilibrium temperature of the background fluid.
Here, the background acts as a thermostat with $\chi_0^2=\zeta (v) k_B T_s / m$.
Applying the first Sonine approximation for this system \cite{resibois1977classical}, one finds that the evolution of temperature is coupled to that of excess of kurtosis, akin to the case of smooth granular hard spheres in Sec.~\ref{SGHS}. 

Another interesting scenario is that of a binary mixture with linear viscous drag \cite{gomez2021mpemba}.  The coupled Boltzmann-Fokker-Planck equations for both species, again assuming homogeneous states, are
\begin{align}
    \left(\partial_t  -\nabla _{\vec{v}}\right)\left(\gamma_i \vec{v} +\frac{\gamma_i T_{\mathrm{ext}}}{m_i}\nabla _{\vec{v}}\right) f_i(\vec{r} , \vec{v};t)=\sum^{2}_{j=1}\mathcal{J}_{i,j}[\vec{v}|f_i(t),f_j(t)],
\label{BEM}
\end{align}
where $\gamma_i$ is the friction coefficient and $T_{\rm ext}$ is the background temperature.
The presence of the external bath couples the time evolution of the total and partial temperatures of each component, allowing the appearance of the Mpemba phenomenon, even when the initial temperature differences are of the same order as the temperatures themselves.

\subsection{Discussion}

There are some crucial differences between the Markovian and kinetic Mpemba effects.
The most important is that all the gas models in the kinetic theory frameworks characterize a driven system that is \textit{intrinsically} nonequilibrium. Again, the driven-gas framework offers an approach where the
system of interest can be treated analytically and is well understood in the thermodynamic limit.
Anomalous relaxation effects are therefore observed when driving the system from one nonequilibrium steady state to another, as opposed to the case described in Sec.~\ref{sec:markovian}, where the system goes from one Boltzmann equilibrium at some initial temperature $T_0$ to another Boltzmann equilibrium at some final temperature $T_b$. Note also the contrast with the Markovian case where the system must be forced to sustain a NESS (see Sec.~\ref{sec:relaxation-to-a-ness} for an example). In the granular case, the system does not obey an equilibrium Boltzmann distribution.

\begin{figure}
    \centering
    \includegraphics[width=0.7\columnwidth]{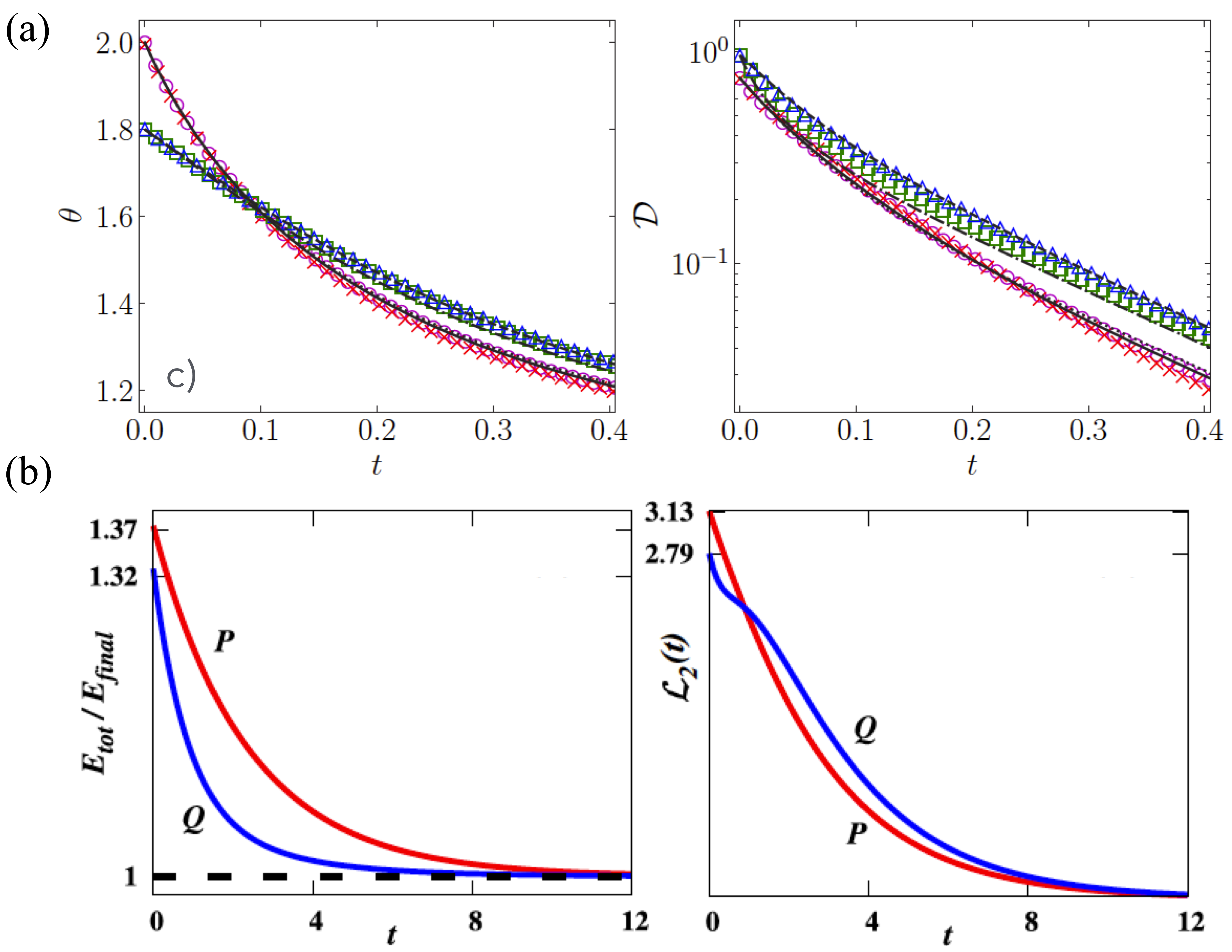}
    \caption{(a) Measure comparison for molecular gases with nonlinear drag: As we discussed in the text,  the relative temperatures of hot and cold states cross (left), but not their relative entropy (right). Lines are theoretical predictions; open symbols are DSMC; and filled symbols are MD. (b) Measure comparison for the case of driven Maxwell model of granular gases. Note that the Mpemba effect is not observed in the total energy for hot and cold states which does not exhibit a crossing (left) whereas the Euclidean distance does (right). \textit{Source:} Reprinted with permission from~\cite{megias2022thermal} and~\cite{biswas2023mpemba_distance} respectively.}
    \label{fig:granularcomparisonmeasures}
\end{figure}
Another difference is that, here, the Mpemba effect was defined based on the relaxation of the granular temperature: a crossing in the temperature implies the existence of the ``thermal" Mpemba effect. However, we also used other measures to define the Mpemba effect, such as relative entropy (Kullback-Leibler divergence) in Sec.~\ref{sec:distance-function}. When using this latter measure, one refers to the  ``entropic" Mpemba effect in the granular-gas community ~\cite{megias2022thermal}. As in the Markovian case \cite{van2024thermomajorization}, for granular gases, the different definitions do not always correspond to the same effects. This was shown in \cite{biswas2023mpemba_distance}, where the two models presented above, the molecular gas with nonlinear drag and anisotropically driven granular models, were considered. It was shown that the ``entropic" and ``thermal" Mpemba effect are not equivalent in these cases. 

An example is given in~\FIG{granularcomparisonmeasures}, which compares the thermal Mpemba effect (kinetic-temperature crossing) with the entropic Mpemba effect (relative-entropy crossing). In the former case, the authors split the relative entropy $D_{\rm KL}(t)=\int \dd \vec{v} f(\vec{v},t) \ln \left( f(\vec{v},t)/f^{\rm eq}(\vec{v}) \right)$ into two terms, $D_{\rm KL}(t)=D^{\rm kin}(t) + D^{\rm SH}(t)$.  The first is associated with the rapid kinetic stage, the second with the slow hydrodynamic stage.  In addition, the authors describe the overshoot Mpemba effect also related with the overshoot that appears in the Kovacs effect; see Sec.~\ref{sec:kovacs}. In \cite{biswas2023mpemba_distance}, the authors compare the occurrence of the Mpemba effect depending on different distance measures used: the total energy $E_{\rm tot}=E_x+E_y$, the Euclidean distance $L_2(t)=\sqrt{(E_{\rm tot}(t)- E_{\rm tot}^{\rm st})^2+(E_{\rm dif}-E_{\rm dif}^{\rm st})^2}$ with $E_{\rm dif}=E_x-E_y$ and $E^{\rm st}$ indicating the stationary value and $E_x$ and $E_y$ the energies along the $x$ and $y$ axes, the Manhattan distance $L_1(t)=|E_{\rm tot}(t)- E_{\rm tot}^{\rm st}|+|E_{\rm dif}-E_{\rm dif}^{\rm st})|$, and the Kullback-Leibler distance $ D_{\rm KL}(t)=\int \dd \vec{v}\ f(\vec{v},t) \ln \left( f(\vec{v},t)/f^{\rm st}(\vec{v}) \right)$. They also derived criteria for the existence of the Mpemba depending on the choice of measure; interestingly, they reached different conclusions using different measures.

Another important issue differentiating the kinetic Mpemba effect from the Markovian one is the apparent symmetry in the direct and inverse effects; see~\FIG{mpembasmooth}.
This symmetry arises from using a rescaled time $\tau$, where $\tau(T)$ emerges from the structure of the rescaled equations; see, for example, Eq.~\eqref{evol-eqs}.

\section{Statistics of anomalous thermal relaxations}
\label{sec:statistics}
\subsection{From one model to an ensemble} 
The Markovian framework developed in~\SEC{markovian-quench} provides a systematic procedure to analyze whether a given temperature-dependent Markovian system has some type of Mpemba effect.  This naturally leads to the question, How often should we expect to observe such effects in various cases? Are these effects a consequence of some fine-tuning in the model parameters, or do they appear generically? A different motivation to ask similar questions is the following: relaxation protocols typically involve a few parameters that we can control and others that we cannot directly control or over which our control is limited (e.g., we can dictate only their range, mean, or variance). In such cases, it is valuable to consider an ensemble of systems with a given probability of observing the setup in a given set of parameters and then estimate the likelihood of observing the Mpemba effect in this ensemble. In addition to their theoretical interest, these questions have experimental implications, such as the probability of observing Mpemba effects in a given ensemble of systems. 

Because of its topological nature, discussed in~\SEC{strong}, the strong Mpemba effect is more amenable to analytical treatment than the weak Mpemba effect. Thus, in what follows, we focus on estimating the occurrence of the strong Mpemba effect. Let us recall from~\SEC{strong} that a lower bound for observing the direct or inverse Mpemba effect is the probability that the parity of the direct or inverse Mpemba index is positive; i.e., that
\begin{align}
    {\rm Prob} \left(\mathcal{P}\left(\mathcal{I}_{\rm M} ^{\rm dir/inv}\right)> 0 \right).  
\end{align} 
The parity of the Mpemba index can be deduced from probability distribution at a few temperatures (see \EQS{Parity-inv}{Parity-dir}), whereas calculating the Mpemba index itself requires scanning over all temperatures. This is an important advantage in a numerical study, although this estimator is only a lower bound on the probability to have a strong Mpemba effect. 

\subsection{How likely is the Mpemba effect in the Random Energy Model?}

The probability to observe a Mpemba effect in an ensemble of systems is a function of the distribution of the parameters in the specific ensemble. In what follows, we illustrate the procedure to estimate such probabilities through a few prototypical examples. 
Specifically, we focus on discrete systems with probability distributions that evolve according to a master equation
\begin{align}
    \partial _t \vec{p} = \textbf{W} \, \vec{p} ,
\end{align}
whose rate matrix $\textbf{W}$ obeys detailed balance (DB). In the specific ensembles we consider~\cite{klich2019mpemba}, the equilibrium distribution is always the Boltzmann distribution, and the physical system is characterized by a set of energies chosen by a stochastic process. 

As an example, we consider the Random Energy Model (REM), one of the simplest models of disordered systems demonstrating a \emph{freezing (condensation) transition}. The REM does not itself describe any specific physical situation; rather, its features are conceptually found in many contexts. The model was introduced by Derrida~\cite{derrida1981random} and later extended by Gross and M{\'e}zard~\cite{gross1984simplest}. In the REM, a set of energies $\{E_1, E_2,\dots,E_L\}$ is drawn from a probability density distribution, $\rho_E(E)$, where $\rho_E(E) \dd{E}$ is the probability of observing energy $E$ in the energy shell $\left(E,E+\dd{E}\right)$. 
The energies represent a \emph{sample} of \emph{quenched disorder}. One can consider these $L = 2^N$ energies as corresponding to the configurations of $N$ Ising spins. 
A common choice for the energy probability density, which we adopt, is the Gaussian distribution
\begin{align}
\label{eq:ProbE}
    \rho_E(E) = \frac{1}{\sqrt{2 \pi N \sigma_E^2}} \, e^{- \frac{E^2}{2 N \sigma_E ^2}}, 
\end{align}
where the variance, $ N\sigma_E ^2$, ensures that the energy and other thermodynamic potentials are extensive.  
Although the above choice is the most common, other distributions have been studied as well. 

The equilibrium distribution for a specific instance of the REM $\{E_i\}$ is given by the corresponding Boltzmann distribution,
\begin{align}
    \pi_i(T,\{E_i\}) = \frac{1}{Z(T,\{E_i\})}e^{- E_i/k_BT}, 
\end{align}
where $Z(T,\{E_i\}) = \sum _{i=1} ^{L} e^{-E_i/k_B T}$ is the partition function of the specific random sample of energy barriers,
$\{E_i\}$. 
Each set of energy barriers has its corresponding Helmholtz free energy 
\begin{align}
    F(T,\{E_i\}) = - T \ln Z(T,\{E_i\}). 
\end{align}
Note that the random choice of energies makes both the partition function and the Helmholtz free energy random variables. The (quenched) average Helmholtz free energy is given by 
\begin{align}
    \langle F(T,\{E_i\}) \rangle _{\rho_E} = \int\hdots\int \prod _{i = 1}^L \dd E_i\ \rho _E(E_i) F(T,\{E_i\})\,,
\end{align}
where the ensemble-average subscript $\langle ... \rangle _{\rho _E}$ 
denotes an average over the distribution of energies. 
Much of the interest in the REM stems from its transition from a delocalized occupancy at high temperatures, where the probability is spread over most states, to a ``condensation'' of the system in several states for temperatures below the critical temperature, $T_{\rm cr}$~\cite{derrida1981random}. 
This ``freezing'' or ``condensation'' phenomenon is a phase transition, since the average free energy $\langle F (T,\{E_i\})\rangle _{E}$ is nonanalytic at the critical temperature~\cite{mezard2009information}. For Gaussian energy statistics, as in ~\EQ{ProbE}, the critical temperature is $T_{\rm cr} = \sigma_E /\sqrt{2 \ln 2}$; see ~\cite{derrida1981random}.

The choice of energy statistics $\rho_E(E)$ determines the equilibrium states of the REM.  To understand the relaxation to equilibrium states, we need to consider the model's dynamics, as well. In what follows, we consider three cases: (i) Glauber dynamics in a 1D chain, (ii)  Random barrier model, and (iii) Isotropic ensemble. 

\subsubsection{The 1D chains with Glauber/Metropolis dynamic ensemble}
\label{sec:1DChainStatistics}

Probably the simplest dynamics that can accompany the REM is to assume that states $\{1,\dots,L\}$ with corresponding energies $\{E_1,\dots,E_L\}$ drawn independently according to~\EQ{ProbE} are located on a 1D chain, and to associate Glauber~\cite{glauber1963time} (or Metropolis~\cite{metropolis1953equation}) transition rates between nearest-neighbor sites. 
In other words, if the system is in state $1<i<L$, it can  jump only to states $i+1$ or $i-1$, with Glauber rates 
\begin{align}
    W_{i+1,\,i} = \frac{\Gamma}{1+e^{(E_{i+1}-E_i)/k_B T_b}}
    \quad \text{and} \quad
    W_{i-1,\,i} = \frac{\Gamma}{1+e^{(E_i-E_{i-1})/k_B T_b}},  
\end{align}
where $\Gamma^{-1}$ has units of time. The transitions from the end sites $1$ and $L$ are only to sites $2$ and $L-1$, respectively, with the corresponding Glauber transitions. 
The same can be done with Metropolis rates, where 
\begin{align}
    W_{i+1,\,i} = \Gamma \min \left(1,e^{-\frac{E_{i+1}-E_i}{k_B T_b}}\right) 
        \quad \text{and} \quad
    W_{i-1,\,i} = \Gamma \min \left(1,e^{-\frac{E_{i-1}-E_i}{k_B T_b}}\right). 
\end{align}

It is worth noting that this ensemble, to some extent, is a discrete version of the Brownian particle in a potential discussed in Sec.~\ref {Sec:Brownian}. The potential is given by the ``random landscape'', whose randomness is provided by the REM.  Notice that the barriers correspond to the ``peaks'' and the energies to the ``valleys'' of a discretized landscape. To relate to the Arrhenius form of transition rates introduced in \EQ{arrhenius}, we note that, in the Glauber case, the barriers are $B_{i\!j} = T_b \ln ((\pi_i(T_b) + \pi _j (T_b))/(\pi_i(T_b)\pi_j(T_b)))$ and, in the Metropolis case, they are $B_{i\!j}= \max(E_i,E_j)$.

\begin{figure}
    \centering
    \includegraphics[width=0.5\linewidth]{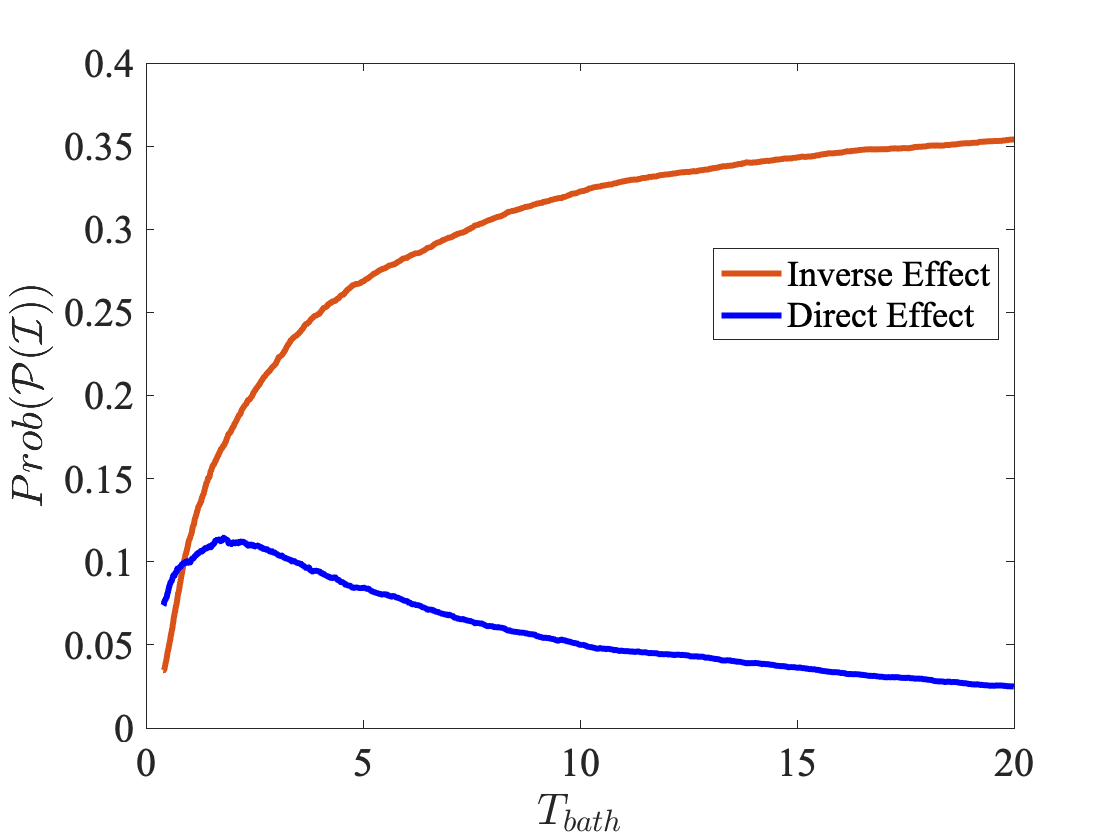}
    \caption{Probability for non-zero parity for the direct (blue) and inverse (red) Mpemba index, in the 1D REM with Metropolis dynamics between nearest neighbors sites, with $L=50$ sites, averaged over 20,000 realizations of the random energies.}
    \label{fig:1DChainStatistics}
\end{figure}

The ensemble that we have introduced can be easily studied numerically, and some limiting cases can be studied analytically.  In \FIG{1DChainStatistics}, we numerically evaluate the probabilities for an even direct and inverse Mpemba index as a function of temperature, for Metropolis dynamics. Repeating the calculation for Glauber dynamics gives indistinguishable results. In both cases, the probability to have a direct Mpemba effect exceeds $0.35$ at high temperatures.

\subsubsection{The random barriers ensemble}
\label{sec:rbe}

The random barrier ensemble was suggested and numerically studied in \cite{klich2019mpemba}. The idea, in this case, is to use the Arrhenius form for the transition rates, specified in~\EQ{arrhenius}, namely
\begin{align}
W_{i\!j}(T_b) = \Gamma e^{-(B_{i\!j}-E_j)/k_B T_b},
\end{align}
where $\Gamma^{-1}$ has units of time and $E_i$ is the energy of site $i$, and $T_b$ is the bath temperature.  The barrier parameter $\{B_{i\!j}\}$  between sites $i$ and $j$ is defined to be symmetric, with $B_{i\!j} = B_{ji}$.
    
For a given system, the energies $\{E_i\}$ are a property of the system itself, regardless of how it is coupled to the environment or the exact nature of the environment. In contrast, the barriers, parameterized by $\{B_{i\!j}\}$, are not just internal properties of the system but rather are functions of the particularities of the environment and its precise coupling to the system. These barriers are often not known or can only be roughly estimated. In the random barrier ensemble, we treat both the energies and the barriers on an equal footing, with the barriers also randomly drawn from a distribution,
\begin{align}
\label{eq:ProbB} 
    \rho_B(B) = \sqrt{\frac{2}{\pi \sigma _B ^2 N}} \, e^{-B^2/2 \sigma _B^2 N} \, \theta (B),
\end{align}
where $\theta$ is the Heaviside step function. Note that to visualize the set of energies and barriers as a landscape of valleys and peaks, one would need to make sure that $B_{i\!j} \geq \{E_i, E_j\}$, and while the above choice does not facilitate that, without loss of generality, one can shift all energies by a constant and ensure that, indeed, $B_{i\!j}$ exceeds $\{E_i, E_j\}$. 

Note that, in contrast to the ensembles discussed in Sec.~(\ref{sec:1DChainStatistics}), the system here has ``mean-field coupling,'' in the sense that state $i$ has non-zero transition rates to all other states. 
Numerical studies~\cite{klich2018solution} of the probability distribution of the parity of the direct Mpemba effect show that it occurs with finite probability both above and below the freezing transition; see~\FIG{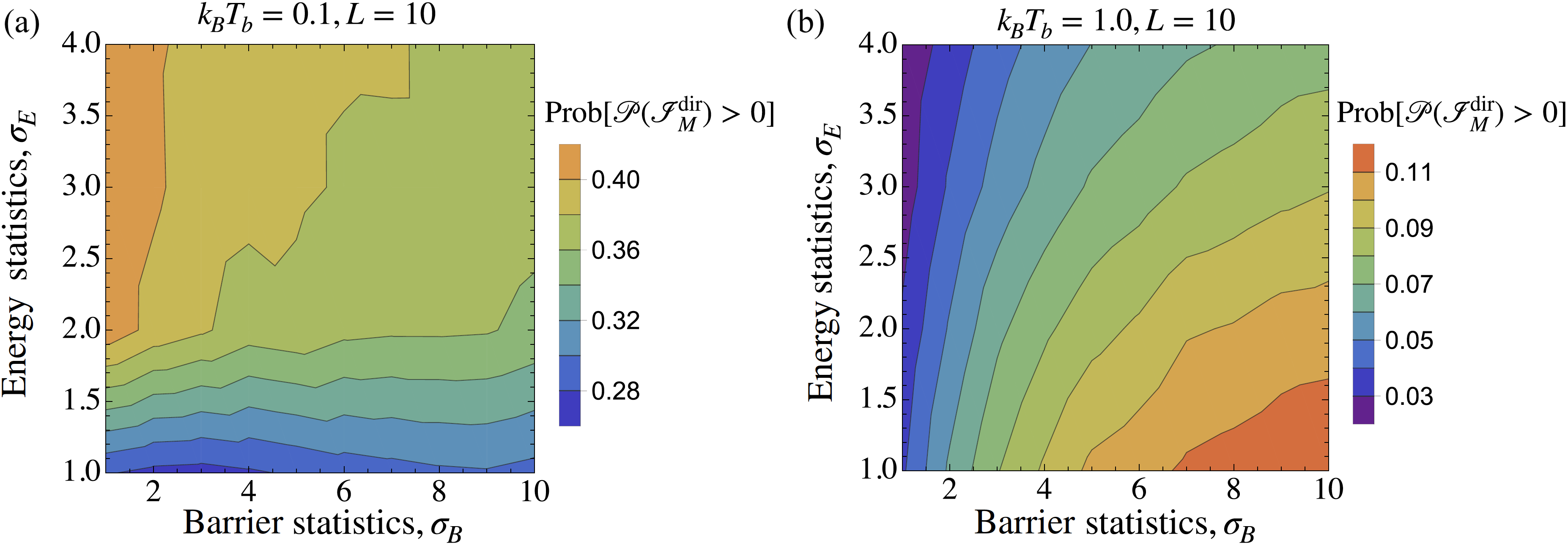}.

\begin{figure}
    \centering
    \includegraphics[width=\columnwidth]{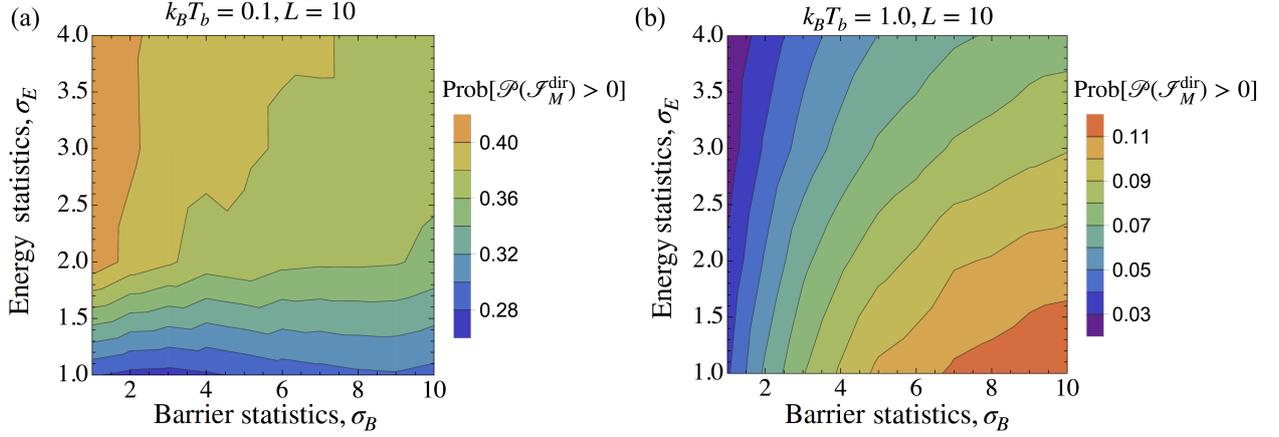}
    \caption{Lower bound for the direct strong Mpemba effect, showing the likelihood to have an odd number of zeros of $a_2$, i.e., for the probability to have a positive Mpemba index parity. The probability is evaluated as a function of the variances of the energy, $\sigma_E^2\log _2 L$, and barrier, $\sigma_B^2 \log _2 L$.  The ensemble statistics are shown for two different bath temperatures: $k_BT_b = 0.1$ on panel (a), and $k_BT_b = 1.0$ on panel (b), for $L = 10$ energy levels. Each point was averaged over $2\times 10^5$ realizations. The freezing transition temperature is $T_{\rm cr} =\sigma _E/\sqrt{2 \ln 2} \approx 0.85 \, \sigma _E$; thus, the figures show that the probability is finite in both phases. \textit{Source:} Adapted with permission from~\cite{klich2019mpemba}.}
    \label{fig:fig-energy-barriers-statistics-lower-bound.png}
\end{figure}

\subsubsection{The isotropic ensemble}

To evaluate the occurrence of the Mpemba effect across a range of systems, it is essential to determine the eigenvector corresponding to the slowest eigenvalue, the second eigenvector, for each instance in the ensemble. Solving analytically for this second eigenvector as a function of the bath temperature is challenging, particularly when dealing with large matrices. However, a method for approximating the second eigenvector was presented in \cite{klich2019mpemba} for a specific ensemble of large random matrices.

To understand the main idea in the isotropic ensemble, let us first consider the two ensembles discussed above in~\SECS{1DChainStatistics}{rbe}. In both cases, the energies are randomly drawn from a Gaussian distribution. By contrast, the barriers are either fixed by the dynamics, as in the Glauber and Metropolis cases, or chosen randomly. The eigenvectors of the relaxation dynamics are themselves random because of the randomness of the energies and, in the random-barrier case, also the barriers. In all these cases, finding the slowest eigenvector for large systems and for a given realization at arbitrary temperature can only be done numerically. In the isotropic ensemble, we avoid this problem by directly specifying the statistics of the slowest eigenvector direction rather than the barrier statistics.

To facilitate this process, it is more convenient to work with the symmetrized relaxation matrix: the relaxation rate matrix $\textbf{W}$ can be transformed to a symmetric matrix $\tilde{\textbf{W}}$ via~\cite{risken1996fokker}
\begin{align}
\label{eq:Wsym}
    \tilde{\textbf{W}} = \textbf{F}^{1/2}\,\textbf{W}\,\textbf{F}^{-1/2},
\end{align}
where $F_{i\!j}(T_b) = \exp(E_i/k_B T_b)\delta_{i\!j}$. The symmetric matrix $\tilde{\textbf{W}}$ has the same eigenvalues as $\textbf{W}$ and has an orthogonal set of eigenvectors $\vec{f}_i = F^{1/2} \vec{v}_i$. The eigenvectors $\{\vec{f} _i\}$ are orthogonal, $\vec{f}_i \cdot \vec{f}_j = ||\vec{f}_i||^2 \delta _{i\!j}$. Working with a symmetric matrix offers other advantages besides orthogonality; numerical diagonalization algorithms are typically more stable. 

\begin{figure}
    \centering
    \includegraphics[width=0.9\linewidth]{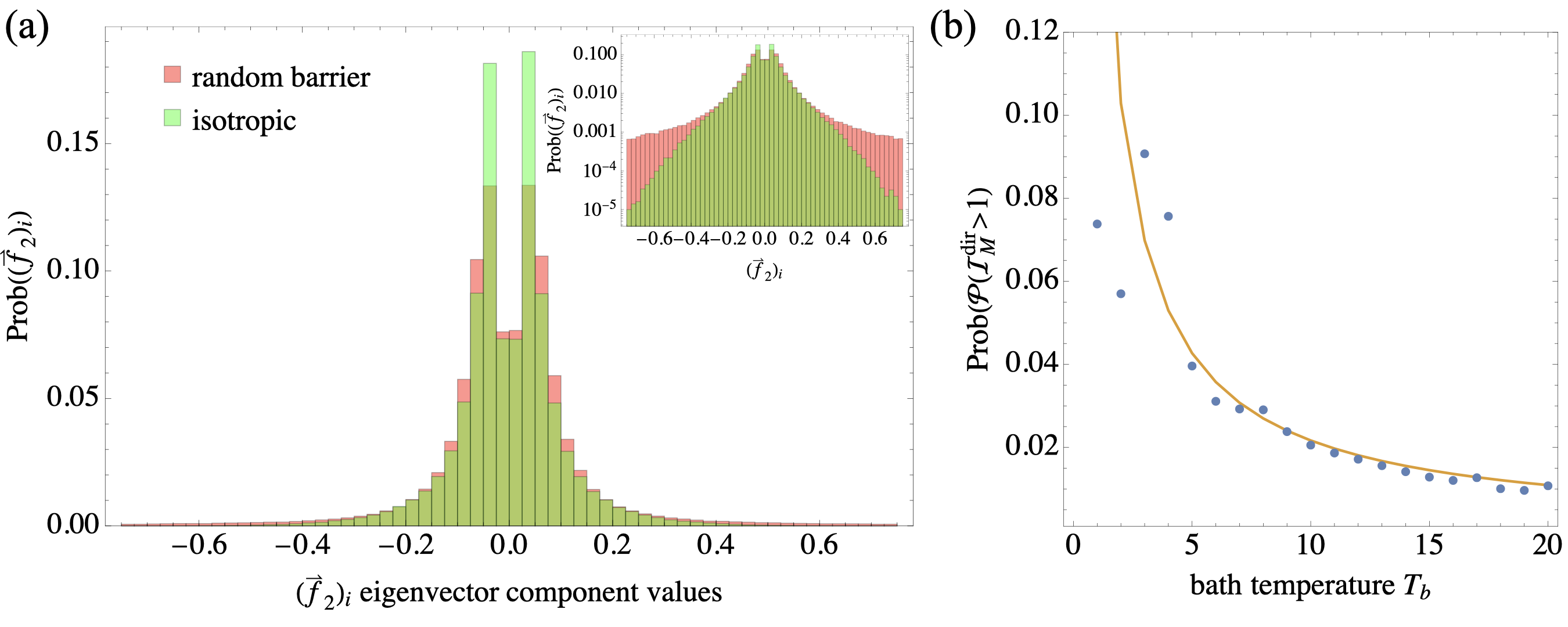}
    \caption{(a) Comparison of the random barrier ensemble and the isotropic ensemble for a quenched energy distribution of $L = 25$ energies chosen from a Gaussian distribution with zero mean and standard deviation $\sigma _E = 1.5$. The bath temperature is $T_b = 10$. The sizes of the isotropic ensemble are $20,000$ random entries of $\vec f ^{\,\,\rm iso} _2$. The size of the random barrier ensemble is $20,000$ random entries of $\vec f _2$. The barrier distribution is a half-normal distribution with $\sigma _B = 75.6$; see~\EQ{ProbB}. For the isotropic ensemble random vector entries, $g_i$ components are i.i.d.s drawn from a Gaussian with zero mean and standard deviation $\sigma _g = 0.19$. The histogram shows the random barrier ensemble probability of $\vec f_2$ components, and the green histogram shows the isotropic ensemble histogram of the components for the ``candidate second eigenvector," $\vec f ^{\,\,\rm iso} _2$. The inset emphasizes the tails of the two histograms, where we see the difference between the two ensembles. (b) Probability of the direct Mpemba index parity as a function of bath temperature $T_b$ for a random barrier ensemble (blue dots) and analytical estimate (orange line). The means and standard deviations of the random vector and the barriers were the same as in a). The quenched sample of $L=25$ energies was different than in a) but obeyed the same statistics (Gaussian ensemble with zero mean and standard deviation 1.5). The two ensembles match well for bath temperature and barrier distributions that are wider than the energy distribution.} 
    \label{fig:ensemble-comparison}
\end{figure}

The zero-eigenvalue eigenvector of $\tilde{\textbf{W}}$ is given by $(\vec f _1)_i \propto e^{-E_i/(2k_B T_b)}$. It is set by choosing the energies $\{ E_i \}$ and the bath temperature $T_b$.  The second eigenvector of $\tilde{\textbf{W}}$ is then chosen as $\vec g$, a vector of i.i.d. (independent and identically distributed) random variables from a Gaussian distribution,  projected onto the space orthogonal to $\vec f _1$: 
\begin{align}
\label{eq:f2approx}
    \vec f _2 ^{\,\,\rm iso} (\vec{g}) \approx \vec g - \frac{\vec g \cdot \vec f _1}{||\vec f _1||^2}\vec f _1.  
\end{align}
However, it is not obvious that the rate matrix having eigenvectors $\vec v_1$ and $\vec v^{\,\rm iso} _2 \equiv  F^{-1/2}\vec f_2 ^{\,\,\rm iso}$ obeys detailed balance. Indeed, for any initial real vector $\vec f ^{\rm iso} _2$ orthogonal to $\vec f_1$, one can find a set of barriers $B_{i\!j}$ that obeys detailed balance~\cite{klich2019mpemba}. Moreover, in the specific ensemble, the probability of choosing $\vec f_1$ and $\vec f_2$ is assumed to be independent of the other eigenvectors. Therefore, as the existence of all types of Mpemba effects is encoded in these vectors, it is enough to average over the probabilities of these vectors. This assumption does not hold for arbitrary ensembles. However, in the random-barrier ensembles considered in Sec.~(\ref{sec:rbe}), the assumption seems to hold in the limit of large $L$, where the mutual dependence between the eigenvectors and their components goes to zero. Below, we will compare the two ensembles on an example, in~\FIG{ensemble-comparison}a. 

To estimate the occurrence of the Mpemba effect, we thus look at the probability of the parity of the Mpemba index. Below, we show the main steps in estimating the occurrence of the direct Mpemba effect. In the same way, one can estimate the probability of having an inverse Mpemba effect. Both direct and inverse Mpemba effects are estimated in~\cite{klich2019mpemba}. 

To estimate the occurrence of the Mpemba effect, we examine the probability associated with the parity of the Mpemba index. Below, we outline the key steps in estimating the occurrence of the direct Mpemba effect. Similarly, one can calculate the probability to have an inverse Mpemba effect. 

From~\EQ{Parity-dir}, the parity of the direct Mpemba index is 
\begin{align}
\label{eq:Parity-dir-sym}
    \mathcal{P}(\mathcal{I}^{\rm dir} _{\rm M}) = 
    \theta\left[\left(\vec f_2\cdot \vec{u}^{\,\rm dir}\right)\left(\vec f_2\cdot \vec{w}\right)\right],
\end{align} 
where $(\vec u^{\,\rm dir}) _i = \exp[E_i/(2 k_B T_b)]$ and  $w_i = \exp[-E_i/(2k_B T_b)] \left(\langle E \rangle _{\vec{\pi}(T_{b})} - E_i \right)$, and $\langle E \rangle _{\vec{\pi}(T_{b})} = \sum _i E_i\, \pi _i (T_b)$ is the average energy with respect to the Boltzmann distribution at $T_b$. It is important to note that the vectors $\vec u^{\,\rm dir}$ and $\vec w$ depend only on the energies and not on $\vec f _2$. Because of this property, the above formula for the parity has a simple geometric meaning. Decomposing $\vec{f}_2$ into $\vec f_{||}$ and $\vec f_{\perp}$, components that are parallel and perpendicular to $\vec{u} ^{\,\rm dir}$, we find
\begin{align}
    \left(\vec f_2 \cdot \vec{u}^{\,\rm dir}\right) \left(\vec f_2 \cdot \vec{w}\right) = f_{||} ^2 \left(\vec u ^{\,\rm dir}\cdot \vec w\right) + f_{||}\,f_{\perp} \left|\vec u^{\,\rm dir} \cdot \vec w\right| K(\vec{u} ^{\,\rm dir}, \vec w),
\end{align}
where 
\begin{align}
\label{eq:Kscalarprod}
    K\left(\vec u ^{\,\rm dir}, \vec w\right) \equiv\sqrt{\frac{\left|\left|\vec u ^{\,\rm dir}\right|\right|^2 \left|\left|\vec w\right|\right|^2}{\left(\vec w \cdot \vec u^{\,\rm dir}\right)^2}-1}. 
\end{align}
Thus, $\mathcal{P}(\mathcal{I}^{\rm dir} _\mathrm{M}) > 0$ is a double wedge, with the boundary of the region given by the lines $f_{\perp} = - f_{||}/K$ and $f_{||} = 0$. The double wedge is illustrated in~\FIG{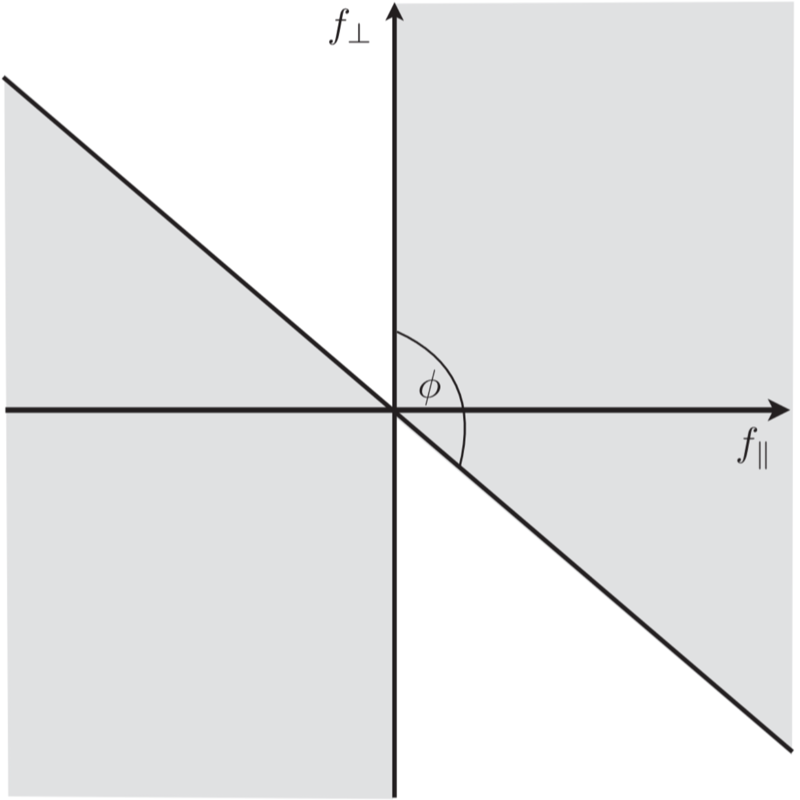}. 

\begin{figure}
\centering 
\includegraphics[width=0.35\columnwidth]{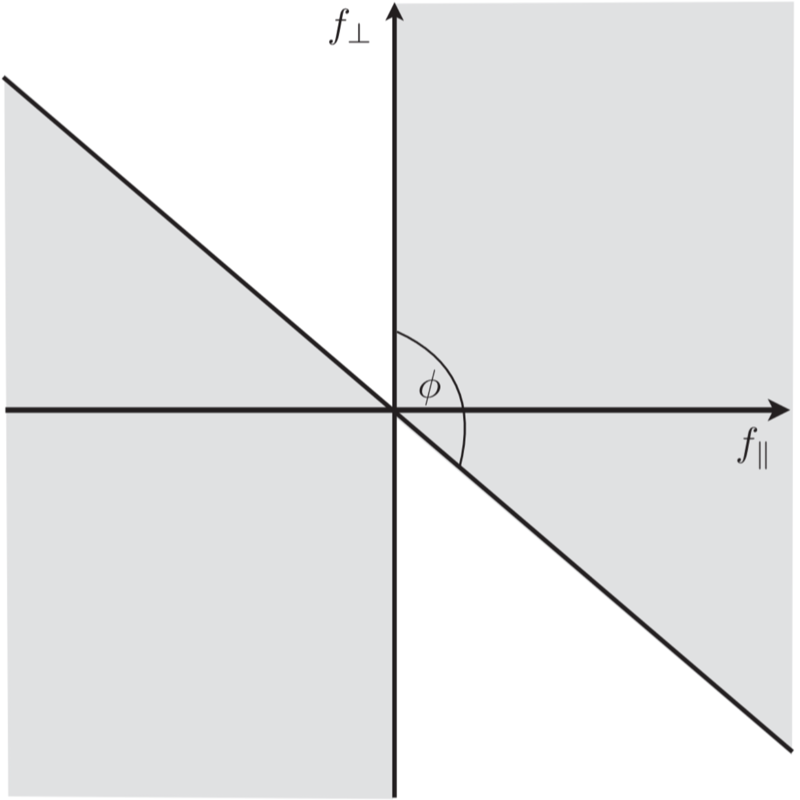}
\caption{Plane of components of the second eigenvector, $\vec f _2$, parallel to $\vec u^{\,\rm dir}$, $f_{||}$, and perpendicular to $\vec u^{\,\rm dir}$, $f_\perp$.  The direct Mpemba index $\mathcal{I}^{\rm dir} _{\rm M}$ is odd in the double-wedge shaded region of the $f_{||}f_{\perp}-$plane if $\vec u^{\,\rm dir} \cdot \vec w >0$, and odd in the white regions if $\vec u^{\,\rm dir}\cdot \vec w <0$. \textit{Source:} Reprinted with permission from~\cite{klich2019mpemba}.}
\label{fig:fig-isotropic-ensemble-v01.png}
\end{figure}

Substituting the approximation for the second eigenvector, ~\EQ{f2approx}, into~\EQ{Parity-dir-sym} gives the parity of the direct Mpemba index,
\begin{align}
   \mathcal{P}(\mathcal{I}^{\rm dir} _{\rm M}) = \theta \left[\left(\vec g \cdot \vec u ^{\,\rm dir} _{\rm iso}\right) \left(\vec g \cdot \vec w\right)\right], 
\end{align}
where 
\begin{align}
    \left(\vec u ^{\,\rm dir} _{\rm iso}\right)_i \equiv e^{E_i/(2k_B T_b)} - \frac{L e^{-E_i/(2k_B T_b)}}{Z(T_b,\{E_i\})}. 
\end{align}
Expressing the vector $\vec g$ in terms of two components, one that is parallel to $\vec u ^{\,\rm dir} _{\rm iso}$ and denoted as $g_{||}$ and another that is orthogonal to $\vec{u}^{\,\rm dir}_{\rm iso}$ and denoted as $g_{\perp}$, we again geometrically get a double wedge where $\mathcal{P}(\mathcal{I}^{\rm dir}_{\rm M}) > 0$. Additionally, when estimating the probability of the Mpemba index, it is essential that the entries of $\vec g$ be distributed isotropically, and thus also the distribution of entries in its orthogonal projection onto $\vec f_1$. Since the components $g_i$ are independent and identically distributed (i.i.d.) Gaussian random variables, the same applies to $g_{||}$ and $g_{\perp}$. Consequently, we conclude that the area corresponding to the double wedge where $\mathcal{P}(\mathcal{I}^{\rm dir}_{\rm M}) > 0$ is defined as 
\begin{align}
\label{eq:ProbParity}
\mathrm{Prob}\left(\mathcal{P}(\mathcal{I}^{\rm dir}_{\rm M}) > 0\right) =
    \frac{1}{2} + \frac{\textrm{sign} \left(\vec u^{\,\rm dir}_{\rm iso}\cdot \vec w\right)}{\pi} \arctan \frac{1}{K \left(\vec u^{\,\rm dir}_{\rm iso}, \vec w\right)}, 
\end{align}
with $K$ defined in~\EQ{Kscalarprod}. 
The above formula estimates the probability that, for a fixed set of energies and bath temperature $T_b$, we have a positive direct Mpemba index. The index gives a lower bound for the occurrence of the direct Mpemba effect. In the high-bath-temperature limit, one finds the asymptotic approximation
\begin{align}
    \mathrm{Prob}\left(\mathcal{P}(\mathcal{I}^{\rm dir} _{\rm M}) > 0\right)\approx \frac{C(\{E_i\})}{k_BT_b}, 
\end{align}
where the constant $C(\{E_i\})$ depends solely on the energies. The explicit form of $C(\{E_i\})$ can be found in Ref.~\cite{klich2019mpemba}. 

The numerical results from the random barrier ensemble are compared to the analytical estimates obtained from the isotropic ensemble in~\FIG{ensemble-comparison}. In panel (a), the histograms depict the components of the second eigenvector, $(\vec{f}_2)_i$, generated from the random barrier ensemble in red, alongside the components of the ``candidate second eigenvector,” $(\vec{f}^{\,\, \rm{iso}}_2)_i$, shown in green. Both ensembles were created for a fixed energy realization comprising $L = 25$ energies drawn from a Gaussian distribution with mean zero and standard deviation $1.5$. Each ensemble consisted of $20,000$ components for the second eigenvector. The barriers were selected from a half-normal distribution with standard deviation of $75.6$. The bath temperature was set to $10$. The components of the random vectors were chosen independently and identically distributed (i.i.d.) from a Gaussian distribution with mean zero and standard deviation $0.19$. As noted in the inset, the two probability distributions exhibit differences at the tails. Figure~\ref{fig:ensemble-comparison}b shows that the expression~\EQ{ProbParity} obtained for the isotropic ensemble (solid line) captures well the probability of the direct Mpemba index for a random barrier ensemble (points) with a fixed energy realization of $L = 25$ energies. This is averaged over $20,000$ randomly selected barrier realizations, based on the~\EQ{ProbB} statistics. The parameters for the statistics of energies, barriers, and the random vector $\vec{g}$ were the same as those used in Fig.~\ref{fig:ensemble-comparison}a, although the energy sample realization was different. Both panels illustrate that the two ensembles align well when the barrier distribution and the bath temperature exceed the energy spread. In general, more work is needed to establish the range over which the average of the random barrier ensemble closely aligns with the isotropic ensemble.

\subsection{Outlook on statistics of the Mpemba effect}

For the ensembles considered above, the Mpemba effect occurs with probabilities $\gtrsim 0.35$, suggesting that the Mpemba effect likely does not have ``self-averaging properties'': the properties of one large system do not determine the typical realization.  Even in large system sizes, different realizations have different Mpemba properties. This fact encourages careful design of relaxation protocols, given an energy landscape that could, in principle, lead to faster relaxation. The REM was introduced as a simple disordered model that shares several qualitative features with spin-glasses. The Mpemba effect has been studied for spin-glasses~\cite{baity2019mpemba} (see Sec.~\ref{Sec:NumericalSpinSystems}), where the connection with the correlation length has been observed. It would be interesting to compare the statistics of the Mpemba effect in spin-glass systems and with those found for the REM. 

Besides annealing, cooling, and heating protocols, questions of efficient relaxation also arise in other contexts, such as the design of efficient samplers. In particular, Markov chain Monte Carlo algorithms use Markov chains to create realizations of the system at its steady state~\cite{sokal1997monte,krauth2007introduction}. It is often desirable to accelerate the relaxation of a Markov chain to a target steady state~\cite{turitsyn2011irreversible,vucelja2016lifting}. One possible approach could be to design relaxation dynamics (barriers) with a Mpemba effect, for a given system and initial condition. For example, although there is no Mpemba effect for a complete graph with Metropolis dynamics, adding barriers that remove some state transitions can induce a Mpemba effect~\cite{klich2018solution}. We discuss more Markov chain Monte Carlo sampling in connection to the Mpemba effect in~\SEC{mcmc}. 

Another example of predicting the Mpemba effect without explicitly calculating the second eigenvector was presented in~\cite{amorim2023predicting}, where the authors use machine-learning techniques to predict the effect in Ising spin systems subject to an external magnetic field. Several methods were employed to perform such tasks, including the decision-tree algorithm, neural networks, linear regression, nonlinear regression with the least absolute shrinkage, and the selection operator (LASSO) method.
For large training datasets, in~\cite{amorim2023predicting}, the authors found that neural networks gave the most accurate prediction of the $a_2$ coefficient, given parameters such as the number of spins, the interaction strength, the external magnetic field, and the initial temperature.

The above methods all have difficulties related to their inability to handle systems characterized by symmetries in the ground state and in the extrapolation of predictions for larger systems. Hence, further investigation in this direction is required to see whether a machine-learning approach can help where methods such as spectral analysis of the evolution operator fall short.

Approximations of the second eigenvector, such as those described above with the isotropic ensemble, or predictions of the Mpemba effect with neural networks without explicit computations of the second eigenvector, provide information about the relaxation of a physical system without knowing exactly the spectra decomposition. They are invaluable when exact solutions are impossible or hard to obtain. Moreover, the results in this section indicate that the Mpemba effect is \emph{non-self-averaging}; that is, the effect's existence cannot be predicted from the average behavior.

We should also mention the study of Meibohm and Knapp on the asymptotic relaxation rate of a system perturbed by quenched disorder~\cite{meibohm2024exponential}. Averaging over disorder, they found that the mean and variance of the asymptotic relaxation rate $\lambda _2$ depends nonmonotonically on the parameters of a broad class of disorder realizations. Thus, the type of disorder affects the asymptotic relaxation rate. It would be interesting to consider further the statistics of spectral changes to the relaxation rate matrix (operator) given such perturbations. And it would be especially interesting to explore them in the context of a potential relaxation speed up caused by the quenched disorder.

\section{Experiments}
\label{sec:experiments} 
In the Introduction (Section~\ref{sec:introduction}), we reviewed historical observations of anomalous behavior in the freezing of water, where, after a sudden temperature quench, hot water is claimed to cool or start freezing more quickly than water at intermediate temperatures.  Modern discussions of this phenomenon date from the experiments by Mpemba and Osborne in 1969~\cite{mpemba1969cool} and the nearly simultaneous, independent work of Kell~\cite{kell1969the}.  We reviewed the complicated history of studies on cooling and freezing water, which continue to this day.  However, although water has played an important historical role in suggesting and providing some evidence for the Mpemba effect, the analysis of individual experiments is complicated.  More importantly, focusing on one system raises the question of the generality of anomalous cooling (and heating) phenomena.  If the time to reach thermal equilibrium is not a monotonic function of the temperature difference between initial and final states, is such behavior observed more generally or only in water?  To persuade the reader that the Mpemba effect is not unique to water, we review in this section experiments in other systems.  

We begin in Sec.~\ref{sec:otherCM} with a survey of observations of Mpemba effects in condensed matter systems other than water.  These are important because they were the first to suggest that an explanation of the Mpemba effect should not be based too firmly on the peculiar features of water.  Unfortunately, the physics involved is at least as complex as in the water-based experiments and is often more poorly understood.  In Sec.~\ref{sec:colloids}, we present a set of recent experiments using a colloidal particle diffusing in water and moving in an external potential.  The experiments were the first to clearly demonstrate the weak (original)  Mpemba scenario, albeit in a simple system.  They also provided the first evidence for more subtle phenomena, such as the strong and inverse Mpemba effects discussed in Sec.~\ref{sec:strong} and \ref{sec:inverse}.  Section~\ref{sec:quantum-experiment} presents the first experiments to demonstrate the Mpemba effect in quantum systems.  

\subsection{Other condensed-matter systems}
\label{sec:otherCM}

The notoriety of the Mpemba effect for cooling and freezing in water has led other researchers to highlight or look explicitly for similar effects in other material systems:

\begin{itemize}

\item \textit{Polymers}.  Hu et al.~studied the crystallization of polylactide (PLA), an environmentally friendly polymer produced from renewable resources~\cite{hu2018conformation}.  In their experimental protocol, they first heated the PLA to 220 $^\circ$C, which completely melts the system and allows it to reach thermal equilibrium in a disordered state.  They then do a first quench into a glassy region of varying temperatures (140--200 $^\circ$C), ``hold" for five minutes, and then quench to 80 $^\circ$C.  They then observe that the fraction of crystallinity is higher for PLA samples held at an initially greater temperature.  Following suggestions from earlier molecular-dynamics simulations of coarse-grained polymer models~\cite{luo2016role}, they conclude that differences in chain conformations at the various holding temperatures accounts for the varying crystallinity.  Qualitatively similar results were observed by Liu et al. in polybutene-I~\cite{liu2023mpemba}.  In this work, the system was heated from 25~$^\circ$C to a higher temperature (120--150 $^\circ$C), held for ten minutes, and then quenched at 30 $^\circ$C/min to 70 $^\circ$C.  They again found that crystallinity increased faster for initially higher holding temperatures.  All of these experiments on polymeric systems differ from the usual Mpemba setting in that the initial states before the final quench are not in thermal equilibrium.

\item \textit{Clathrates}.  Ahn et al.~observed the Mpemba effect experimentally in clathrate hydrates consisting of hydrogen-bonded water frameworks and enclathrated tetrahydrofuran (THF) molecules~\cite{ahn2016experimental}.  They observed that THF solutions display Mpemba-like behavior in the temperature range 278--318 K, where warmer solutions form hydrates faster than colder solutions. At an initial temperature above 318 K, the formation time was delayed, as some THF molecules in gas form in solution at high temperature must condense before they can form hydrates.  Ahn et al. defined the freezing temperature as the temperature at which the THF solution structurally transforms into THF hydrates and measured it via differential scanning calorimetry (DSC).  They checked that they could observe a Mpemba effect in a pure-water system using DSC and found a nonmonotonic, roughly parabolic dependence of time to freeze on the initial temperature $T_i$ that is characteristic of the Mpemba effect.  THF solutions also showed a nonmonotonic relationship between time to freeze and $T_i$ (roughly a cubic).  Ahn et al.~then studied the cooperative relationship between the intramolecular polar-covalent bonds (O--H) and intermolecular hydrogen bonds (O:H) for the THF system.  Using Raman spectra, they observed that O:H stretching phonons of water in THF showed a blue shift (stiffened) when cooled, whereas the O--H stretching mode showed a redshift (softened).   Such trends are also found in pure water and thus at least qualitatively support a role for hydrogen bonding in anomalous cooling, as suggested by Zhang et al.~\cite{zhang2014hydrogen,sun2023the}.

\item \textit{Ionic liquids}.  Ionic liquids have potential applications as environmentally friendly solvents and as energy-storage media, using the latent heat of phase transitions.  The latter motivated a recent study of crystallization of 1-ethylpyridinium triflate (epy) where the Mpemba effect was observed in an aqueous epy solution~\cite{chorazewski2023the}.  In particular, quenching an epy-water solution did speed up the solidification time for higher initial temperatures, whereas slow heating and cooling using a DSC did not show such anomalies.

\item \textit{AgAu nanoshells}. Ferbonik et al. studied the pump-probe response of hollow AgAu nanoparticles of diameter 20--30 nm~\cite{ferbonink2020stochastic}.  The sudden absorption of high laser power excites acoustic-phonon modes that can be detected by a probe laser.  The authors find that the phonon-response decay time at lower powers increases monotonically with pump-laser power; however, at intermediate powers, the decay time starts to decrease with pump power.  One might naturally speculate that such a nonmonotonic response would imply a Mpemba effect, since the initial temperature should correlate with the energy injected.  A subsequent theoretical analysis partially backs up this intuition.  Using molecular dynamics (MD) simulations, the authors deduce an effective potential landscape for the excited system, simulated a Fokker-Planck equation for the evolution of the excited system, and examined the decay of a ``distance from equilibrium function.'' The authors do not observe a Mpemba effect in their simulated system; rather the distance function and equilibration times saturate with increasing initial energy (pump power).  Still, the saturation is ``close" to a Mpemba effect, which would require a decrease in equilibration time.

\item \textit{Magnetic system}.  Chaddah et al. studied the magnetic, not thermal, relaxation of polycrystalline manganite La$_{0.5}$Ca$_{0.5}$MnO$_3$ (LCMO)~\cite{chaddah2010overtaking}.  A sample was cooled from 320 K to 5 K in different external fields.  Then, at 5 K, the field is quenched to a smaller value, and the authors measure the decay of the initial magnetization.  They find that states that are initially farther from equilibrium can overtake states that start closer to equilibrium, in a magnetic analog of the Mpemba effect.
\end{itemize}

As a final note, while all these experiments exhibit at least some aspects of the  ``standard'' Mpemba effect, where a system starts in equilibrium and relaxes to a cooler equilibrium (perhaps with a phase transition to a new state) more quickly for initially hotter samples, they all have complicating aspects.  The polymer samples do not start in equilibrium states.  The clathrate samples depend on differential scanning calorimetry (DSC).  Although the authors do claim to check that they see a Mpemba effect in water via DSC, the technique involves slower cooling (5 K/min.) than typical Mpemba experiments, where the environmental temperature is changed by 10s of degrees K in a few seconds.  The smaller sample size used in DSC measurements (20 mg, about 1000 times smaller than other experiments) might explain the discrepancy, but that needs to be demonstrated.  The hollow nanoparticles and magnetic systems showed nonmonotonic effects in other variables---these could be characterized as anomalous non-thermal relaxations.  The ionic-liquid experiment is perhaps closest to the water version of the Mpemba effect, but only a handful of cooling curves are examined.  Still, the experiments collectively make the case that anomalies in relaxation are present in other systems, making it all the more likely that any explanations need not be based on the particular physics of water.

\subsection{Colloidal systems}
\label{sec:colloids}

In influential experiments, Kumar, Ch\'etrite, and Bechhoefer introduced a colloidal system as a simple setting to study the Mpemba effect and close variants~\cite{kumar2020exponentially,chetrite2021metastable,kumar2022inverse,bechhoefer2021fresh}. The work is distinctive for three reasons: first, it was directly inspired by and makes direct contact with the recent theoretical advances discussed in this Review (especially Refs.~\cite{lu2017nonequilibrium} and \cite{klich2019mpemba}); second, the experiments showing the weak Mpemba effect were significantly more quantitative and reproducible than earlier work; and third, the experiments gave the first observations of the strong and inverse Mpemba effects.

Unlike previous work that focused on materials made from macroscopic numbers of atoms, the colloidal system is a single, micron-scale particle at position $x(t)$ that diffuses in a carefully designed potential, $U(x)$. The relevant particle dynamics are well described by a one-dimensional, overdamped Langevin equation of the form
\begin{align}
\label{eq:langevin-colloid}
	\gamma \dot{x}(t) = - \left. \pdv{U}{x} \right|_{x(t)} + \xi(t) \,,
\end{align}
where the friction coefficient $\gamma$ is approximately equal to the Stokes drag, $6\pi \eta r$, with $r$ the particle radius and $\eta$ the fluid viscosity. Note that when the particle is near a solid surface,  the drag increases. The mass of the particle has been set to unity. The stochastic force $\xi(t)$ describes the irregular impacts of water molecules on the particle and has moments $\langle \xi \rangle = 0$ and $\langle  \xi(t) \, \xi(t')  \rangle = 2 k_BT \gamma \, \delta (t-t')$, which is an expression of the fluctuation-dissipation relation. The potential $U(x)$ has a tilted double-well shape that is inserted into a box with steep sidewalls, as illustrated in Fig~\ref{fig:strong_sketch.png}b.

\begin{figure}[ht]
    \centering
    \includegraphics[width=0.6\columnwidth]{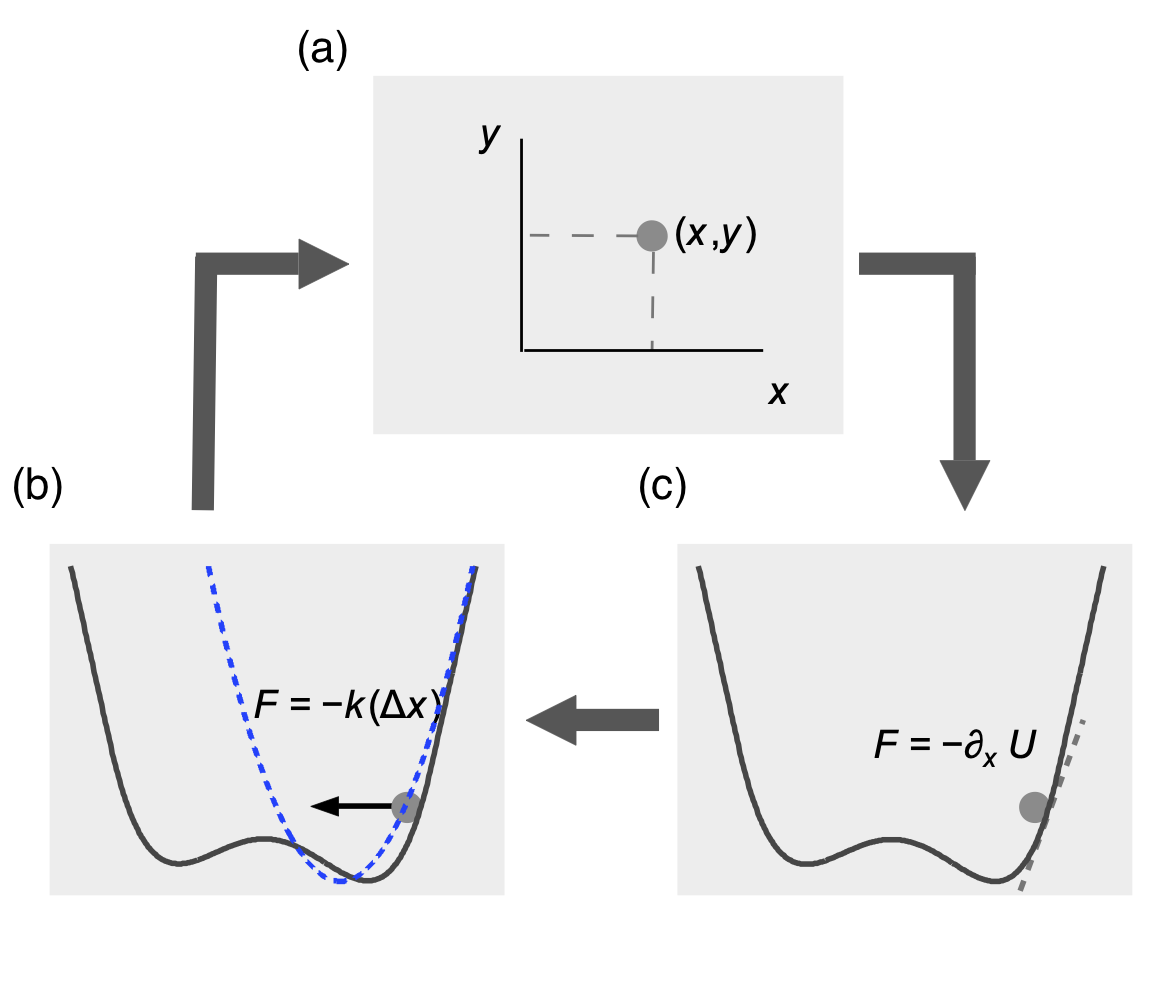}
    \caption{Creation of a virtual potential in three steps:  (a) Measure the particle position;  (b)  calculate the force as the negative gradient of the desired potential, $U(x)$; shift the center of the optical tweezer by an amount calculated to produce the desired force.  The arrows depict one cycle of a feedback loop.
    \textit{Source:}  Figure reproduced from~\cite{kumar2020exponentially}.}
    \label{fig:virtualPotential}
\end{figure}

As we shall see, controlling the shape and spatial scale of the potential are important requirements.  In the work reported here, the potentials were ``virtual potentials'' that were created by a feedback loop.  The basic procedure is illustrated in Fig.~\ref{fig:virtualPotential} and adapts the feedback (or ``ABEL") trap created for biophysical applications~\cite{cohen2005control} to the study of larger-scale colloidal particles~\cite{jun2012virtual}.  The Mpemba experiments used a second-generation version of feedback trap that had special hardware to reduce the feedback loop time and applied forces based on the displacements of an optical tweezer (dashed-blue potential in Fig.~\ref{fig:virtualPotential}b).  The latter was more stable than earlier feedback traps that depended on electrokinetic forces and showed attendant chemical drifts. 

Feedback traps have two key advantages for the Mpemba experiments:  First, one can create a nearly arbitrary potential shape, limited only by the maximum force (limit on potential slope) and feedback loop time (limit on potential curvature)~\cite{jun2012virtual}.  Second, because the method needs only to measure the position of an isolated particle, the localization and hence potential shape can be smaller than the wavelength of light used in the optical tweezers.  In the experiments reported here, for example, the typical separation between the two potential wells was $\ell = 80$~nm.  With diffusive dynamics, the overall  of experiments is proportional to $\ell^2$.  One could create double-well potentials using dual tweezers or spatial-light modulators (SLMs), but their length scale would be $\ell \gtrsim 1$~\textmu m.  The ratio of length scales implies that experiments done using a feedback trap can be more than $100$ times faster than similar experiments done using dual-tweezers or SLMs.

Using the feedback trap, Kumar and Bechhoefer reported experiments on the standard Mpemba effect and the strong Mpemba effect in Ref.~\cite{kumar2020exponentially}.  Then Ref.~\cite{kumar2022inverse} reported first observations of the inverse Mpemba effect.  The procedure for all these experiments is to draw an initial condition for a particle from a Boltzmann distribution $\pi(x;T)$ that corresponds to a system in equilibrium at a temperature $T$ that, in general, differs from the bath temperature $T_b$.  Because the particle moves in water at temperature $T_b$, the effective quench from $T$ to $T_b$ is instantaneous.  This trick of using the initial position of the particle as a substitute for a physical temperature eliminates any complication arising from finite-time transients. 

After the particle is released, its trajectory is measured for a time that is long enough to thermalize.  The small energy barriers (typically 2 to 3 $k_BT_b$) and the small separation between the wells meant that the equilibration times were short, even in the ``longest" cases.  Thus, the dynamics was typically followed for 60 \textmu s.  Because each run was short, it was possible to have a large number of trials for each condition considered (from 1000 to 30,000). The large number of trials greatly improved the reproducibility of the results. Recall from Section~\ref{sec:historical} that experiments on water have all fewer than 10 trials per condition.

\begin{figure}[ht]
    \centering
    \includegraphics[width=0.6\columnwidth]{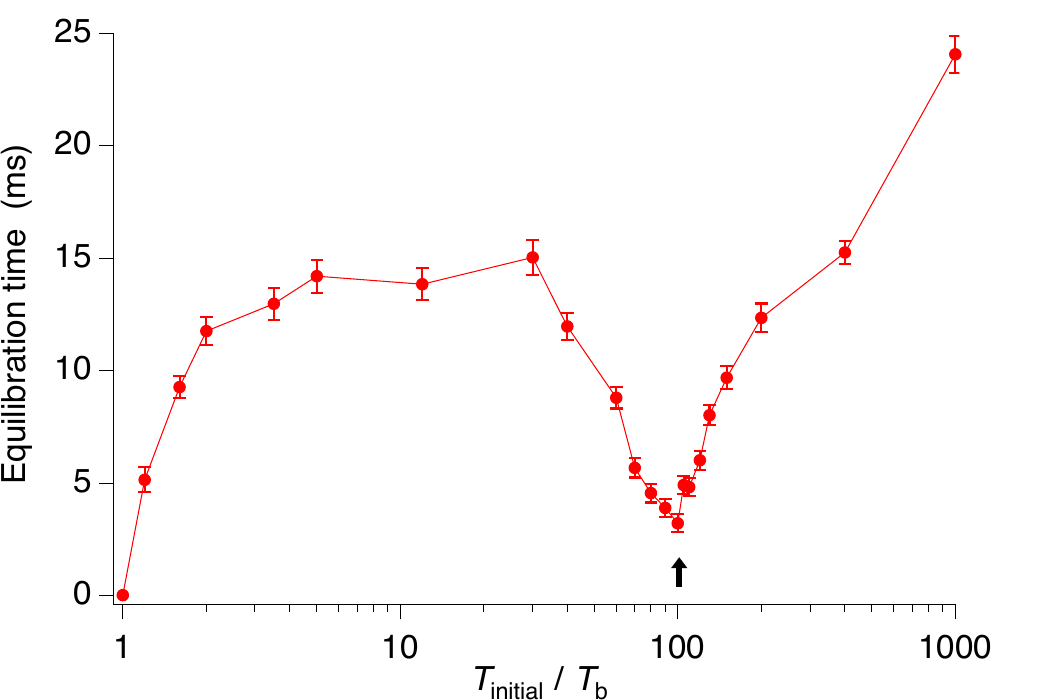}
    \caption{Equilibration time of ensemble of colloidal particles as a function of the initial temperature, for an asymmetry parameter $\alpha=9$. Arrow indicates the strong Mpemba effect. \textit{Source:} Reprinted with permission from~\cite{kumar2020exponentially}.}
    \label{fig:strongMpemba}
\end{figure}

The trajectories $x(t)$ for repeated trials are grouped to form a statistical ensemble, from which, at each time $t$, one can estimate the probability density $p(x,t)$ of beads.  This probability density then relaxes to the equilibrium Boltzmann distribution, $\pi(x,T_b)$. Figure~\ref{fig:strongMpemba} shows an example of the evolution of the equilibration time as a function of the initial temperature.  The time is by definition zero when $T_\textrm{initial} = T_b$.  As $T_\textrm{initial}$ increases, the time to equilibrate also increases.  Then, for $T_\textrm{initial} / T_b \approx 10$, there is a local maximum in equilibration time and further increases in initial temperature \textit{decrease} the equilibration time.  This decrease is a signature of the Mpemba effect.  The local minimum in equilibration time at $T_\textrm{initial} / T_b \approx 100$ (indicated by the black arrow) then reflects the \textit{strong Mpemba effect}.  To identify the local minimum with the strong Mpemba effect, the authors used measurements of the relaxation of $p(x,t)$ to equilibrium, to extract the (modulus of the) $a_2$ coefficient discussed in Sec.~\ref{sec:phase-transitions}.  The strong Mpemba effect corresponds to the vanishing of $|a_2|$ and, as a result, an exponential speedup of relaxation.

To understand the physical origin of the Mpemba effect, we first note that, as suggested by Lu and Raz~\cite{lu2017nonequilibrium} (see Fig.~\ref{fig:brownian_system} and also the discussion in Ref.~\cite{chetrite2021metastable}), relaxation proceeds in two stages:  In the first stage, there is a rapid relaxation from the initial position $x_0$ where the particle (usually) falls into the nearest well.   The fraction of particles in each well will usually differ from the equilibrium fraction.  In the second stage of relaxation, particles hop back and forth across the barrier and equilibrate the populations in the two wells.   

The initial fraction of particles that fall into each well is influenced by the placement of the double-well potential within the confining box.  In particular, in the limit of high initial temperatures, the distribution of conditions approaches uniform, so that an asymmetric placement (illustrated in Fig.~\ref{fig:strong_sketch.png}b and quantified by the asymmetry parameter $\alpha = x_\textrm{max} / x_\textrm{min}$ defined in that figure) implies a simple relationship between the asymmetry $\alpha$ and the fraction of particles in each well after the first stage.  If one adjusts $\alpha$ so that the fraction after Stage 1 equals the equilibrium partition, then there is no need for Stage 2.  This situation corresponds to the strong Mpemba effect.  The intuitive argument given here has recently been made more rigorous and generalized~\cite{walker2022mpemba}.

The same general set up was also used to study the inverse Mpemba effect~\cite{kumar2022inverse}.  Here, a colder initial condition can heat up more quickly than a warmer one.  The physics and mathematics describing the inverse effect is qualitatively similar to that describing the usual cooling scenarios and, indeed, both weak and strong versions of the inverse Mpemba effect were observed.  One significant difference between the two settings, though, is that the effects were harder to observe.  A generic reason is that the target high-temperature state had very low energy barriers and the motion was almost a pure diffusion.  In such a system, the ratio of eigenvalues $\lambda_3 / \lambda_2 \approx 4$.  This ratio gives a measure of how much data is needed to separate the $e^{-\lambda_3 t}$ decay component from the $e^{-\lambda_2 t}$ component that dominates at long times.  The larger the ratio, the easier the separation and the fewer the trials required.  In the original forward-Mpemba experiments, the energy barriers present at the bath temperature led to an exponential enhancement of the ratio $\lambda_3 / \lambda_2 \approx 15$.  As a consequence, the inverse-Mpemba experiments needed 5000 trials to infer $a_2$ and $a_3$, whereas only 1000 trials were needed for corresponding accuracy in the forward-Mpemba experiments.

Taken collectively, these experiments on a colloidal system give convincing demonstrations of the basic phenomena associated with Mpemba effect in a very simple setting.  The good agreement between model calculations based on Fokker-Planck and Langevin equations bolsters the argument that these kinds of anomalous relaxation are generic in the sense that they can occur in a wide range of systems and are not linked to the special physics of one particular system such as water.

\subsection{Experiments on quantum Mpemba effects}
\label{sec:quantum-experiment}

In January, 2024, three preprints posted to arXiv gave evidence for the Mpemba effect in systems where quantum effects are important. The experimental apparatus in all three works is based on trapped ions, but the three works are quite different: \cite{aharony2024inverse} measured an inverse Mpemba effect of the type discussed in~\SEC{QuantumMarkov};  \cite{zhang2024observation} measured exponential acceleration in the relaxation rate by first applying a unitary transformation on the initial state to annihilate its overlap with the slowest relaxation mode, as discussed in~\SEC{QuantumUnitary}; and  \cite{joshi2024observing} measured the entanglement asymmetry quantum Mpemba effect (discussed in~\SEC{EntanglementAsymmetry}) in a system of 12 ions.

In all three experiments, the system is composed of one~\cite{aharony2024inverse,zhang2024observation} or many~\cite{joshi2024observing} ions, trapped in a Paul trap, and laser cooled. This type of setup has become standard, with applications ranging from accurate clocks \cite{burt2021demonstration} to precise magnetometry \cite{kotler2011single} to quantum computing \cite{bruzewicz2019trapped}. See \cite{leibfried2003quantum,duan2010colloquium,haffner2008quantum} for reviews describing the experimental setups.  
Interestingly, an additional experimental manuscript that aims to measure the quantum Mpemba effect was also uploaded to the arXiv in January 2024 \cite{edo2024study}. However, this manuscript differs from the other three manuscripts in two major points: (i) Instead of trapped ions, the authors used a publically available IBM quantum computer that is based on superconducting qubits. The authors considered a single-qubit thermal relaxation that exists because of the coupling between the system and its environment. Rather than initializing the qubit in a hot thermal state, it was initialized in a pure state, and the thermal state relaxation trajectory was calculated by averaging over these states. More importantly, (ii) no Mpemba effect of any type, experimental or theoretical, was observed in this work. We therefore do not discuss this manuscript further. 

\subsubsection{Inverse Mpemba effect in a single strontium ion} 
In \cite{aharony2024inverse}, a nonequilibrium strong inverse quantum Mpemba effect was demonstrated experimentally on a single trapped $^{88}$Sr$^+$ ion. The relevant degrees of freedom in the system are the ion's internal energy levels, not the motion degrees of freedom. Of all the internal states of the ion, two states are considered as ``the qubit system," whereas other states are used as auxiliary states~\cite{aharony2024inverse}. The transitions between the two states have several sources: first, there is a coherent driving generated by an oscillating magnetic field with a frequency that is tuned to the specific transition. In addition, there is an incoherent decay transition between the two states, mediated by a thermal bath, which is generated by coupling the system to several auxiliary states in a way that can effectively be analyzed as a thermal bath. Finally, there are dephasing processes, also generated by the thermal bath, namely by the coupling to the auxiliary states. As these different processes cannot be coupled simultaneously to the system, they are done sequentially, i.e., by Trotterization. Because there is coupling both to a thermal bath and to a coherent source, the steady state of the system is not an equilibrium but rather a nonequilibrium steady state. Nonetheless, since coherent driving is kept fixed  while the bath temperature varies, the result is an equivalent to the ``equilibrium locus'' in the classical Mpemba effect. 

To demonstrate a strong inverse Mpemba effect, one prepares the system with a steady-state density matrix corresponding to a low temperature and then quenches to a high temperature. The dynamics of the density matrix then follows~\EQ{LindbladianDyna}, relaxing towards the new steady state, with the slowest decay mode controlling the long-time limit of the relaxation. For this specific system, only a strong inverse Mpemba effect exists; there is no strong direct effect.   

\subsubsection{Unitary transformation in a single calcium ion}
\label{sec:unitary-quantum-experiment}
As in the above experiment, in~\cite{zhang2024observation} the internal energy states of $^{40}$Ca$^+$ ions were used as the relevant degrees of freedom for the relaxation process. Here, three energy levels  coupled via lasers having the relevant frequencies. The density matrix describing the system evolves under Lindbladian dynamics as described in Eq.~\eqref{eq:LindbladianDyna}, and the steady state of the system is a nonequilibrium steady state, not a Boltzmann distribution.

To observe the desired effect, the system is initiated in a pure state that corresponds to one of the energy levels and is denoted by $|0\rangle$. Its evolution towards the steady state is then observed. The relaxation is compared to a system that, after starting from the same initial condition, evolves according to a carefully designed unitary Hamiltonian dynamics to a state that has no overlap with the slowest relaxation mode of the Lindbladian; as a result, the relaxation is exponentially faster than generic relaxation. An additional effect in this system is the Liouvillian Exceptional Point (LEP), where two real eigenvalues coalesce to form a pair of complex-conjugate eigenvalues. In this case, there are two decaying modes with identical real values; the overlap of initial state with both must vanish in order to obtain a faster relaxation rate. 

\subsubsection{Entanglement-asymmetry Mpemba effect in a linear string  of twelve calcium ions}
\begin{figure}
    \centering
    \includegraphics[width=0.65\linewidth]{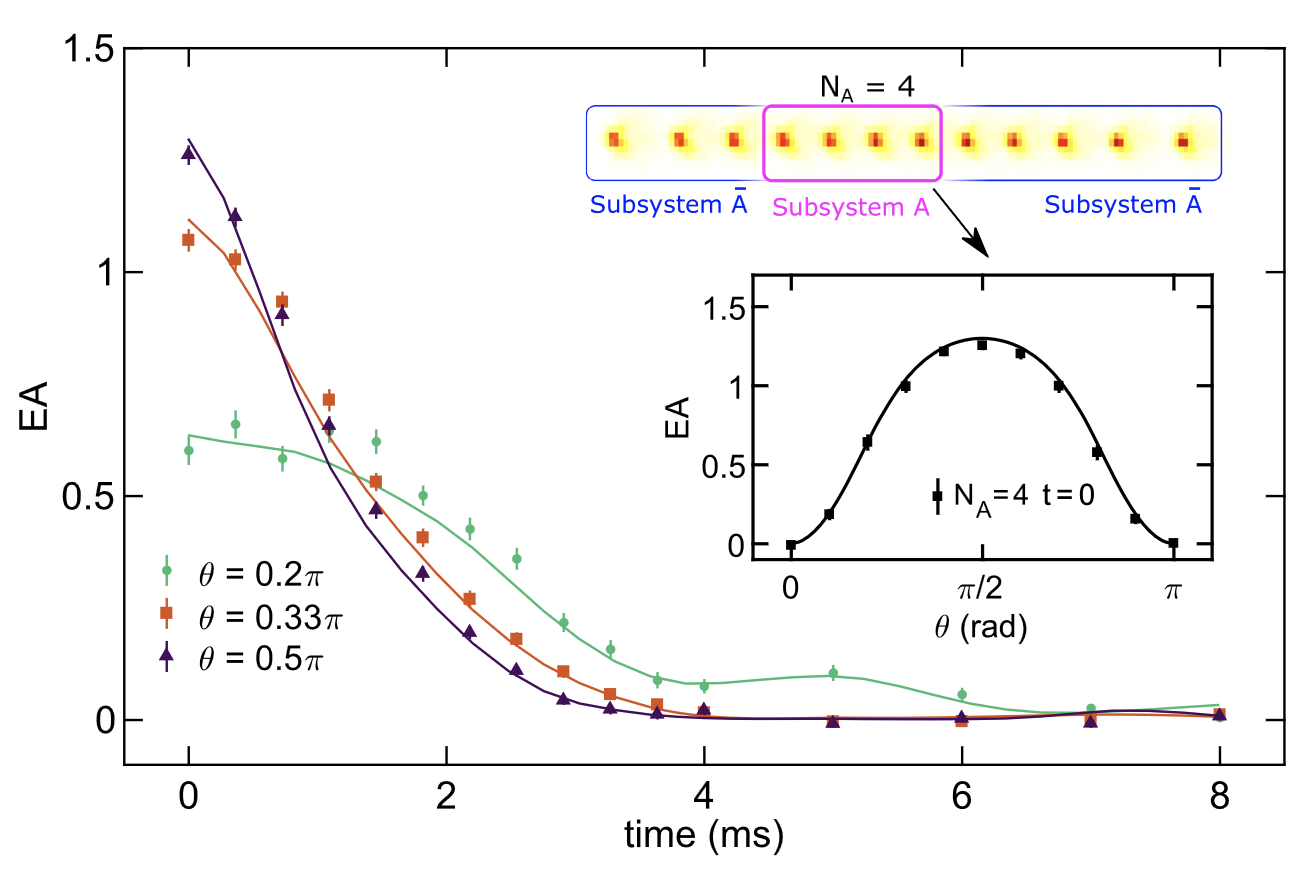}
    \caption{Experimental demonstration of the entanglement asymmetry quantum Mpemba effect. The system (upper inset) is a 1D system of twelve trapped ions, of which four constitute ``the system" (subsystem $A$), and the rest the ``bath" (subsystem $\tilde A$). The entanglement asymmetry (EA) of subsystem $A$ changes as a function of a global rotation angle, $\theta$ (lower inset). Initiating the system at different values of $\theta$ and observing the entanglement asymmetry evolving in time shows a clear crossing between the relaxation trajectories. \textit{Source:} Reprinted with permission from~\cite{joshi2024observing}.} 
    \label{fig:quantum_asymmetry}
\end{figure}

In \cite{joshi2024observing}, a string of 12 $^{40}$Ca$^+$  ions was cooled to the motional ground state. In each ion, two spin states are used as the relevant degrees of freedom. The spin-$\frac{1}{2}$ of each ion interacts with all other ions, and the interaction strength changes with the distance between the ions. The system is initialized in the state where all spins point in the $z$ direction. Then, they are all tilted by an angle $\theta$ from the $z$-axis. After initialization, the system evolves according to an engineered $XY$ Hamiltonian \cite{monroe2021programmable}. This dynamics possesses a $U(1)$ symmetry, with a conserved quantity of the total magnetization in the $z$ direction: $Q=\sum_i \sigma^z_i$. However, initial states with $\theta\notin\{0,\pi\}$ explicitly break the $U(1)$ symmetry. The broken symmetry can be quantified using the entanglement asymmetry, which is defined as
\begin{align}
    \Delta S_A = \ln\left[{\rm Tr}(\rho_A ^2)\right] - \ln\left[{\rm Tr}(\rho_{A,Q}^2)\right],
\end{align}
where subsystem $A$ is defined as ions 4--7, the density matrix $\rho_A$ describes the state of the system traced over all spins that do not belong to subsystem $A$, and $\rho_{A,Q}$ is the density matrix $\rho_A$ projected on the total magnetization of subsystem $A$. A plot of entanglement asymmetry as a function of initial angle $\theta$ is given in the inset of \FIG{quantum_asymmetry}. 

For a 1D system, it is clear that in the thermodynamic limit any small subsystem should relax to a Boltzmann distribution that would restore the $U(1)$ symmetry after a long time, since there are no phase transitions in the system. Thus, the above entanglement asymmetry should decay as a function of time. Indeed, as shown in \FIG{quantum_asymmetry}, the entanglement asymmetry decays with time for all initial conditions. Moreover, the same figure demonstrates the existence of anomalous relaxation in the system, where initial conditions with higher entanglement asymmetry restore the symmetry faster than initial conditions with lower initial values of the entanglement asymmetry.

\section{Numerical observations}
\label{sec:numerical-observations}
Numerical simulations are useful and powerful, if not always insightful: as in experiments, the underlying mechanism cannot be directly observed. In the context of anomalous relaxations, numerical investigations have been used in several ways. For example, they have accompanied theoretical calculations and demonstrated abstract concepts and physical mechanisms relevant to both the Markovian and kinetic approaches discussed above. Even though such numerical results are important---they can be reassuring and helpful in developing intuitions---we do not focus on them here. Rather, we mostly discuss numerical studies that demonstrate anomalous effects that cannot be directly addressed using our current theoretical tools, either because these numerical results demonstrate a physical phenomenon that is different from those described by our current theoretical models, or because we cannot yet connect the numerical results to any of the theoretical results.

In what follows, we discuss three classes of numerical results: In Sec.~\ref{Sec:NumericalSpinSystems}, we discuss various results on spin systems; in Sec.~\ref{Sec:NumericalWater}, we discuss dynamical models of water or other liquids; and in Sec.~\ref{Sec:NumericalKinetic}, we discuss numerical results pertaining to the driven granular-gas systems discussed in Sec.~\ref{sec:kinetic-framework}. These classes differ both in their numerical aspects and in the observed physical phenomena: the aim of dynamical models is not necessarily to demonstrate the Mpemba effect directly but rather to find precursors of the effect, such as the density of nucleation sites, or rotation vs. translation thermal relaxation. In spin systems and driven granular gases, however, the relevant effects (direct and inverse Mpemba effects, optimal heating protocols that are not monotonic in temperature, etc.) are directly observed. 

\subsection{Numerical results in spin systems}
\label{Sec:NumericalSpinSystems}

Spin systems show a rich set of anomalous phenomena, ranging from the Markovian framework discussed in~\SEC{markovian} in 1D~\cite{lu2017nonequilibrium} to mean-field systems \cite{klich2018solution}, direct glassy Mpemba effect in spin glass systems~\cite{baity2019mpemba}, and other types of effects in ferromagnetic systems~\cite{vadakkayil2021should}. In what follows, we discuss these various phenomena.  

\subsubsection{Spin glass system}

The most extensive numerical study on a spin system was done in~\cite{baity2019mpemba}. In this work, a cubic system of $160\times 160\times 160$ spins was studied using the Janus II spin glass simulator \cite{baity2014janus}. Janus II is a computer implemented on field programmable gate arrays (FPGA), configured
to optimize Monte Carlo simulations of spin systems. Using special algorithms that take full advantage of its customized architecture, huge simulations can be efficiently implemented. The Hamiltonian of the system was chosen as
\begin{align}
    \mathcal{H} = -\sum_{{\rm n.n.\,} (x,y)}J_{x,y}\,\sigma_x \sigma_y,
\end{align}
where $\sigma_{x}\in\{+1,-1\}$ are spin degrees of freedom. Each of the coupling constants $J_{x,y}$ was randomly chosen from $\{+1,-1\}$ with equal probability, with only nearest neighbors coupling different from zero. This system is known to have a glass transition at the critical temperature $T_c\approx 1.102$. Even above the glass transition, equilibrating the system takes an extremely long time; it was thus not possible to sample initial conditions from the equilibrium distributions. Instead, the initial conditions for the study were created by first connecting the spins to a heat bath of the chosen temperature and then letting the system evolve under Metropolis Monte Carlo dynamics for a long time (although one that is shorter than the corresponding relaxation time). In particular, the system is subject to a first direct quench to a fixed temperature from a completely disordered initial state, i.e., one that corresponds to an infinite temperature, during a given waiting time.   In~\cite{baity2019mpemba}, the bath temperatures were $T_1=1.3$ for the initially hotter system and $T_2=1.2$ for the initially colder system. To ensure that the initially hot system is indeed ``hotter," the authors calculated the energy of each system at an initial time that was chosen by coupling to the bath long enough that the initially hotter system had a significantly higher mean energy than the one coupled to a cooler system. 
Thus, the initial conditions in the study were actually nonequilibrium distributions.  In this respect, the observed anomalous relaxation effect differs from other frameworks.

After the preparation stage, the systems were quenched to the spin-glass phase by coupling to a bath at $T_{final} = 0.7\approx0.64\ T_c$. As the final temperature is below the glass transition, the relaxation is extremely slow, and even $10^{11}$ simulation sweeps were not enough to observe the final stage of relaxation, as expected in a glassy system. To find a Mpemba effect, the energy per spin, averaged over both trajectories and the quenched disorder $J_{x,y}$, was calculated through the relaxation dynamics. Crossings of the energy relaxation curves versus time were taken to indicate a Mpemba effect. Such crossings, for initial temperatures above and below the glass transition and with subsequent relaxation below the glass transition temperature, can be seen on~\FIGS{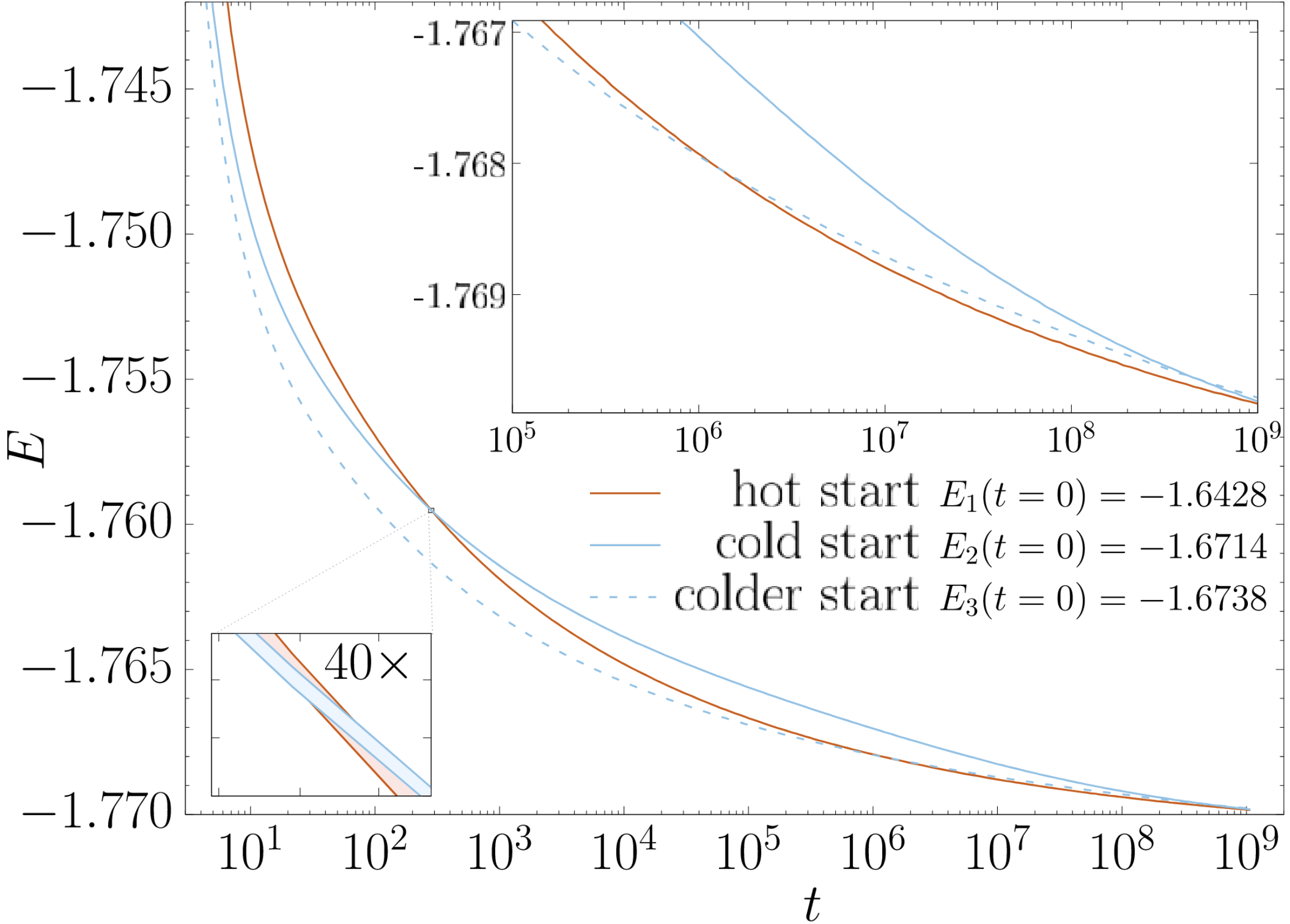}{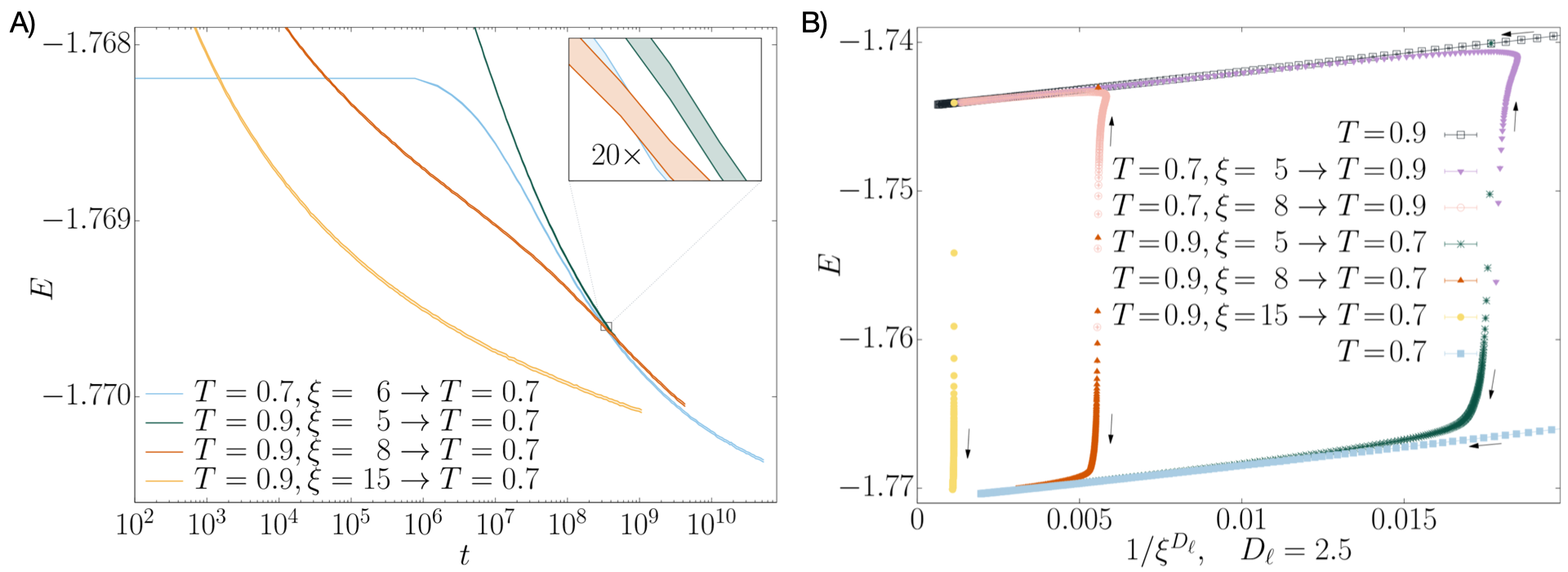}. An inverse Mpemba effect was also observed; in this case, all three temperatures were below the glass temperature of the model.   

\begin{figure}[tbh]
    \centering
    \includegraphics[width=0.5\linewidth]{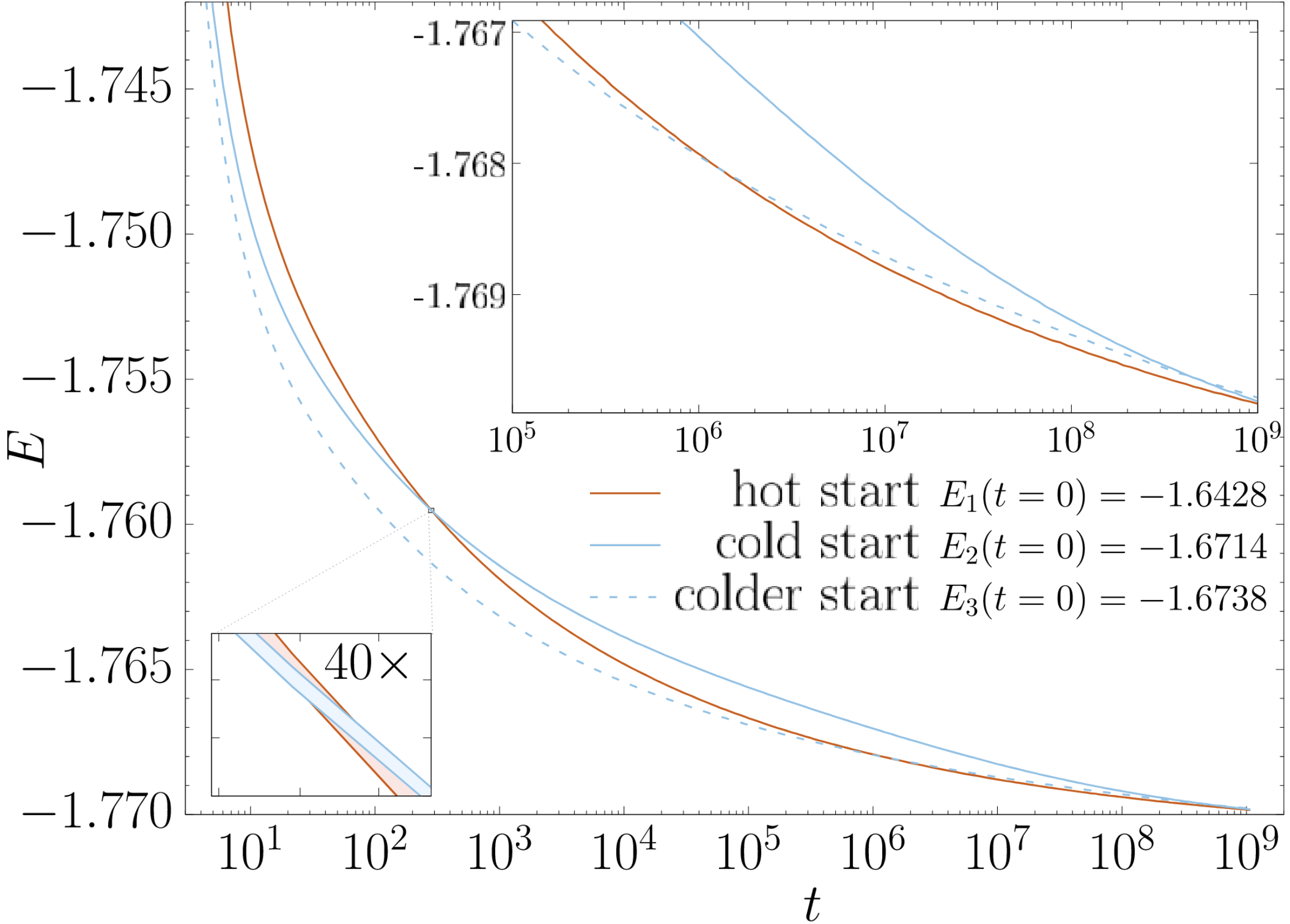}
    \caption{Mpemba effect in the spin-glass system starting from a paramagnetic phase, which is quenched below the critical temperature, $T_c = 1.102$, to the spin-glass phase, with $T_{\rm final} = 0.7 \approx 0.64\ T_c$. The hot start, ($T_1 = 1.3$, $\xi _1 = 12$, $E_1(t=0) =-1.6428$), overtakes the two cold starts ($T_2 = 1.2$, $\xi_{2}=5$, $E_2(t=0) = -1.6714$) and ($T_3 = 1.2$, $\xi _3=8$, $E_3(t=0) = -1.6738$). The correlation length grows for temperatures close to $T_c$; thus, the hot start, since it has the longest correlation length, is the ``closest" to the state the system is relaxing to. The order of the two crossings implies the importance of the correlation length. The error bars are depicted in the thickness of the lines. \textit{Source:} Reprinted with permission from~\cite{baity2019mpemba}.} 
    \label{fig:fig-Baity-spin-glass-Mpemba.png}
\end{figure}

In addition to finding indications for direct and inverse Mpemba effects in a specific spin-glass system, the authors also gained insight about the mechanism behind these effects~\cite{baity2019mpemba}. In particular, they found that it is not enough to record the bath temperature and not enough to also include the average energy at the start of the quench.  Rather, the correlation length $\xi$
seems to be a key ``hidden parameter'' that is also important for understanding the relaxation dynamics. Note that, for a given temperature, different values of the evolving initial correlation length can be achieved by changing the waiting time of the first direct quench.  To see this, the authors considered two different relaxation dynamics.  Both began by connecting the system to a bath of $T_2 = T_3 = 1.3$, and both evolved in contact with that bath to have approximately equal energies at the time of the quench ($E_2 \approx E_3 = -1.67$); nonetheless, the relaxation curves crossed the $T_1$ relaxation curve at very different times.  The relevant difference turned out to be in their correlation lengths ($\xi_2 = 5$ and $\xi_3=8$).  The greater value of the latter suggests that it is closer to the final state, and thus its relaxation curve crosses that of the $T_1$ curve at a later time. 

\begin{figure}[tbh]
    \centering
    \includegraphics[width=\linewidth]{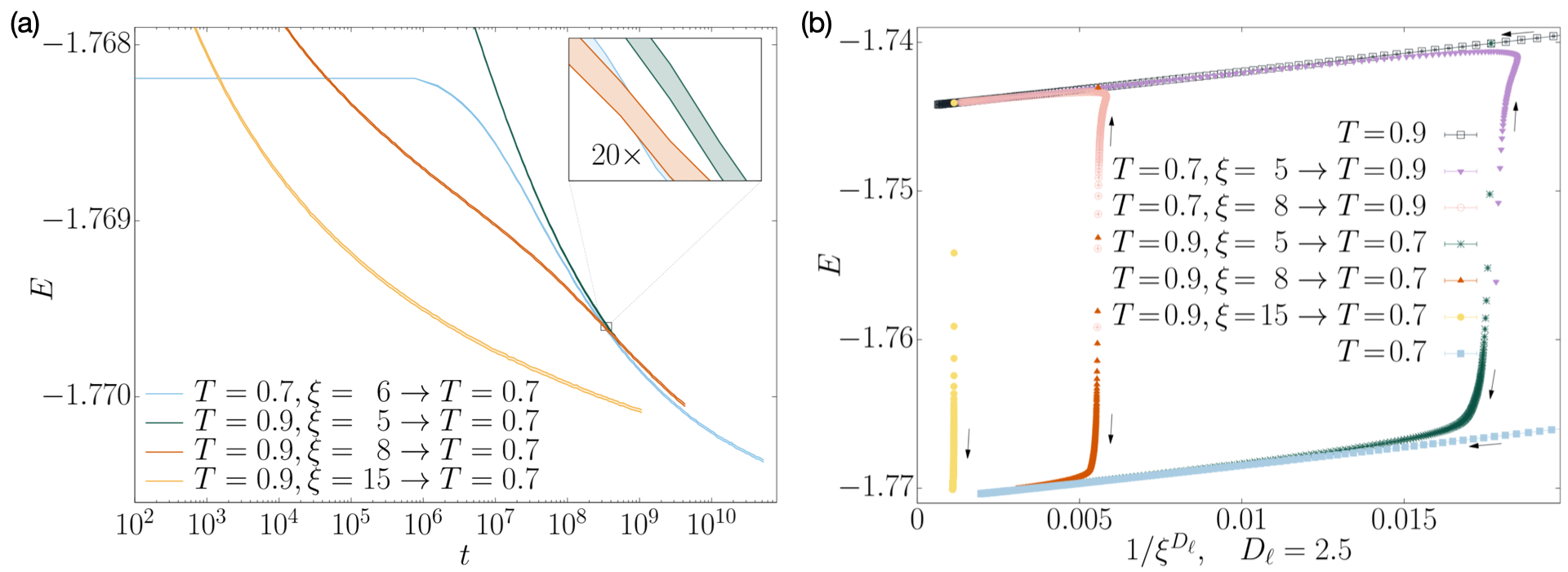}
    \caption{(a) Mpemba effect in the spin-glass system below the glass transition, $T_c = 1.102$, for several correlation lengths. The energy relaxation curves show that the coherence length plays a major role in the relaxation. (b) Relationship between energy and coherence length $\xi$. A thermal protocol is responsible for the vertical moves in (b) that produce the Mpemba effect. \textit{Source:} Reprinted with permission from~\cite{baity2019mpemba}.}
    \label{fig:fig-spin-glass-below-transition.png}
\end{figure}

Below the critical temperature, the nonequilibrium time evolution of the internal energy follows that of the coherence length, with a linear dependence, $E(t)=E_{\infty}(T)+E_1 / \xi^{D_l}(t)+\dots$, where $D_l\approx 2.5$ is the lower critical dimension; see~\FIG{fig-spin-glass-below-transition.png}.  Both the crossing dependence and the energy evolution suggest that coherence-length dynamics are a major factor in determining the relaxation. 

\subsubsection{Ferromagnetic systems}
Many works have considered the Markovian framework for anomalous relaxations in antiferromagnetic systems, and a rich phase diagram of anomalous effects was demonstrated in many variants of antiferromagnetic systems, e.g., \cite{lu2017nonequilibrium,gal2020precooling,klich2018solution,teza2023eigenvalue,teza2022far}. It is therefore natural to consider also the case of a ferromagnetic systems. Recently, a series of investigations \cite{vadakkayil2021should,chatterjee2023mpemba,das2023perspectives,chatterjee2024mpemba} reports numerical observations of a different type of anomalous effect in ferromagnetic systems.  The ferromagnetic systems surveyed include the 2D Ising model \cite{vadakkayil2021should}, the q-Potts model in 2D and with varying interaction lengths \cite{chatterjee2023mpemba}, and the 3D Ising model along with the 2D Ising model with long-range order \cite{chatterjee2024mpemba}. In these numerical results, the system is initiated at equilibrium corresponding to a temperature above the phase transition (but with the specific constraint that half the spins are up and half down~ \cite{vadakkayil2021should,das2023perspectives}). After the preparation, the system is instantaneously quenched to a temperature below the phase transition. The quenching means that the system starts to follow a single spin flip dynamics, with Glauber or Metropolis flipping rates, where the temperature corresponds to the bath temperature. Rather than looking for the phase transition point, the energy per spin is monitored as a function of the simulation time.  It is numerically observed that although the initially higher temperature system has a higher energy per spin, after long enough evolution with dynamic corresponding to bath temperature below the phase transition, the initially hotter system has a lower energy per spin. In addition to observing the energy per spin, in \cite{vadakkayil2021should} crossing between average domain length was also observed.

Interestingly, the exact nature of the observed effect is not understood at this point. For example, it seems to be important for the observed effect that the initial condition have an exactly equal mix of up and down spins (and not the Boltzmann distribution of up and down spins expected in equilibrium), but the reason is unclear. Note that even though above the phase transition point the mean magnetization is expected to be zero on average, the variance scales as the square root of the number of spins; hence, having exactly the same number of up and down spins is a very strong constraint. It is quite possible that this effect is related to the Markovian Mpemba effect discussed in~\SEC{markovian}. Unfortunately, finding the slowest decaying mode in a large 2D Ising model (ref. \cite{vadakkayil2021should} used an array of $256 \times 256$ spins) is not possible analytically or even numerically, and no trace for Markovian Mpemba effect was found for smaller systems where analytic calculations are possible. One route that might determine the existence of a Mpemba effect without the need to evaluate $a_2$ could, in principle, come from the study of the relaxation of relevant observables, as outlined in the following section as well as in~\SEC{strong}.

\subsubsection{Numerical observations of the Mpemba effect using Monte Carlo simulations}
\label{SubSec:NumericalMonteCarlo}
The Markovian framework, discussed in~\SEC{markovian}, prescribes a way to find Mpemba effects: the coefficient of the slowest relaxation mode of the final temperature Markovian operator on the initial conditions contains the relevant information. This prescription is very useful in cases where the coefficient of the slowest dynamics can be found analytically (as in simple potentials of overdamped particle \cite{walker2021anomalous,biswas2023mpemba}) or numerically. However, for large many-body systems, this is very often impractical. Consider, for example, the 2D Ising model: a small system of $100\times 100$ spins has 10,000 spins. Since each spin can be in one of two possible states, the number of micro-configurations in the system is $2^{10,000}$, and there are no known macrostates that can be used to reduce this number, for antiferromagnets in the presence of a magnetic field. In other words, the probability distribution is a vector in a $2^{10,000}$ dimensional space, and the Markovian operator is a $2^{10,000}\times2^{10,000}$ matrix. It is clearly impossible to find numerically the relevant slowest eigenvector, and since there are very few systems for which the relaxation eigenvalues are known analytically, a different approach is needed.  

\begin{figure}[tbh]
    \centering
    \includegraphics[width=0.8\linewidth]{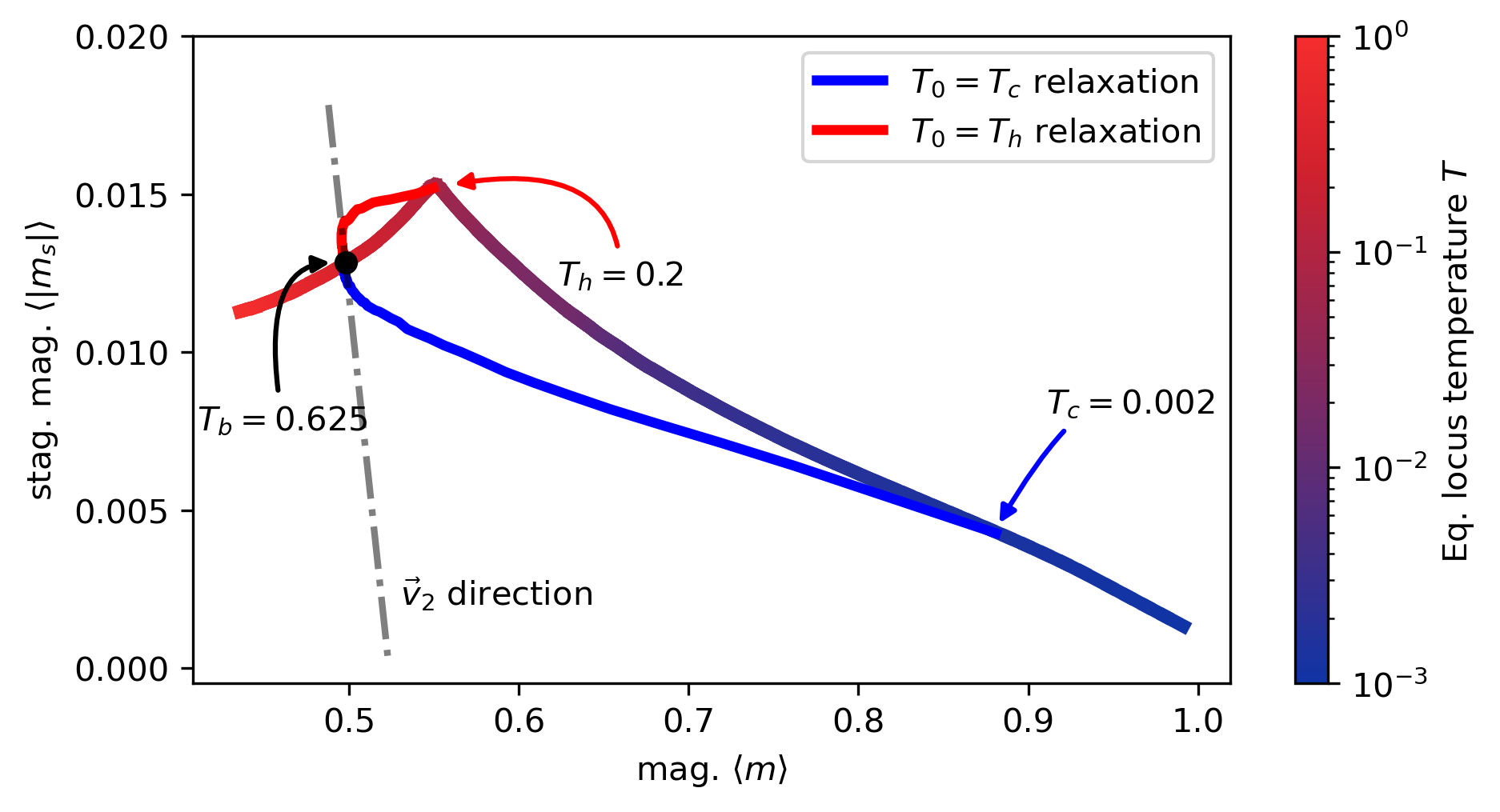}
    \caption{Strong inverse Mpemba effect in the 2D Ising model. The system consists of $100\times 100$ spins on a rectangular lattice, with periodic boundary conditions. The Hamiltonian is given in~\EQ{2DIsingHamiltonian}, with $J=-1$ and $h=1.0025$. The gradient-colored line is the equilibrium line, projected on the $(\langle m\rangle,\langle |m_{s}|\rangle)$ plane, which exhibits a nonmonotonic dependence on temperature for the $|m_{s}|$ observable (see \SEC{eigenvalue-crossing}).
    The red (blue) line is the projection of the relaxation trajectory of a system prepared at an initial temperature $T_{h} = 0.2$ ($T_2 = 0.005$) relaxing towards a final bath temperature $T_{b} = 0.625$ (highlighted with a black dot). As can be seen, the two trajectories approach equilibrium along the same vector (parallel to $\vec{v}_2$) but from opposite directions, corresponding to opposite signs of $a_2$.
    }
    \label{fig:2DIsingInverse}
\end{figure}

One numerical approach to observe strong Mpemba effects in Markovian systems using Monte Carlo simulations was proposed in~\cite{gal2020precooling}. The main idea in this approach is to detect a zero crossing in $a_2$, which implies a strong Markovian Mpemba effect. Let us denote the temperature to which the system relaxes as $T_{f}$, and the initial temperature at which there is a strong (direct or inverse) Mpemba effect as $T_{s}$. We also denote the coefficient along $\vec v_2$ of the Boltzmann distribution at temperature $T$ as $a_2(T)$. In the very high-dimensional space of probabilities, the relaxation process $\vec p(t)$ simply follows linear dynamics in probability space, with the final equilibrium distribution, $\vec \pi(T_{f})$ a fixed point of the dynamics. The probability distributions with $a_2=0$ form a co-dimension 1 hyperplane, namely a linear subspace with dimension $N-1$ where $N$ is the dimension of the probability space (see Fig. \ref{fig:strong_effect_geometry}). This hyperplane separates the probability space into two disjoint sets, one with $a_2>0$ and the other\footnote{Note that the sign of $a_2$ is arbitrary: if $\vec v_2$ is an eigenvector of some matrix or an operator, then $-\vec v_2$ is also an eigenvector of the same operator. The sign choice of $\vec v_2$ affects the sign of $a_2$.} with $a_2<0$. The existence of a strong Mpemba effect implies that the relaxation trajectory from $\vec \pi(T_{s})$ towards $\vec \pi(T_{f})$ is confined to the $a_2=0$ hyperplane. All initial conditions that are not associated with a strong Mpemba effect approach the final equilibrium asymptotically from the $\vec v_2$ direction. However, the sign change in $a_2$ corresponds to relaxation along $\vec v_2$ from opposite directions. The approach of relaxation trajectories from opposite sides of the hyperplane $a_2=0$ can be detected using a simple Monte Carlo simulation, as follows:

A Monte Carlo simulation does not track the evolution of the probability distribution; instead, it tracks the evolution of a function of several observables, e.g., $\{x_1,x_2,\dots,x_n\}$, averaged using the time-dependent probability $\vec p(t)$. We denote this average by $\langle f(x_1,x_2,\dots,x_n)\rangle_{\vec p(t)}$.  As an example of such averages, consider the 2D antiferromagnetic Ising model in the presence of an external magnetic field. The Hamiltonian of the system is given by 
\begin{align}\label{eq:2DIsingHamiltonian}
    \mathcal{H}(\{\sigma\}) = - \frac{J}{4}\sum_{{\rm n.n.}\,(i,j;\,k,l) }\sigma_{i,j}\sigma_{k,l} - h\sum_{i,j}\sigma_{i,j},
\end{align}
where $\sigma_{i,j}$ is the spin at the $i$-th column and $j$-th row, $J=-1$ is the antiferromagnetic coupling constant and $h$ is the external magnetic field and with $\{\sigma\}$ we denote a specific spin-configuration.
The first sum is restricted to nearest neighbor spins, and the second sum is over all spins in the system. Examples for averages that a Monte Carlo simulation tracks include the mean magnetization and the mean staggered magnetization, defined for a given spin configuration $\{\sigma\}$ as
\begin{align}
    m(\{\sigma\}) = \frac{1}{N}\sum_{i,j}\sigma_{i,j} \quad\text{and}\quad m_{s}(\{\sigma\}) = \frac{1}{N^2} \left({\sum_{i,j}(-1)^{i+j}\,\sigma_{i,j}}\right)^2,
\end{align}
where the sign of $(-1)^{i+j}$ specifies the sublattice, while the square over the sum in $m_{s}$ is used, since the two sublattices are symmetric.  Averaging these observables over the distribution $\vec p(t)$ can be viewed as giving an unnormalized linear projection of the high-dimensional probability vector onto a low-dimensional space. The equilibrium average $\langle(m,m_{s})\rangle_{\vec\pi(T_{f})}$ is a fixed point, and relaxation trajectories $\langle(m,m_{s})\rangle_{\vec p(t)}$ are converging towards the fixed point. Unless $a_2=0$, they relax from the same direction, which is the projection of $\vec v_2$. However, trajectories with opposite values of $a_2$ relax from opposite directions towards the fixed point. The trajectory initiated at the equilibrium of $T_{s}$ relaxes from a different direction, corresponding to the $\vec v_3$ projection. This idea is demonstrated in~\FIG{2DIsingInverse}.

\subsection{Numerical results on water and other fluids}\label{Sec:NumericalWater}
In this part, we discuss numerical efforts to explain the Mpemba effect in water by studying microscopic processes. The studies include molecular dynamics simulations in various models for water or other liquid-solid models (the Lenard-Jones and Yukawa potentials), and a specific model for the skin effect in water. We do not cover simulations of macroscopic systems. At this stage, it is not clear at all whether the numerical results reported below and the conclusions taken from these relate to one of the frameworks discussed in this review, nor with experimental results on water. However, some of these works motivated the search for a Mpemba effect in clathrate hydrates \cite{ahn2016experimental}, which demonstrate their influence. 

\subsubsection{Molecular dynamics simulations}

Four works \cite{jin2015mechanisms,tao2017different,gijon2019paths,ghosh2024simulations} used molecular dynamics simulations to try to shed light on the Mpemba effect in water. 
The first, \cite{jin2015mechanisms}, considered molecular dynamics simulations of up to 1000 molecules. The initial configurations were sampled from equilibrium distributions corresponding to the temperature before the quench, from 280~K to 390~K. Three different force field models were used (Molinero with a Stillinger-Weber angular function, SPC rigid water model and the F3C flexible water model), all in thermostated NPT simulation (Nose-Hoover thermostat), to a final temperature of 100~K. In each case, the vibrational density of states (DOS) was calculated, and the population of states in the region 80--160~cm$^{-1}$, which is known to be important for water crystallization, was recorded. It was found that quenching from warm water (above 370~K) led to a higher population in the relevant density of states at this specific energy interval, in comparison to ice.  By contrast, a system quenched from cool water (below 300~K) has a lower population in this energy range. The observations might suggest that the relaxation of vibrational states to that of ice is faster in the initially hot system. This work also explored the importance of the specific range of DOS and found that the specific regime is associated with internal vibrations of water hexagons or perhaps fused decagons.

A different mechanism was explored in \cite{tao2017different}: First, an extensive study of small water molecule clusters of 50-mers was performed, where the strongest 16 H-bond configurations were identified. Then, molecular dynamics simulations of 1000 water molecules were performed in a fixed volume, with periodic boundary conditions, using the TIP5P force field in the NPT ensemble at 1 bar, at temperatures 283~K and 363~K and for a simulation time of 2~ns. The temperature was changed by rescaling the velocities of the particles. 
At each temperature, the density of H-bonds of each type was calculated. Although the total density of H-bonds decreases with increasing temperature, a careful look on their geometry and strength showed that hot water is actually more suitable to ice formation: in cold water, there are more electrostatic H-bonds, whereas in hot water these are more covalent H-bonds. This suggests a plausible mechanism for a Mpemba effect in water.

In a different approach~\cite{gijon2019paths}, a molecular dynamics simulation of 64,000 water molecules was performed. The force fields used were the SPC, TIP3P, and TIP4P rigid models, as well as a flexible version of the TIP3P model; all gave the same qualitative results. In this research, the molecular dynamics imposed energy conservation; that is, the simulation was micro-canonical.  As a consequence, direct thermal relaxation was not explored. Instead, the initial atomic velocities were generated by sampling the Maxwell-Boltzmann distribution at the desired target temperature. Once the given system has equilibrated at the desired temperature, a new, out-of-equilibrium initial condition was generated by modifying the velocity distribution of the molecules so as to study the system’s evolution back to equilibrium. This was done four ways: (1) the velocities
were unaffected (normal equilibrium case); (2) the kinetic energy was placed only in translational degrees of freedom, whereas the molecular rotations were frozen at the initial condition; (3) the translational degrees of freedom were initially frozen, with the kinetic energy placed only in rotational degrees of freedom, and (4) both translation and rotation degrees of freedom were frozen, and the kinetic energy was contained in internal molecular vibrations alone. These initial conditions are clearly out of equilibrium, as equipartition does not hold. The idea is then to examine the relaxation rates of the different degrees of freedom (rotational, translational and vibrational) towards their equilibrium value. Indeed, the authors find that the relaxation rates of the different degrees of freedom do differ, in accord with the mechanism proposed in \cite{torrente2019large}, where energy transfer between rotational and translational degrees of freedom generates huge anomalous relaxations (See section \ref{Sec:BreakingEP}). Lastly, configurations from an equilibrium at 330~K were sampled and heated by setting the kinetic energy to a value corresponding to 370~K, or cooled by setting the kinetic energy to a value corresponding to 290~K. In this case it was observed that when the initial kinetic energy does not obey equipartition, the instantaneous temperature approaches the equilibrium value from below, both in the cases of cooling and heating.

\begin{figure}[tbh]
    \centering
    \includegraphics[width=0.5\linewidth]{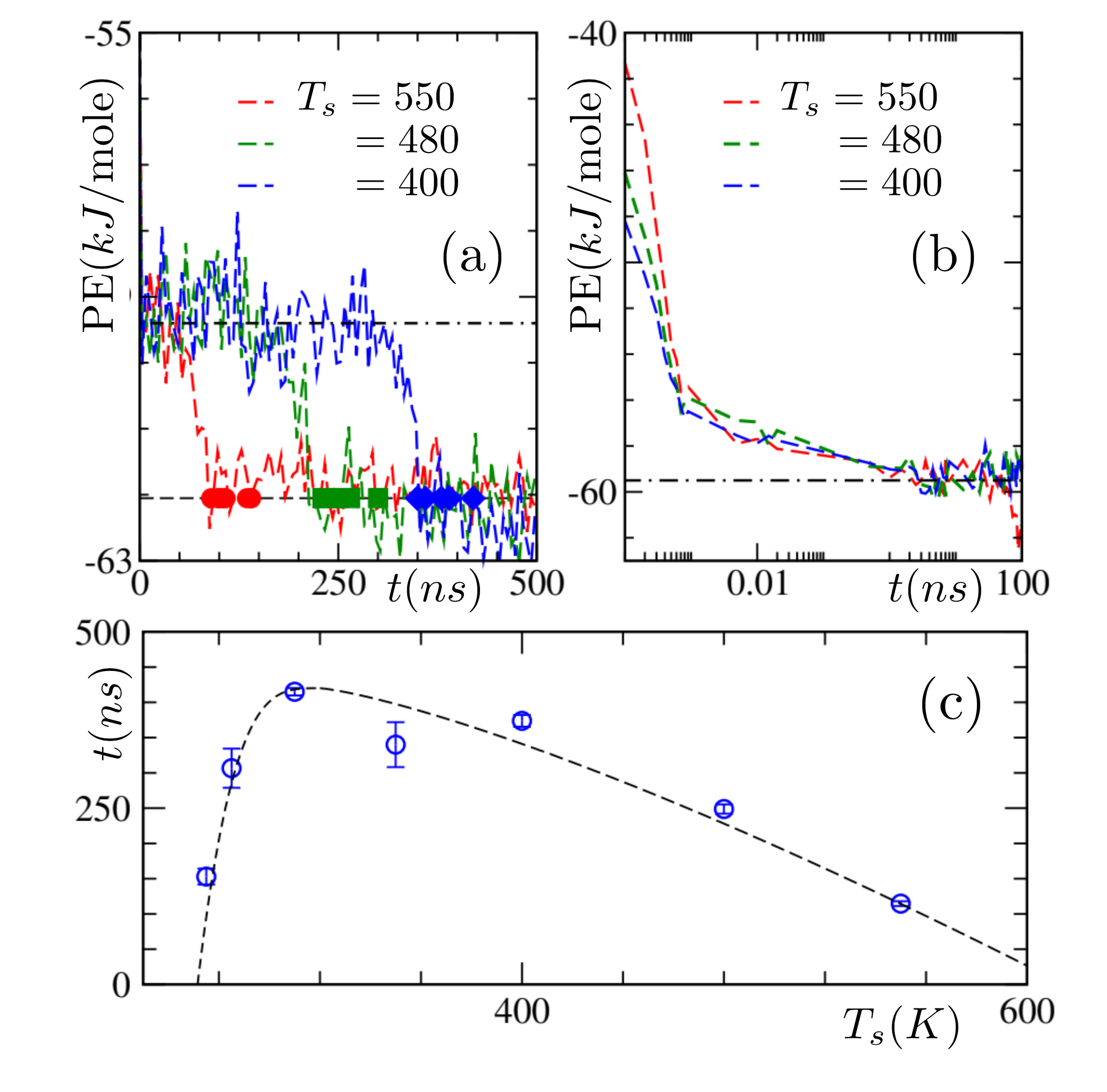}
\caption{Numerical results of the TIP4P/Ice model with 96 molecules. The upper-left figure shows the potential energy as a function of time for various initial temperatures, all quenched to $T_{f}=230\,{\rm K}$. The symbols show the locations of the jump based on which the phase transition time is declared (with additional symbols for other trajectories that are not plotted). The initial part of the relaxation is shown in the upper right figure. The lower figure shows that the mean freezing time is a nonmonotonic function of the initial temperature for the chosen set of parameters. \textit{Source:} Reprinted with permission from~\cite{ghosh2024simulations}.}
    \label{fig:LJ-Water}
\end{figure}

A different type of results were reported in \cite{ghosh2024simulations}. These simulations addressed directly the liquid-solid phase transition, in the TIP4P/Ice model and in a Lenard-Jones potential. For the Lenard-Jones potential, a molecular dynamics simulation in 2D with a truncated, shifted, and force-corrected potential was used in a constant NVT ensemble with a Nose-Hoover thermostat. For the water system, the TIP4P/Ice model was used in an NPT molecular dynamic simulation, for 96 molecules. In both water and Lenard-Jones, the system was initiated at various temperatures above the freezing temperature, and then quenched to a temperature well below the freezing point. After the quench, the potential energy in the system was monitored, and a phase transition was declared when a sharp decrease in the potential energy was observed; see Fig.~\ref{fig:LJ-Water}. By averaging the time to phase transition, the authors calculated the mean freezing time as a function of the initial temperature. They found a nonmonotonic behavior.

Lastly, Ref.~\cite{barba2024anomalous} reported numerical simulations in Yukawa fluids that show anomalous thermal relaxations. The authors performed numerical simulations of screened ion fluid, where the interaction potential between the ions is a Yukawa potential, $V(r) \propto e^{-r/\lambda}/r$, with $\lambda$ a characteristic screening distance. A molecular dynamics simulation for an ensemble of $N=375$ such particles was carried out using generalized Langevin dynamics. In order to induce the quenching effect, the kinetic energy was lowered through dynamical scattering with the rigid box walls, which are assumed to have a temperature $T_\mathrm{wall}$. Specifically, the authors considered systems with ionic density $nI=1.1\times10^{-4}$~fm$^{-3}$ that were initiated at temperatures $T_\mathrm{init} = 6$, 9, and 15~MeV. Additional important parameters were (i) the ratio between the screening length $\lambda$ and the box size, $L$; (ii) the charge spread of the particles. For some specific values of these parameters, a direct Mpemba effect was reported. 

It is worth mentioning that, in all the above works, the timescale associated with the observed dynamics is at most a few microseconds, a time interval that is many orders of magnitude shorter than any relevant experimentally observed macroscopic phenomena. It is therefore not at all clear whether these results correspond to any of the experimental observations. 
Moreover, all these quenching studies carry out the quench via numerical methods (e.g. rescaling velocities) that are convenient but do not represent the dynamics of a real physical process.

\subsubsection{Nonlinear Fourier with supersolidity skin}
A different approach was taken in \cite{zhang2013mpemba} and \cite{zhang2014hydrogen}, where the authors considered 1D, macroscopic, nonlinear Fourier-law dynamics.  The material and heat transport properties are proposed to differ between the bulk of the system and the edge, where a supersolidity skin effect occurs. For a range of parameters, these works predict the existence of a Mpemba effect, with a crossing in the system temperature, even before the phase transition point. Although this line of works is based on macroscopic dynamics (Fourier law for a system of a few millimeters in size), the basic argument behind it---a supersolidity skin effect that changes significantly the macroscopic parameters at the boundary of the system---is of microscopic origin. Therefore, it is somewhat different from other numerical works on macroscopic systems that did not take this effect into account, e.g.~\cite{vynnycky2010evaporative,vynnycky2012axisymmetric,vynnycky2015convection}. It will be of great interest to combine the suggested supersolidity skin effect with 3D coupled fluid-mechanics heat-convection equations. Exploring the effect in such a simulation has never been done, as far as we know.

\subsection{Simulations in kinetic theory} \label{Sec:NumericalKinetic}
Observing the Mpemba effect in the kinetic-theory framework is difficult: Except for the case of breaking equipartition, the Mpemba effect is small, and many trajectories are needed to have enough statistics. Two kinds of simulations were extensively used to measure anomalous relaxation effects: Event-driven Molecular Simulation and Direct Simulation Monte Carlo Method.

Event-Driven Molecular Simulation (EDMS) is based on solving Newton's equations of motion for the system of interest. In particular, for granular particles that interact only during a collision, an inelastic version of the classical algorithm of Allen and Tidesley~\cite{allen2017computer} is used. In EDMS, one computes the collision time between a pair of particles by integrating the equation of motion of particles; after choosing the smallest collision time, one moves each particle accordingly and calculates the outcome of the specific collision, based on the specific collision rule in the model (see~\SEC{kinetic-framework}).

The direct simulation Monte Carlo (DSMC) method, due originally to Bird~\cite{bird1994molecular}, solves the Boltzmann equation and can be applied to simulate rarefied gas dynamics. The basic ideas can be adapted to a wide range of models of dilute granular gases; see~\SEC{kinetic-framework}. The space region is divided into a series of cells whose dimension must be smaller than the mean free path, so that flows are approximately constant inside the cell cavities. The flow of particles between collisions is considered uncoupled over a time interval $\Delta t$, which must be smaller than the average time between collisions. Next, each particle is moved a distance $v_i\Delta t$, where $v_i$ is the velocity of the $i$-th particle. Then the collision part of the algorithm is resolved. When the collision process is computed, only particles inside the same cell can collide; otherwise, the positions of particles play no role. A crucial aspect is that, in each cell, pairs of particles are chosen randomly, and their collision is accepted or rejected according to a probability that is proportional to the relative velocity between them.  If the collision is accepted, the post-collisional velocities are computed using an impact parameter chosen at random from a uniform distribution of incident particles. The time counter is advanced after collisions and the process repeated in each cavity until its time counter exceeds the time step $\Delta t$~\cite{brey1999direct}.  The DSMC method is a faster, more efficient way to simulate the dilute granular dynamics.  See~\FIGS{mpembasmooth}{mpembarough} for a comparison between methods and theoretical predictions.

\section{Applications of the Mpemba effect}
\label{sec:applications}
In this section, we discuss two broad classes of application of the Mpemba effect.  The first might be termed ``direct" and involves situations where speeding up the relaxation to equilibrium has immediate consequences.  The second class is more inspirational and starts from an understanding of basic mechanisms underlying the Mpemba effect and uses similar ideas in somewhat different contexts.  We begin with the direction applications.

\subsection{Direct applications}
To cool an object rapidly is a common need, and heat transfer is a well-developed topic of engineering~\cite{lavine2018fundamentals}. Even Aristotle's first mention of anomalous cooling quoted in~\SEC{introduction} cites how the inhabitants of Pontus start with previously heated water to produce ice more quickly. And Mpemba, in his introductory statement in Ref.~\cite{mpemba1969cool}, relates that ice-cream vendors in the city of Tanga use hot milk to speed the production of ice cream.

A modern application of such ideas has recently been presented by Yang et al.~\cite{yang2024exploring}. Piezoceramic materials are important for applications such as gas lighters, sonar, and nanomechanics (as actuators for subnanometer displacements). To form them, the raw materials first need to be compressed to increase the piezoelectric response. The compression is usually done by mechanical force. In Ref.~\cite{yang2024exploring}, the authors instead use the expansion of freezing ice. Squeezing by ice rather than a conventional press produces higher pressures and is simple to implement. In exploring the parameters of their experiment, the authors found that freezing begins sooner if hot water (60 $^\circ$C) is used, rather than warm water (30 $^\circ$C), and ascribe this difference to the Mpemba effect. The speedup is important not only for saving time but also for reducing the grain size of the piezoceramic material, a consequence of the more rapid freezing. The resulting material has a density that is as much as 11\% higher.

\subsubsection{Optimal heating and cooling protocols}
\label{sec:optimal-heating-cooling-protocols}
A natural application for many variants of the Mpemba effect is optimal cooling, or heating in the case of an inverse Mpemba effect. 
If cooling a system initiated at $T_\mathrm{hot}$ is faster than cooling the same system initiated at $T_\mathrm{medium}$ when the cooling mechanism is coupling the system to a bath at $T_\mathrm{cold}$, then perhaps cooling a system initiated at $T_\mathrm{medium}$ by first heating it towards $T_\mathrm{hot}$ and then coupling it to $T_\mathrm{cold}$ is faster than coupling it to a thermal bath at $T_\mathrm{cold}$ without the pre-heating stage. But if pre-heating takes longer than the time gained during the cooling stage, the extra step will not reduce the overall protocol time.

Optimal heating/cooling protocols and their relations to the Mpemba effect were studied in \cite{gal2020precooling}, where the problem was carefully defined, and a framework to address it was suggested. The main conclusions of this manuscript were (i) the phenomenon exists---there are cases where pre-heating (pre-cooling) speeds cooling (heating); (ii) if there is a strong direct Mpemba effect in the system, namely there is some temperature $T_\mathrm{strong}$ for which the coefficient of the slowest relaxation mode vanishes (in the definitions of~\SEC{markovian} this means $a_2=0$), then for every initial temperature $T_\mathrm{initial}$ such that $T_\mathrm{cold}<T_\mathrm{initial}<T_\mathrm{strong}$, the cooling process can be improved by coupling the system to some bath at $T_\mathrm{hot}>T_\mathrm{strong}$ for some time (the exact range of heating time depends on $T_\mathrm{hot}$); (iii) optimal cooling (heating) protocols that have a pre-heating (pre-cooling) stage exist even in systems that have no direct (inverse) Mpemba effect at the target temperature. These results were demonstrated on a few concrete systems. 

Defining ``optimal heating” (or optimal cooling) protocols is tricky:
A system that is initiated at some $T_\mathrm{init}$ and is coupled to $T_\mathrm{cold}$ relaxes towards the bath temperature, but at least formally never reaches the exact Boltzmann distribution associated with $T_\mathrm{cold}$. The same is true for most systems, where reaching the Boltzmann distribution exactly at a finite time is impossible, even for complicated protocols with a time-dependent temperature that is 
bounded
from below by $T_\mathrm{cold}$. Thus, one cannot define the optimal protocol as ``the fastest way to reach the Boltzmann distribution exactly." There are a few ways to nevertheless define optimal protocols. The approach chosen in \cite{gal2020precooling} is to assume that in the very last stage of the protocol, it is nevertheless coupled to a bath with the target temperature, $T_\mathrm{cold}$. Since the distribution is not the corresponding Boltzmann distribution, the relaxation then continues through this last stage too. The optimal protocol is therefore the one at which the relaxation in this final stage is optimal. This is achieved when as many as possible of the slowest relaxation modes of the Markovian operator corresponding to coupling the system to a bath at $T_\mathrm{cold}$ have no coefficients, and minimizing the slowest mode that exists. In other words, assuming that for any time  $t>\tau$ the system is coupled to $T_\mathrm{cold}$, we want the probability distribution $P(\tau)$ to have $a_2=a_3= \ldots = a_n=0$ for the largest possible $n$, and minimize over the absolute value of $a_{n+1}$. 

In most cases, finding the exact optimal protocol is very difficult, even numerically. However, for systems with only a few states, this is sometimes possible: in optimal-control theory, finding $T_b(t)$ that minimizes $a_{n+1}$ subject to $a_2 = a_3 = \ldots = a_n=0$ is a quadratic end-point minimization problem with linear end-point constraints, a bounded domain and no path cost. Such minimization problems can be addressed numerically using Pontryagin’s maximum principle~\cite{bechhoefer2021control}. Since there is no path cost, they can also be solved numerically with methods such as sequential quadratic programming~\cite{nocedal2006numerical}. In \cite{gal2020precooling}, an example for optimal heating in a four-state system where at a given finite time $a_2 = a_3 = 0$ and $|a_4|$ is minimal was demonstrated. Unfortunately, when the number of states become large, this optimization problem becomes intractable. Nevertheless, it is often possible to find a protocol that sets $a_2=0$ or even $a_2=a_3=0$ at a finite time. These protocols are not expected to be optimal, but they are superior to the naive protocol of coupling only to the final temperature. Examples for such protocols were given for a Brownian particle diffusing in a potential, and for the 2D nearest-neighbor-coupling antiferromagnetic Ising model.

Why might heating before cooling be faster, overall?  
To develop some intuition,
consider the experiment done in \cite{kumar2020exponentially}, namely a Brownian particle diffusing in the potential given in \FIG{brownian_system}b. The slowest dynamic in cooling this system is hopping from the left well to the right well (or vice versa), to balance the accumulated probabilities in the two wells. The existence of an inverse Mpemba effect indicates that for the specific $T_{\rm b}$, there exist $T_\mathrm{strong}$ for which the corresponding Boltzmann probability is balanced between the two wells such that there is no coefficient along the slowest dynamics. For temperatures exceeding $T_\mathrm{strong}$, the balance between the wells is inverted in comparison to initial conditions at $T_\mathrm{cold}<T_\mathrm{init}<T_\mathrm{strong}$, namely, the equilibrium distribution corresponding to $T>T_\mathrm{strong}$ has more probability in the well which is depleted in $T_\mathrm{init}$. Therefore, preheating the system to $T>T_\mathrm{strong}$ can balance the two wells at a finite time, setting $a_2=0$.

\begin{figure}[htb]
    \centering
    \includegraphics[width=0.7\linewidth]{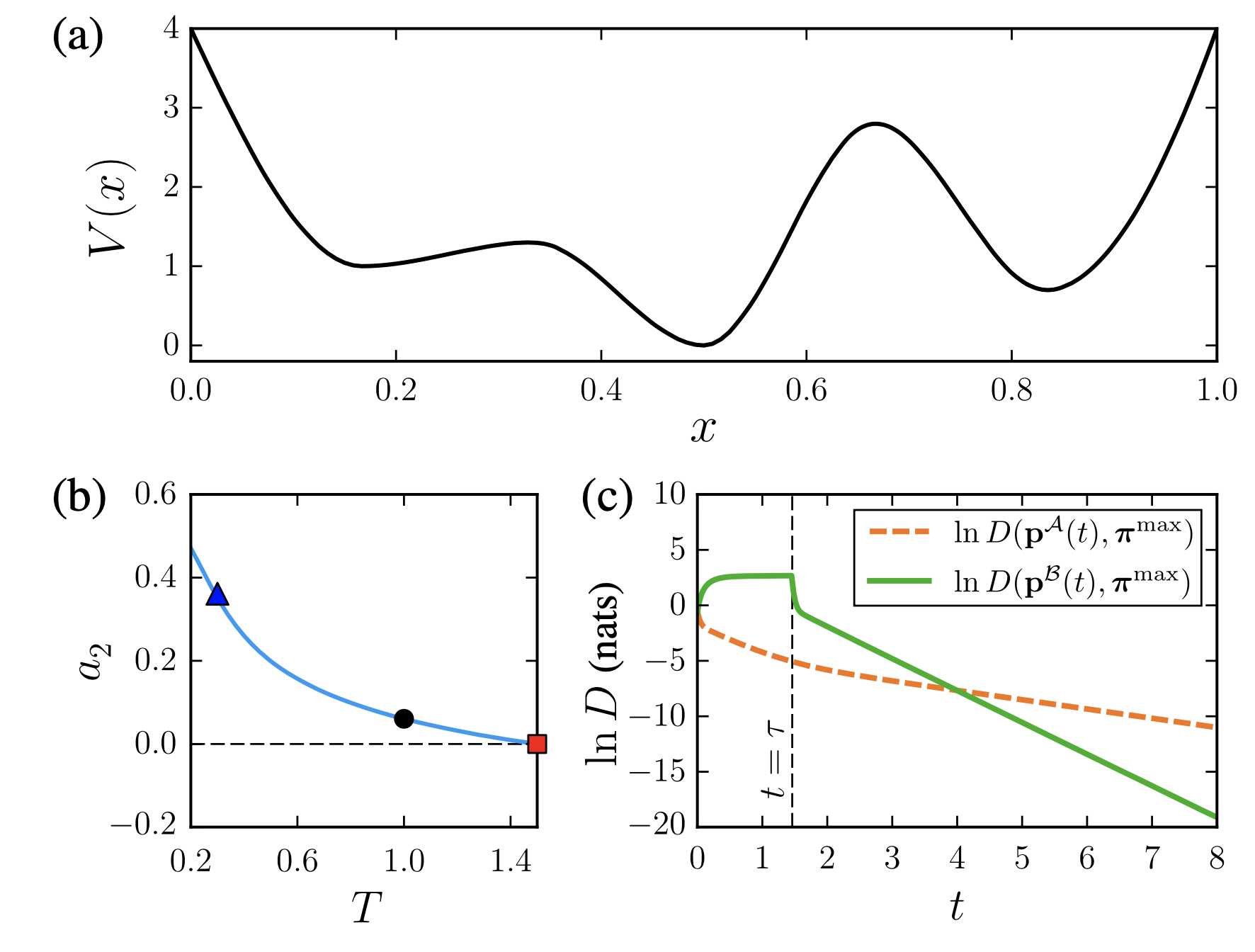}
    \caption{Improving heating by first cooling the system (figure taken from \cite{gal2020precooling}). (a) The system is a Brownian particle diffusing in the plotted 1D potential and reflecting boundary conditions. (b) A plot of $a_2$ as a function of temperature for $T<T_\mathrm{hot}$. The final temperature $T_\mathrm{hot}$ is represented by the red square, the initial temperature $T_\mathrm{init}$ by the black circle, and the cold temperature $T_\mathrm{cold}$ to which the system is coupled in the pre-cooling state by the blue triangle. The plot of $a_2$ demonstrates that there is no inverse Mpemba effect, weak or strong, for $T_\mathrm{hot}$. (c) The  Kullback-Leibler divergence of the distribution $p(t)$ from the equilibrium at $T_\mathrm{hot}$ for the naive heating protocol where the system is coupled to $T_\mathrm{hot}$ (dashed orange line), and the pre-cooling protocol, where the system is first coupled to $T_\mathrm{cold}$ for some finite time, and only then it is coupled to $T_\mathrm{hot}$ (solid green line). \textit{Source:} Reprinted with permission from~\cite{gal2020precooling}.}
    \label{fig:OptimalProtocols}
\end{figure}

The above intuitive explanation does not give the full story: in~\FIG{OptimalProtocols}, a Brownian particle diffuses in a different potential, as shown in \FIG{OptimalProtocols}a.  The system is initiated at equilibrium for $T_\mathrm{init}$ and then cooled by coupling to $T_\mathrm{cold}<T_\mathrm{init}$ for a time $\tau$, until $a_2=0$.  (Here, $a_2$ is calculated for the heating process resulting from coupling the system to $T_\mathrm{hot}$.) This is possible even though there is no inverse Mpemba, strong or even weak effect, at $T_\mathrm{hot}$; see \FIG{OptimalProtocols}b. The exact conditions for the existence of protocols that can improve the heating/cooling time remain unknown.

\begin{figure}
    \centering
    \includegraphics[width=0.65\linewidth]{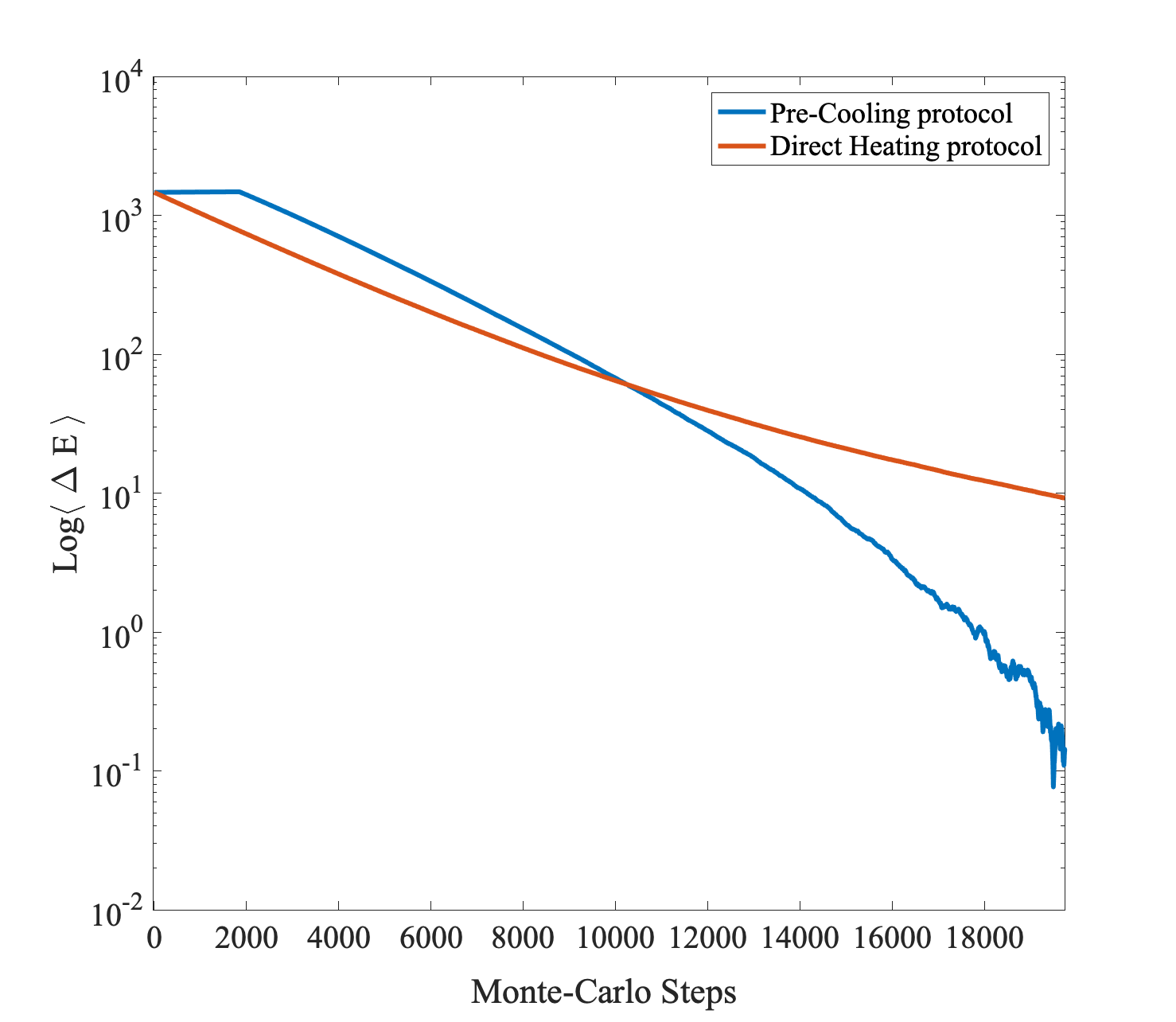}
    \caption{Improving heating by first cooling the system in the 2D nearest-neighbor Ising model with an external magnetic field. The energy differences between the two cases is clearly visible.}
    \label{fig:OptimalProtocolsIsing}
\end{figure}

Lastly, one has to verify that setting $a_2=0$ indeed improves the cooling/heating time in a meaningful way: although formally, the relaxation never gets to the Boltzmann distribution, in practice, at some point, the differences between $p(t)$ and the Boltzmann distribution cannot be detected. Thus, improving the protocol is relevant only if the differences between the improved protocol and the naive protocol are detectable and important for the relevant application. An example where this is the case is the 2D antiferromagnetic nearest-neighbor Ising model with an external magnetic field. For a $70 \times 70$ spin lattice with periodic boundary conditions, a pre-cooling protocol that sets $a_2=0$ was found numerically using the method discussed in Sec.~\ref{SubSec:NumericalMonteCarlo}. In this case, the difference between the two protocols is not only detectable but also quite useful: it takes about five additional Monte-Carlo swipes for the naive protocol to get to the same accuracy as the pre-cooling protocol (see Fig.\ref{fig:OptimalProtocolsIsing}).

\textit{Application to AFM imaging}.  
The idea that one can manipulate heating and cooling by using nonequilibrium protocols that set $a_2$---and even higher-order coefficients---equal to zero has been carried out experimentally in a recent study by Pottier et al.~\cite{pottier2023accelerating}. In this work, the goal was to speed up AFM imaging by controlling more rapidly the temperature of an atomic force microscope (AFM) cantilever.   To understand the connection between cantilever and imaging speed, we note that many AFM imaging modes involve exciting one of the lower-frequency mechanical modes of the cantilever. A very clean way to excite just the cantilever and not other mechanical parts of the instrument is via photothermal excitation, where the temperature of the cantilever is varied periodically at the resonance frequency, and thermal expansion couples the varying temperature to mechanical motion. In more fundamental applications, model heat engines at small scales have also been created using AFM cantilevers as the ``working engine." In both cases, the ability to change rapidly the temperature of an AFM is important. In Ref.~\cite{pottier2023accelerating}, the proxies for temperatures are nonequilibrium steady states and are controlled by varying the power of a laser that is focused on the cantilever.  

In~\cite{pottier2023accelerating}, the transition between two nonequilibrium steady states uses an approach that is analogous to (and inspired by) the work of Gal and Raz~\cite{gal2020precooling}.    The protocol has two stages. The first drives the system into a desired nonequilibrium state; the second is the relaxation, which is to be accelerated. In the first stage, the laser power $P(t)$ is varied over a protocol of duration $t_f$. The variation is done in such a way that at the end ($t= t_f$), the nonequilibrium spatial temperature profile $T(x,t_f)$ has a shape with zero projection onto the eigenfunction describing the slowest relaxation to the nonequilibrium steady state corresponding to the desired end-time laser power $P(t_f)$. The relaxation then shows the same exponential speedup as in the strong Mpemba effect. Indeed, the flexibility allowed by controlling the laser power continuously for $0 \le t \le t_f$ allows one to systematically eliminate as many orders as one likes, subject to limitations of laser power (from below, $P=0$ and from above, by capabilities of the laser hardware) and the melting temperature of the cantilever itself. The authors demonstrate cancellation of the slowest four eigenmodes, with more than 30-fold speedup of the relaxation time in the second stage of the protocol, for $t > t_f$. The work by Pottier et al.~\cite{pottier2023accelerating} is an exciting extension of Mpemba-related ideas with a clear practical application, speeding up AFM imaging.

\subsubsection{Optimal transport and heat engine efficiency}
\label{sec:optimal-transport-heat-engine-efficiency}
Possible connections between the conditions for minimal dissipation and the Mpemba effect were explored in Ref.~\cite{walker2023optimal}. The main question considered was whether relaxation can be both ``fast" (shows the Mpemba effect) and ``optimal" (minimal entropy dissipation). To explore this option, the authors analyze a continuous system---specifically, an over damped particle diffusing in a potential landscape, which can be described by overdamped Langevin dynamics. Such a model was introduced in~\SEC{markovian} and later discussed in relation to colloidal experiments in~\SEC{colloids}. For a one-dimensional potential, the authors find that the occurrence of the strong Mpemba effect most often corresponds to high total entropy production. Thus ``fast" and ``optimal" do not typically occur simultaneously. 

Subsequently, the authors examined discrete cases of Markov jump processes, where the rates obey detailed balance and are parameterized by energies $\{E_j\}$ and barriers $\{B_{i\!j}\}$ such that $W_{i\!j} \propto \exp\left(-(E_j - B_{i\!j})/k_BT_b\right)$, as noted in ~\EQ{arrhenius}. In these discrete cases, they identify and analyze scenarios where the same dynamics produces a strong Mpemba effect and minimal total entropy dissipation. It would be intriguing to generalize these findings on optimal transport and the Mpemba effect to larger networks or more complex physical systems and to determine the aspects of relaxation protocols that simultaneously yield a Mpemba effect and minimal total entropy dissipation.

The efficiency of a mechanical Maxwell demon in the presence of a Mpemba effect was discussed in two studies, \cite{bera2023effect, bao2023designing}. In 1867, James C. Maxwell posed a \emph{gedanken experiment}, imagining a ``demon'' that could harness information and thereby extract work from a thermal bath~\cite{maxwell1871theory}. The paradox is that the demon seemingly violates the second law of thermodynamics. After over a century and a half of debate and inspiring discussions, the paradox was resolved mainly by Landauer~\cite{landauer1961irreversibility}, Bennett~\cite{bennett1982the} and Penrose~\cite{penrose1970foundations}, whose work led to a comprehensive theory of information thermodynamics~\cite{parrondo2015thermodynamics}. In particular, there is no violation of the second law of thermodynamics if one accounts for the cost of information storage. A fully mechanical version of the Maxwell's demon was proposed by Mandal and Jarzynsky in~\cite{mandal2012work}. Their analytically solvable model consists of a memory tape and a device (Maxwell's demon). By exploiting the information on the tape, the demon can extract work from thermal equilibrium.  

\begin{figure}
     \centering
     \includegraphics[width=0.85\linewidth]{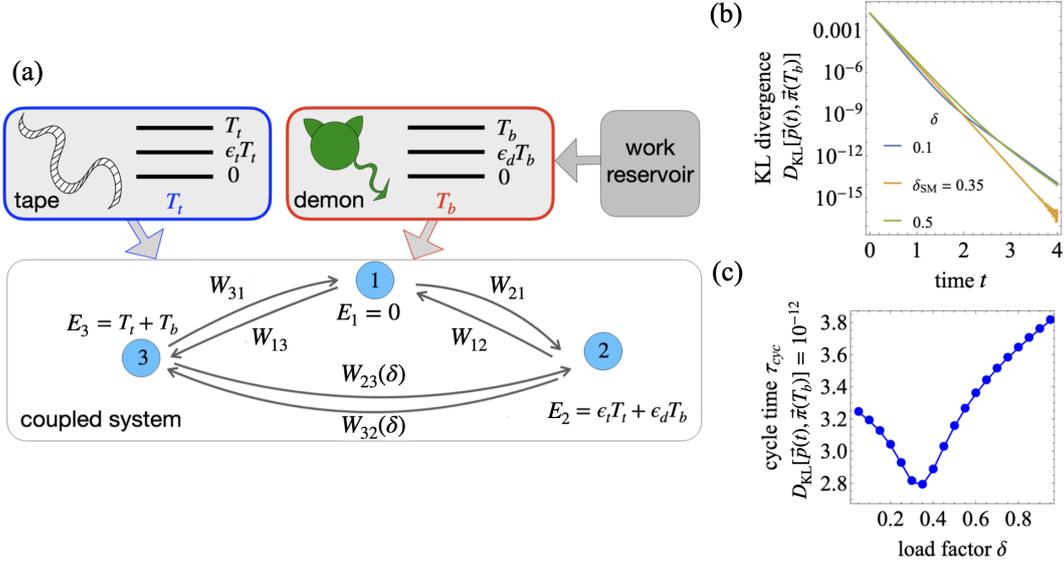}
     \caption{Mpemba effect in a Maxwell demon. (a) The device (mechanical Maxwell's demon here) interacts with tape and work reservoir. The device and the tape are modeled as a three-level Markov jump system. They form a coupled system during the working cycle. (b) Kullback-Leibler divergence as a function of time for the same initial conditions and three different load distribution factors. The system has a strong Mpemba effect at load factor $\delta _{\rm SM}=0.35$ for parameters: $T_t = 0.824$, $T_b = 2$, $\epsilon _t = 0.9$ and $\epsilon_d = 0.4$. (c) Cycle time, $\tau _{\rm cyc}$ is chosen as the time when the KL divergence has a cutoff (here $10^{-12}$). The cycle time has a minimum at $\delta_{\rm SM} = 0.35$, where we also have the strong Mpemba effect. \textit{Source:} Adapted with permission from~\cite{bera2023effect}. }
     \label{fig:fig-Maxwell-demon.png}
\end{figure}

The efficiency of such mechanical Maxwell demon was studied in~\cite{bera2023effect}. Specifically, the authors considered a ``demon model'' with a three-level Markov jump process and an information-carrying tape, see Ref.~\cite{bera2023effect}. The tape is also modeled as a three-level Markov jump process. During a cycle time, denoted as $\tau_{\mathrm{cyc}}$, the two systems interact and form a coupled system; see~\FIG{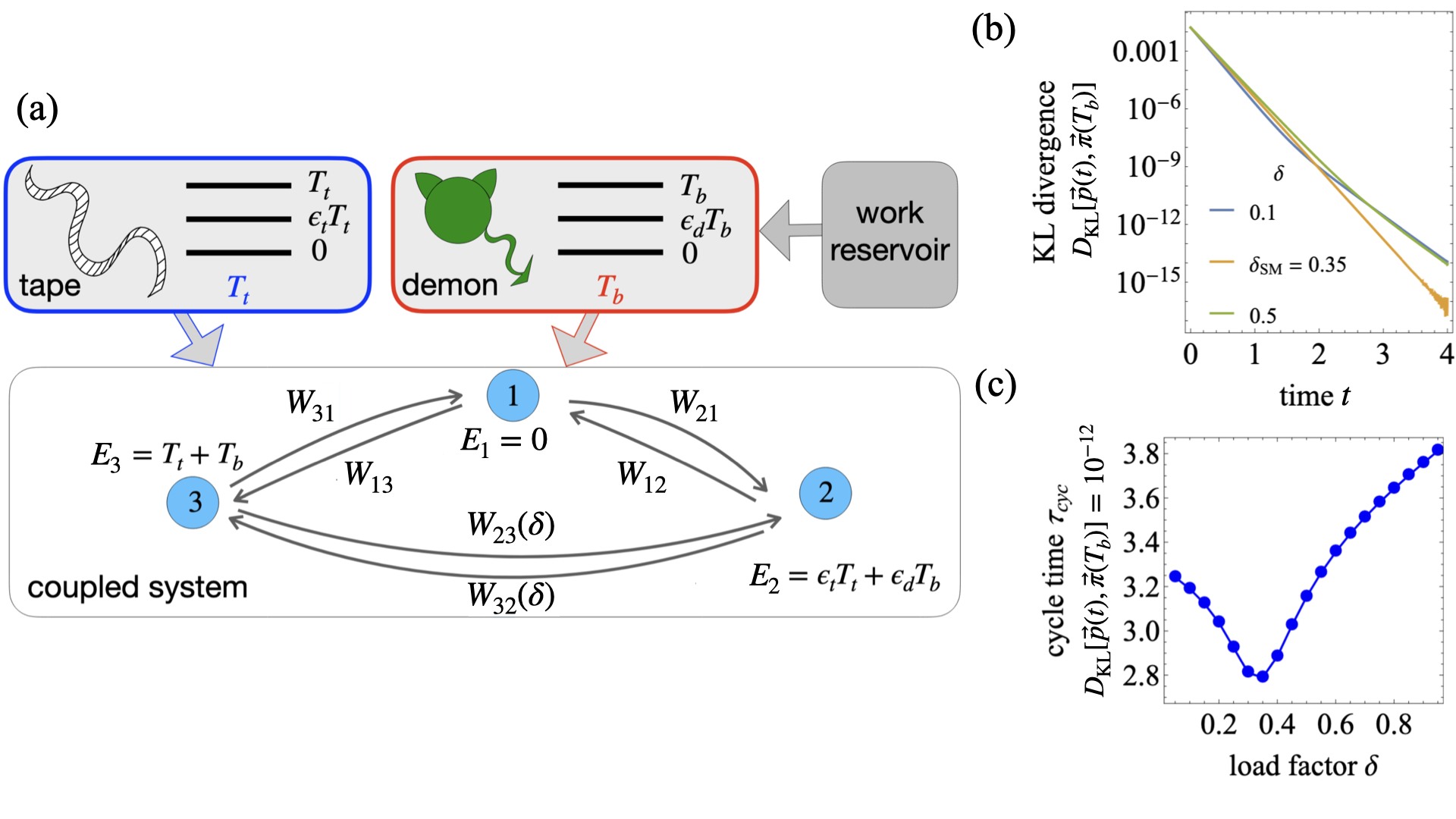}a. The transition rates are parameterized by the energies of the coupled system $\{E_i\}$ and the load factor $\delta \in [0,1]$, as follows:
\begin{align}
    W_{21}(T_b) \propto e^{-(E_2 - E_1)/k_B T_b}, \quad
    W_{13}(T_b) \propto e^{-(E_1 - E_3)/k_B T_b},  \quad \text{and}\quad 
    W_{32}(T_b, \delta) \propto e^{-(E_3 - E_2) \delta /k_B T_b}.
\end{align}
The reverse rates are determined by detailed balance, $W_{i\!j} \pi_j(T_b) = W_{ji} \pi_i(T_b)$. For simplicity,  only the rates between two states (here $2$ and $3$) depend on the load factor. The authors consider different dynamics with the same initial and steady-state conditions. They consider cases whereby modifying the dynamics, specifically by changing the value of $\delta$, one can observe the strong Mpemba effect. The same distance from equilibrium, quantified by the Kullback-Leibler divergence between the initial and final states, $D_{\mathrm{KL}}(\vec{p}(t), \vec{\pi}(T_b))$, is achieved in a shorter time when the load factor is set to yield the strong Mpemba effect ($\delta _{\rm SM} = 0.35$ on~\FIG{fig-Maxwell-demon.png}b). This results in a reduction of the cycle time, $\tau _{\rm cyc}$, defined as the period at which the KL divergence reaches a specific long-time cutoff value. An example with a 
chosen cut-off value of$10^{-12}$ is shown in~\FIG{fig-Maxwell-demon.png}c. A shorter cycle time results in a greater average power output. It should be noted that this power output is obtained without a reduction in the device's efficiency and remains stable. 

Enhanced performances of an autonomous Maxwell’s demon, in the sense of reaching the functional state faster, was considered in~\cite{bao2023designing}. The authors, in addition to the Maxwell-demon setup, introduce \emph{stochastic resetting} to the dynamics. We mention their work when we discuss connections between stochastic resetting and the Mpemba effect; see~\SEC{stochastic-resetting}. 

Closely related are work-heat engines, which played an important role in the development of thermodynamics. A simple heat-engine model where a three-state Markov jump process in which the temperature is periodically switched between two temperatures. It thus serves as the ``working fluid'' of a cyclic Otto engine~\cite{lin2022power}. In this work, the authors compare the evolution of the system from the same initial conditions, both with and without the Mpemba effect. They demonstrate that the Mpemba effect can enhance the machine's performance by decreasing the time required to reach steady state. The shorter cycle increases power output for a given efficiency,  while maintaining stability. The findings align with previous results in ~\cite{bera2023effect} and~\cite{bao2023designing}.  An analogous  quantum version of the Otto engine with a quantum Mpemba effect was studied in \cite{liu2024speeding}. 

Integrating the Mpemba effect into optimal protocols for efficient heat engines and erasers is still in its early stages. The studies described above touch upon some of the potentially promising applications.

\subsubsection{Quantum state preparation}
\label{sec:quantum-state-preparation}
For quantum systems, a natural application of anomalous relaxation is to accelerate the preparation of quantum states that are useful for quantum computing, quantum simulations, and experiments in quantum thermodynamics. 
The goal is typically to engineer the connection between system and bath to generate fast relaxation into the desired state \cite{chen2024boosting,verstraete2009quantum,harrington2022engineered}. Indeed, this was the motivation for the discussions in Secs.~\ref{sec:QuantumUnitary} (theoretical aspects) and \ref{sec:unitary-quantum-experiment} (experimental demonstrations). In those sections, Mpemba ideas were used in a quantum setup to increase the rate at which the system relaxes to its target (equilibrium) state, by first applying fast unitary operations that annihilate the coefficient of the slowest relaxation process; in our notation, this corresponds to setting $a_2=0$. 

\subsubsection{Rate of chemical reactions}
\label{sec:rate-of-chemical-reactions}
A surprising application to chemical systems is the discovery that exposing a sensor molecule to a high concentration of target ligand molecules and then, briefly, to a low-concentration environment can increase overall detection sensitivity
~\cite{pagare2024mpemba}. The boosted performance exceeds that of a sensor starting from an initial low-concentration environment. This result is based on the same formalism and ideas used to understand the Mpemba effect in Markovian systems~\cite{lu2017nonequilibrium}.

In a related study, Hatakeyama~\cite{hatakeyama2024enzymatic} examined enzymatic reactions and found, unusually, that higher enzyme concentrations can actually slow down these reactions. The author draws parallels with the Mpemba effect in the Markovian setup discussed in~\SEC{markovian}. Most biochemical reactions involve enzymes as catalysts, with reaction kinetics described by Michaelis-Menten kinetics~\cite{michaelis1913kinetik}. The reaction rate $w$ is then expected to be 
proportional to enzyme concentration:
\begin{align}
\label{eq:MichaelisMentenkinetics}
    w = \frac{k[E][S]}{K + [S]}, 
\end{align}
where $k$ is the rate constant for the elemental reaction, $[E]$ and $[S]$ the enzyme and substrate concentrations, and $K$ the dissociation constant between enzyme and substrate. The Michaelis-Menten kinetics of \EQ{MichaelisMentenkinetics} suggests that increasing enzyme concentrations typically accelerate enzymatic reactions. Hatakeyama considers a biomolecule with multisite modifications catalyzed by enzymes. For a single enzyme, the conventional Michaelis-Menten scenario holds; however, for two enzymes, an increase in the concentration of one can slow relaxation to a steady state, even if both enzymes individually act as catalysts. As a paradigm, Hatakeyama uses an allosteric dimer, where each monomer represents a modification site, and where monomers can form two different structures. The equilibrium distribution is independent of the number of enzymes, but the relaxation times depend on the population of each particular enzyme. The two enzymes catalyze reactions by changing the modification site. They show a branched network structure, with one of the branches having a local energy minimum (a metastable state). The substrates in the metastable state sequestrate the enzymes; these molecules remove all enzymes,  and the system stays in the metastable state for a long time. In this case, two enzymes have different affinities for substrates, and one preferentially binds to a substrate.  Together, they increase the flux to the metastable state and slow the relaxation toward the target state. The role of temperature here is replaced by the enzyme concentration. Hatakeyama's work shows an example of a system where the relaxation of the system toward a steady state is a nonmonotonic function of the enzyme concentration. It was also noted in this work that two or more enzymes of different affinities for substrates are needed for the result.  

\begin{figure}[ht]
    \centering
    \includegraphics[width=0.5\linewidth]{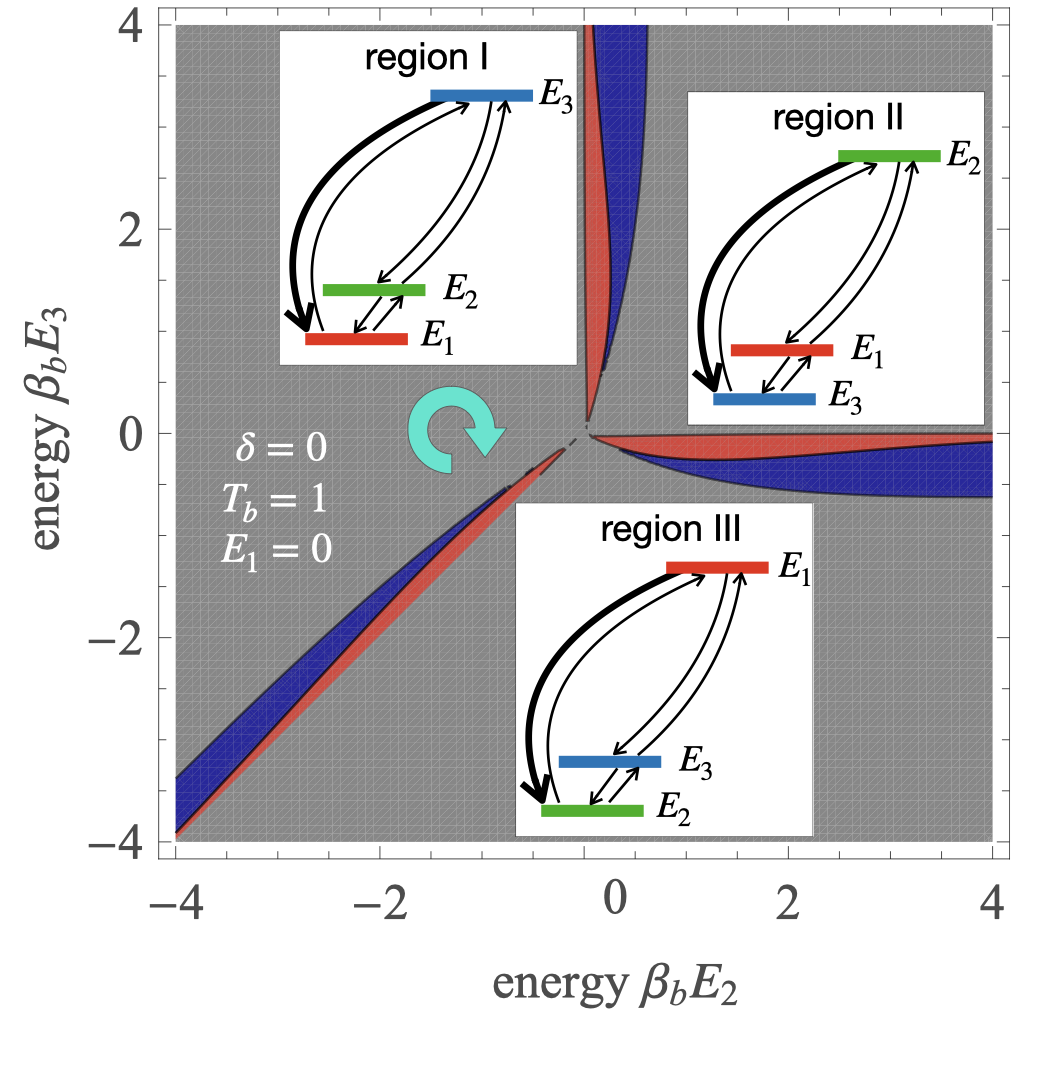}
    \caption{Phase space of a fully connected three-level system characterized by energies $E_1 = 0$, $E_2$, and $E_3$ and situated within a thermal bath at a temperature $T_b = 1$. The load distribution factor is set to $\delta = 0$, as introduced in ~\EQ{delta-rates}. In this space, the gray regions indicate areas where there is no strong Mpemba effect. The red regions represent areas with a strong Mpemba effect during heating, which have a parity index given by $\mathcal{P}\left(\mathcal{I}_M^{\textrm{inv}}\right) = 1$. Conversely, the blue regions indicate the existence of the strong Mpemba effect during cooling, with a parity index described by $\mathcal{P}\left(\mathcal{I}_M^{\textrm{dir}}\right) = 1$. The definitions of the parity indices can be found in~\EQS{Parity-inv}{Parity-dir}. Moreover, the red and blue regions have the highest rate (the global maximum of rates) between the highest and lowest energy levels. The swirling cyan arrow illustrates the chirality of the arms associated with the strong Mpemba effect. \textit{Source:} Adapted with permission from~\cite{bera2023effect}.}
    \label{fig:fig-3level-cyclic-SM-regions-v03-m.png}
\end{figure}
A Mpemba effect in linear reaction networks, which has the same flavor as the above work, was studied in ~\cite{bera2023effect}. The authors show examples where changing the dynamics allows one to choose, for a given initial condition, dynamics with anomalous thermal relaxation. They provide analytical results and insights regarding when the strong Mpemba effect occurs in unimolecular reactions involving three and four species, depending on the dynamics.  To quantify these dynamics, the authors introduce a load-distribution factor, $\delta$, which modifies the forward and backward rates while maintaining detailed balance. Specifically, the transition rates are defined as
\begin{align}
\label{eq:delta-rates}
    W_{i\!j} = \Gamma e^{-(E_j-E_i)\delta/k_B T_b}\quad \text{and} \quad W_{\!ji} = \Gamma e^{-(E_j-E_i)(1-\delta)/ k_B T_b}, 
\end{align}
with $\Gamma^{-1}$ having unit of time, and with load factor $\delta \in [0,1]$. The load-distribution factor has been studied in molecular motors~\cite{kolomeisky2007molecular,kolomeisky2013motor}, negative differential mobility~\cite{teza2020rate}, and Markov jump processes~\cite{remlein2021optimality}. Once again, as in other cases, such as~\SEC{metastability} and~\SEC{statistics}, the strong Mpemba effect is particularly amenable to analytical treatment. Results include characterizations of the regions where both the direct and inverse strong Mpemba effect occur. For example, in~\FIG{fig-3level-cyclic-SM-regions-v03-m.png}, we examine the phase space of a three-level system. Here, the energy of state 1 is fixed at $E_1 = 0$, while the energies of the other two states are varied. The bath temperature is set to unity, $T_b = 1$, and the load factor is $\delta = 0$. The authors identify the regions that exhibit the strong Mpemba effect via the parity index introduced in~\SEC{strong} and defined in~\EQS{Parity-inv}{Parity-dir}. This index provides a lower bound on the strong Mpemba effect. The gray regions in the figure indicate areas where the strong Mpemba effect is absent, while the red regions illustrate where it is in heating and the blue where the effect is in cooling. In this system, the strong Mpemba effect occurs when two energy levels are close to each other, specifically when $|E_i - E_j| = \mathcal{O}(T_b)$, and when the transition rate from the highest to the lowest energy level is the global maximum among the rates. The regions where we see the effect resemble ``propeller arms" and have chirality, which on ~\FIG{fig-3level-cyclic-SM-regions-v03-m.png} is emphasized with the guide to the eye: cyan arrow. The chirality is present due to the mirror symmetry about $\delta = 1/2$ due to the system's inherent symmetry. Specifically, for $E_1 = 0$, the system's behavior with parameter $\delta$ is equivalent to that with $1 - \delta$ when $E_2 \rightleftharpoons E_3$. Moreover, the regions exhibiting the strong Mpemba effect as a function of $\delta$ change as follows: At $\delta = 0$, the ``propeller arms" where we have the effect are the thickest and have chirality, gradually thinning while maintaining chirality, as $\delta$ approaches $1/2$. For $\delta=1/2$, the strong Mpemba effect entirely disappears. Because the system upholds this chiral symmetry (mirror symmetry about $\delta = 1/2$) at $E_2 = E_3$, the only way to satisfy both left and right chirality is for the effect to vanish everywhere. For more details, see~\cite{bera2023effect}. 

It was also shown in this work that the direct and inverse strong Mpemba effect regions do not overlap in a three-level cycle, as measured with the parity of the strong Mpemba effect,~\EQS{Parity-inv}{Parity-dir}. However, both effects can be present simultaneously in the four-level system~\cite{bera2023effect}. 

The above examples focus on the role of dynamics in chemical systems undergoing thermal relaxation and the connections between the details of the dynamics and anomalous thermal relaxations. General aspects of the role of the dynamics in anomalous thermal relaxations are also discussed in~\SEC{dynamics}. 

\subsection{Mpemba-inspired applications and generalizations}
\subsubsection{Relaxation to a NESS}
\label{sec:relaxation-to-a-ness}
Motivated by water, the original formulation of the Mpemba effect is a thermal quench between Boltzmann equilibria at different temperatures. However, one can exploit the Markovian framework outlined in~\SEC{markovian} to explore relaxations towards nonequilibrium steady states (NESS). This extension provides a first step towards the creation of more general optimal heating / cooling protocols, in which the system can be driven to explore a much wider configurational space beyond those accessible when quenching only towards equilibrium steady states~\cite{gal2020precooling}. The boundary setup introduced in~\cite{teza2023relaxation} naturally introduces the possibility to explore such a scenario, by coupling, e.g., different degrees of freedom to two (or more) thermal baths set at different temperatures.

\begin{figure}[tbh]
    \centering
    \includegraphics[width=0.75\linewidth]{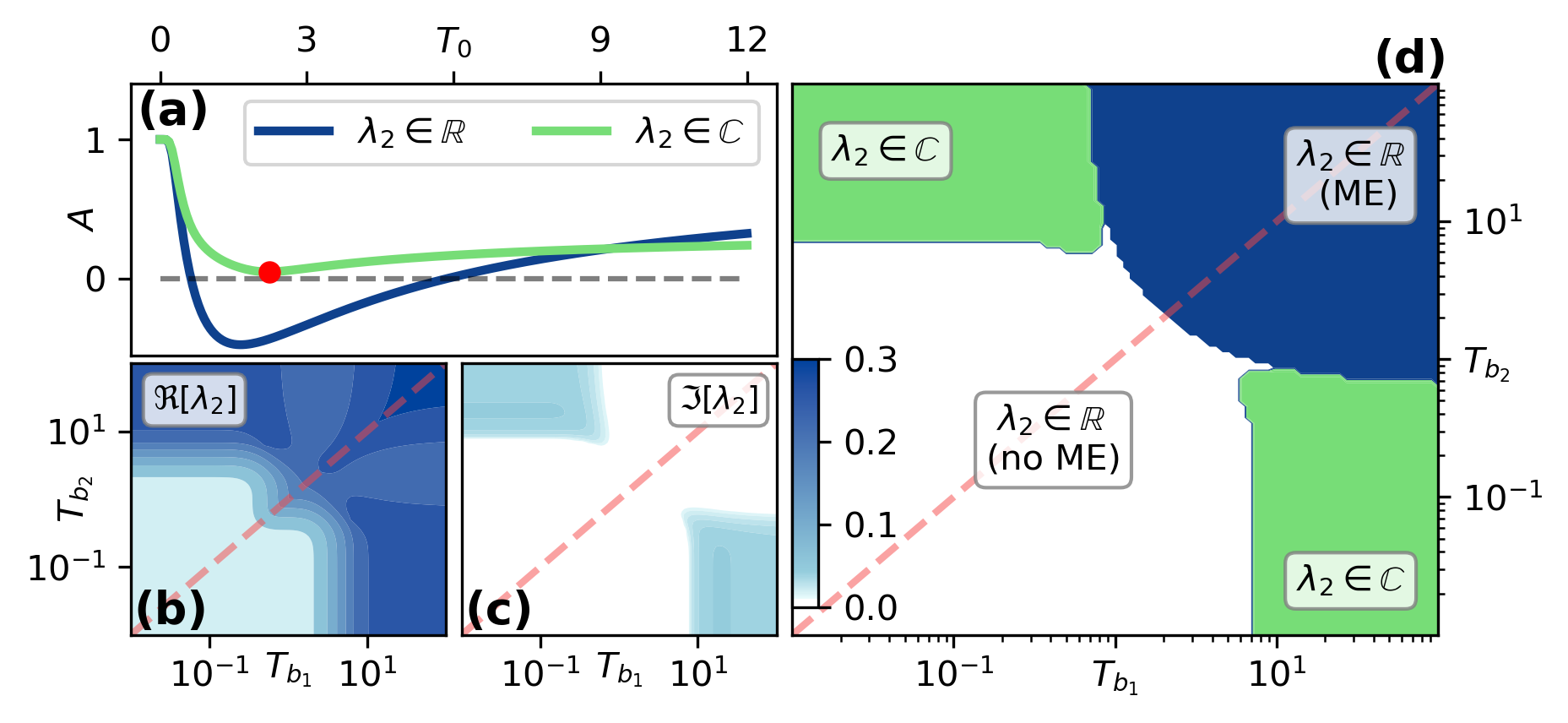}
    \includegraphics[width=0.20\linewidth]{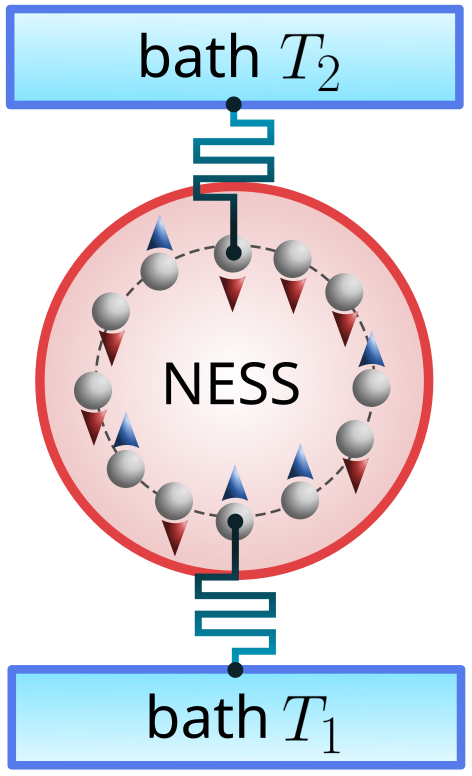}
    \caption{(a) Amplitude of the slowest relaxation coefficients to a NESS (Eq. \ref{eq:ness_relax}) relaxing from an initial equilibrium at $T_0$ for a real (blue) and complex (green) $\lambda_2$.
    (b,c) Real and imaginary parts of $\lambda_2$ as a function of the two bath temperatures $T_{b_1}$ and $T_{b_2}$.
    (d) Phase diagram of an $N=10$ antiferromagnetic spin chain in a magnetic field $H=\sqrt{2}$ with two opposite spins coupled ($C=1$) to different thermal baths.
    The green area corresponds to complex-valued $\lambda_2$, while in the white and blue regions it is real-valued (only in the blue area an analogous of the Mpemba effect can be observed).
    The bisectors $T_{b_1}=T_{b_2}$ (red dashed lines) represent the equilibrium case. \textit{Source:} Reprinted with permission from the supplementary material of~\cite{teza2023relaxation}.  
    }
    \label{fig:ness}
\end{figure}

Taking the 1D antiferromagnetic spin chain introduced in~\SEC{boundary_coupling} as an example, we can connect each of two opposing spins to thermal baths with different temperatures $T_{b_1}\neq T_{b_2}$, respectively (see sketch in~\FIG{ness}). In this scenario, the evolution of the system is regulated by a transition matrix with rates breaking the detailed balance condition, inducing a relaxation to a steady state $\vec{\pi}(T_{b_1},T_{b_2})$ that does not correspond to any Boltzmann equilibrium \cite{katz1983phase,teza2020rate}. Additionally, the intrinsic phenomenology of the relaxation process is somewhat different: while the Perron-Frobenius theorem still ensures the dominant eigenvalue to be identically zero and non-degenerate, the others can be complex-valued and are therefore sorted according to the real part $0=\lambda_1>\Re \lambda_2 \geq \Re \lambda_3 \geq \dots$. Complex eigenvalues come in conjugate pairs, implying periodic oscillations in time of the $a_i$ coefficients. 

Ultimately, two possible scenarios emerge for the slowest relaxation mode; see~\FIG{ness}: (i) $\lambda_2\in\mathbb{R}$, where a characterization of the Mpemba effect as for an equilibrium steady state; or (ii) $\lambda_2\in\mathbb{C}$ implying that $\lambda_3$ is its complex-conjugate, and consequently the probability distribution for long enough times is
\begin{align}
\label{eq:ness_relax}
    \vec{p}(t) \approx \vec{\pi}(T_{b_1},T_{b_2}) + 2 A e^{\Re(\lambda_2)t} 
    \left( \cos(\Im(\lambda_2) t +\phi) \Re(\vec{v}_2) - \sin(\Im(\lambda_2) t+\phi)\Im(\vec{v}_2)\right),
\end{align}
where the amplitude $A=\sqrt{(\Re a_2)^2+(\Im a_2)^2}$ and phase $\phi=\tan^{-1}\left( \Im a_2 /\Re a_2 \right)$ retain the information about the initial condition, which we set here to a Boltzmann equilibrium at temperature $T_0$. The system approaches the NESS via damped sinusoidal oscillations between the real and imaginary parts of $\vec{v}_2$, which can result in a spiraling approach to equilibrium when $\Im \lambda_2 /\Re \lambda_2 \approx 1$. In such a scenario, even if $A$ never crosses zero, there must always exist an initial temperature $T_0\in[0,\infty)$ that minimizes its value (red marker in~\FIG{ness}a): this determines which initial equilibrium condition allows one to reach the NESS more quickly. In \FIG{ness}d, we illustrate the phase diagram for an $N=10$ chain subject to external magnetic field $H=\sqrt{2}$ and with opposing spins coupled ($C=1$) to thermal baths at temperatures $T_{b_1}$ and $T_{b_2}$, respectively. The oscillating relaxations can be observed (green region) only with a considerable gap between the two bath temperatures, with the smallest not exceeding a certain critical value. Consistently, with the relaxation to equilibrium analysis, we also here find that Mpemba-like effects are always observable when both temperatures are above such threshold (blue area).

Although both the scenario presented here and the driven granular gas framework discussed in~\SEC{kinetic-framework} consider relaxations between NESS, they have a very different nature: the latter case is nonlinear.

\subsubsection{Isothermal analogs}
\label{sec:isothermal-analogs}
Section~\ref{sec:relaxation-to-a-ness} considers the case of relaxation to a NESS; here,
we consider an isothermal analog of the Mpemba effect and study the relaxation from a NESS to thermal equilibrium, as suggested in~\cite{degunther2022anomalous}.
The role of temperature in the original version of the effect, as in~\cite{lu2017nonequilibrium}, is taken up by a driving force. 
The 
prototypical 
example is that of a particle diffusing in a 1D periodic potential landscape, $V(x) = V(x+L)$, while also subject to a non-conservative force, $f$; see~\FIG{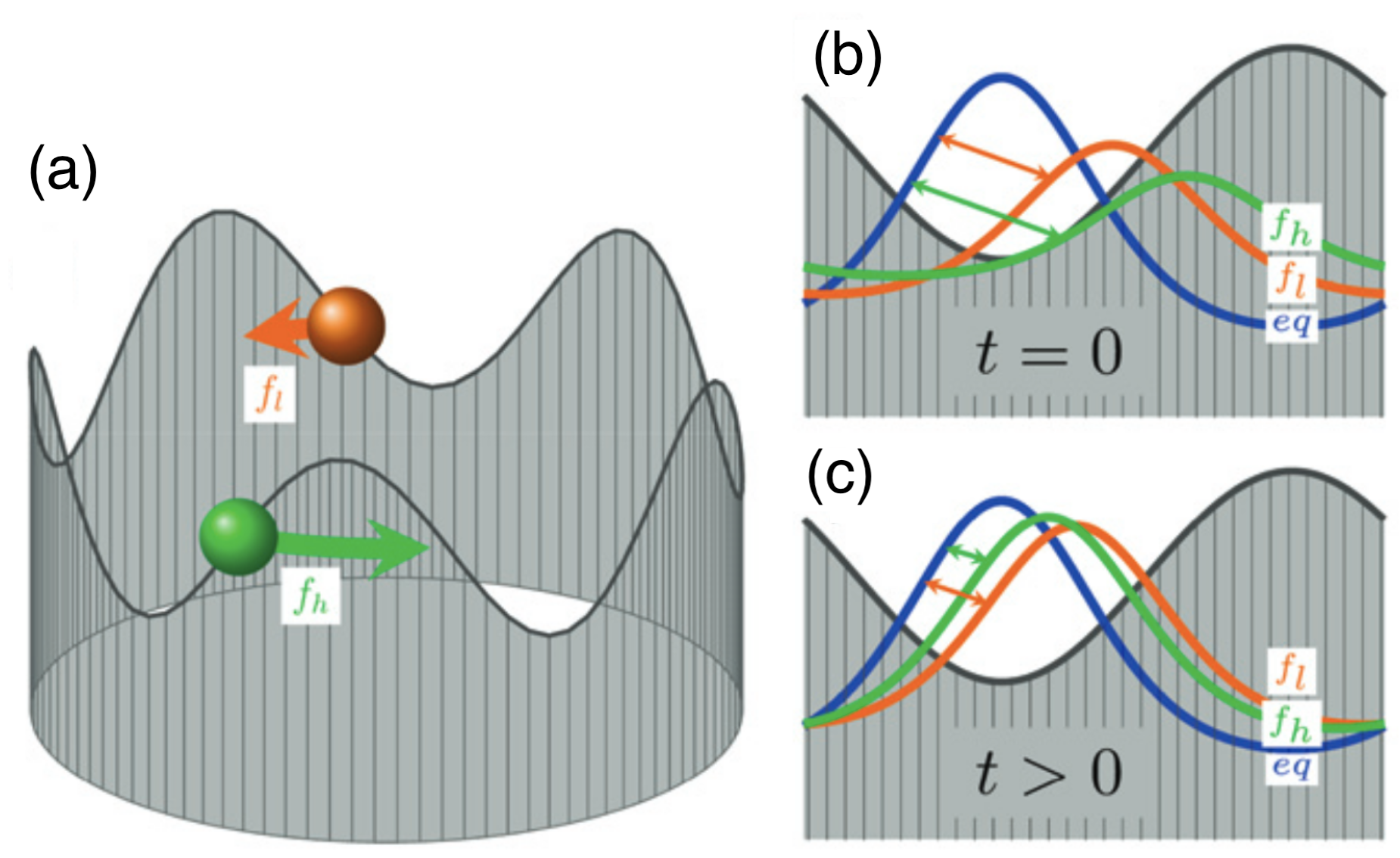}.  

\begin{figure}[tbh]
    \centering
\includegraphics[width=0.5\linewidth]{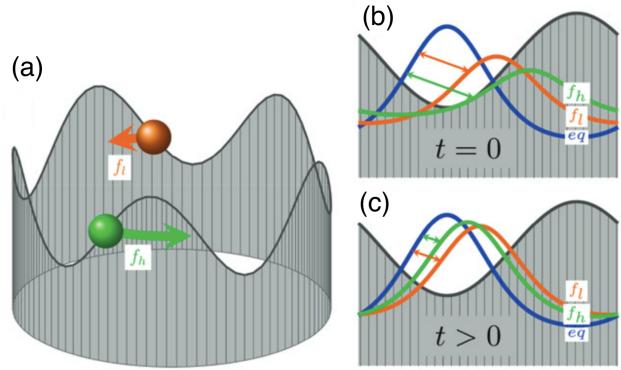}
    \caption{Particle in a periodic potential, $V(x) = V(x+L)$, subject in addition to a nonconservative driving force $f$. (a) Two different initial conditions: a particle driven with a high force $f_h$ and a low force $f_l$. (b) The particle driven with a high force $f_h$ has an initial probability distribution that is further from the equilibrium than the one driven by a low force $f_l$. (c) After some time, the particle driven by $f_h$ has a probability distribution closer to equilibrium. \textit{Source:} Reprinted with permission from~\cite{degunther2022anomalous}.}
    \label{fig:fig-degunther-seifert-isothermal-Mpemba.png}
\end{figure}

The main results are that the Mpemba-like effect is generic for asymmetric potentials but absent for symmetric ones. The second eigenvector is obtained perturbatively for potentials whose amplitudes are smaller than the diffusion scale. 
The authors compute the projection coefficient $a_2$ explicitly and analyze the occurrence of the Mpemba effect as a function of the Fourier amplitudes characterizing the periodic potential. Using topological arguments, the authors extend their perturbative results to periodic potentials with large amplitudes.

\subsubsection{Active-particle systems}
\label{sec:active}
The Mpemba effect has also been predicted in systems containing active, self-propelled particles~\cite{bechinger2016active}. In a first effort, Schwarzendahl and L\"owen studied active particles in a one-dimensional double-well potential placed asymmetrically in a box~\cite{schwarzendahl2022anomalous}. This is the same potential used in the colloidal studies discussed in Sec.~\ref{sec:colloids}. In the study, active particles mimic bacterial run-and-tumble motion by reversing direction after a characteristic time $\tau_p$, drawn from an exponential distribution modeling the first-order kinetics of the transition. In contrast to the passive case,  the authors predict (and study theoretically via coupled Fokker-Planck equations for left- and right-moving particles) that an ``overcooling" can occur. To see this, the authors define an effective temperature $T_{\rm eff}$ using the $L_1$ distance measure 
(Sec.~\ref{sec:distance-function}),
\begin{align}
	D_{T_{\rm eff}}(t) = \sum_i \left| p_i(t) - \pi_{i,T_{\rm eff}} \right| \,,
\label{eq:Teff}
\end{align}
where, at each time $t$, one searches through all equilibrium (here, steady-state) distributions, as parameterized by the temperature $T_{\rm eff}$. Intuitively, one is projecting the distribution onto the closest equilibrium distribution, and associate the temperature of this equilibrium distribution to $\vec p(t)$. As one might expect, the effective temperature description generally breaks down for strongly nonequilibrium regimes, except in some cases in mean-field systems \cite{klich2019mpemba}. Nevertheless, the description makes sense near equilibrium, even in complicated systems. This allows the authors to study the approach to a steady state in the long-time limit. With this definition, the effective temperature can transiently dip below the bath temperature during the cooling process, so that the asymptotic dynamics has an increasing effective temperature. A similar reversal (overheating) is predicted for heating quenches and the inverse Mpemba effect, as well as in the thermodynamic limit of the mean-field antiferromagnetic Ising model \cite{klich2019mpemba}.

Biswas and Pal have recently studied similar dynamics on discrete-state systems (three states and two velocities)~\cite{biswas2024mpemba}. They found that the eigenvalues of the rate-transition matrix of the master equation can become complex, leading to oscillations in the relaxational distance dynamics. The same authors also studied a continuous system similar to that of Ref.~\cite{schwarzendahl2022anomalous} but having an asymmetric V-shaped potential with no metastable state~\cite{biswas2023mpemba_a}. In that case, they found that large spontaneous velocities tend to suppress the Mpemba effect, but smaller velocities can sometimes enhance it~\cite{biswas2024mpemba1}. Given the wide range of realizations of active-particle systems~\cite{bechinger2016active}, it should be possible to test such predictions experimentally.
 
\subsubsection{Stochastic resetting}
\label{sec:stochastic-resetting}
Stochastic processes that occasionally reset the state to a specific value occur in many biological and chemical scenarios. Examples include a population that stabilizes at a specific size before a rare and catastrophic event reduces its numbers, diffusion of a particle that occasionally resets to its original position, visual search for a face in a crowd in psychology, animal foraging, and backtrack recovery by RNA polymerases. Resetting is crucial in simulated-annealing algorithms, which are used to find deep minima in optimization problems. Typically, the resetting process is stochastic. 

Research on stochastic resetting began with a seminal paper by Evans and Majumdar, which examined a diffusing Brownian particle that resets to the origin after intervals determined by a Poisson distribution~\cite{evans2011diffusion}. In this scenario, the overdamped Langevin diffusion with stochastic resetting reaches a nonequilibrium steady state (NESS) and exhibits a dynamical phase transition. The rate of stochastic resetting can be used to ensure that the average search time for a target is both finite and optimized. This area of research, focusing on the stochastic resetting of random processes, is actively explored both theoretically and experimentally, as reviewed by Evans and Majumdar in~\cite{evans2020stochastic} and in a recent special collection~\cite{kundu2024stochastic}.

The relations between stochastic resetting and the Mpemba effect within discrete Markov systems was studied in~\cite{busiello2021inducing}. In this work, the probability distribution of a discrete Markov system evolves according to the master equation,~\EQ{master_equation}, with rates specifically chosen to satisfy detailed balance and follow the Arrhenius form,~\EQ{arrhenius}. The system evolves by 
\begin{align}
    \partial _t \vec{p} = \left(\textbf{W} - r\, \theta (t_r - t) \, \textbf{I} \right) \vec{p} + r \, \theta(t_r - t)\, \vec{\Delta}, 
\end{align}
where $\theta$ is the Heaviside step function, $\textbf{I}$ is the identity matrix, and $\vec \Delta$ is a vector with all zeros except at the resetting location where it is unity. The solution to the master equation with stochastic resetting yields 
\begin{align}
    \vec p (t) = \begin{cases}
\vec \pi (T_b) + \displaystyle\sum _{i \geq 2} a _i ^{(r)} (t, T_{\rm init}, T_b) \vec{v}_i e^{\lambda_i t} & t \leq t_r,\\ 
\\
\vec \pi (T_b) + \displaystyle \sum _{i \geq 2} a _i ^{(r)} (t_r, T_{\rm init}, T_b) \vec{v}_i e^{\lambda_i t} & t > t_r.
\end{cases}
\end{align}
The eigenvector projection coefficients, $a_i ^{(r)}$, are time dependent and have the form 
\begin{align}
\label{eq:acoef-reset}
    a^{(r)} _i (t,T_{\rm init},T_b) = \left[\frac{r d_i}{r - \lambda_i} + \left(a_i(T_{\rm init},T_b) - \frac{r d_i}{r - \lambda_i}\right) e^{(\lambda_i - r)t}\right]e^{- \lambda_i t}. 
\end{align}
Here, $a_i(T_{\rm init},T_b)$ is the projection coefficient without resetting, given with~\EQ{ai-coef}, and $d_i$ are the projection coefficients of $\vec{\Delta}$ in the basis of $\{\vec \pi (T_b),\vec{v}_i\}$, as 
\begin{align}
\label{eq:d-coef}
    \Delta _j = \pi _j (T_b) + \sum _{i \geq 2} d_i \left(\vec v^{\,(i)}\right)_j. 
\end{align}

The authors of \cite{busiello2021inducing} showed that stochastic reset can induce a strong Mpemba effect in a system lacking the effect without the reset. Moreover, the stochastic reset can provide driving, independent of the initial condition, leading to the long-time dynamics being dominated by $\lambda_3$ instead of $\lambda_2$. 
Imposing the conditions for a strong Mpemba effect, $a^{(r)}(t_{\rm SM}) =0$, a time $ t = t_{\rm SM}$, in~\EQ{acoef-reset}, gives
\begin{align}
    t_{\rm SM}(r) = \frac{1}{r + \lambda_2} \ln \left[1 - \frac{(r + \lambda_2)}{r}\frac{a_2(T_{\rm init},T_b)}{d_2}\right]. 
\end{align}
Thus, provided that 
\begin{align}
\label{eq:Bussielo-result}
\frac{a_2(T_{\rm init},T_b)}{d_2} \leq 0, 
\end{align} 
there is a strong Mpemba effect, where $d_2$ depends on resetting states probabilities, i.e. $\vec \Delta$. Moreover, $t_{\rm SM}(r)$ is smaller the closer to zero the ratio of $a_2(T_{\rm init},T_b)/d_2$ or the bigger the resetting rate $r$ is. An example of a stochastic-resetting-induced strong Mpemba effect is shown in~\FIG{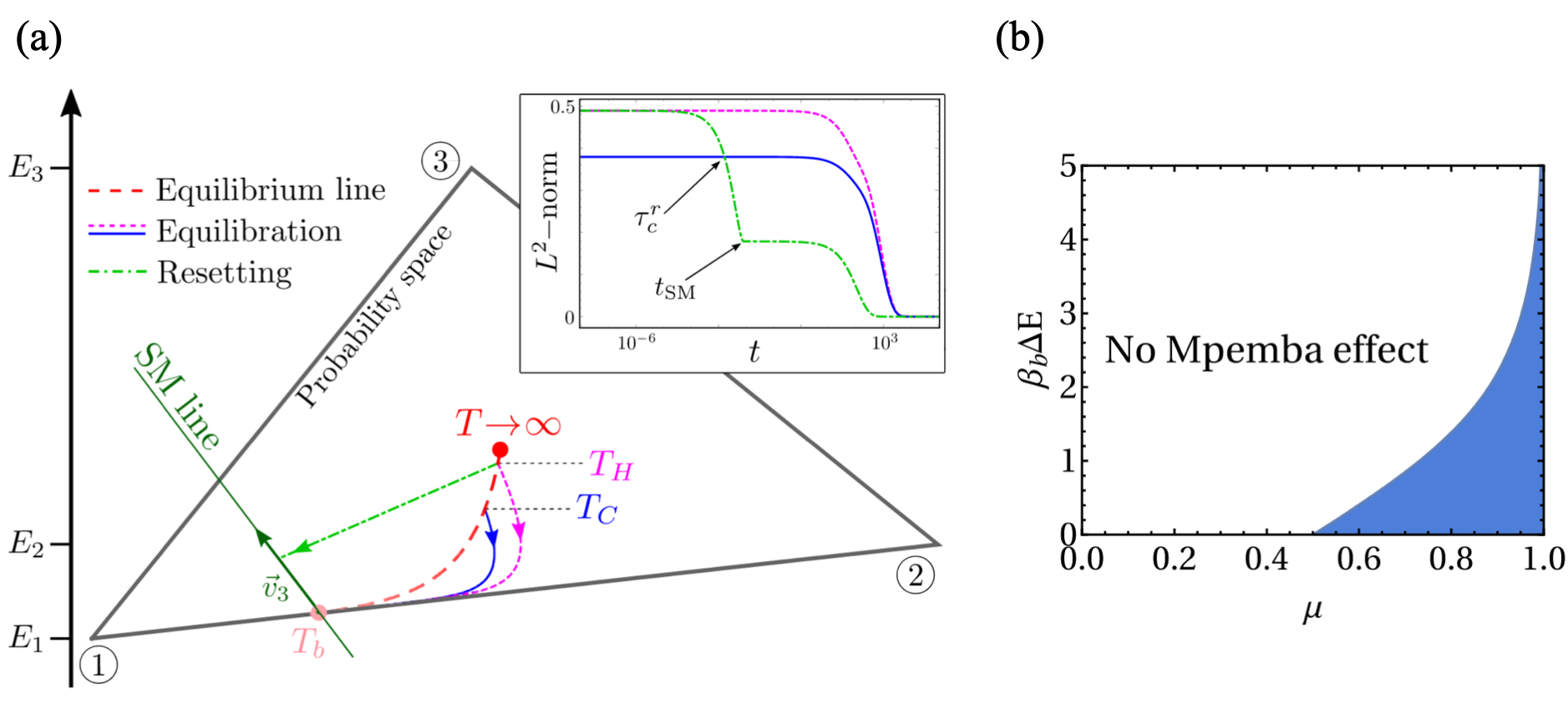}a. 
\begin{figure}
    \centering
    \includegraphics[width=0.85\linewidth]{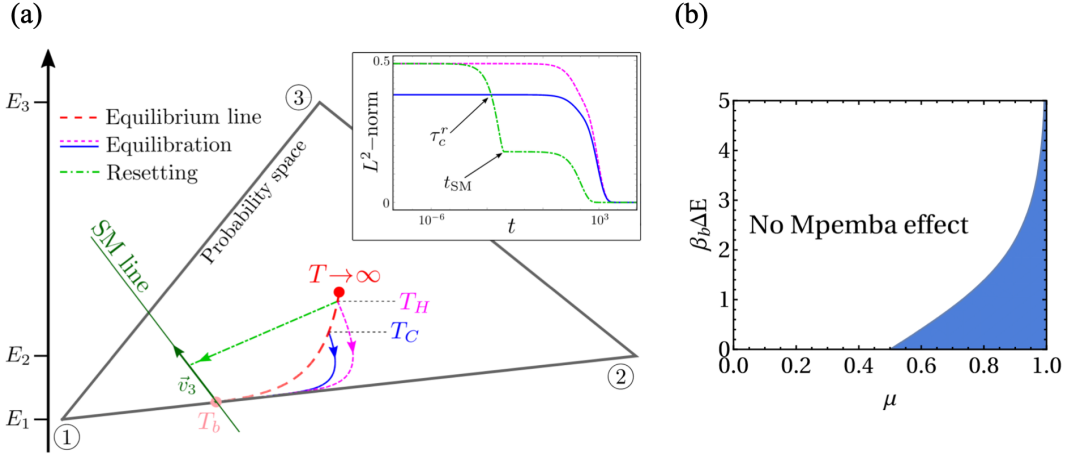}
    \caption{(a) Probability space of a three-state system. The dashed-red line shows the equilibrium locus. The relaxation with Arrhenius rates,~\EQ{arrhenius}, to the bath temperature, $T_b$, is shown with dashed-magenta (initial temperature $T_H$) and solid-blue lines (initial temperature $T_C$). The dot-dashed-green line shows the relaxation to $T_b$ in the presence of stochastic resetting. Here $\tau^r _c$ is the crossing time for the $L^2$ norm, which quantifies the onset of the Mpemba effect, and $t_{\rm SM}$ is the time at which the strong Mpemba effect occurs; i.e., $a_2^{(r)}(t_{\rm SM}) = 0$. (b) Phase space of the existence of the Mpemba effect in a two-state model with stochastic resetting, with the probability of resetting to State $1$ being $\mu$, and State 2 being $1-\mu$. The energy difference $\Delta E = E_2 - E_1$. The shaded area shows the region where the Mpemba effect can be observed. \textit{Source:} Adapted with permission from~\cite{busiello2021inducing}.}
    \label{fig:fig-stochastic-resetting.png}
\end{figure}

To appreciate how resetting affect the Mpemba effect, we recall from Sec.~\ref{Sec:Example-3-states} that the Mpemba effect does not occur for a two-state model.  However, with stochastic resetting, one can also have a Mpemba effect in the two-state system. Moreover, it is enough to consider resetting to a single state. Generalizing to resetting to both states and introducing the probability of resetting to State 1 is $\Delta _1 = \mu$ and to State 2 is $\Delta_2 = 1-\mu$ where $\mu \in [0,1]$ we have the following condition for the occurrence of the Mpemba effect
\begin{align}
    \frac{\Delta E}{k_B T_b} < \ln \frac{\mu}{1-\mu}, 
\end{align}
where $\Delta E \equiv E_2 - E_1$. The phase space $(\mu,\Delta E/k_B T_b)$ is shown on~\FIG{fig-stochastic-resetting.png}b. 

Lastly, the authors derive an optimal driving protocol for simultaneously optimizing the appearance time of the Mpemba effect and the total energy dissipation into the environment. This is captured by a Pareto front in~\FIG{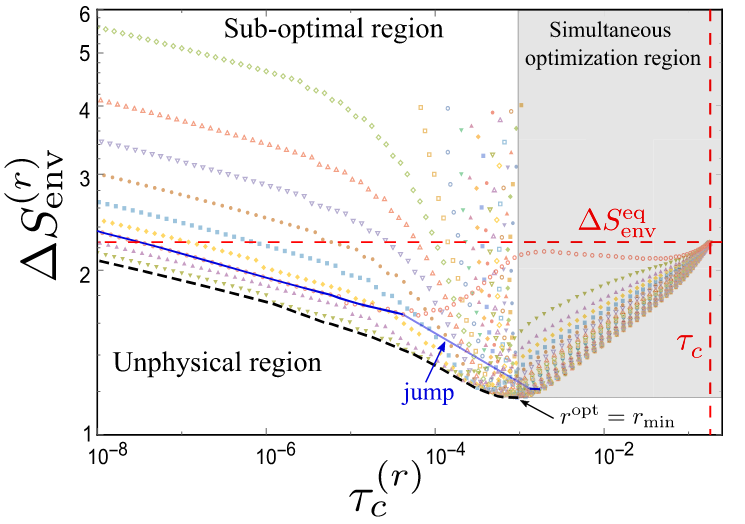}. 
\begin{figure}[tbh]
    \centering
    \includegraphics[width=0.55\linewidth]{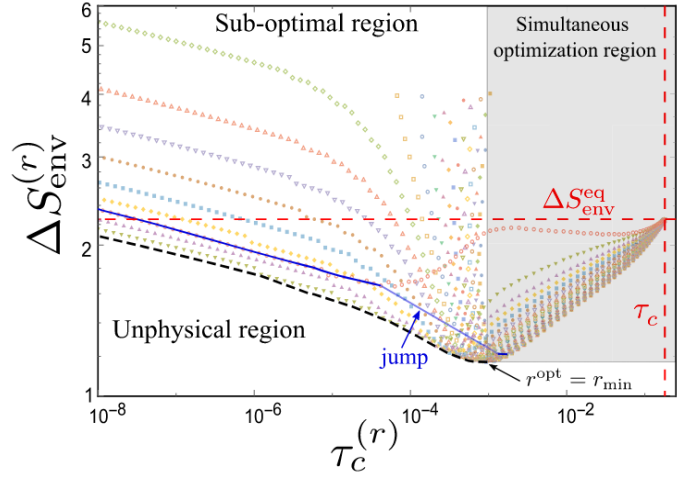}
    \caption{Pareto front showing the optimal protocol that minimizes the crossing time $\tau_{\rm c}^{(r)}$ and the dissipation to the environment, $\Delta S^{(r)} _{\rm env}$ for a fully connected four-state Markov system. The solid blue line represents resetting to a single state, while the dashed-black line represents resetting to two states (1 and 4 here) with $\Delta = \mu \delta_{i,1} + (1 - \mu) \delta_{i,4}$. Suboptimal choices of resetting probability (to state 1) $\mu$ are above the dashed black line. For protocols in the shaded area, both the crossing time $\tau_c ^{(r)}$ and the dissipation to the environment $\Delta S^{(r)} _{\rm en}$ can be simultaneously optimized, hence corresponding to a single optimal point $r^{\rm opt} = r_{\rm min}$. \textit{Source:} Adapted with permission from~\cite{busiello2021inducing}.} 
    \label{fig:fig-Pareto-front.png}
\end{figure}

\begin{figure}
    \centering
    \includegraphics[width=0.65\linewidth]{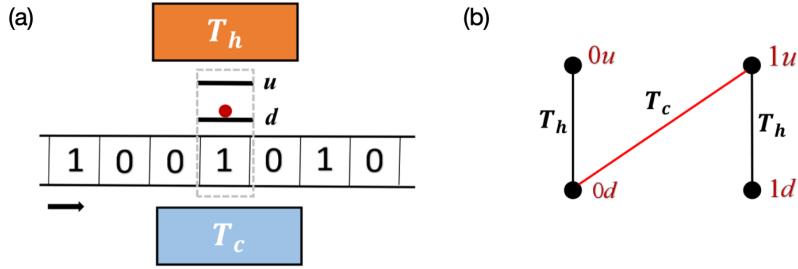}
    \caption{(a) Two-state Maxwell demon interacting with a tape (sequence of bits) and two reservoirs at temperatures $T_h$ and $T_c$. (b) A composite four-state system consisting of demon and tape. The black edges correspond to transitions at high temperature $T_h$, and the red edge corresponds to the transition at low temperature $T_c$. \textit{Source:} Adapted with permission from~\cite{bao2023designing}. }
    \label{fig:fig-Maxwell-refrigerator.png}
\end{figure}

Additional connection between stochastic resetting and Mpemba were discussed in ~\cite{bao2023designing}, where stochastic resetting was utilized to optimize the relaxation process of an autonomous Maxwell's demon by inducing a strong Mpemba effect. Maxwell's demon is introduced and discussed in~\SEC{optimal-transport-heat-engine-efficiency}. Specifically, the manuscript us the exactly solvable demon model proposed in~\cite{mandal2012work}. By introducing a discrete-time stochastic resetting mechanism and applying it to the demon setup, the authors showed that the time for the autonomous demon to relax to its periodic steady state can be reduced by a protocol inspired by a strong Mpemba effect. The design strategy is illustrated for the two-state Maxwell demon shown in~\FIG{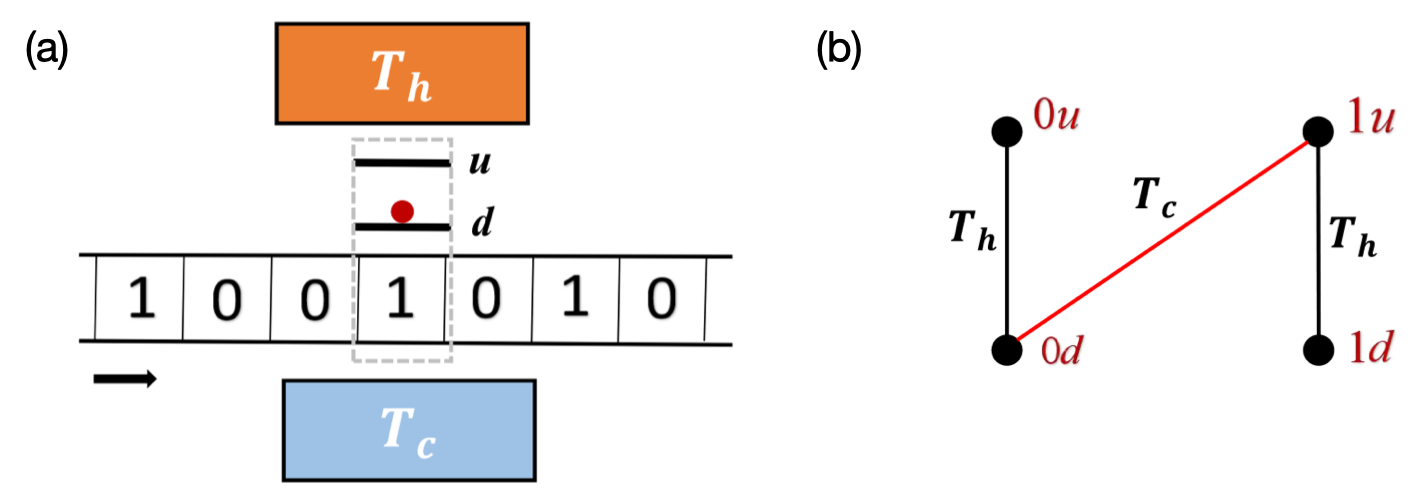}a. This model system consists of a demon with two states and an infinitely long tape, a stream of information with bits $0$ and $1$. The combined system is shown on~\FIG{fig-Maxwell-refrigerator.png}b.  The demon starts the cycle at a temperature of one of the reservoirs, and, during a time $\tau$, interacts with the memory tape. There is no friction, and the demon interacts with only a single bit at a time. During the interaction, the demon undergoes an intrinsic transition between its pairs of states. The bit can also change its state. The two transitions can occur simultaneously and lead to anomalous work production. The demon's disordered transitions can be rectified by incoming bits. The combined system of the demon and the tape obeys a master equation. After a certain time, it reaches a periodic steady state. 

The authors introduce a strategy of resetting the demon with a constant probability in discrete time for a fixed amount of time $t_c = N_c \tau$, such that the resulting evolution does not have the coefficient to the slowest mode, that is $a_2^{(r)}(N_c) = 0$. They show that the sufficient condition to find a physical critical number of periods, $N_c \geq 1$ is $a_2/d_2 \leq 0$, where $d_2$, as before, is the coefficient of the resetting state $\vec \Delta$ in front of the second right eigenvector $\vec v _2$ and $a_2$ is the coefficient to the second eigenvector in the absence of resetting. This agrees with the above-described result from \cite{busiello2021inducing} c.f.~\EQ{Bussielo-result}. The authors show how stochastic resetting can notably enhance the performance of an autonomous Maxwell's demon in two ways: the speed of reaching their functional state and the range of their effective work regions. Namely, they show that one can drive the autonomous demon system to its functional periodic steady state with a shortened relaxation time by resetting the demon at a predetermined critical time and then stopping the resets. They also show how, by continuing with the resets, one can increase the effective working area of Maxwell's demon. Lastly, they discover a dual-functioning region of the demon with resetting, where it can both produce work and erase information at the same time. The dual region comes at the cost of resetting. 

A Maxwell demon model interacting with a tape in connection with the Mpemba effect was discussed in~\cite{bera2023effect}. Similar to previous research, the authors identify conditions under which the relaxation to the functioning state is accelerated. In~\cite{bera2023effect}, the tape is maintained at a finite temperature, and the dynamics of relaxation are explored as a function of a control parameter (load factor, see also~\SEC{rate-of-chemical-reactions}). We reference this work in the context of optimizing the performance of the device in~\SEC{optimal-transport-heat-engine-efficiency}. Finally, for quantum systems, the acceleration of relaxation in Markovian open quantum systems through quantum reset processes was considered in ~\cite{bao2022accelerating}.

\subsubsection{Markov chain Monte Carlo sampling}
\label{sec:mcmc}
Monte Carlo algorithms have played a crucial role in investigating the Mpemba effect. For instance, Markov chain Monte Carlo algorithms were utilized to study the Mpemba effect in Ising spin systems, as discussed in Section \ref{SubSec:NumericalMonteCarlo}. Interestingly, this effect may potentially also enhance existing sampling algorithms, particularly the Markov chain Monte Carlo sampling algorithm itself.

In physics, the Markov chain Monte Carlo (MCMC) algorithm is frequently used for calculating high-dimensional integrals in quantum field theory and statistical physics, as well as for generating samples from a target probability distribution, often the equilibrium probability distribution. This algorithm was first introduced in 1953 in a groundbreaking paper by Metropolis, Rosenbluth, Rosenbluth, Teller, and Teller~\cite{metropolis1953equation}. Beyond the realm of physics, Monte Carlo algorithms are widely applied in fields such as biology, chemistry, and various other natural and social sciences. Numerous reviews and books on this topic have been published, such as those by Krauth and Sokal~\cite{krauth2006statistical, sokal1997monte}.

For Monte Carlo sampling to be efficient, significant effort is dedicated to designing Markov chains that quickly converge to the target probability distribution and generate independent samples. To accelerate convergence to the steady state, it is often beneficial to increase the spectral gap, ${\rm Re} \lambda_2$. This can be attempted in various ways, such as introducing auxiliary variables—like momentum in Hamiltonian Monte Carlo~\cite{neal2011mcmc}, or by generally increasing the phase space in a method called \emph{lifting Monte Carlo}~\cite{chen1999lifting, diaconis2000analysis, turitsyn2011irreversible, vucelja2016lifting, zhao2022nonreversible}. Another approach involves modeling a collection of systems at different temperatures and allowing swaps between them, as seen in \emph{parallel tempering}~\cite{ballard2009replica, machta2009strengths}. Other methods include breaking detailed balance, as mentioned with lifting, introducing piecewise deterministic processes~\cite{bierkens2019zig}, or using event-chain Monte Carlo~\cite{krauth2021event}. Likewise, there are approaches that employing neural networks to optimize the rate matrix~\cite{song2017nice, wang2018meta}.

One potential application of anomalous thermal relaxations is in selecting initial conditions, $p_{\rm init}$, or the properties of the dynamics, $\mathbf{W}$, so that the relaxation displays a Mpemba effect, enabling a quicker approach to the target equilibrium state. For instance, the study in~\cite{gal2020precooling} demonstrated that an optimal heating strategy could involve cooling a system for a certain period before heating it. This method allows the system to relax to the target distribution from a different trajectory compared to when there is no precooling stage. By introducing a different ``target" temperature for a finite duration, the system can approach the equilibrium from an alternate direction: $-\vec v_2$ rather than $\vec v_2$. This indicates that a strong Mpemba effect occurred at some point during this process. Therefore, it could be beneficial to utilize a different "target" distribution for a limited time if it enables one to minimize the projection along $\vec{v}_2$, i.e., minimize $a_2$, and thus speed up the relaxation. 

The relaxation of Metropolis dynamics on dense graphs was studied in~\cite{klich2018solution}, where it was discovered that while no Mpemba effect occurs on a complete graph, it can in non-complete graphs.

Research on the potential applications of the Mpemba effect in designing efficient Markov chain Monte Carlo sampling is still in its early stages, and it will be interesting to see future developments in this area.

\subsubsection{Controlled state transformations}
\label{sec:controlled}
In the Mpemba effect, a system initially in equilibrium with a bath at temperature $T_{i}$ is suddenly put in contact (``quenched'') with a bath at temperature $T_{b}$.  The surprise is that the time to relax is not necessarily a monotonic function of the temperature difference $T_{i}-T_{b}$.  An alternate point of view is that of the system, where one can say that the bath temperature is switched at time $t=0$.  That is, the bath temperature $T_{b}(t)$ follows
\begin{align}
	T_{b}(t) = T_{i} +(T_{b}-T_{i}) \Theta(t) \,,
\label{eq:temp-switch}
\end{align}
where $\Theta(t)$ is the step function.  In this view, the Mpemba effect is the result of subjecting the system to an external protocol $T_{b}(t)$, and the relaxation follows as a response to the imposed protocol.  Similarly, we can view the Kovacs effect as the result of a protocol with a third, intermediate temperature, see Sec.~\ref{sec:kovacs}.  

Continuing in this framework, one can view the steady-state Mpemba effect describing isothermal transitions between nonequilibrium steady states (Sections~\ref{sec:zoology} and \ref{sec:isothermal-analogs}) as simply the result of a step protocol in some external  parameter (other than temperature), $\theta(t)$, that drives the system out of equilibrium. An example for such a case was considered in~\cite{degunther2022anomalous}, where $\theta(t)$ is a non-conservative driving force that pushes particles around in a ring; c.f. Fig.~\ref{fig:fig-degunther-seifert-isothermal-Mpemba.png}.  In a finite-state systems, the control parameter $\theta$ alter transition rates between states, breaking detailed balance.  Each value $\theta$ would generate a nonequilibrium steady state with circulating currents, and one could look for a Mpemba effect involving abrupt transitions of driving strength.  To our knowledge, this has not been done so far.

Since simple protocols such as steps in temperature or some driving force lead to interesting results, it is natural to consider more general protocols $\vec\theta(t)$ that may include multiple (even an infinite number) of parameters and may or may not involve temperature.  This is the domain of \textit{control theory}, a subject of increasing interest to physicists~\cite{bechhoefer2021control}. In the framework of control theory, these protocols all aim to implement \textit{state-to-state} transitions.  The protocols are typically finite-time protocols, lasting a time $\tau$.  The states either reach the desired state at the end of the protocol or reach it after a passive relaxation that starts at time $t=\tau$. For the latter case, one can still define a notion of protocol relaxation time using the magnitude of the $a_2$ coefficient, as discussed in~\SEC{markovian-quench}.

In general, there are many protocols $\vec\theta(t)$---often, infinitely many---that transform a given state into a specified final state. Sometimes finding a single feasible protocol, a $\vec\theta(t)$ that takes a system from its initial state $x(0)$ to a desired final state $x(\tau)$, is enough. Usually, though, the transformation should satisfy additional considerations.  One that is nearly always relevant is that the allowed values of a control parameter are typically limited: for example, applied forces must be below some value. Beyond constraints on the values of a control parameter, one might seek to make the transformation in minimum time, to minimize dissipation, or to follow some prescribed path.  

One goal that has been intensely pursued, first in quantum contexts~\cite{gueryodelin2019shortcuts} and then in classical~\cite{gueryodelin2023driving} is to find ``shortcuts to adiabaticity.'' The goal generalizes notions of adiabatic transformations in classical and quantum mechanics, where a system is transformed from the ground state corresponding to $\vec\theta(0)$ to the ground state corresponding to $\vec\theta(\tau)$ over the time interval $(0,\tau)$.  At each intermediate time $t$, the system should be in its instantaneous equilibrium value.  If $\vec\theta(t)$ were the bath temperature $T_{b}(t)$, this would amount to requiring that a relaxing system be characterized by a Boltzmann distribution $\pi(x,T_{b}(t))$.  Of course, this is just what occurs in the long-time limit, but now the time $\tau$ is comparable to or shorter than $\lambda_2^{-1}$.

One way to realize a shortcut to adiabaticity is to add a precisely specified \textit{counterdiabatic} force, an external potential $U_{\rm c}(x)$. Unfortunately, adding such a force is difficult in general, as it amounts to requiring a continuous infinity of parameters (the potential or force at each state $x$), a situation known as ``full control.'' Moreover, in the under-damped case, often the counterdiabatic term is also a function of the momentum, which is even more difficult to achieve \cite{patra2017shortcuts}.  Other goals, such as the minimization of dissipation require a similar level of control. Alternatively, techniques such as \emph{Engineered Swift Equilibration} (ESE) give feasible solutions depending on a small number of parameters~\cite{martinez2016engineered}.  The parameters can be chosen to optimize a particular goal, such as minimizing the required work to carry out the transition.  The values achieved will in general be below what is possible with full control.

In between the simple step control of equilibrium or steady-state Mpemba and the full control required to minimize dissipation or enforce shortcuts to adiabaticity, there are simpler controls that can generalize the Mpemba effect in interesting ways.  We have already mentioned the Kovacs effect, as an example of a three-temperature protocol.  Another possibility is to ask for continuous-time temperatures $T_{b}(t)$ that minimize heating or cooling times.  The temperatures should be constrained by upper and lower bounds to make the problem physically realizable. As previously introduced in Sec.~\ref{sec:optimal-heating-cooling-protocols}, Ref.~\cite{gal2020precooling} showed that a technique from optimal control, the \emph{Pontryagin Maximum Principle}~\cite{bechhoefer2021control}, applies and leads to ``bang-bang" control where the fastest transitions are achieved by alternating the temperature between the allowed limits.  Their analysis shows that the strong Mpemba effect implies the existence of an optimal nonmonotonic protocol, but even if there is no Mpemba effect of any type, the optimal protocol might still be nonmonotonic in temperature.  

More recently, Patr\'on et al. studied optimal bang-bang protocols in harmonic potentials~\cite{patron2022thermal}, which they referred to as leading to the ``thermal brachistochrone,'' (minimum-time trajectory).  The term refers to the classic problem posed by Johann Bernoulli and solved by Isaac Newton of the shape with the fastest transition time for a ballistic particle~\cite{boyer2011history}.  For harmonic potentials, the nonequilibrium distributions are always Gaussian at all times in the protocol, which allows for transitions where the equilibrium is reached in finite time.  This is generally not possible for non-Gaussian potentials when controlling only the temperature.  Furthermore, the same authors have generalized their calculation to the nonequilibrium steady states created in a Brownian gyrator, where a particle trapped in a two-dimensional harmonic oscillator is subject to different noise strengths along different coordinate axes and rotates as a result~\cite{patron2024minimum}. In~\cite{boubakour2024dynamical} dynamics invariants, based on shortcuts to adiabaticity, are used to speedup the equilibration of an open quantum system. The speedup is compared to that resulting from a Mpemba effect.

Finally, we discuss a more sophisticated control extension of the Mpemba effect proposed by Chittari and Lu~\cite{chittari2023geometric}.   They seek to establish conditions for the existence of rapid (``hasty'') protocols that speed up operations that transform a system from some initial state to some desired end state in a finite time.  The basic idea is to manipulate the location of manifolds describing fast and slow relaxations such that a trajectory can always proceed along fast segments; hence, the term ``hasty.'' Chittari and Lu argue that Mpemba effect-like shortcuts only constitute a small fraction of the diverse categories of hasty shortcuts and refer to the Mpemba effect as a ``one-step'' hasty shortcut since the system is ``steered away'' from slow dynamics via a single control step (by setting the initial condition). Moreover, that single control step is what the authors call ``counter-shooting only,'' as the controls go beyond the range of the initial and final bounds.  That is, they start from a hotter initial temperature to facilitate faster cooling to the environment.

Thus, we see that more actively controlling the dynamics of a particle can open up a wide range of generalizations.  The Mpemba effect represents one extreme of a continuum of control possibilities---free relaxation after ``release'' from an initial state---whereas techniques such as shortcuts to adiabaticity require full control, where one chooses an applied force at every moment of time.  Within these limits, the challenge is to balance the greater range of behavior provided by more control against ease of implementation in physical systems.

\section{Related effects}
\label{sec:other-related}
In this section, we describe several kinds of anomalous effects and concepts related with the Mpemba Effect. In particular, we show how different strategies and protocols can affect the relaxation rhythm of a given system. In Sec.~\ref{sec:kovacs}, we discuss the Kovacs effect, where internal states affect the dynamics when the protocol is generalized to more than one temperature quench. The Kovacs effect appears when a naive strategy with two quenches is considered, for example to try to reach faster a given cold temperature by going first to a much colder one. A first quench is to a very cold temperature. Then, suddenly, when the system is at the desired temperature, a second quench is applied at that temperature. In Sec.~\ref{sec:asymmetric}, we consider various types of asymmetries between heating and cooling quenches, that is, heating can be faster than cooling and vice versa. Specifically, we discuss different results and ways to characterize the existence of an asymmetry between cooling and heating in systems far from equilibrium.   In Sec.~\ref{sec:eigenvalue-crossing}, we describe how changes in the ordering of eigenvalues (and their associated eigenvectors) impact the Mpemba effect.  And finally, in Sec.~\ref{sec:relaxation-speedup}, we show some other mechanisms that can lead to speedups in the relaxation to thermal equilibrium. The first is to gain leverage using the non-monotonicity of equilibrium magnitudes and a weak version of the master equation. Using that approach, asymmetry between heating and cooling, preheating strategy (analogous to the precooling strategy introduced in \cite{gal2020precooling}) and the Mpemba effect are all reproduced.  The second addresses how to use the evolution of the internal structure of magnetic materials to reach faster a target temperature. In this case, it is shown that the size of internal magnetic domains affects to the velocity of the relaxation after a quench.

\subsection{Kovacs effect and nonmonotonic temporal relaxation}
\label{sec:kovacs}
\begin{figure}
    \centering
    \includegraphics[width=5.0in]{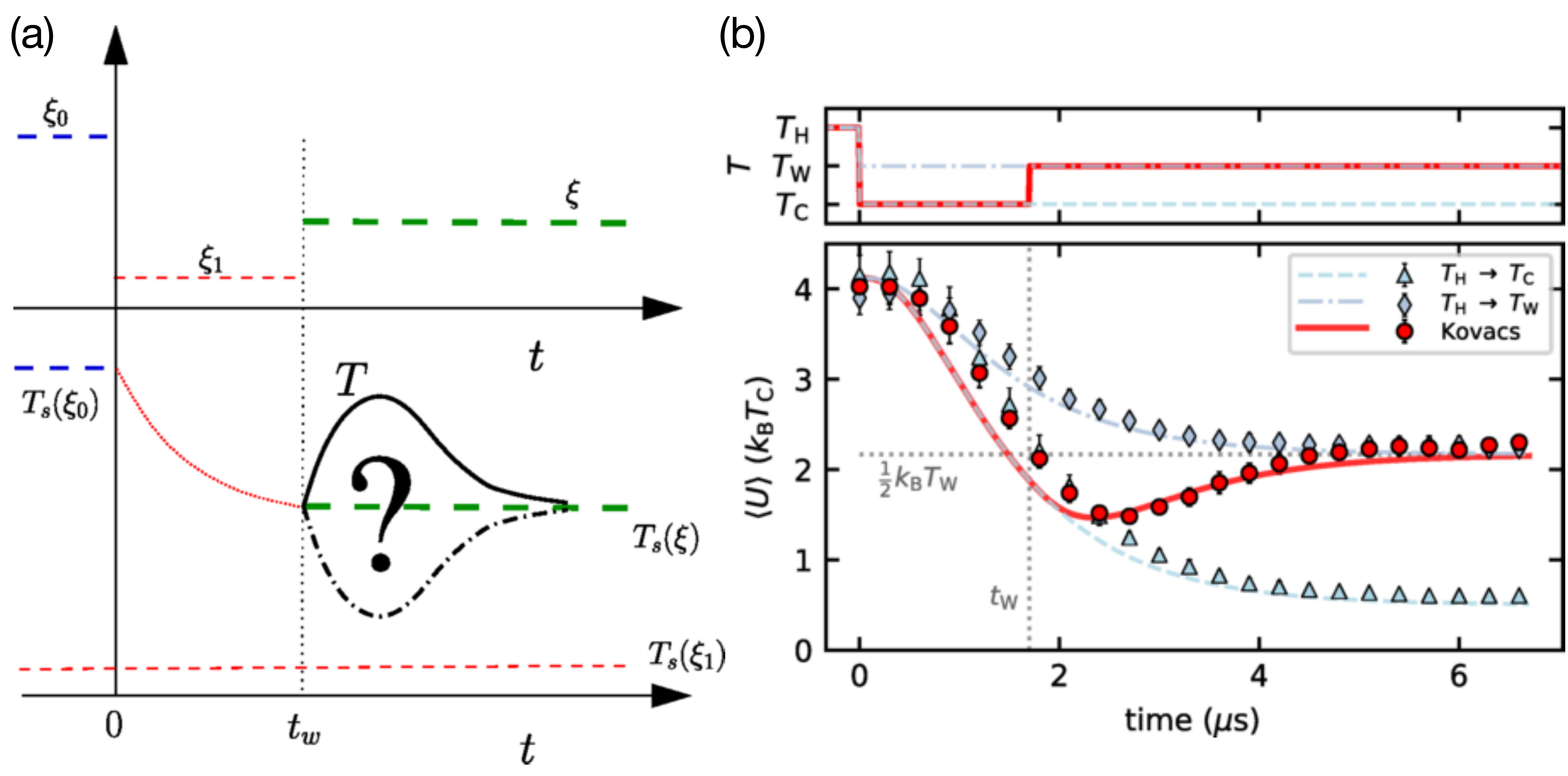}
    \caption{Illustration of Kovacs effect. (a) Sketch of the normal and anomalous  Kovacs response produced in a driven granular gas; see Sec. \ref{sec:kinetic-framework}.  Top: Driving protocol has three forcing levels, $\xi_0$, $\xi_1$, and $\xi_2$.  Bottom:   Kinetic temperature time evolution. Initially, the system is at the hot temperature  $T_s(\xi_0)$, in a nonequilibrium steady state resulting from the driving $\xi_0$. Then, the driving level is suddenly highly decreased to $\xi_1$, and the system would asymptotically reach the cold temperature $T_s(\xi_1)$. However, at $t=t_w$, the granular temperature is measured and the driving modified to $\xi$. The new value would lead to a warm temperature, $T_s(\xi)=T(t_w)$. The question mark illustrates the two possible Kovacs hump, a positive hump with a maximum (normal behavior, solid line) and a negative anomalous hump (dot-dashed line).  (b) Experimental observation of the Kovacs effect in an optically trapped particle. Top: Temperature protocol of the effective bath during the cooling protocol from $T_H$ to $T_C$ (dashed light blue line), cooling protocol from $T_H$ to $T_W$ (dash-dotted silver line), and Kovacs protocol (red solid line). Bottom panel: Time evolution of average potential energy during the protocols described in (a). Discrete symbols (light blue triangles, silver diamonds, and red circles) are used for experimental data, whereas lines are used for theory curves. The horizontal dotted line shows the equilibrium value at the warm temperature. The vertical dotted line marks the time $t_w$ at which the temperature is changed to $T_w$. The plot shows an anomalous (downward) ``Kovacs hump." \textit{Source:} Reprinted with permission from~\cite{prados2014kovacs} and~\cite{militaru2021kovacs}.}
    \label{fig:kovacs}
\end{figure}

Our discussion of the Mpemba effect has considered situations where a single sudden temperature step, or quench, is applied to a system. Although the relaxation to equilibrium is monotonic in time for a given initial temperature (Fig.~\ref{fig:kovacs}a), the overall time to relax to equilibrium is---surprisingly---not necessarily a monotonic function of the temperature difference between initial and equilibrium temperatures; see Fig.~\ref{fig:strongMpemba}. A natural generalization is to ask whether more complicated temperature protocols also lead to new effects. Indeed, at roughly the same time that Mpemba and Osborne were studying water, A. J. Kovacs~\cite{kovacs1964transition} published an influential study of polymer relaxation showing that a temperature protocol with \textit{two} temperature steps could lead to nonmonotonic relaxation of the specific volume as a function of \textit{time}; see Fig.~\ref{fig:kovacs}b.  In the original work, the response led to a positive overshoot that has become known as the ``Kovacs hump." Later, Prados and Trizac \cite{prados2014kovacs} showed in granular gases that there also can exist an ``anomalous" downward Kovacs hump; see  Fig.~\ref{fig:kovacs}a for a sketch of both types of Kovacs response, and see  Fig.~\ref{fig:kovacs}b for an anomalous hump observed in an optically trapped colloidal-particle experiment).

In glassy materials such as glucose or polyvinyl acetate, Kovacs considered a protocol where the system is prepared in equilibrium at temperature $T_\mathrm{hot}$, quenched for a short time at a temperature $T_\mathrm{cold}$ and then quenched for a long time at temperature $T_\mathrm{warm}$. The switch from cold to warm temperatures is done when the measured response equals the equilibrium value for $T_\mathrm{warm}$.\footnote{
In the original Kovacs protocol, the time spent at $T_\mathrm{cold}$ was only approximately the time for the response to equal its value at  $T_\mathrm{warm}$. The observed phenomena are qualitatively similar but depend on the exact dwell time. Modern studies simplify the analysis and interpretation by choosing the dwell time as the time for the response to cross the equilibrium value at $T_{\rm warm}$.}
Naively, if the system is close enough to equilibrium to be described by a single relaxation rate and if the switch to $T_\mathrm{warm}$ occurs when the time-dependent response crosses the equilibrium value at $T_\mathrm{warm}$, it should stay at that value. However, in systems with multiple degrees of freedom, the previous history---here, the first step---can leave other variables out of equilibrium. The result is an over- or undershoot of the response before a final relaxation to equilibrium. Measuring the response via the fractional volume, Kovacs found overshoots (a single oscillation) in the relaxation for certain combinations of temperatures and times. With colleagues, he proposed an explanation in terms of the multiple rates of relaxation encountered in these highly viscous, glassy systems~\cite{kovacs1979isobaric}. A general explanation using the master-equation formalism \cite{van1992stochastic} was given in \cite{prados2010the,ruiz2014kovacs}, where it was shown that the normal Kovacs effect is produced in the linear-response regime.  

Since the original work of Kovacs, investigations have developed in ways reminiscent of the development of our understanding of the Mpemba effect and have established two important conclusions: first, nonmonotonic relaxation is a generic behavior of complex systems; second, even very simple systems can nonetheless display Kovacs effects. The overall picture arising from these studies emphasizes the important role of memory and multiple rates.

\textit{Observations in other systems.}  The original experiments by Kovacs were in polyvinyl acetate, a glassy polymer, and in glycerol, a small molecule that shows highly viscous, glassy behavior at low temperatures. It was then natural to look for the Kovacs effect in other complex materials and systems. For example, Mossa and Sciortino found and studied the Kovacs effect in a molecular-dynamics model of orthoterphenyl~\cite{mossa2004crossover}, a small molecule whose relaxation is qualitatively similar to that of glycerol, and Berthier and Bouchaud found the effect numerically in an Ising spin glass \cite{berthier2002geometrical}. Later, Chen et al.~applied the Kovacs protocol to a colloidal glass of nanospheres with temperature-dependent, attractive interactions~\cite{chen2023memory}. Using X-ray photon correlation spectroscopy (XPCS), they found nonmonotonic responses in quantities such as the elastic and viscous response functions.
 
If one allows for other driving variables, there are many more examples. Several investigations have used an applied force as the control parameter instead of temperature and measured a stress response. Systems where nonmonotonic stress relaxation was observed include crumpled Mylar sheets and elastic foams~\cite{lahini2017nonmonotonic},  the contact area of two solids pressed together~\cite{dillavou2018nonmonotonic}, rock salt~\cite{he2019nonmonotonic}, a simulation of an athermal jammed solid consisting of a binary mixture of soft harmonic particles~\cite{mandal2021memory}, and a globular protein stretched using magnetic tweezers~\cite{morgan2020glassy}. 

The Kovacs effect has also been observed in still other settings. Josserand et al.~observed overshoot in the relaxation of the level of a granular fluid (glass beads in a vertically vibrated cylinder) after a Kovacs protocol in the vertical acceleration~\cite{josserand2000memory}. Prados and Trizac showed for the first time the existence of the normal and anomalous Kovacs effect in granular gases \cite{prados2014kovacs,trizac2014memory};  Lasanta et al.~studied numerically a granular gas of rough spheres, finding not only unusually large, order-one Kovacs response but also responses with two and even three extrema during the relaxation~\cite{lasanta2019emergence}. Motivated by the results in granular gases, K\"ursten et al.~studied a 2D model of interacting particles inspired by studies of bacterial swarms~\cite{kursten2017giant}. They applied a temperature-step protocol and monitored the response of the orientational order of the particles.  Interestingly, Momp\'o et al. \cite{mompo2021memory} showed in a gas of viscoelastic granular particles a unified picture of the Mpemba and Kovacs effects based on the evolution of kurtosis; see Sec. \ref{sec:kinetic-framework}.

\textit{Kovacs effect in simpler settings.}  The above experiments and simulations showing the Kovacs effect were all in complex materials and systems that have many similar relaxation rates. In 2020, Peyrard and Garden showed analytically and numerically that a three-state Markov model described by a master equation displayed a Kovacs effect for the mean energy~\cite{peyrard2020memory}. Interestingly, the same model is also the simplest setting for observing the Mpemba effect~\cite{lu2017nonequilibrium}.  More recently, a study of Newton's law (single exponential) cooling with a delay also showed Kovacs-like (and Mpemba-like) relaxation effects that highlight the role of memory in these effects~\cite{santos2024mpemba}. 

A 2021 experiment by Militaru et al.~observed experimentally an anomalous Kovacs effect, also in mean energy response, for the relaxation of a nanoparticle trapped by optical tweezers in air~\cite{militaru2021kovacs}; see Fig. \ref{fig:kovacs}b). In contrast to previous observations of the Kovacs effect, there was just a single particle with a single relaxing mode. The pressure (near ambient) created a nearly critical damping level, where the natural oscillation frequency and relaxation rate were comparable. In the experiment, an effective temperature was manipulated by subjecting the bead to random fluctuating electrostatic forces~\cite{martinez2013effective}. The ensemble-averaged potential energy relaxation response showed a nonmonotonic response for nearly critical damping levels; shaded region in~\FIG{kovacs}b. Indeed, at the time of the switch to $T_\mathrm{warm}$, the average potential energy equalled its equilibrium value, but the average kinetic energy did not. The hump reflects the subsequent equilibration of both kinetic and potential energies. Interestingly, the effect vanishes when the vibration and damping rates are mismatched, i.e., in the limits of weak damping or strong overdamping.

Given the complexity of the materials where the Kovacs effect has been typically studied, it is not surprising that there have been many theoretical studies, e.g., Refs.~\cite{kovacs1979isobaric,bouchbinder2010nonequilibrium,prados2010the,godreche2022the}. A common feature is the role of multiple, comparable rates and of memory~\cite{patron2023non}. Indeed, the Kovacs effect has been considered as a simple example of the importance of memory effects (initial conditions and history) as a means of creating and engineering matter with desirable new properties~\cite{keim2019memory}. Now that we understand that the Kovacs effect can arise in systems with a few comparable rates, it is perhaps not surprising that it is so widespread in Nature and in theoretical models.

\textit{Application of the Kovacs effect.}  An ``inverse'' Kovacs effect observed in a dipolar glass has inspired a temperature protocol that alters the nonlinear optical response of a ferroelectric material so that nonlinear diffraction cancels diffractive broadening of a light beam propagating through the solid material~\cite{parravicini2012equalizing}. The ferroelectric material, lithium-enriched potassium-niobate-tantalate (KTN:Li), is subjected to a temperature protocol resembling that used to produce the Kovacs effect. The result is a state whose desired nonlinear optical properties are set by quenched disorder, with material properties stable over time. Successfully designed materials support ``scale-free propagation," where light beams propagate through the material without broadening. Importantly, such propagation can occur independent of beam size and intensity, which is often not the case when material properties depend on nonlinear response. Although the connections to the Kovacs effect are somewhat loose, the result is a new way to design materials with useful optical properties.

\subsection{Asymmetric heating and cooling}
\label{sec:asymmetric}
In 2020, Lapolla and Godec argued that heating in a wide range of systems should occur generically faster than cooling~\cite{lapolla2020faster}. In particular, they establish a protocol where the relaxation time of a given system from a cold and a hot temperatures to an intermediate warm one, respectively, are compared. Naively, the claim is somewhat surprising, because linear nonequilibrium thermodynamics suggests that relaxation times should be equal~\cite{degroot1984nonequilibrium}. For example, in a macroscopic system, finite-size bodies in a uniform heat bath of temperature $T_b$ will, in the long-time limit, decay as to equilibrium following Newton's law, 
\begin{align}
	\dv{T(t)}{t} = - \frac{1}{\tau} \left[T(t) - T_b \right] \,,
\label{eq:macroTempRelax}
\end{align}
where the timescale $\tau$ is proportional to the heat capacity of the body. In Eq.~\ref{eq:macroTempRelax}, the relaxation rate $\tau^{-1}$ is the same for initial temperatures $T(0)$ that are greater or lesser than $T_b$, and the relaxation time would be equal when $T_h(0) - T_b = T_b - T_c(0)$. 

For the described protocol, the situation is similar in a mesoscopic system whose state $x$ is described by a probability density $p(x,t)$. At long times ($t \gg |\lambda_3|^{-1}$) and for initial temperatures $T_0$ very close to the bath temperature $T_b$, the eigenfunction expansion for $p(x,t)$ given in Eq.~\ref{eq:eigenfunc_continuous} reduces to
\begin{align}
	p(x,t) \approx \pi(x,T_b) + (T_0-T_b) \, a \, e^{\lambda_2(T_b)t}v_2(x,T_b) \,,
\label{eq:eigenfunction_a2continuous_neareq}
\end{align}
where we have Taylor expanded $a_2(T_0,T_b)$ about $T_b$, where $a_2(T_0,T_0)=0$, since the system is in equilibrium.  Here, $a \equiv \left[ \partial_T \left. a_2(T,T_b) \right|_{T=T_b} \right]$, and we again expect equal relaxation times for $T_0$ symmetrically above or below $T_b$. 

Although the above discussion suggests that heating and cooling times are identical for initial temperatures that are infinitesimally different from the bath, relaxation times might differ when the initial states are far from equilibrium. Mathematically, the function $a_2(T_0,T_b)$ need not be odd about $T_b$, even though it must vanish at $T_0=T_b$. Following this line of thought, Lapolla and Godec used a nonequilibrium free energy difference $F(t)$ to quantify the distance between the initial state and final equilibrium. In particular, if we consider three temperatures $T_c$, $T_w$, and $T_h$ (cold, warm, hot) and let $i \in \{ h,c \}$, then~\cite[Sec. 5.2]{peliti2021stochastic} 
\begin{align}
	F_i (t) 
		& = \left\langle \Delta E_i(t) \right\rangle - T_w \left\langle \Delta S_i(t) \right\rangle  = T_w \int_{-\infty}^\infty \dd{x} p_i(x,t) 
			\ln  \left( \frac{p_i(x,t)}{\pi(x,T_w)} \right) \nonumber \\
		& = T_w D_{\rm KL}\left[ p_i(x,t), \pi(x,T_w)\right] \ge 0
\label{eq:noneq_free_energy}
\end{align}
measures the nonequilibrium free energy difference between the relaxing hot or cold state $p_i(x,t)$ and the (warm) equilibrium bath state characterized by $\pi(x,T_w)$, which is related to the Kullback-Leibler divergence $D_{\rm KL}$ and defines a good way to compare two different states of the system by means of its probability distributions \cite{jordan1997free}. 

The ensemble averages $\langle \Delta E_i(t) \rangle$ and $\langle \Delta S_i(t) \rangle$ are the mean differences of energy and entropy between the state $p_i(x,t)$ and the equilibrium at $T_w$.  At time zero, the hot and cold temperatures $T_h$ and $T_c$ are chosen so that they are ``equidistant" in the relative entropy measure, namely, $F_{\rm h}(0) = F_{\rm c}(0)$.  Under these conditions and with a potential $U(x)$ having a single quadratic minimum, the time to relax to $T_w$ starting from $T_c$ is always shorter than the time to relax from $T_h$.  Consistent with the above discussion of linear response, the difference in relaxation times goes to zero as $\{T_h, T_c \} \to T_w$. The systems studied theoretically in Ref.~\cite{lapolla2020faster} included ones with multidimensional states $x$, such as a chain of $N+1$ beads connected by harmonic springs and $N$ hard-core Brownian point particles diffusing in a tilted, one-dimensional box. In these more complicated examples, the effect was studied using variables that project the full state down to a low-dimensional quantity, such as the end-to-end distance for a chain~\cite{lapolla2019manifestations}.

\begin{figure}
    \centering
    \includegraphics[width=5.0in]{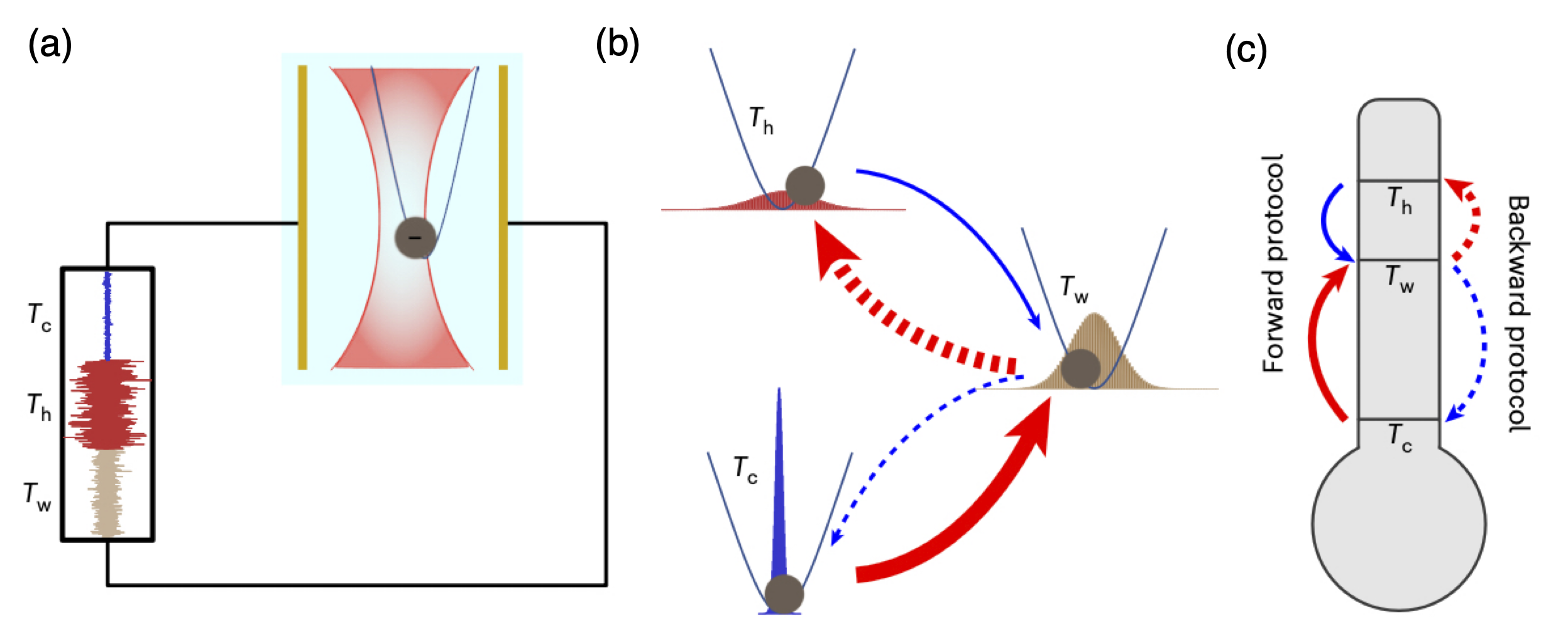}
    \caption{Heating is faster than cooling for equidistant quenches. (a) Schematic of experimental setup: A bead in an optical trap with harmonic potential and stochastic electrical forces whose strength quantifies the effective temperature of the bead. (b) Evolution of Gaussian position distributions after quenches from $T_h$ or $T_c$ to $T_w$ (forward-protocol) or vice versa (backward protocol). Thicker lines denote faster relaxation pathways. (c) Schematic of forward and reverse protocols. \textit{Source:} Reprinted with permission from~\cite{ibanez2024heating}.}
    \label{fig:asymmetric}
\end{figure}

\begin{figure}
    \centering
    \includegraphics[width=1.0\columnwidth]{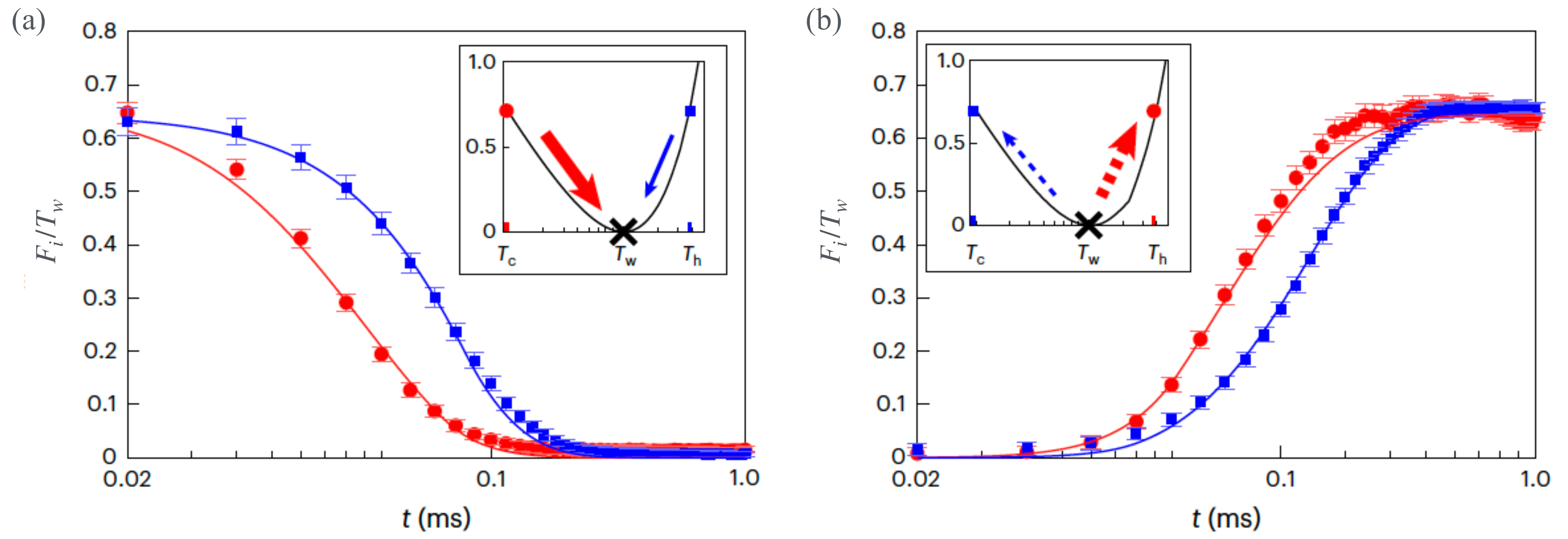}
    \caption{Heating is faster than cooling for equidistant quenches.  Red arrows represent heating whereas blue ones cooling.  Solid lines correspond to the theoretical predictions.  Solid and dashed arrows stand for the forward and backward protocols, respectively, and thicker lines indicate the fastest evolution. (a) and (b) show time evolution of the generalized excess free energy for equidistant quenches, (a) are results for the forward protocol and (b) for the backward protocol. Insets represent the initial value of the equidistant relative entropy 
 as a function of the temperature. \textit{Source:} Adapted with permission from~\cite{ibanez2024heating}. 
 }
    \label{fig:asymmTE}
\end{figure}
Recently, Iba\~nez et al. showed experimentally that the expected asymmetry is present in a simple system consisting of a colloidal particle in water in an optical trap that created a harmonic potential;  see~\FIG{asymmetric}a. The colloidal system is modeled by the Langevin equation in the overdamped limit, Eq.~\ref{eq:langevin-colloid}, with the potential $U(x)=kx^2/2$, where $k$ is the stiffness of the optical trap. The effective temperature of the particle was varied by imposing fluctuating voltages on two electrodes and, hence, fluctuating electroosmotic forces on the particle. The effective temperature is $T_{\rm eff} = T+\sigma^2/(2\gamma)$, where $T$ is the room temperature, $\sigma^2 = q^2V_0^2/d^2$, with $V_0^2$ the variance of the noisy voltage output, $q$ the charge and $d$ the microparticle diameter.  Not only did the authors observe that heating was faster than cooling for equidistant quenches from $T_h$ and $T_c$ to $T_w$, they also observed the same asymmetry for the reverse process and for a simpler system comparing quenches from $T_c$ to $T_h$ with reverse quenches from $T_h$ to $T_c$. 

Thus, in this simple colloidal system, heating and cooling always followed distinct kinetic pathways. In order to explain these observations, the authors developed a ``thermal kinematics''~\cite{ibanez2024heating} based on geometric information theory \cite{ito2020stochastic}.

A natural question to ask is how universal is this asymmetry. Already, Lapolla and Godec noted counterexamples when the potential had metastable states where the relaxation required thermal hops over barriers~\cite{lapolla2020faster}. Then Meibohm found counterexamples for single-well potentials with non-quadratic forms $U(x) = |x|^\alpha$, for $\alpha \gtrsim 3$~\cite{meibohm2021relaxation}. A small additional quadratic contribution to the potential does not alter the result, meaning that there can be counterexamples for wells with strong-enough anharmonic shapes. Van Vu and Hasegawa examined relaxation in classical discrete-state systems~\cite{van2021toward}. They confirmed the asymmetry for two-state systems but found counterexamples in systems with more than two states. In particular, they found and demonstrated that in more than two-level quantum systems, for a three-temperature protocol with a warm intermediate temperature and for some fitting parameters, the asymmetry can be reversed, and the cooling process can be faster than the heating one. Even for degenerate two-state systems, the asymmetry could be reversed beyond a critical threshold for the energy gap. Also, Manikandan showed that in discrete-state quantum systems, asymmetries are not generic and can have either sign, depending on various system parameters~\cite{manikandan2021equidistant}. 

\begin{figure}
    \centering
    \includegraphics[width=0.9\columnwidth]{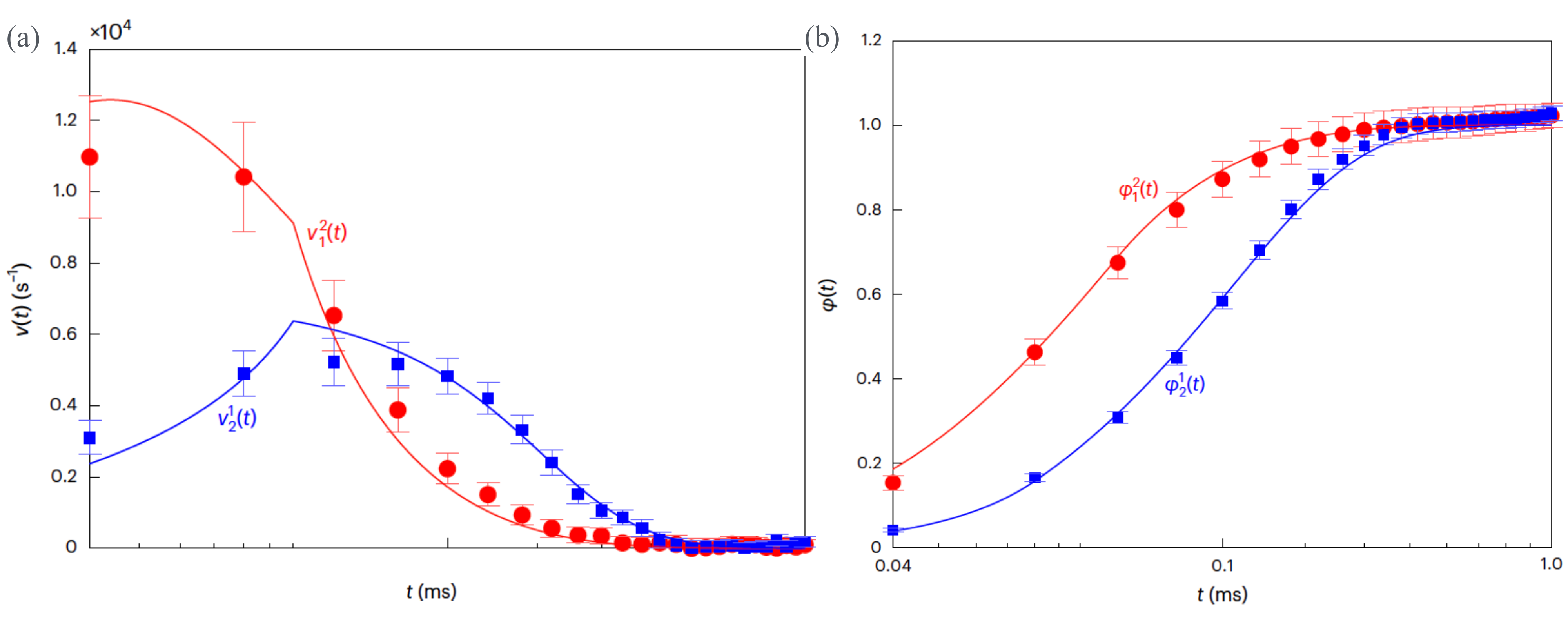}
    \caption{Thermal kinematics of heating and cooling between any pair of temperatures. Red circles denote heating, blue squares cooling. Solid lines are theoretical predictions. (a) Instantaneous statistical velocity as a function of time. (b) Degree of completion as a function of time. \textit{Source:} Adapted with permission from~\cite{ibanez2024heating}.
    }
    \label{fig:asymmTT}
\end{figure}
In order to unravel the question, two more protocols were designed in~\cite{ibanez2024heating}. First, there is the ``backward process,'' defined with respect to the direct process, where there is a free evolution of the system from a common warm temperature to first a hot and then a cold temperature.  The hot and cold temperatures are chosen to be ``equidistant'' from the warm temperature, as measured by the relative entropy.  Note that, in the backward process, $\lambda_2(T_c) \neq \lambda_2 (T_h)$ and $a_2(T_b,T_c) \neq a_2(T_b,T_c)$; see Fig.~\ref{fig:asymmTE}b. That contrasts with the ``equidistant'' case, where the bath temperature is identical and thus $\lambda_2(T_w) = \lambda_2 (T_w)$. Second, because the heating and cooling protocols end up at two different temperatures,  $\lambda_2(T_c) \neq \lambda_2 (T_h)$ and $a_2(T_c,T_h) \neq a_2(T_h,T_c)$; see Fig.~\ref{fig:asymmTT}. We emphasize that, in the former protocol, a different measure must be used, since the relative entropy is not symmetric and is not a metric. Instead, one can use the Fisher information $I(t)=\int \dd x\, (\partial_t P(x,t))^2/P(x,t)$, which is related to the Kullback-Leibler entropy by $D_{\mathrm{KL}}\left[P(x,t+\dd t), P(x,t)\right] = I(t) \dd t^2 + \mathcal{O}(\dd t^4)$. From the previous equation, one can define a ``statistical'' velocity~\cite{ito2020stochastic} at a given time $t$ as $v(t):= \sqrt{I(t)}$; see Fig.~\ref{fig:asymmTT}a. By analogy to classical kinematics, a statistical length traced by  $P(x,\tau )$ is computed as ${\mathcal{L}}(t)=\int\nolimits_{0}^{t}\dd \tau \,{v}(\tau)$ \cite{ito2020stochastic}. In turn, the total distance traveled between the initial and the final states is defined to be ${\mathcal{L}}(\infty )$. From that quantity, one can define the completion degree ${\varphi }(t)\equiv {\mathcal{L}}(t)/{\mathcal{L}}(\infty )$, which quantifies the thermal relaxation in the present framework and goes from $0$ to $1$; see Fig.~\ref{fig:asymmTT}b.

There have been generalizations of the original work, too. Dieball et al. studied thermal relaxation in driven systems where the relaxation is to a nonequilibrium steady state (NESS)~\cite{dieball2023asymmetric}. Focusing on the example of a 2D harmonically confined Rouse polymer with hydrodynamic interactions and internal friction that is driven out of equilibrium by a force. The same type of asymmetry noted in the equilibrium case persists in the driven case, for both conservative and non-conservative forcing. 

\begin{figure}
    \centering
    \includegraphics[width=0.9\columnwidth]{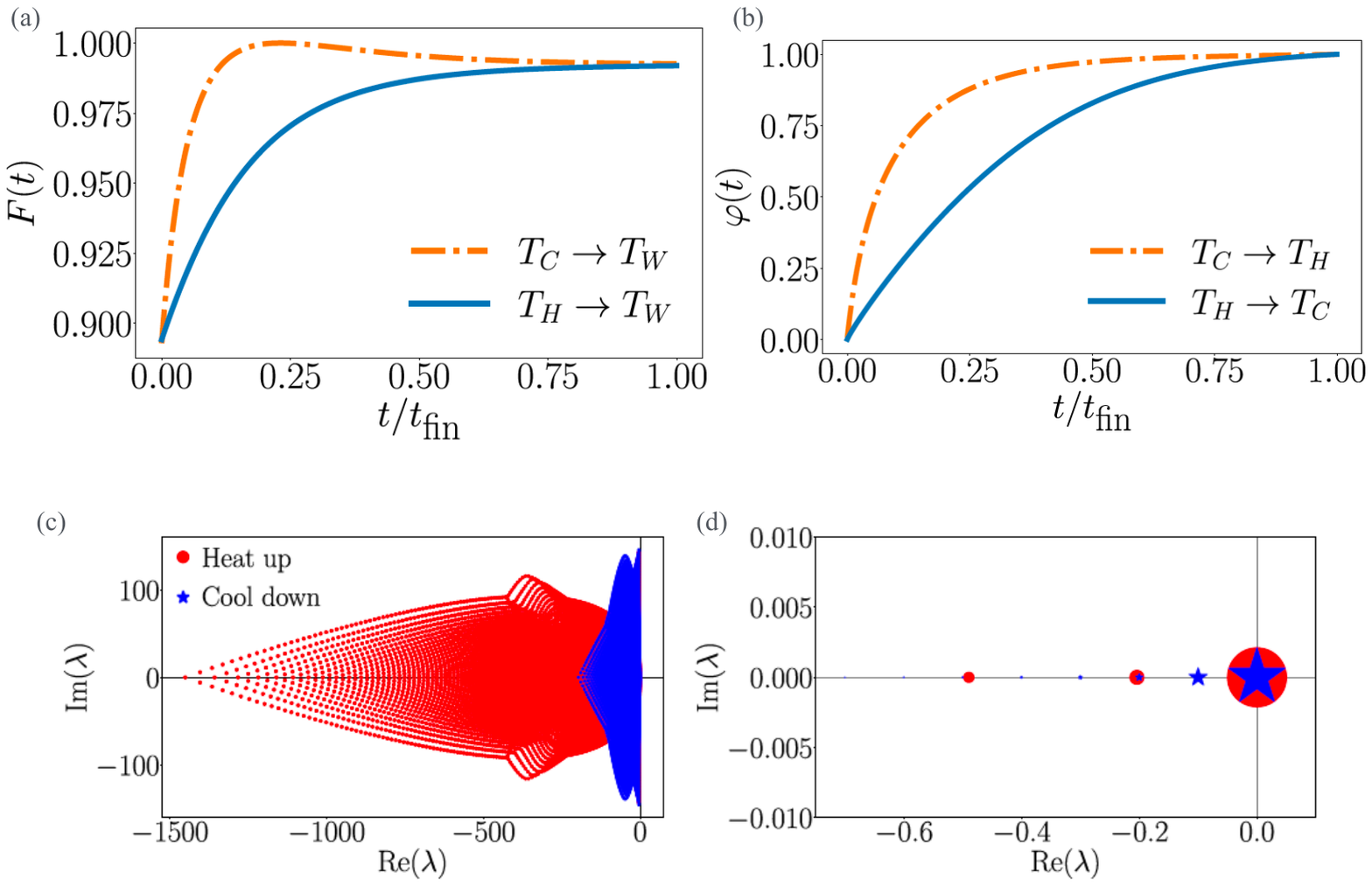}
    \caption{Heating is faster than cooling in a quantum Brownian particle. (a) Simulation of the evolution of the fidelity with respect to the thermal state $T_w$ for equidistant initial states. (b) Degree of completion computed for the heating and cooling processes in the two-temperature protocol. For both panels the $x$ axes represent a rescaled time with respect the total time the system takes to reach the final state. (c) Eigenvalues of the operator $\mathcal{L}$ from ~\EQ{Lind}. (d) First eigenvalues of (c); marker size is proportional to the overlap with the thermal state at the opposite temperature. \textit{Source:} Reprinted with permission from~\cite{tejero2024asymmetries}.}
    \label{fig:asymmQ}
\end{figure}

Results similar to the ones found in~\cite{ibanez2024heating} can be generalized to quantum systems~\cite{tejero2024asymmetries}. In particular, the authors study a quantum two-level system, the quantum harmonic oscillator, and a trapped quantum Brownian particle. They extend the thermal kinematics analysis to the quantum realm and offer a general explanation based on the spectral decomposition of the Liouvillian and the spectral gap of reciprocal processes of cooling and heating. The systems are described by the formalism of~\SEC{quantum-framework}. The studied magnitudes are the Bures distance $\left(D_{B} [\rho_1,\rho_2]\right)^2 \equiv 2\left(1 -  F[\rho_1,\rho_2 ]\right)$, where $F[\rho_1,\rho_2]\equiv \Tr\sqrt{\sqrt{\rho_1}\rho_2 \sqrt{\rho_1}}$ is the \emph{fidelity}; see Fig.~\ref{fig:asymmQ}a. Note that fidelity, like the Kullback-Leibler divergence, is not a metric but a symmetric measure~\cite{nielsen2010quantum}. In analogy with the classical case, in the context of thermal relaxation, an infinitesimal statistical line element may be defined as $\left(D_{B} \left[\rho(t), \rho(t+\dd t) \right]\right)^2 = \mathcal{I}_{\rm Q} [\rho(t)]  \dd t^2/4 + \mathcal{O}(\dd t^4)$ being $\mathcal{I}_{\rm Q}$ the quantum Fisher information \cite{tejero2024asymmetries,bettmann2024information} $ \mathcal{I}_{\rm Q} [\rho(t)]\equiv \text{Tr} \left[L_t^2 \rho(t) \right]$; see Fig.~\ref{fig:asymmQ}b, where $L_t$ is the logarithmic time-derivative operator defined by $\dot{ \rho}(t):= \left(L_t \rho(t) + \rho(t) L_t \right)/2$. Interestingly, by studying the Liouvillian spectrum of the system, the authors
find that the eigenvalues spread towards the negative real line as temperature increases. This indicates that, for thermal baths at higher temperatures, there are more fast-decaying modes, making the evolution faster; see Fig.~\ref{fig:asymmQ}c. Additionally, the overlap between the initial state and the fast decaying modes is larger for the heating-up processes; see Fig.~\ref{fig:asymmQ}d.

In a different kind of generalization, asymmetries are interpreted in relaxation through the lens of information geometry~\cite{bravetti2024asymmetric}. In particular, they propose to characterize the asymmetry using a differential-geometry formalism for endoreversible processes. First, they define generalized coordinate and thermodynamic potentials, and then measure differences between probability densities via the Bregman divergence, which is analogous to the Kullback-Leibler divergence. Defining the dissipation as the length of the trajectory followed for the system in a manifold, they propose a criterion using the Amari-Chentsov tensor for predicting the appearance of a heating-cooling  asymmetry in endoreversible evolution   of thermodynamics systems \cite{hoffmann2008introduction} and show several examples. Endoreversible thermodynamics deals with irreversibilities in the evolution of a system and considers each intermediate state as an equilibrium one, with irreversibility manifested only at the boundary between the bath and the system. In other words, an endoreversible system can be viewed as internally in equilibrium and interchanges energy with the environment in an irreversible way. In particular, Bravetti et al. show that there exists asymmetry, but not that heating is always faster than cooling. Sometimes, cooling can be faster than heating. 

Finally, it is worth comparing this asymmetry between equidistant temperature quenches and the Mpemba effect. The latter also highlights differences between relaxation times, but these are from two temperatures that are each higher (or lower) than the final equilibrium temperature. Nonetheless, asymmetries between the forward Mpemba effect (cooling) and the inverse Mpemba effect (heating) have been noted. In their experimental study of the inverse Mpemba effect, Kumar and Bechhoefer noted that the inverse effect involving heating was weaker than the forward Mpemba effect involving cooling~\cite{kumar2022inverse}. ``Weaker" here means harder to measure, in the sense that more trials are required for similar statistical noise levels in measurements. This traces back to the differing ratios of relaxation rates, $\lambda_3/\lambda_2$, which was about $16$ for the forward effect but only about $4$ for the inverse effect. The difference in eigenvalue ratios in turn traces back to the dominant influence of the energy landscape at low temperatures and entropic effects at high temperatures 
(corresponding to the two terms in the expression for the free energy). A quench to low temperatures then involves modes that correspond to barrier hopping and have exponentially smaller $\lambda_2$ and thus high values of $\lambda_3/\lambda_2$. For quenches to high temperatures, the motion is close to pure diffusion, and the eigenvalue ratio is smaller. Thus, there is also an asymmetry between forward and inverse Mpemba effects, but its origin is somewhat different.

\subsection{Eigenvalue crossing}
\label{sec:eigenvalue-crossing}
\begin{figure}
    \centering
    \includegraphics[width=0.75\linewidth]{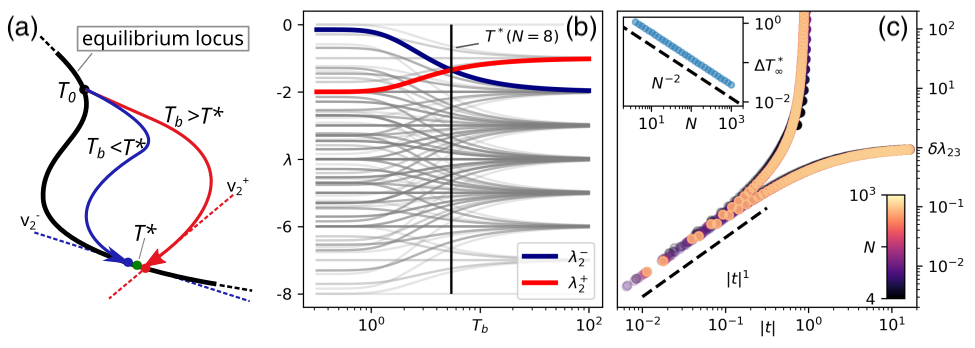}
    \caption{(a) A small change in the bath temperature $T_b$ drastically changes the approach to equilibrium direction in the presence of a level crossing like the one observed in the four-wells setup shown in panel (b), representing the spectrum of an $N=8$ Ising antiferromagnet at zero magnetic field with Glauber dynamics \cite{glauber1963time}.
    Highlighted in blue and red are the first two dominant eigenvalues, which are relevant for the dynamics. They cross at  $T^*(N=8)\sim5.45$.
    (c) Eigenvalue difference $\delta\lambda_{23}$ as a function of the {reduced} temperature $t=(T-T^*)/T^*$, exhibiting an excellent finite-size collapse $\propto|t|^1$ around the crossing temperature.
    Inset: the distance from the asymptotic crossing temperature $T^*_\infty=2/\mathrm{arctanh }(1/3)$ decays quadratically. \textit{Source:} Adapted with permission from~\cite{teza2023eigenvalue}.}
    \label{fig:ev_crossing_1}
\end{figure}

Another relaxation phenomenon that is deeply intertwined with the Mpemba effect is  \textit{eigenvalue crossing} \cite{teza2023eigenvalue}.
It consists in a crossing of the eigenvalues of the Markovian evolution operator with respect to a change of a system's parameter, which can be either an external parameter  (e.g., an external driving or field) or the bath temperature $T_b$ regulating the thermal quench.
If at some temperature $T^*$ there is a crossing between the second and third eigenvalues, it induces a singularity in the long-time-limit approach to equilibrium: the system relaxes along two distinct directions when quenched to $T=T^*+\delta$ and $T^*-\delta$, no matter how small $\delta$ is. This is schematically depicted in~\FIG{ev_crossing_1}a.

Eigenvalue crossing can be viewed as a phase transition in the relaxation dynamics, which occurs in the long-time limit $t\to\infty$, in analogy with an equilibrium phase transition, which can be observed in the thermodynamic limit $N\to\infty$.
From the equation regulating the last stages of the evolution in a thermal quench, \EQ{longtime_a2}, one sees how the timescale is regulated by an exponential decay with characteristic time $\propto 1/|\lambda_2|$.
To highlight this, we can formally rearrange the dynamics equation, obtaining
\begin{align}
\label{eq:app_dir_p}
    e^{-\lambda_2t}\partial_t\vec{p}(t) = a_2\lambda_2\vec{v}_2 + \sum_{n>2}a_n\lambda_n e^{-\Delta\lambda_{2,n} t}\vec{v}_n,
\end{align}
where we introduced the eigenvalue gaps $\Delta\lambda_{2,n} = \lambda_2-\lambda_n\geq0$.
If $\lambda_2$ is not degenerate and if $a_2\neq 0$, the final stage of the relaxation is in the direction of $\vec{v}_2$ and changes continuously with $T_b$. However, $\vec{v}_2(T_b)$ can abruptly change at some temperature $T^*$ if at this temperature there is an eigenvalue crossing, namely $\lambda_2(T^*) = \lambda_3(T^*)$, as in~\FIG{ev_crossing_1}b.

Referring to the eigenvalues and eigenvectors that dominate the long-time dynamics before and after $T^*$ as $\lambda_2^{\pm}$ and $\vec{v}_2^{\pm}$ (Fig. \ref{fig:ev_crossing_1}b), one can characterize the singular behavior in the long-time limit of~\EQ{app_dir_p} as
\begin{align}
\vec{v}_2 = 
    \begin{cases}
    \vec{v}^{-}_2 & T_b<T^* \\
    a_2^-\vec{v}^{-}_2 +  a_2^+\vec{v}^{+}_2 & T_b=T^* \\
    \vec{v}^{+}_2, & T_b>T^* \\
    \end{cases}
\end{align}
where $a_2^{\pm}$ are coefficients determined by the initial conditions.
This singularity is not pronounced in the equilibrium state, but rather in the relaxation towards equilibrium, and this is why eigenvalue crossing can be linked with anomalous phenomena arising in the relaxation process, such as the Mpemba effect.

\paragraph{Connection of the eigenvalue crossing with the Mpemba effect}
While eigenvalue crossing can be observed in very small systems (e.g., in Ref.~\cite{teza2023eigenvalue}, where the example of a 4-state Markovian system is presented), it can also be observed in complex many-body interacting systems as Ising antiferromagnets in zero external magnetic fields $H=0$.
For a 1D periodic chain of $N$ spins with Glauber dynamics and $H=0$, there is an exact solution of the spectrum $\lambda_k(T_b)$ (with $k=1,\dots,2^N$)  \cite{felderhof1971spin}.
This solution allows visualization (see \FIG{ev_crossing_1}b for the case $N=8$) of how the spectrum evolves as a function of the bath temperature $T_b$, highlighting a high number of crossing among the eigenvalues.
We highlighted in red and blue the relevant eigenvalues $\lambda_2^{\pm}$, which are those that are non-orthogonal to all Boltzmann distributions. The higher eigenvalues are orthogonal to all Boltzmann distributions. Thus, equilibrium initial conditions have zero projection on these and are hence irrelevant to our discussion.
Their analytic expressions for finite $N$ are given by 
\begin{align}
    \lambda_2^-&=-2+2\tanh (2J/k_B T_b) \cos \left( \frac{N-1}{N} \pi \right), 
    \\
    \lambda_2^+&=-1+\tanh (2J/k_B T_b),
\end{align}
where $J<0$ is the antiferromagnetic pairwise spin interaction for nearest-neighboring spins on the chain.
The analytic expressions of the eigenvalues enabled a formally study of the phenomenon in the thermodynamic limit: imposing $\lambda_2^-=\lambda_2^+$, one finds that the crossing survives the $N\to\infty$ limit, asymptotically approaching $T^*_{\infty}=2/\mathrm{arctanh }(1/3)$ (Fig. \ref{fig:ev_crossing_1}b).

The presence of the above eigenvalue crossing at $H=0$ is directly connected \cite{teza2023relaxation} with the emergence of a Mpmeba effect in antiferromagnets for $H \ll 1$ above a certain temperature (see Fig. \ref{fig:ising_1dAF_boundary_sketch}b).
Indeed, such a crossing dramatically changes the coefficient of the second eigenvector in the initial condition and hence affects its monotonicity through which the effect is defined.
The crossing becomes progressively sharper in the antiferromagnetic system as the magnetic field $H$ goes to zero.  
This feature was observed transversely across different Ising antiferromagnets systems and different dynamics, for which the effect does not seem to appear for any value of the two external parameters---bath temperature $T_b$ and magnetic field $H$---but rather in a specific domain (e.g., compare the phase diagrams for a mean-field model \cite{klich2018solution} and a 1D model subject to boundary coupling \cite{teza2023relaxation}).

\begin{figure}
    \centering
    \includegraphics[width=0.75\linewidth]{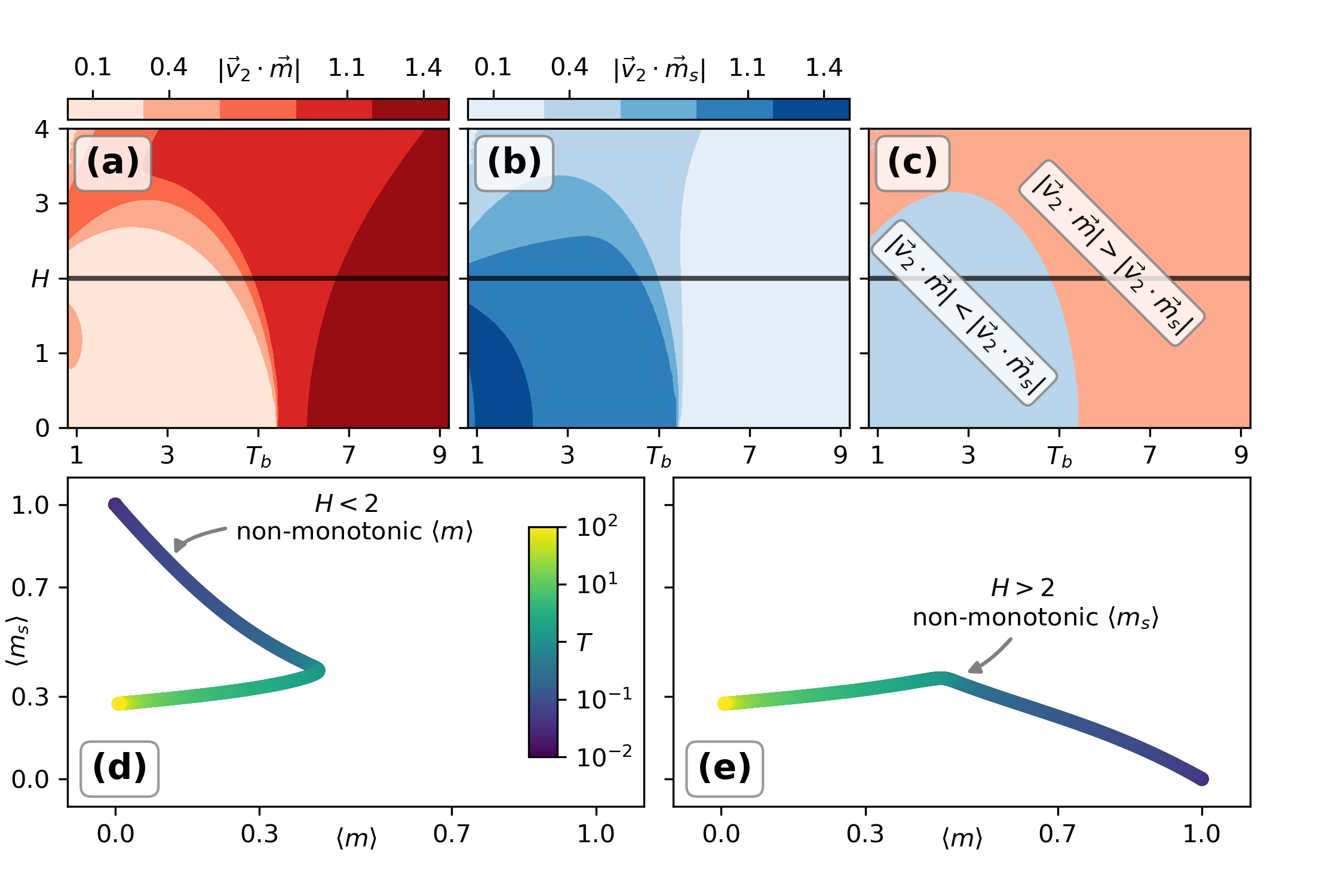}
    \caption{Projection of the slowest relaxation vector $\vec{v}_2$ on the magnetization (a) and staggered magnetization (b) for an $N=8$ 1D antiferromagnetic spin chain; see the main text for the specifics of the model.
    (c) Highlight of the areas in which the projection of $\vec{v}_2$ is larger along the magnetization (red area) and staggered magnetization (blue area).
    (d,e) Equilibrium loci showing the values of magnetization and staggered magnetization of the chain in equilibrium at different temperatures $T_b$ below ($H=1.95$) and above ($H=2.05$) the critical value of the magnetization that determines the phase transition in the groundstate of the Ising chain. \textit{Source:} Reprinted with permission from~\cite{teza2023eigenvalue}.}
    \label{fig:v2_vs_m_ms}
\end{figure}

To better understand the connection between these two effects, an important observation is related to the fact that the eigenvectors associated with the crossing eigenvalues carry a physical meaning: $\vec{v}_2$ is parallel to the staggered magnetization ($\vec{m}_{\rm s}$) before the crossing, where no Mpemba effect is observed, while it is parallel to the magnetization vector ($\vec{m}_{\rm s}$) after the crossing, where a strong inverse Mpemba effect can be observed~\cite{teza2023eigenvalue}.
For small magnetic fields (such that the antiferromagnetic interaction dominates), the magnetization of an Ising antiferromagnet at equilibrium is a nonmonotonic observable with respect to the temperature, while the staggered magnetization at equilibrium, on the other hand, is monotonic with respect to the temperature; see~\FIG{v2_vs_m_ms}d.
In Figs.~\ref{fig:v2_vs_m_ms}a and \ref{fig:v2_vs_m_ms}b,  the projection of $\vec{v}_2$ on $\vec{m}$ and $\vec{m}_{\rm s}$ respectively are plotted, highlighting the magnitude of the projections, while in Fig. \ref{fig:v2_vs_m_ms}c we color in red the area in which $|\vec{v}_2\cdot\vec{m}|>|\vec{v}_2\cdot\vec{m}_{\rm s}|$ and blue the one in which $|\vec{v}_2\cdot\vec{m}|>|\vec{v}_2\cdot\vec{m}_{\rm s}|$.
This explains geometrically why, having the slowest relaxation occurring along the magnetization for $H<2$, is a prerequisite in order to observe a strong Mpemba effect, and consequently why there is no effect before the crossing.

For magnetic fields that are large enough to overcome the antiferromagnetic ordering, the opposite holds: the staggered magnetization becomes nonmonotonic while the magnetization is now monotonic; see Fig. \ref{fig:v2_vs_m_ms}e.
Indeed, strong Mpemba effect can be observed in this case only in the blue area above $H>2$; see Refs. \cite{teza2023relaxation,klich2019mpemba}.
This demonstrates clearly that the Mpemba effect is a phenomenon intrinsically related to the \textit{dynamics}, as discussed in Sec.~\ref{sec:dynamics}: indeed, the geometric shortcut that is enabled by the nonmonotonicity of an observable as in \FIG{v2_vs_m_ms} (see also Fig. \ref{fig:relax_path} in the case of a weakly coupled system) is not a two-way road, as the spectrum of the rate matrix explicitly depends on $T_b$ and can therefore be dramatically different in the two cases.

\subsection{Relaxation speedup}
\label{sec:relaxation-speedup}
The following sections discuss two methods for accelerating relaxation processes. These methods use the time evolution of physical observables and the complex internal structures of physical systems near a phase transition. The first approach, outlined in~\SEC{speed-up-equil-phys-magnitudes}, shifts the focus from the evolution of the probability vector (or density) to the evolution of physical observables. This exploits the relationship between two formulations of the dynamics, one in which vectors carry the time dependence and the other in which time dependence is expressed through observables. A quantum-mechanical analogy of this is switching from a formulation with time-dependent states to one with time-dependent observables. The second approach, presented in~\SEC{internal-structure-speed-up}, leverages the complex internal structure near a phase transition to reduce the time required to reach a target state. We illustrate both approaches using 1D and 2D models of Ising spins with antiferromagnetic interactions.

\subsubsection{Speedup 
equilibrium properties}
\label{sec:speed-up-equil-phys-magnitudes}
\begin{figure}[ht]
	\centering
	\includegraphics[width=0.7\columnwidth]{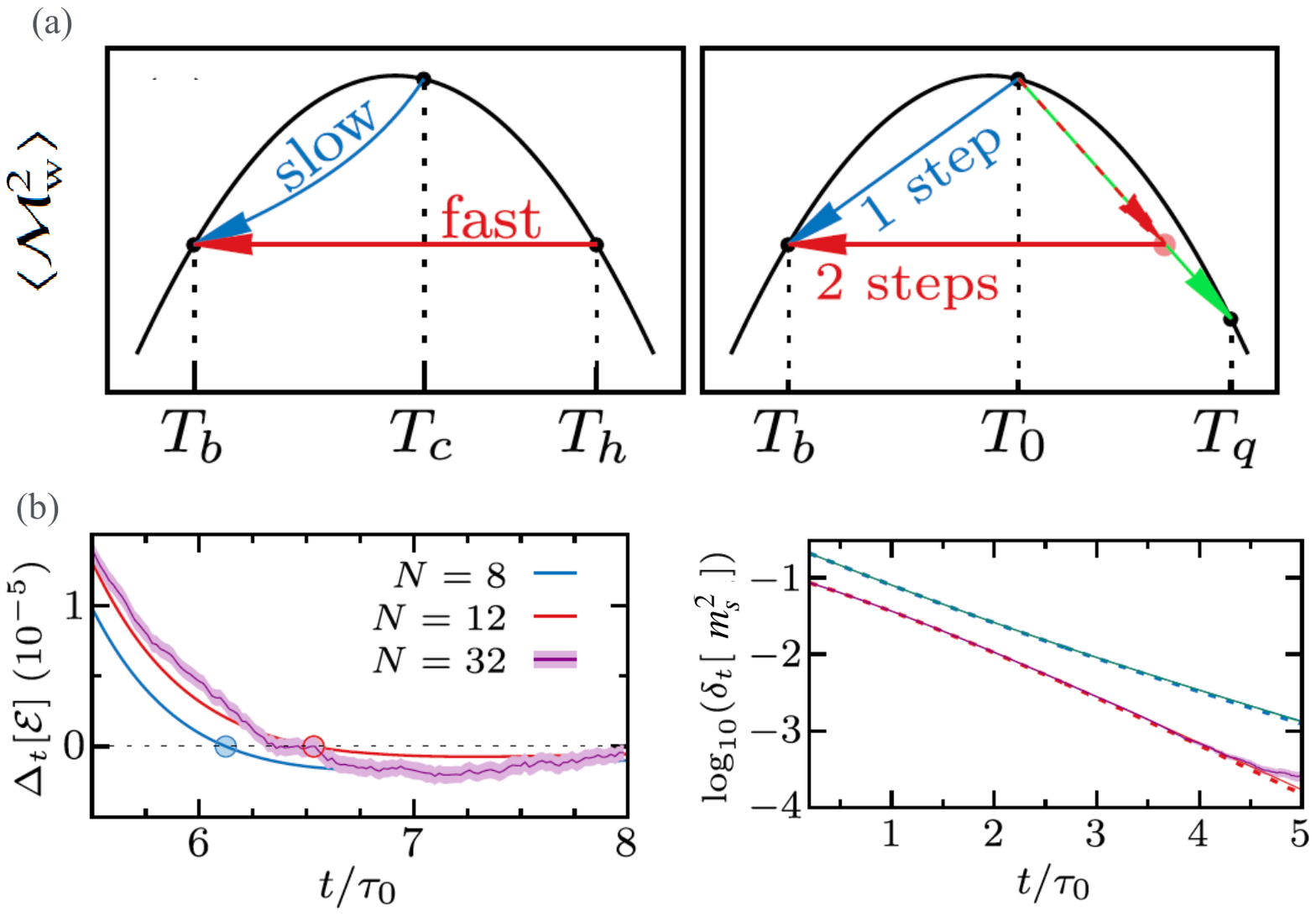}
	\caption{
    An observable whose equilibrium expected value exhibits a nonmonotonic dependence on temperature can be utilized to optimize its value near the maximum. Two illustrative examples are presented in the panels. Panel (a), on the left, depicts the Mpemba effect, while the right side illustrates the speedup achieved by briefly heating the system before cooling it. The red lines represent the hotter system, while the blue lines indicate the colder system. For a 1D Ising antiferromagnetic model, panel (b) left reproduces the Mpemba effect in the change of sing of $\Delta_t  \langle \varepsilon \rangle =(\langle \varepsilon (t) \rangle^{h}-\langle \varepsilon (t) \rangle ^{c})/N$ (with $\varepsilon$ defined in~\EQ{ham_AF} and $c$ and $h$ indicates cold and hot respectively). Panel (b) right shows the supercooling produced by the preheating strategy for $\delta_t  \langle m_s^2 \rangle =(\langle m_s^2 (t) \rangle-\langle m_s^2 (t) \rangle ^{b})/N$ the squared staggered magnetization, where the superindex $b$ indicates the bath. \textit{Source:} Reprinted with permission from~\cite{pemartin2024shortcuts}.
}
\label{fig:paradoxical_protocols}
\end{figure}
As we have seen throughout the review, the various Mpemba effects can reduce the time needed to cool down (or heat up) a system, as has been shown, for example, in Markovian systems that satisfy detailed balance (\SEC{markovian}). The analysis of the speed up relies on the diagonalization of the transition-rate matrix of the master~\EQ{master_equation} and finding cases in which the probability vector $\vec p(t)$ has no contribution from the eigenvector $\vec v_2$ (which describes the slowest evolving direction in probability space), thus forcing an exponential speed up towards equilibrium. In ~\cite{pemartin2024shortcuts}, the authors propose studying the weak version of the master~\EQ{master_equation}. The idea is similar in spirit, although not in detail, to the numerical method introduced in Sec.~\ref{SubSec:NumericalMonteCarlo}. We focus on how different observables evolve in time instead of focusing on the probability vector itself. Related to the eigenbasis $(\vec \pi,\vec v_2,\vec v_3,\ldots)$, there is a dual basis of observables $\left (\boldsymbol{1},\mathcal{O}^\mathrm{b}_2,\mathcal{O}^\mathrm{b}_3,\ldots\right)$. Analogously, this allows one to write the evolution of the expected value of an observable $\mathcal{A}$ as
\begin{align}
\label{eq:spectral_decomposition}
	\langle \mathcal{A} (t)\rangle
    &= \langle \mathcal{A} \rangle_{T_{\text{b}}} + \sum_{k\geq2} \alpha_k^{(t=0)} \beta_{k}^{\mathcal{A}}\, \mathrm{e}^{-|\lambda_k|t/\tau_0}\,,\\
	\beta_{k}^{\mathcal{A}} &= \langle \, {\mathcal O}_k^{\text{b}}\,|\,\mathcal{A}\,\rangle\,, \quad 
    \alpha_k^{(t=0)}\;=\;\sum_{\boldsymbol{x}\in\Omega}\,
	P_{\boldsymbol{x}}^{(t=0)}\,{\cal O}_k^{\text{b}}(\boldsymbol{x})\,.
\end{align}
Note that~\EQ{spectral_decomposition} is akin to~\EQ{m} but using an observable instead of a probability---much like the differences between the Schr\"odinger and Heisenberg representations in quantum mechanics~\cite{sakurai2020modern}. An observable shows an exponential speedup if it initially has no contribution from $\mathcal{O}^\mathrm{b}_2$, as that is the slowest evolving observable mode. However, the observables $\mathcal{O}^\mathrm{b}_k$ are mere mathematical constructs. That is, they might not represent macroscopic physical observables in the thermodynamic limit. To overcome this shortcoming, the authors propose general requirements for the system based on timescale separation and nonmonotonic temperature evolution of an important state function. These features are generic near a first-order phase transition.

Consider any general physical observable $\mathcal{M}_{\text{w}}^2$,  having an equilibrium expected value $\langle \mathcal{M}_{\text{w}}^2\rangle_T$ that attains a maximum at $T=T^*$.  Then, for thermal bath temperatures $T_\mathrm{b}$ near that maximum, the fluctuating part of $\mathcal{M}_{\text{w}}^2$, given by $\mathcal{M}_{\text{w}}^2-\boldsymbol{1}\,\langle \mathcal{M}_{\text{w}}^2\rangle_{T_\mathrm{b}}$, will be a proxy for the slowest evolving observable, \(\mathcal{O}_2^{\text{b}}\). This allows one to connect to the requirement that $\alpha^{(t=0)}_{2}=0$ in~\eqref{eq:spectral_decomposition}, since
\begin{align}\label{eq:astucia}
	\alpha^{(t=0)}_{2}\approx \frac{1}{\Lambda}\Big(\langle \mathcal{M}^2_{\text{w}}\rangle_{T^*}\ -\ \langle \mathcal{M}^2_{\text{w}}\rangle_{T_{\text{b}}}\Big)\,.
\end{align}
With this in mind, the authors explain the appearance of several paradoxical behaviors near that maximum; cf.~\FIG{paradoxical_protocols}. This also serves the purpose of finding effective strategies to devise such phenomena in mesoscopic systems. In particular, the model system studied in~\cite{pemartin2024shortcuts} is the antiferromagnetic 1D Ising model given in~\EQ{ham_AF}. In addition, the equilibrium observable used by the authors is the squared staggered magnetization (described in Sec.~\ref{sec:phase-transitions} as $m_s= \sum_{k=1}^{N} (-1)^k \sigma_k$). The three found anomalous effects are (i) the Mpemba effect, (ii) the Preheating strategy, and (iii) two asymmetries in heating and cooling processes, namely, heating faster than cooling and vice versa, as discussed in Sec.~\ref{sec:asymmetric}. In~\FIG{paradoxical_protocols} two of these anomalous effects are shown.

\subsubsection{Internal structure speedup}
\label{sec:internal-structure-speed-up}
Often, a system has a complex internal structure near a (first- or second-order) phase transition. In~\cite{pemartin2021slow}, the authors show that the slow growth of magnetic domains helps to shorten the time it takes the system to reach a desired state. To this end, they consider a 2D Ising model without a magnetic field, described by the Hamiltonian
\begin{align}
\mathcal{H}=-J\displaystyle\sum_{{\rm n.n. }\, x,y} \sigma_x \sigma_y,
\end{align}
where spin $\sigma_x$ has position $x$ in a lattice $L \times L$. To demonstrate their idea, the authors study the 2D Ising Model (with $L = 4096$) using extensively two simulation methods, the Metropolis and the heat-bath algorithms. This model has no conserved magnitudes and belongs to the \emph{model A universality class}, where the dynamics has no conserved quantities~\cite{hohenberg1977theory}. A second-order phase transition between the ferromagnetic and paramagnetic phase appears at $T=T_{\rm c}$. It is well known that in the ferromagnetic phase and excluding fast initial relaxations, the energy $E(t)$ and the correlation length $\xi(t)$ are related because the excess energy is located at the boundaries of the magnetic domains.

\begin{figure}[htb]
    \centering 
    \includegraphics[width=0.8\columnwidth]{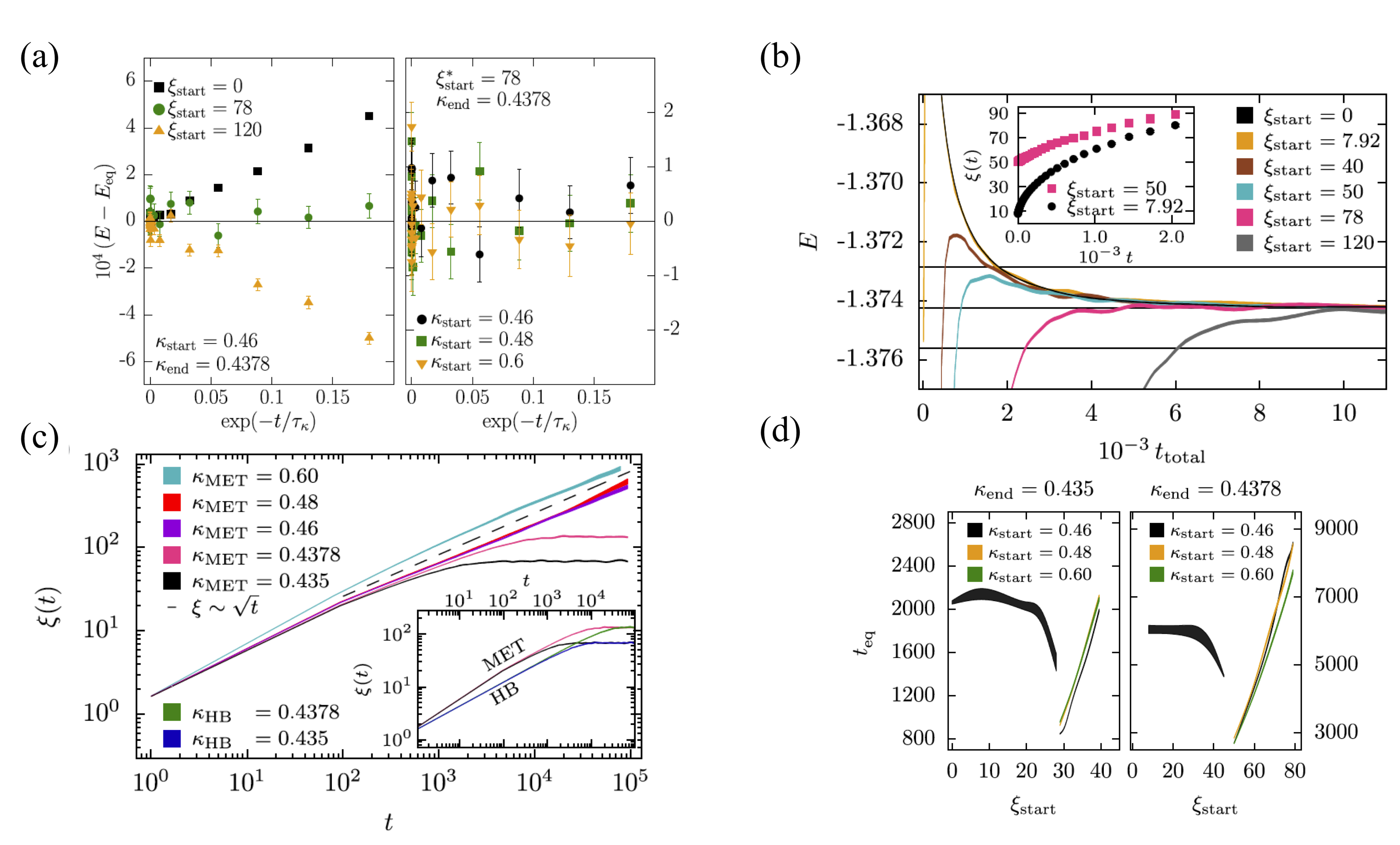}
    \caption{Example of anomalous relaxation in a 2D Ising model, using a two-step protocol for $\kappa_{\rm start} = 0.46$ and $\kappa_{\rm end} = 0.4378$.   (a) Change of sign of $\mathcal{\xi}$ and an example of $\xi^*_{\rm start}$.  (b) Examples of the speedup using a two-step protocol.  (c) Evolution of $\xi(t)$ in the one-step protocol. Note that the largest timescale exists as a finite-size effect in the ferromagnetic phase. Indeed, when $\kappa > \kappa_{\rm c}$ and keeping leading terms, domains grow as $\xi(t) \approx \sqrt{t}$ until $\xi \approx L$. (d) Equilibration time $t_{\rm eq}$ as a function of $\xi_{\rm start}$. Note that the equilibration time $t_{\rm eq}$ here is defined as the time such that $E_{\rm total}$ differs from $E_{\rm eq}$ by less than $0.1 \%$ for any $t_{\rm total}>t_{\rm eq}$. MET and HB indicate the use of Metropolis and Heat Bath algorithms, respectively. 
    \textit{Source:} Adapted with permission from~\cite{pemartin2021slow}.}
    \label{fig:slowgrowth}
\end{figure}

The paper focuses on the time evolution of energy. In the paramagnetic phase, the leading term can be expressed as
\begin{align} 
\label{eq:Eslowgrowth}
    E(t)-E_{\rm eq}= \mathcal{A}(\xi) e^{-t/\tau_k},
\end{align}
where $\tau_k$ is the largest relaxation timescale. Here, $\mathcal{A}(\xi)$ is an amplitude that depends only on the coherence length $\xi$. Interestingly, $\mathcal{A}(\xi)$ change its sign as the initial coherence length grows, and $\mathcal{A}(\xi^*_{\rm start})=0$ at some point; see~\FIG{slowgrowth}a. Given that $\xi(t)$ grows faster in the ferromagnetic phase (see \FIG{slowgrowth}a), it is likely that $E(t)$ could equilibrate faster in the paramagnetic phase through an excursion to the ferromagnetic one. In addition, an exponential speedup is possible if the optimal $\xi^*_{\rm start}$ is found. Two thermal protocols are designed, with one and two steps, to study that phenomenon. In the one-step protocol, a disordered spin configuration, corresponding to infinite temperature, is put in contact with a thermal bath at inverse-temperature $\kappa=J/k_BT$; see~\FIG{slowgrowth}c. The equilibrium value of $\xi$ is reached only in the paramagnetic phase. In the two-step protocol, a fully disordered spin configuration is initially placed at an inverse temperature in the ferromagnetic phase, $\kappa_{\rm start} > \kappa_{\mathrm{c}}$, where it evolves until $\xi(t)$ reaches a target value $\xi_{\rm start}(t)$. Just at that point, the bath is changed to a new $\kappa_{\rm end} < \kappa_{\mathrm{c}}$. Thus, the second change must be done when $\mathcal{A}(\xi^*_{\rm start})=0$, where $\xi^*_{\rm start}$ can be estimated numerically. From~\FIG{slowgrowth}a and \FIG{slowgrowth}b, the speedup due to the choice of $\xi^*_{\rm start}=78$ can be clearly observed. Interestingly, for $\xi^*_{\rm start}=7.92$, the Kovacs effect is observed; cf.~\SEC{kovacs}.

\section{Perspectives}
\label{sec:perspectives}
This review centers on the Mpemba effect, beginning with the original research conducted by Mpemba and Osborne in 1969 \cite{mpemba1969cool} and providing a historical context that traces back to Aristotle. Our focus, however, is on the significant advances made in this field over the past decade. After discussing the various challenges and efforts made to explain the Mpemba effect in water, highlighting its unique properties in Sec.~\ref{sec:historical}, we examine the range of observations and characterizations of the Mpemba effect and closely related phenomena. The different types of Mpemba effects defined in Sec.~\ref{sec:markovian} that can be understood within Markovian frameworks are crucial for understanding the effect. They have sparked considerable interest in identifying similar anomalous-relaxation phenomena across various systems, from single-particle interactions to many-body physics, in both classical and quantum settings.

The common theme in all Mpemba phenomena is a nonmonotonicity in the time a system takes to relax from an initial probability distribution to the final probability distribution or phase transition.  We start by discussing the effect for a discrete set of states with Markovian rates, overdamped Langevin dynamics, and antiferromagnetic systems.  In Sec.~\ref{sec:phase-transitions}, we discuss various approaches to accurately characterize the Mpemba effect in the presence and vicinity of first- or second-order phase transitions. Whereas a clear picture was obtained for mean-field models, not much is understood about this effect in 2D or 3D systems, which are of main practical interest. The kinetic theory framework covered in Sec.~\ref{sec:kinetic-framework} complements these concepts by providing a means to study related effects in driven systems, with an effective description of the thermodynamic limit. Beyond the scope of the kinetic theory framework are many interesting questions about the Mpemba effect in the thermodynamic limit, where issues such as vanishing spectral gaps and continuous spectra are largely open problems.  

We address general questions regarding the likelihood of encountering a Mpemba-related effect in these systems in Sec.~\ref{sec:statistics}. Generalizing this approach to other ensembles of random models is of high interest. At this stage, the relevant properties of the ensemble are unclear. Next, we provide an overview of the experiments in Sec.~\ref{sec:experiments} and the numerical simulations described in Sec.~\ref{sec:numerical-observations}, which validate our theoretical frameworks.  A potential area for further exploration on the experimental side is the search for the inverse Mpemba effect in macroscopic systems. Aside from the colloidal experiment conducted by Kumar and Bechhoefer~\cite{kumar2020exponentially} and a quantum analog effect~\cite{aharony2024inverse}, the inverse Mpemba effect has not yet been observed in other systems, whereas the direct effect has, most notably in water. At the start of 2024, three experiments demonstrated a Mpemba effect in quantum systems~\cite{aharony2024inverse,joshi2024observing,zhang2024observation}. More quantum experiments are needed, particularly those involving open quantum systems.

The information-geometry approach has theoretically proven useful in explaining the asymmetry between heating and cooling ~\cite{tejero2024asymmetries}. This method was also applied in ~\cite{bettmann2024information} to study classical stochastic thermodynamics, particularly the Mpemba effect. Further exploration into anomalous thermal relaxation using this approach would be of interest. Additionally, in studies addressing entanglement asymmetry and the Mpemba effect, the nonmonotonic relaxation of entanglement entropy can be interpreted as a form of symmetry restoration~\cite{ares2023entanglement}.

Sections~\ref{sec:applications} and~\ref{sec:other-related} highlight several applications and related effects, hinting at future prospects. Practical perspectives will also face experimental challenges related to the impossibility of resolving the system with infinite precision and waiting an arbitrarily long time for the relaxation process to terminate. These issues are particularly relevant in quantum simulations \cite{joshi2024observing,aharony2024inverse}, where experimental noise and decoherence set strong bounds in both directions. Another common issue that could arise in these settings is that in open quantum settings, the spectrum of the Lindbladian can be complex \cite{carollo2021exponentially}, as in the classical case with broken detailed balance. Various perspectives on ``state preparation'' highlight a rich research avenue for many-body quantum simulations, potentially based on the results discussed in this review.

The question of how different effects, frameworks, and experiments are related poses a significant challenge for the coming years. For instance, can the Markovian framework discussed in Sec.~\ref{sec:markovian} explain the phase transition effects covered in Sec.~\ref{sec:phase-transitions}? Additionally, can the intuitive single-particle perspective from experiments mentioned in Sec.~\ref{sec:experiments} be applied to macroscopic systems such as water, magnetic alloys, and others? Furthermore, is it possible to extend the Boltzmann-Fokker-Planck equation and the framework outlined in Sec.~\ref{sec:kinetic-framework} to make predictions about macroscopic materials such as water? It is our hope that a framework encompassing all these phenomena can be found, one that goes all the way from making ice cream in Tanzania to preparing states for quantum computing, and explains why---sometimes---the longer path can nevertheless be faster.

\section{Acknowledgements}
MV would like to express gratitude to Massimo Vergassola for advising that it is time to write a review of the Mpemba effect. We greatly appreciate the contributions of our coauthors and the vibrant scientific community surrounding the Mpemba effect. Special thanks to Siddharth Sane for generating~\FIG{average-energy}. Also, we appreciate the critical reading and comments of the manuscript by Antonio Prados and especially Andr\'es Santos.  This material is based on work supported by the National Science Foundation under Grant No. DMR-1944539. MV would like to thank the Isaac Newton Institute for Mathematical Sciences, Cambridge, for support and hospitality during the programme ``Stochastic systems for anomalous diffusion", where part of the work on this paper was undertaken. This work was supported by EPSRC grant no EP/R014604/1. JB was supported by Discovery Grants from the National Science and Engineering Research Council (NSERC, Canada). 
OR acknowledges financial support from the ISF, grant no. 232/23, and by the Minerva foundation with funding from the Federal German Ministry for Education and Research. AL was supported by the Project PID2021-128970OA-I00 financed by MICIN/AEI/10.13039/501100011033 and FEDER a way to make Europe and by C-EXP-251-UGR23 European Regional Development Fund, Junta de Andaluc\'ia-Consejer\'ia de Econom\'ia y Conocimiento.

\bibliography{references-arXiv-without-url-doi} 

\begin{thebibliography}{100}
\expandafter\ifx\csname url\endcsname\relax
  \def\url#1{\texttt{#1}}\fi
\expandafter\ifx\csname urlprefix\endcsname\relax\def\urlprefix{URL }\fi
\expandafter\ifx\csname href\endcsname\relax
  \def\href#1#2{#2} \def\path#1{#1}\fi

\bibitem{mpemba1969cool}
E.~B. Mpemba, D.~G. Osborne, \href{https://dx.doi.org/10.1088/0031-9120/4/3/312}{Cool?}, Phys. Educ. 4 (1969) 172.

\bibitem{ball2006does}
P.~Ball, \href{https://iopscience.iop.org/article/10.1088/2058-7058/19/4/32}{Does hot water freeze first?}, Phys. World 19 (2006) 19.

\bibitem{lu2017nonequilibrium}
Z.~Lu, O.~Raz, \href{https://www.pnas.org/doi/abs/10.1073/pnas.1701264114}{Nonequilibrium thermodynamics of the {M}arkovian {M}pemba effect and its inverse}, Proc.~Natl.~Acad.~Sci. (USA) 114~(20) (2017) 5083--5088.

\bibitem{lasanta2017hotter}
A.~Lasanta, F.~Vega~Reyes, A.~Prados, A.~Santos, \href{https://link.aps.org/doi/10.1103/PhysRevLett.119.148001}{When the hotter cools more quickly: {M}pemba effect in granular fluids}, Phys.~Rev. Lett. 119~(14) (2017) 148001.

\bibitem{klich2019mpemba}
I.~Klich, O.~Raz, O.~Hirschberg, M.~Vucelja, \href{https://link.aps.org/doi/10.1103/PhysRevX.9.021060}{Mpemba index and anomalous relaxation}, Phys.~Rev. X 9~(2) (2019) 021060.

\bibitem{kumar2020exponentially}
A.~Kumar, J.~Bechhoefer, \href{https://doi.org/10.1038/s41586-020-2560-x}{Exponentially faster cooling in a colloidal system}, Nature 584 (2020) 64--68.

\bibitem{nava2019lindblad}
A.~Nava, M.~Fabrizio, \href{https://link.aps.org/doi/10.1103/PhysRevB.100.125102}{Lindblad dissipative dynamics in the presence of phase coexistence}, Phys. Rev. B 100 (2019) 125102.

\bibitem{aharony2024inverse}
S.~Aharony~Shapira, Y.~Shapira, J.~Markov, G.~Teza, N.~Akerman, O.~Raz, R.~Ozeri, \href{https://link.aps.org/doi/10.1103/PhysRevLett.133.010403}{Inverse {M}pemba effect demonstrated on a single trapped ion qubit}, Phys. Rev. Lett. 133 (2024) 010403.

\bibitem{zhang2024observation}
J.~Zhang, G.~Xia, C.-W. Wu, T.~Chen, Q.~Zhang, Y.~Xie, W.-B. Su, W.~Wu, C.-W. Qiu, P.-X. Chen, W.~Li, H.~Jing, Y.-L. Zhou, \href{https://doi.org/10.1038/s41467-024-54303-0}{Observation of quantum strong {M}pemba effect}, Nat. Commun. 16~(1) (2025) 301.

\bibitem{joshi2024observing}
L.~K. Joshi, J.~Franke, A.~Rath, F.~Ares, S.~Murciano, F.~Kranzl, R.~Blatt, P.~Zoller, B.~Vermersch, P.~Calabrese, C.~F. Roos, M.~K. Joshi, \href{https://link.aps.org/doi/10.1103/PhysRevLett.133.010402}{Observing the quantum {M}pemba effect in quantum simulations}, Phys.~Rev.~Lett. 133~(1) (2024) 010402.

\bibitem{debenedetti2001supercooled}
P.~G. Debenedetti, F.~H. Stillinger, \href{https://doi.org/10.1038/35065704}{Supercooled liquids and the glass transition}, Nature 410 (2001) 259--267.

\bibitem{gueryodelin2019shortcuts}
D.~Gu{\'e}ry-Odelin, A.~Ruschhaupt, A.~Kiely, E.~Torrontegui, S.~Mart{\'\i}nez-Garaot, J.~G. Muga, \href{https://link.aps.org/doi/10.1103/RevModPhys.91.045001}{Shortcuts to adiabaticity: {C}oncepts, methods, and applications}, Rev. Mod. Phys. 91 (2019) 045001.

\bibitem{webster1923aristotle}
Aristotle, Meterologica, {Transl. E. W. Webster, Book 1, Part 12} Edition, Oxford: Clarendon Press, 1923.

\bibitem{bacon1900opus}
R.~Bacon, The Opus Majus of Roger Bacon, Vol.~3, Williams and Norgate, 1900.

\bibitem{clagett1941giovanni}
M.~Clagett, Giovanni Marliani and Late Medieval Physics, Columbia Univ. Press, New York, 1941.

\bibitem{bacon1878novum}
F.~Bacon, \href{https://hdl.handle.net/2027/mdp.39015026483944}{{N}ovum {O}rganum}, Clarendon {P}ress, 1878.

\bibitem{descartes1637discours}
R.~Descartes, \href{https://gallica.bnf.fr/ark:/12148/btv1b86069594/f5.item}{Discours de la M\'ethode}, Ian Maire, Leyden, 1637, Ch. Les M\'et\'eores.

\bibitem{descartes1638mersenne}
R.~Descartes, The Philosophical Writings of Descartes, Vol. III: The Correspondence, Cambridge Univ. Press, 1984.

\bibitem{barker1775the}
R.~Barker, \href{https://royalsocietypublishing.org/doi/10.1098/rstl.1775.0023}{The process of making Ice in the {E}ast {I}ndies}, Philos. Trans. Roy. Soc. (London) 65 (1775) 252--257.

\bibitem{black1775the}
J.~Black, \href{https://royalsocietypublishing.org/doi/10.1098/rstl.1775.0014}{The Supposed Effect of Boiling upon Water, in Disposing It to Freeze More Readily, Ascertained by Experiments}, Philos. Trans. Roy. Soc. (London) 65 (1775) 124--128.

\bibitem{kumar1980Mpemba}
K.~Kumar, \href{https://dx.doi.org/10.1088/0031-9120/15/5/101}{{M}pemba effect and 18th century ice-cream}, Phys. Educ. 15 (1980) 268.

\bibitem{jeng2006mpemba}
M.~Jeng, \href{https://doi.org/10.1119/1.2186331}{The {M}pemba effect: {W}hen can hot water freeze faster than cold?}, Am. J. Phys. 74 (2006) 514--522.

\bibitem{groves1962now}
L.~R. Groves, {Now It Can Be Told: The Story of the Manhattan Project}, Harper \& Row, New York (1962).

\bibitem{rsc2013oral-history}
{Unsigned press release from the Royal Society of Chemistry}, \href{https://edu.rsc.org/resources/the-mpemba-effect/1018.article}{The {M}pemba effect: {C}ompetition and resources} (2013).

\bibitem{burridge2016questioning}
H.~C. Burridge, P.~F. Linden, \href{https://doi.org/10.1038/srep37665}{Questioning the {M}pemba effect: {H}ot water does not cool more quickly than cold}, Sci. Rep. 6 (2016) 37665.

\bibitem{kell1969the}
G.~S. Kell, \href{https://doi.org/10.1119/1.1975687}{The Freezing of Hot and Cold Water}, Am. J. Phys. 37 (1969) 564--565.

\bibitem{vynnycky2010evaporative}
M.~Vynnycky, S.~Mitchell, \href{https://doi.org/10.1007/s00231-010-0637-z}{Evaporative cooling and the {M}pemba effect}, Heat Mass Transf. 46 (2010) 881--890.

\bibitem{vynnycky2012axisymmetric}
M.~Vynnycky, N.~Maeno, \href{https://doi.org/10.1016/j.ijheatmasstransfer.2012.07.060}{Axisymmetric natural convection-driven evaporation of hot water and the {M}pemba effect}, Int. J. Heat Mass Transf. 55 (2012) 7297--7311.

\bibitem{mirabedin2017numerical}
S.~M. Mirabedin, F.~Farhadi, \href{https://doi.org/10.1016/j.ijrefrig.2016.09.006}{Numerical investigation of solidification of single droplets with and without evaporation mechanism}, Int. J. Refrig. 73 (2017) 219--225.

\bibitem{brownridge2011when}
J.~D. Brownridge, \href{https://doi.org/10.1119/1.3490015}{When does hot water freeze faster then cold water? {A} search for the {M}pemba effect}, Am. J. Phys. 79 (2011) 78--84.

\bibitem{deeson1971cooler}
E.~Deeson, \href{https://iopscience.iop.org/article/10.1088/0031-9120/6/1/311}{Cooler-Lower Down.}, Phys. Educ. 6 (1971) 42--44.

\bibitem{freeman1979cooler}
M.~Freeman, \href{https://iopscience.iop.org/article/10.1088/0031-9120/14/7/314}{Cooler still---an answer?}, Phys. Educ. 14 (1979) 417.

\bibitem{wojciechowski1988freezing}
B.~Wojciechowski, I.~Owczarek, G.~Bednarz, \href{https://onlinelibrary.wiley.com/doi/10.1002/crat.2170230702}{Freezing of aqueous solutions containing gases}, Cryst. Res. Technol. 23 (1988) 843--848.

\bibitem{auerbach1995supercooling}
D.~Auerbach, \href{https://doi.org/10.1119/1.18059}{Supercooling and the {Mpemba} effect: {When} hot water freezes quicker than cold}, Am. J. of Phys. 63~(10) (1995) 882--885.

\bibitem{katz2009hot}
J.~I. Katz, \href{https://doi.org/10.1119/1.2996187}{When hot water freezes before cold}, Am. J. Phys. 77 (2009) 27--29.

\bibitem{zimmerman2021towards}
W.~B. Zimmerman, \href{https://doi.org/10.1016/j.ces.2021.116618}{Towards a microbubble condenser: {D}ispersed microbubble mediation of additional heat transfer in aqueous solutions due to phase change dynamics in airlift vessels}, Chem. Eng. Sci. 238 (2021) 116618.

\bibitem{zimmerman2022in}
W.~B. Zimmerman, \href{https://doi.org/10.1016/j.ces.2021.117043}{In search of a {M}pemba effect protocol: {S}ome hot water does cool and freeze faster than cold}, Chem. Eng. Sci. 247 (2022) 117043.

\bibitem{zimmerman2022mediating}
W.~B. Zimmerman, \href{https://doi.org/10.1016/j.applthermaleng.2022.118720}{Mediating heat transport by microbubble dispersions: {T}he role of dissolved gases and phase change dynamics}, Appl. Thermal Eng. 213 (2022) 118720.

\bibitem{lifshitz1981physical}
E.~M. Lifshitz, L.~P. Pitaevskii, {P}hysical {K}inetics, Vol.~10 of L. D. Landau and E. M. Lifshitz Course of Theoretical Physics, Elsevier, 1981.

\bibitem{dorsey1948the}
N.~E. Dorsey, \href{https://doi.org/10.2307/1005602}{The Freezing of Supercooled Water}, Trans. Am. Phil. Soc. 38 (1948) 247--328.

\bibitem{esposito2008mpemba}
S.~Esposito, R.~{De Risi}, L.~Somma, \href{https://www.sciencedirect.com/science/article/pii/S0378437107010722}{Mpemba effect and phase transitions in the adiabatic cooling of water before freezing}, Physica A 387~(4) (2008) 757--763.

\bibitem{burridge2020observing}
H.~C. Burridge, O.~Hallstadius, \href{https://doi.org/10.1098/rspa.2019.0829}{Observing the {M}pemba effect with minimal bias and the value of the {M}pemba effect to scientific outreach and engagement}, Proc. R. Soc. A 476 (2020) 20190829.

\bibitem{maciejewski1996evidence}
P.~K. Maciejewski, \href{https://doi.org/10.1115/1.2824069}{Evidence of a Convective Instability Allowing Warm Water to Freeze in Less Time Than Cold Water}, J. Heat Transf. 118 (1996) 65--72.

\bibitem{vynnycky2015convection}
M.~Vynnycky, S.~Kimura, \href{http://doi.org/10.1007/s00231-010-0637-z}{Can natural convection alone explain the {M}pemba effect?}, Int. J. Heat Mass Transf. 80 (2015) 243--255.

\bibitem{zhang2014hydrogen}
X.~Zhang, Y.~Huang, Z.~Ma, Y.~Zhou, J.~Zhou, W.~Zheng, Q.~Jiang, C.~Q. Sun, \href{https://doi.org/10.1039/C4CP03669G}{Hydrogen-bond memory and water-skin supersolidity resolving the {M}pemba paradox}, Phys. Chem. Chem. Phys. 16~(42) (2014) 22995--23002.

\bibitem{sun2023the}
C.~Q. Sun, Y.~Huang, X.~Zhang, Z.~Ma, B.~Wang, \href{https://doi.org/10.1016/j.physrep.2022.11.001}{The physics behind water irregularity}, Phys. Rep. 998 (2023) 1--68.

\bibitem{jin2015mechanisms}
J.~Jin, W.~A. Goddard~III, \href{https://doi.org/10.1021/jp511752n}{Mechanisms underlying the {M}pemba effect in water from molecular dynamics simulations}, J. Phys. Chem. C 119~(5) (2015) 2622--2629.

\bibitem{tao2017different}
Y.~Tao, W.~Zou, J.~Jia, W.~Li, D.~Cremer, \href{https://doi.org/10.1021/acs.jctc.6b00735}{Different ways of hydrogen bonding in water---why does warm water freeze faster than cold water?}, J. Chem. Theory Comput. 13~(1) (2017) 55--76.

\bibitem{ahtee1969investigation}
M.~Ahtee, \href{https://doi.org/10.1088/0031-9120/4/6/114}{Investigation into the freezing of liquids}, Phys. Educ. 4 (1969) 379--380.

\bibitem{katz2017reply}
J.~I. Katz, \href{https://doi.org/10.48550/arxiv.1701.03219}{{Reply to Burridge \& Linden: H}ot water may freeze sooner than cold}, arXiv:1701.03219 (2017).

\bibitem{tang2022direct}
Z.~Tang, W.~Huang, Y.~Zhang, Y.~Liu, L.~Zhao, \href{https://doi.org/10.1002/inf2.12352}{Direct observation of the {M}pemba effect with water: {P}robe the mysterious heat transfer}, InfoMat. 5 (2022) e12352.

\bibitem{elton2021pathological}
D.~C. Elton, P.~D. Spencer, \href{https://doi.org/10.1007/978-3-030-67227-0_8}{Biomechanical and Related Systems}, in: A.~Gadomski (Ed.), Biologically Inspired Systems, Vol.~17 of Biologically Inspired Systems, Springer, 2021, Ch. Pathological water science -- Four examples and what they have in common, pp. 155--169.

\bibitem{ortega2024thermographic}
A.~B. Ortega, E.~Hern{\'a}ndez-Figueroa, J.~A. del R{\'\i}o, \href{https://doi.org/10.31349/revmexfis.70.050601}{Thermographic study of freezing water drops: {A}n insight on {M}pemba effect}, Rev. Mex. de Fis. 70 (2024) 050601.

\bibitem{turnbull1956phase}
D.~Turnbull, \href{https://www.sciencedirect.com/science/article/pii/S0081194708601344}{Phase Changes}, in: F.~Seitz, D.~Turnbull (Eds.), Solid State Physics, Vol.~3 of Solid State Physics, Academic Press, 1956, pp. 225--306.

\bibitem{ball1999h2O}
P.~Ball, H$_2$O: A Biography of Water, Weidenfeld and Nicolson Ltd, 1999.

\bibitem{sun2016the}
C.~Q. Sun, Y.~Sun, \href{https://doi.org/10.1007/978-981-10-0180-2}{The Attribute of Water: Single Notion, Multiple Myths}, Springer, 2016.

\bibitem{holtzman2022landau}
R.~Holtzman, O.~Raz, \href{https://doi.org/10.1038/s42005-022-01063-2}{Landau theory for the {M}pemba effect through phase transitions}, Commun. Phys. 5~(1) (2022) 280.

\bibitem{zhang2022theoretical}
S.~Zhang, J.-X. Hou, \href{https://doi.org/10.1103/PhysRevE.106.034131}{Theoretical model for the {M}pemba effect through the canonical first-order phase transition}, Phys.~Rev. E 106~(3) (2022) 034131.

\bibitem{baity2019mpemba}
M.~Baity-Jesi, E.~Calore, A.~Cruz, L.~A. Fernandez, J.~M. Gil-Narvi{\'o}n, A.~Gordillo-Guerrero, D.~Iñiguez, A.~Lasanta, A.~Maiorano, E.~Marinari, V.~Martin-Mayor, J.~Moreno-Gordo, A.~M. Sudupe, D.~Navarro, G.~Parisi, S.~Perez-Gaviro, F.~Ricci-Tersenghi, J.~J. Ruiz-Lorenzo, S.~F. Schifano, B.~Seoane, A.~Taranc{\'o}n, R.~Tripiccione, D.~Yllanes, \href{https://doi.org/10.1073/pnas.1819803116}{The {M}pemba effect in spin glasses is a persistent memory effect}, Proc.~Natl.~Acad.~Sci. (USA) 116~(31) (2019) 15350--15355.

\bibitem{van2024thermomajorization}
T.~Van~Vu, H.~Hayakawa, \href{https://link.aps.org/doi/10.1103/PhysRevLett.134.107101}{Thermomajorization {M}pemba Effect}, Phys. Rev. Lett. 134 (2025) 107101.

\bibitem{biswas2020mpemba}
A.~Biswas, V.~Prasad, O.~Raz, R.~Rajesh, \href{https://doi.org/10.1103/PhysRevE.102.012906}{Mpemba effect in driven granular {M}axwell gases}, Phys. Rev. E 102 (2020) 012906.

\bibitem{nava2024mpemba}
A.~Nava, R.~Egger, \href{https://link.aps.org/doi/10.1103/PhysRevLett.133.136302}{Mpemba effects in open nonequilibrium quantum systems}, Phys. Rev. Lett. 133 (2024) 136302.

\bibitem{chatterjee2023quantum}
A.~K. Chatterjee, S.~Takada, H.~Hayakawa, \href{https://link.aps.org/doi/10.1103/PhysRevLett.131.080402}{Quantum {M}pemba effect in a quantum Dot with Reservoirs}, Phys. Rev. Lett. 131 (2023) 080402.

\bibitem{kumar2022inverse}
A.~Kumar, R.~Ch{\'e}trite, J.~Bechhoefer, \href{https://doi.org/10.1073/pnas.2118484119}{Anomalous heating in a colloidal system}, Proc. Natl. Acad. Sci. USA 119 (2022) e2118484119.

\bibitem{gal2020precooling}
A.~Gal, O.~Raz, \href{https://doi.org/10.1103/PhysRevLett.124.060602}{Precooling strategy allows exponentially faster heating}, Phys.~Rev.~Lett. 124~(6) (2020) 060602.

\bibitem{zwanzig2001nonequilibrium}
R.~Zwanzig, \href{https://doi.org/10.1093/oso/9780195140187.001.0001}{Nonequilibrium Statistical Mechanics}, Oxford {U}niv. {P}ress, 2001.

\bibitem{horowitz2009exact}
J.~M. Horowitz, C.~Jarzynski, \href{https://doi.org/10.1007/s10955-009-9818-x}{Exact formula for currents in strongly pumped diffusive systems}, J. Stat. Phys. 136 (2009) 917--925.

\bibitem{schnakenberg1976network}
J.~Schnakenberg, \href{https://doi.org/10.1103/RevModPhys.48.571}{Network theory of microscopic and macroscopic behavior of master equation systems}, Rev. Mod. Phys. 48~(4) (1976) 571--585.

\bibitem{mandal2011proof}
D.~Mandal, C.~Jarzynski, \href{https://doi.org/10.1088/1742-5468/2011/10/P10006}{A proof by graphical construction of the no-pumping theorem of stochastic pumps}, J. Stat. Mech: Theory Exp. 2011~(10) (2011) P10006.

\bibitem{teza2023relaxation}
G.~Teza, R.~Yaacoby, O.~Raz, \href{https://doi.org/10.1103/PhysRevLett.131.017101}{Relaxation shortcuts through boundary coupling}, Phys.~Rev. Lett. 131~(1) (2023) 017101.

\bibitem{biswas2023mpemba_a}
A.~Biswas, R.~Rajesh, \href{https://link.aps.org/doi/10.1103/PhysRevE.108.024131}{Mpemba effect for a {B}rownian particle trapped in a single well potential}, Phys. Rev. E 108 (2023) 024131.

\bibitem{horodecki2013fundamental}
M.~Horodecki, J.~Oppenheim, \href{https://doi.org/10.1038/ncomms3059}{Fundamental limitations for quantum and nanoscale thermodynamics}, Nat. Commun. 4~(1) (2013) 2059.

\bibitem{vadakkayil2021should}
N.~Vadakkayil, S.~K. Das, \href{https://doi.org/10.1039/D1CP00879J}{Should a hotter paramagnet transform quicker to a ferromagnet? {M}onte {C}arlo simulation results for {I}sing model}, Phys. Chem. Chem. Phys. 23~(19) (2021) 11186--11190.

\bibitem{ohga2024microscopic}
N.~Ohga, H.~Hayakawa, S.~Ito, \href{https://doi.org/10.48550/arxiv.2410.06623}{Microscopic theory of {M}pemba effects and a no-{M}pemba theorem for monotone many-body systems}, arXiv preprint arXiv:2410.06623 (2024).

\bibitem{chetrite2021metastable}
R.~Ch{\'e}trite, A.~Kumar, J.~Bechhoefer, \href{https://doi.org/10.3389/fphy.2021.654271}{The Metastable {M}pemba Effect Corresponds to a Non-monotonic Temperature Dependence of Extractable Work}, Front. Phys. 9 (2021) 141.

\bibitem{walker2022mpemba}
M.~R. Walker, M.~Vucelja, \href{https://doi.org/10.48550/arxiv.2212.07496}{Mpemba effect in terms of mean first passage times of overdamped {L}angevin dynamics on a double-well potential}, arXiv preprint arXiv:2212.07496 (2022).

\bibitem{malhotra2024double}
I.~Malhotra, H.~L{\"o}wen, \href{https://pubs.aip.org/aip/jcp/article/161/16/164903/3317716/Double-Mpemba-effect-in-the-cooling-of-trapped}{Double {M}pemba effect in the cooling of trapped colloids}, J. Chem. Phys. 161 (2024) 164903.

\bibitem{teza2020exact}
G.~Teza, A.~L. Stella, \href{https://link.aps.org/doi/10.1103/PhysRevLett.125.110601}{Exact Coarse Graining Preserves Entropy Production out of Equilibrium}, Phys. Rev. Lett. 125 (2020) 110601.
\newblock \href {https://doi.org/10.1103/PhysRevLett.125.110601} {\path{doi:10.1103/PhysRevLett.125.110601}}.

\bibitem{glauber1963time}
R.~J. Glauber, \href{https://doi.org/10.1063/1.1703954}{Time-dependent statistics of the {I}sing model}, J. Math. Phys. 4~(2) (1963) 294--307.

\bibitem{felderhof1971spin}
B.~Felderhof, \href{https://doi.org/10.1016/S0034-4877(71)80006-X}{Spin relaxation of the {I}sing chain}, Rep. Math. Phys. 1~(3) (1971) 215--234.

\bibitem{hamming1950error}
R.~W. Hamming, \href{https://doi.org/10.1002/j.1538-7305.1950.tb00463.x}{Error detecting and error correcting codes}, Bell Labs Tech. J. 29~(2) (1950) 147--160.

\bibitem{teza2023eigenvalue}
G.~Teza, R.~Yaacoby, O.~Raz, \href{https://doi.org/10.1103/PhysRevLett.130.207103}{Eigenvalue crossing as a phase transition in relaxation dynamics}, Phys.~Rev. Lett. 130~(20) (2023) 207103.

\bibitem{kramers1940brownian}
H.~A. Kramers, \href{https://doi.org/10.1016/S0031-8914(40)90098-2}{Brownian motion in a field of force and the diffusion model of chemical reactions}, Physica 7~(4) (1940) 284--304.

\bibitem{chandrasekhar1943stochastic}
S.~Chandrasekhar, \href{https://doi.org/10.1103/RevModPhys.15.1}{Stochastic problems in physics and astronomy}, Rev. Mod. Phys. 15~(1) (1943) 1.

\bibitem{wang1945theory}
M.~C. Wang, G.~E. Uhlenbeck, \href{https://doi.org/10.1103/RevModPhys.17.323}{On the theory of the {B}rownian motion {II}}, Rev. Mod. Phys. 17~(2-3) (1945) 323.

\bibitem{van1992stochastic}
N.~Van~Kampen, \href{https://doi.org/10.1016/B978-0-444-52965-7.X5000-4}{Stochastic Processes in Physics and Chemistry}, North-Holland Publishing Co, 1992.

\bibitem{landauer1961frequency}
R.~Landauer, J.~Swanson, \href{https://doi.org/10.1103/PhysRev.121.1668}{Frequency factors in the thermally activated process}, Phys. Rev. 121~(6) (1961) 1668.

\bibitem{langer1969statistical}
J.~S. Langer, \href{https://doi.org/10.1016/0003-4916(69)90153-5}{Statistical theory of the decay of metastable states}, Ann. Phys. 54~(2) (1969) 258--275.

\bibitem{risken1996fokker}
H.~Risken, \href{https://doi.org/10.1007/978-3-642-61544-3}{The {F}okker-{P}lanck {E}quation}, Springer, 1996.

\bibitem{gavrilov2017direct}
M.~Gavrilov, R.~Ch{\'e}trite, J.~Bechhoefer, \href{https://doi.org/10.1073/pnas.1708689114}{Direct measurement of weakly nonequilibrium system entropy is consistent with {G}ibbs-{S}hannon form}, Proc.~Natl.~Acad.~Sci. (USA) 114~(42) (2017) 11097--11102.

\bibitem{walker2021anomalous}
M.~R. Walker, M.~Vucelja, \href{https://doi.org/10.1088/1742-5468/ac2edc}{Anomalous thermal relaxation of {L}angevin particles in a piecewise-constant potential}, J. Stat. Mech: Theory Exp. 2021~(11) (2021) 113105.

\bibitem{biswas2023mpemba}
A.~Biswas, R.~Rajesh, A.~Pal, \href{https://doi.org/10.1063/5.0155855}{Mpemba effect in a {L}angevin system: {P}opulation statistics, metastability, and other exact results}, J.~{C}hem.~{P}hys 159~(4) (2023).

\bibitem{biswas2025mpemba}
A.~Biswas, R.~Rajesh, \href{https://doi.org/10.1063/5.0246857}{Mpemba effect in the relaxation of an active {B}rownian particle in a trap without metastable states}, J. Chem. Phys. 162~(3) (2025).

\bibitem{kumar2022anomalous}
A.~Kumar, \href{https://doi.org/10.1007/978-3-031-13280-3}{Anomalous relaxation in colloidal systems}, Springer Nature, 2022.

\bibitem{chen2023memory}
Y.~Chen, Q.~Zhang, S.~Ramakrishnan, R.~L. Leheny, \href{https://pubs.aip.org/aip/jcp/article-abstract/158/2/024906/2868297/Memory-in-aging-colloidal-gels-with-time-varying?redirectedFrom=fulltext}{Memory in aging colloidal gels with time-varying attraction}, J. Chem. Phys. 158 (2023) 024906.

\bibitem{klich2018solution}
I.~Klich, M.~Vucelja, \href{https://doi.org/10.48550/arxiv.1812.11962}{Solution of the {M}etropolis dynamics on a complete graph with application to the {M}arkov chain {M}pemba effect}, arXiv preprint arXiv:1812.11962 (2018).

\bibitem{bera2023effect}
S.~Bera, M.~R. Walker, M.~Vucelja, \href{https://doi.org/10.48550/arxiv.2308.04557}{Effect of dynamics on anomalous thermal relaxations and information exchange}, arXiv preprint arXiv:2308.04557 (2023).

\bibitem{mitra18quantum}
A.~Mitra, \href{https://doi.org/10.1146/annurev-conmatphys-031016-025451}{Quantum quench dynamics}, Annu. Rev. Condens. Matter Phys. 9 (2018) 245--259.

\bibitem{keller2018quenches}
T.~Keller, V.~Torggler, S.~B. Jäger, S.~Schütz, H.~Ritsch, G.~Morigi, \href{https://dx.doi.org/10.1088/1367-2630/aaa161}{Quenches across the self-organization transition in multimode cavities}, New J. Phys. 20~(2) (2018) 025004.

\bibitem{ares2025quantum}
F.~Ares, P.~Calabrese, S.~Murciano, \href{{https://arxiv.org/abs/2502.08087}}{The quantum {M}pemba effects}, arXiv preprint arXiv:2502.08087 (2025).

\bibitem{chruscinski2017brief}
D.~Chru{\'s}ci{\'n}ski, S.~Pascazio, \href{https://doi.org/10.1142/S1230161217400017}{A brief history of the {GKLS} equation}, Open Systems \& Information Dynamics 24~(03) (2017) 1740001.

\bibitem{graf2024role}
J.~Graf, J.~Splettstoesser, J.~Monsel, \href{https://doi.org/10.48550/arxiv.2412.18456}{Role of electron-electron interaction in the {M}pemba effect in quantum dots}, arXiv preprint arXiv:2412.18456 (2024).

\bibitem{zhou2023accelerating}
Y.-L. Zhou, X.-D. Yu, C.-W. Wu, X.-Q. Li, J.~Zhang, W.~Li, P.-X. Chen, \href{https://link.aps.org/doi/10.1103/PhysRevResearch.5.043036}{Accelerating relaxation through {Liouvillian} exceptional point}, Phys. Rev. Res. 5~(4) (2023) 043036.

\bibitem{wang2024mpemba}
X.~Wang, J.~Wang, \href{https://link.aps.org/doi/10.1103/PhysRevResearch.6.033330}{Mpemba effects in nonequilibrium open quantum systems}, Phys. Rev. Res. 6 (2024) 033330.

\bibitem{chatterjee2024multiple}
A.~K. Chatterjee, S.~Takada, H.~Hayakawa, \href{https://link.aps.org/doi/10.1103/PhysRevA.110.022213}{Multiple quantum {M}pemba effect: {E}xceptional points and oscillations}, Phys. Rev. A 110 (2024) 022213.

\bibitem{kheirandish2024mpemba}
F.~Kheirandish, N.~Cheraghpour, A.~Moradian, \href{https://doi.org/10.48550/arxiv.2412.03943}{The {M}pemba effect in quantum oscillating and two-level systems}, arXiv preprint arXiv:2412.03943 (2024).

\bibitem{ivander2023hyperacceleration}
F.~Ivander, N.~Anto-Sztrikacs, D.~Segal, \href{https://doi.org/10.1103/PhysRevE.108.014130}{\href{}{Hyperacceleration of quantum thermalization dynamics by bypassing long-lived coherences: An analytical treatment}}, Phys.~Rev. E 108~(1) (2023) 014130.
\newblock \href {https://doi.org/10.1103/PhysRevE.108.014130} {\path{doi:10.1103/PhysRevE.108.014130}}.
\newline\urlprefix\url{https://doi.org/10.1103/PhysRevE.108.014130}

\bibitem{moroder2024thermodynamics}
M.~Moroder, O.~Culhane, K.~Zawadzki, J.~Goold, \href{https://link.aps.org/doi/10.1103/PhysRevLett.133.140404}{Thermodynamics of the Quantum {Mpemba} Effect}, Phys. Rev. Lett. 133 (2024) 140404.

\bibitem{bettmann2024information}
L.~P. Bettmann, J.~Goold, \href{https://link.aps.org/doi/10.1103/PhysRevE.111.014133}{Information geometry approach to quantum stochastic thermodynamics}, Phys. Rev. E 111 (2025) 014133.

\bibitem{wang2024mpembaa}
X.~Wang, J.~Su, J.~Wang, \href{http://arxiv.org/abs/2410.06669}{Mpemba meets quantum chaos: {A}nomalous relaxation and {M}pemba crossings in dissipative {S}achdev-{Y}e-{K}itaev models}, arXiv preprint arXiv:2410.06669 (2024).

\bibitem{strachan2024non}
D.~J. Strachan, A.~Purkayastha, S.~R. Clark, \href{https://doi.org/10.48550/arxiv.2402.05756}{Non-{M}arkovian quantum {M}pemba effect}, arXiv preprint arXiv:2402.05756 (2024).

\bibitem{chang2024imaginary}
W.-X. Chang, S.~Yin, S.-X. Zhang, Z.-X. Li, \href{https://doi.org/10.48550/arxiv.2409.06547}{Imaginary-time {M}pemba effect in quantum many-body systems}, arXiv preprint arXiv:2409.06547 (2024).

\bibitem{wang2024going}
Z.-M. Wang, S.~Wu, M.~S. Byrd, L.-A. Wu, \href{https://doi.org/10.48550/arXiv.2411.17197 }{Going beyond quantum {M}arkovianity and back to reality: {A}n exact {M}aster equation study}, arXiv preprint arXiv:2411.17197 (2024).

\bibitem{dong2411quantum}
J.~W. Dong, H.~F. Mu, M.~Qin, H.~T. Cui, \href{https://link.aps.org/doi/10.1103/PhysRevA.111.022215}{Quantum {M}pemba effect of localization in the dissipative mosaic model}, Phys. Rev. A 111 (2025) 022215.

\bibitem{furtado2024strong}
J.~Furtado, A.~C. Santos, \href{https://doi.org/10.48550/arxiv.2411.04545}{Strong Quantum {M}pemba Effect with Squeezed Thermal Reservoirs}, arXiv preprint arXiv:2411.04545 (2024).

\bibitem{longhi2024mpemba}
S.~Longhi, \href{https://arxiv.org/abs/2411.09589}{Mpemba effect and super-accelerated thermalization in the damped quantum harmonic oscillator}, arXiv preprint arXiv:2411.09589 (2024).

\bibitem{qian2024intrinsic}
D.~Qian, H.~Wang, J.~Wang, \href{https://doi.org/10.48550/arxiv.2411.18417}{Intrinsic Quantum {M}pemba Effect in {M}arkovian Systems and Quantum Circuits}, arXiv preprint arXiv:2411.18417 (2024).

\bibitem{longhi2024photonic}
S.~Longhi, \href{https://opg.optica.org/ol/abstract.cfm?URI=ol-49-18-5188}{Photonic {M}pemba effect}, Opt. Lett. 49~(18) (2024) 5188--5191.

\bibitem{longhi2024bosonic}
S.~Longhi, \href{https://doi.org/10.1063/5.0234457}{Bosonic {M}pemba effect with non-classical states of light}, APL Quantum 1~(4) (2024).

\bibitem{medina2024anomalous}
I.~Medina, O.~Culhane, F.~C. Binder, G.~T. Landi, J.~Goold, \href{https://doi.org/10.48550/arxiv.2412.13259}{Anomalous discharging of quantum batteries: {T}he ergotropic {M}pemba effect}, arXiv preprint arXiv:2412.13259 (2024).

\bibitem{carollo2021exponentially}
F.~Carollo, A.~Lasanta, I.~Lesanovsky, \href{https://doi.org/10.1103/PhysRevLett.127.060401}{Exponentially accelerated approach to stationarity in {M}arkovian open quantum systems through the {M}pemba effect}, Phys.~Rev. Lett. 127~(6) (2021) 060401.

\bibitem{kochsiek2022accelerating}
S.~Kochsiek, F.~Carollo, I.~Lesanovsky, \href{https://doi.org/10.1103/PhysRevA.106.012207}{Accelerating the approach of dissipative quantum spin systems towards stationarity through global spin rotations}, Phys.~Rev. A 106~(1) (2022) 012207.

\bibitem{ares2023entanglement}
F.~Ares, S.~Murciano, P.~Calabrese, \href{https://doi.org/10.1038/s41467-023-37747-8}{Entanglement asymmetry as a probe of symmetry breaking}, Nat. Commun. 14~(1) (2023) 2036.

\bibitem{rylands2024microscopic}
C.~Rylands, K.~Klobas, F.~Ares, P.~Calabrese, S.~Murciano, B.~Bertini, \href{https://doi.org/10.1103/PhysRevLett.133.010401}{Microscopic origin of the quantum {M}pemba effect in integrable systems}, Phys.~Rev.~Lett. 133~(1) (2024) 010401.

\bibitem{murciano2024entanglement}
S.~Murciano, F.~Ares, I.~Klich, P.~Calabrese, \href{https://doi.org/10.1088/1742-5468/ad17b4}{Entanglement asymmetry and quantum {M}pemba effect in the {XY} spin chain}, J. Stat. Mech: Theory Exp. 2024~(1) (2024) 013103.

\bibitem{caceffo2024entangled}
F.~Caceffo, S.~Murciano, V.~Alba, \href{https://doi.org/10.1088/1742-5468/ad4537}{Entangled multiplets, asymmetry, and quantum {M}pemba effect in dissipative systems}, J. Stat. Mech: Theory Exp. 2024~(6) (2024) 063103.

\bibitem{liu2024symmetry}
S.~Liu, H.-K. Zhang, S.~Yin, S.-X. Zhang, \href{https://link.aps.org/doi/10.1103/PhysRevLett.133.140405}{Symmetry Restoration and Quantum Mpemba Effect in Symmetric Random Circuits}, Phys. Rev. Lett. 133 (2024) 140405.

\bibitem{turkeshi2024quantum}
X.~Turkeshi, P.~Calabrese, A.~De~Luca, \href{https://doi.org/10.48550/arxiv.2405.14514}{Quantum {M}pemba Effect in Random Circuits}, arXiv preprint arXiv:2405.14514 (2024).

\bibitem{chalas2024multiple}
K.~Chalas, F.~Ares, C.~Rylands, P.~Calabrese, \href{https://dx.doi.org/10.1088/1742-5468/ad769c}{Multiple crossings during dynamical symmetry restoration and implications for the quantum Mpemba effect}, J. Stat. Mech.: Theory Exp. 2024~(10) (2024) 103101.

\bibitem{ares2024quantum}
F.~Ares, V.~Vitale, S.~Murciano, \href{https://doi.org/10.48550/arxiv.2405.08913}{The quantum {M}pemba effect in free-fermionic mixed states}, arXiv preprint arXiv:2405.08913 (2024).

\bibitem{foligno2024non}
A.~Foligno, P.~Calabrese, B.~Bertini, \href{https://link.aps.org/doi/10.1103/PRXQuantum.6.010324}{Nonequilibrium dynamics of charged dual-unitary circuits}, PRX Quantum 6 (2025) 010324.

\bibitem{yang2020non}
Z.-Y. Yang, J.-X. Hou, \href{https://doi.org/10.1103/PhysRevE.101.052106}{Non-{M}arkovian {M}pemba effect in mean-field systems}, Phys.~Rev. E 101~(5) (2020) 052106.

\bibitem{rylands2024dynamical}
C.~Rylands, E.~Vernier, P.~Calabrese, \href{https://doi.org/10.1088/1742-5468/ad97b3}{Dynamical symmetry restoration in the {H}eisenberg spin chain}, J. Stat. Mech.: Theory Exp. 2024~(12) (2024) 123102.

\bibitem{yamashika2024quenching}
S.~Yamashika, P.~Calabrese, F.~Ares, \href{https://doi.org/10.48550/arxiv.2410.14299}{Quenching from superfluid to free bosons in two dimensions: entanglement, symmetries, and quantum {M}pemba effect}, arXiv preprint arXiv:2410.14299 (2024).

\bibitem{yamashika2024entanglement}
S.~Yamashika, F.~Ares, P.~Calabrese, \href{https://doi.org/10.1103/PhysRevB.110.085126}{Entanglement asymmetry and quantum {M}pemba effect in two-dimensional free-fermion systems}, Phys. Rev. B 110~(8) (2024) 085126.

\bibitem{ferro2024non}
F.~Ferro, F.~Ares, P.~Calabrese, \href{https://doi.org/10.1088/1742-5468/ad138f}{Non-equilibrium entanglement asymmetry for discrete groups: {T}he example of the {XY} spin chain}, J. Stat. Mech: Theory Exp. 2024~(2) (2024) 023101.

\bibitem{klobas2024translation}
K.~Klobas, C.~Rylands, B.~Bertini, \href{https://doi.org/10.48550/arxiv.2406.04296}{Translation symmetry restoration under random unitary dynamics}, arXiv preprint arXiv:2406.04296 (2024).

\bibitem{banerjee2024entanglement}
T.~Banerjee, S.~Das, K.~Sengupta, \href{https://doi.org/10.48550/arxiv.2412.03654}{Entanglement asymmetry in periodically driven quantum systems}, arXiv preprint arXiv:2412.03654 (2024).

\bibitem{liu2408quantum}
S.~Liu, H.~Zhang, S.~Yin, S.~Zhang, H.~Yao, \href{https://doi.org/10.48550/arxiv.2408.07750}{Quantum {M}pemba effects in many-body localization systems (2024)}, arXiv preprint arXiv:2408.07750 39 (2024).

\bibitem{hu2018conformation}
C.~Hu, J.~Li, S.~Huang, H.~Li, C.~Luo, J.~Chen, S.~Jiang, L.~An, \href{https://doi.org/10.1021/acs.cgd.8b01250}{Conformation directed {M}pemba effect on polylactide crystallization}, Cryst. Growth Des. 18 (2018) 5757--5762.

\bibitem{ahn2016experimental}
Y.-H. Ahn, H.~Kang, D.-Y. Koh, H.~Lee, \href{https://doi.org/10.1007/s11814-016-0029-2}{Experimental verifications of {M}pemba-like behaviors of clathrate hydrates}, Korean J. Chem. Eng. 33 (2016) 1903--1907.

\bibitem{yang2022mpemba}
Z.-Y. Yang, J.-X. Hou, \href{https://doi.org/10.1103/PhysRevE.105.014119}{Mpemba effect of a mean-field system: {T}he phase transition time}, Phys.~Rev. E 105~(1) (2022) 014119.

\bibitem{patron2023non}
A.~Patr{\'o}n, B.~S{\'a}nchez-Rey, C.~Plata, A.~Prados, \href{https://doi.org/10.1209/0295-5075/acf7e5}{Non-equilibrium memory effects: Granular fluids and beyond}, Europhys. Lett. 143~(6) (2023) 61002.

\bibitem{torrente2019large}
A.~Torrente, M.~A. L{\'o}pez-Casta{\~n}o, A.~Lasanta, F.~V. Reyes, A.~Prados, A.~Santos, \href{https://doi.org/10.1103/PhysRevE.99.060901}{Large {M}pemba-like effect in a gas of inelastic rough hard spheres}, Phys.~Rev. E 99~(6) (2019) 060901.

\bibitem{megias2022mpemba}
A.~Meg{\'\i}as, A.~Santos, \href{https://doi.org/10.3389/fphy.2022.971671}{Mpemba-like effect protocol for granular gases of inelastic and rough hard disks}, Front. Phys. 10 (2022) 971671.

\bibitem{santos2020mpemba}
A.~Santos, A.~Prados, \href{https://doi.org/10.1063/5.0016243}{Mpemba effect in molecular gases under nonlinear drag}, Phys. Fluids 32~(7) (2020).

\bibitem{megias2022thermal}
A.~Meg{\'\i}as, A.~Santos, A.~Prados, \href{https://doi.org/10.1103/PhysRevE.105.054140}{Thermal versus entropic {M}pemba effect in molecular gases with nonlinear drag}, Phys.~Rev. E 105~(5) (2022) 054140.

\bibitem{takada2021mpemba}
S.~Takada, H.~Hayakawa, A.~Santos, \href{https://doi.org/10.1103/PhysRevE.103.032901}{Mpemba effect in inertial suspensions}, Phys.~Rev. E 103~(3) (2021) 032901.

\bibitem{mompo2021memory}
E.~Momp{\'o}, M.~L{\'o}pez-Casta{\~n}o, A.~Lasanta, F.~V. Reyes, A.~Torrente, \href{https://doi.org/10.1063/5.0050804}{Memory effects in a gas of viscoelastic particles}, Phys. Fluids 33~(6) (2021).

\bibitem{gijon2019paths}
A.~Gij{\'o}n, A.~Lasanta, E.~Hern{\'a}ndez, \href{https://doi.org/10.1103/PhysRevE.100.032103}{Paths towards equilibrium in molecular systems: {T}he case of water}, Phys.~Rev. E 100~(3) (2019) 032103.

\bibitem{brey1996homogeneous}
J.~J. Brey, M.~Ruiz-Montero, D.~Cubero, \href{https://doi.org/10.1103/physreve.54.3664}{Homogeneous cooling state of a low-density granular flow}, Phys.~Rev. E 54~(4) (1996) 3664.

\bibitem{gomez2021mpemba}
R.~G{\'o}mez~Gonz{\'a}lez, N.~Khalil, V.~Garz{\'o}, \href{https://doi.org/10.1063/5.0050530}{Mpemba-like effect in driven binary mixtures}, Phys. Fluids 33~(5) (2021).

\bibitem{brush1965kinetic}
S.~G. Brush, Kinetic Theory. 1. The Nature of Gases and of Heat, Pergamon Press, 1965.

\bibitem{cercignani1988boltzmann}
C.~Cercignani, \href{https://doi.org/10.1007/978-1-4612-1039-9}{The Boltzmann Equation and Its Applications}, Springer, 1988.

\bibitem{puglisi2014transport}
A.~Puglisi, \href{https://doi.org/10.1007/978-3-319-10286-3}{Transport and fluctuations in granular fluids: From Boltzmann equation to hydrodynamics, diffusion and motor effects}, Springer, 2014.

\bibitem{vega2015steady}
F.~Vega~Reyes, A.~Santos, \href{https://doi.org/10.1063/1.4934727}{Steady state in a gas of inelastic rough spheres heated by a uniform stochastic force}, Phys. Fluids 27~(11) (2015).

\bibitem{resibois1977classical}
P.~Resibois, Classical Kinetic Theory of Fluids, Wiley, 1977.

\bibitem{van1998velocity}
T.~Van~Noije, M.~Ernst, \href{https://doi.org/10.1007/s100350050009}{Velocity distributions in homogeneous granular fluids: the free and the heated case}, Granular Matter 1~(2) (1998) 57--64.

\bibitem{brey1998hydrodynamics}
J.~J. Brey, J.~W. Dufty, C.~S. Kim, A.~Santos, \href{https://doi.org/10.1103/physreve.58.4638}{Hydrodynamics for granular flow at low density}, Phys.~Rev. E 58~(4) (1998) 4638.

\bibitem{carnahan1969equation}
N.~F. Carnahan, K.~E. Starling, \href{https://doi.org/10.1063/1.1672048}{Equation of state for nonattracting rigid spheres}, J.~{C}hem.~{P}hys 51~(2) (1969) 635--636.

\bibitem{prados2010the}
A.~Prados, J.~J. Brey, \href{https://doi.org/10.1088/1742-5468/2010/02/P02009}{The {K}ovacs effect: {A} {M}aster equation analysis}, J. Stat. Mech 2010 (2010) P02009.

\bibitem{brilliantov1996model}
N.~V. Brilliantov, F.~Spahn, J.-M. Hertzsch, T.~P{\"o}schel, \href{https://doi.org/10.1103/PhysRevE.53.538}{Model for collisions in granular gases}, Phys.~Rev. E 53~(5) (1996) 5382.

\bibitem{biswas2022mpemba_a}
A.~Biswas, V.~Prasad, R.~Rajesh, \href{https://dx.doi.org/10.1209/0295-5075/ac2d54}{Mpemba effect in an anisotropically driven granular gas}, Europhys. Lett. 136~(4) (2022) 46001.

\bibitem{biswas2022mpemba_b}
A.~Biswas, V.~Prasad, R.~Rajesh, \href{https://doi.org/10.1007/s10955-022-02891-w}{Mpemba effect in anisotropically driven inelastic {M}axwell gases}, J. Stat. Phys. 186~(3) (2022) 45.

\bibitem{biswas2023mpemba_distance}
A.~Biswas, V.~Prasad, R.~Rajesh, \href{https://link.aps.org/doi/10.1103/PhysRevE.108.024902}{Mpemba effect in driven granular gases: {R}ole of distance measures}, Phys. Rev. E 108 (2023) 024902.

\bibitem{derrida1981random}
B.~Derrida, \href{https://doi.org/10.1103/PhysRevB.24.2613}{Random-energy model: An exactly solvable model of disordered systems}, Phys.~Rev. B 24~(5) (1981) 2613.

\bibitem{gross1984simplest}
D.~J. Gross, M.~M{\'e}zard, \href{https://doi.org/10.1016/0550-3213(84)90237-2}{The simplest spin glass}, Nucl. Phys. B 240~(4) (1984) 431--452.

\bibitem{mezard2009information}
M.~Mezard, A.~Montanari, \href{https://doi.org/10.1093/acprof:oso/9780198570837.001.0001}{{I}nformation, {P}hysics, and {C}omputation}, Oxford {U}niv. {P}ress, 2009.

\bibitem{metropolis1953equation}
N.~Metropolis, A.~W. Rosenbluth, M.~N. Rosenbluth, A.~H. Teller, E.~Teller, \href{https://doi.org/10.1016/j.physd.2010.10.003}{Equation of state calculations by fast computing machines}, J.~{C}hem.~{P}hys. 21~(6) (1953) 1087--1092.

\bibitem{sokal1997monte}
A.~Sokal, \href{https://doi.org/10.1007/978-1-4899-0319-8_6}{Monte {C}arlo Methods in Statistical Mechanics: Foundations and New Algorithms}, in: Functional Integration: Basics and Applications, Springer, 1997, pp. 131--192.

\bibitem{krauth2007introduction}
W.~Krauth, \href{https://doi.org/10.1007/BFb0105457}{Introduction to {M}onte {C}arlo Algorithms}, in: Advances in Computer Simulation: Lectures Held at the E{\"o}tv{\"o}s Summer School in Budapest, Hungary, 16--20 July 1996, Springer, 2007, pp. 1--35.

\bibitem{turitsyn2011irreversible}
K.~S. Turitsyn, M.~Chertkov, M.~Vucelja, \href{https://doi.org/10.1016/j.physd.2010.10.003}{Irreversible {M}onte {C}arlo algorithms for efficient sampling}, Physica D 240~(4-5) (2011) 410--414.

\bibitem{vucelja2016lifting}
M.~Vucelja, \href{https://doi.org/10.1119/1.4961596}{Lifting---a nonreversible {M}arkov chain {M}onte {C}arlo algorithm}, Am. J. Phys. 84~(12) (2016) 958--968.

\bibitem{amorim2023predicting}
F.~Amorim, J.~Wisely, N.~Buckley, C.~DiNardo, D.~Sadasivan, \href{https://link.aps.org/doi/10.1103/PhysRevE.108.024137}{Predicting the {M}pemba effect using machine learning}, Phys.~Rev. E 108~(2) (2023) 024137.

\bibitem{meibohm2024exponential}
J.~Meibohm, S.~H.~L. Klapp, \href{https://link.aps.org/doi/10.1103/PhysRevLett.134.087101}{Exponential Change of Relaxation Rate by Quenched Disorder}, Phys. Rev. Lett. 134 (2025) 087101.

\bibitem{luo2016role}
C.~Luo, J.-U. Sommer, \href{https://pubs.acs.org/doi/10.1021/acsmacrolett.5b00668}{Role of Thermal History and Entanglement Related Thickness Selection in Polymer Crystallization}, ACS MacroLett. 5 (2016) 30--34.

\bibitem{liu2023mpemba}
J.~Liu, J.~Li, B.~Liu, I.~W. Hamley, S.~Jiang, \href{https://doi.org/10.1039/D3SM00309D}{Mpemba effect in crystallization of polybutene-1}, Soft Matter 19 (2023) 3337--3347.

\bibitem{chorazewski2023the}
M.~Chorazewski, M.~Wasiak, A.~V. Sychev, V.~I. Korotkovskii, E.~B. Postnikov, \href{https://doi.org/10.1007/s10953-023-01268-1}{The curious case of 1-ethylpyridinium triflate: {I}onic liquid exhibiting the {M}pemba effect}, J. Solution Chem. (2023).

\bibitem{ferbonink2020stochastic}
G.~F. Ferbonink, T.~S. Rodrigues, P.~H.~C. Camargo, R.~Q. de~Albuquerque, R.~A. Nome, \href{https://doi.org/10.26434/chemrxiv.12846695.v1}{Stochastic thermodynamics analysis of ultrafast {AgAu} nanoshell dynamics in the nonlinear response regime}, chemRxiv: 12846696.v1 (2020).

\bibitem{chaddah2010overtaking}
P.~Chaddah, S.~Dash, K.~Kumar, A.~Banerjee, \href{https://doi.org/10.48550/arXiv.1011.3598}{Overtaking while approaching equilibrium}, arXiv:1011.3598 (2010).

\bibitem{bechhoefer2021fresh}
J.~Bechhoefer, A.~Kumar, R.~Ch{\'e}trite, \href{https://doi.org/10.1038/s42254-021-00349-8}{A fresh understanding of the {M}pemba effect}, Nat. Rev. Phys. 3 (2021) 534--535.

\bibitem{cohen2005control}
A.~E. Cohen, \href{https://link.aps.org/doi/10.1103/PhysRevLett.94.118102}{Control of nanoparticles with arbitrary two-dimensional force fields}, Phys. Rev. Lett. 94 (2005) 118102.

\bibitem{jun2012virtual}
Y.~Jun, J.~Bechhoefer, \href{https://link.aps.org/doi/10.1103/PhysRevE.86.061106}{Virtual potentials for feedback traps}, Phys. Rev. E 86 (2012) 061106.

\bibitem{burt2021demonstration}
E.~Burt, J.~Prestage, R.~Tjoelker, D.~Enzer, D.~Kuang, D.~Murphy, D.~Robison, J.~Seubert, R.~Wang, T.~Ely, \href{https://doi.org/10.1038/s41586-021-03571-7}{Demonstration of a trapped-ion atomic clock in space}, Nature 595~(7865) (2021) 43--47.

\bibitem{kotler2011single}
S.~Kotler, N.~Akerman, Y.~Glickman, A.~Keselman, R.~Ozeri, \href{https://doi.org/10.1038/nature10010}{Single-ion quantum lock-in amplifier}, Nature 473~(7345) (2011) 61--65.

\bibitem{bruzewicz2019trapped}
C.~D. Bruzewicz, J.~Chiaverini, R.~McConnell, J.~M. Sage, \href{https://doi.org/10.1063/1.5088164}{Trapped-ion quantum computing: {P}rogress and challenges}, Appl. Phys. Rev. 6~(2) (2019).

\bibitem{leibfried2003quantum}
D.~Leibfried, R.~Blatt, C.~Monroe, D.~Wineland, \href{https://doi.org/10.1103/RevModPhys.75.281}{Quantum dynamics of single trapped ions}, Rev.~Mod.~Phys. 75~(1) (2003) 281.

\bibitem{duan2010colloquium}
L.-M. Duan, C.~Monroe, \href{https://doi.org/10.1103/RevModPhys.82.1209}{Colloquium: {Q}uantum networks with trapped ions}, Rev.~Mod.~Phys. 82~(2) (2010) 1209--1224.

\bibitem{haffner2008quantum}
H.~H{\"a}ffner, C.~F. Roos, R.~Blatt, \href{https://doi.org/10.1016/j.physrep.2008.09.003}{Quantum computing with trapped ions}, Phys. Rep. 469~(4) (2008) 155--203.

\bibitem{edo2024study}
M.~E. Edo, L.-A. Wu, \href{https://doi.org/10.48550/arXiv.2403.14630}{Study on quantum thermalization from thermal initial states in a superconducting quantum computer}, arXiv preprint arXiv:2403.14630 (2024).

\bibitem{monroe2021programmable}
C.~Monroe, W.~C. Campbell, L.-M. Duan, Z.-X. Gong, A.~V. Gorshkov, P.~W. Hess, R.~Islam, K.~Kim, N.~M. Linke, G.~Pagano, P.~Richerme, C.~Senko, N.~Y. Yao, \href{https://doi.org/10.1103/RevModPhys.93.025001}{Programmable quantum simulations of spin systems with trapped ions}, Rev.~Mod.~Phys. 93~(2) (2021) 025001.

\bibitem{baity2014janus}
M.~Baity-Jesi, R.~Baños, A.~Cruz, L.~Fernandez, J.~Gil-Narvion, A.~Gordillo-Guerrero, D.~Iñiguez, A.~Maiorano, F.~Mantovani, E.~Marinari, V.~Martin-Mayor, J.~Monforte-Garcia, A.~{Muñoz Sudupe}, D.~Navarro, G.~Parisi, S.~Perez-Gaviro, M.~Pivanti, F.~Ricci-Tersenghi, J.~Ruiz-Lorenzo, S.~Schifano, B.~Seoane, A.~Tarancon, R.~Tripiccione, D.~Yllanes, \href{https://doi.org/10.1016/j.cpc.2013.10.019}{Janus {II}: {A} new generation application-driven computer for spin-system simulations}, Comput. Phys. Commun. 185~(2) (2014) 550--559.

\bibitem{teza2022far}
R.~Yaacoby, O.~Raz, G.~Teza, \href{https://doi.org/10.48550/arXiv.2203.11644}{Far from equilibrium relaxation in the weak coupling limit}, arXiv preprint arXiv:2203.11644 (2022).

\bibitem{chatterjee2023mpemba}
S.~Chatterjee, S.~Ghosh, N.~Vadakkayil, T.~Paul, S.~K. Singha, S.~K. Das, \href{https://doi.org/10.48550/arXiv.2309.03709 }{Mpemba Effect and Superuniversality across Orders of Magnetic Phase Transition}, arXiv preprint arXiv:2309.03709 (2023).

\bibitem{das2023perspectives}
S.~K. Das, \href{https://doi.org/10.1021/acs.langmuir.3c00668}{Perspectives on a Few Puzzles in Phase Transformations: {W}hen Should the Farthest Reach the Earliest?}, Langmuir 39~(31) (2023) 10715--10723.

\bibitem{chatterjee2024mpemba}
S.~Chatterjee, S.~Ghosh, N.~Vadakkayil, T.~Paul, S.~K. Singha, S.~K. Das, \href{https://link.aps.org/doi/10.1103/PhysRevE.110.L012103}{Mpemba effect in pure spin systems: {A} universal picture of the role of spatial correlations at initial states}, Phys.~Rev. E 110~(1) (2024) L012103.

\bibitem{ghosh2024simulations}
S.~Ghosh, P.~Pathak, S.~Chatterjee, S.~K. Das, \href{https://doi.org/10.48550/arXiv.2407.06954}{Simulations of {M}pemba Effect in water, {L}ennard-{J}ones and {I}sing Models: {M}etastability vs Critical Fluctuations}, arXiv preprint arXiv:2407.06954 (2024).

\bibitem{barba2024anomalous}
D.~Barba-Gonz{\'a}lez, C.~Albertus, M.~P{\'e}rez-Garc{\'\i}a, \href{https://doi.org/10.48550/arXiv.2406.01700}{Anomalous thermal relaxation in {Y}ukawa fluids}, arXiv preprint arXiv:2406.01700 (2024).

\bibitem{zhang2013mpemba}
X.~Zhang, Y.~Huang, Z.~Ma, C.~Q. Sun, \href{https://doi.org/10.1021/jp407836n}{Mpemba Paradox Revisited: Numerical Reinforcement}, J. Phys. Chem. B 117 (2013) 13639--45.

\bibitem{allen2017computer}
M.~P. Allen, D.~J. Tildesley, \href{https://doi.org/10.1093/oso/9780198803195.001.0001}{Computer Simulation of Liquids}, Oxford {U}niv. {P}ress, 2017.

\bibitem{bird1994molecular}
G.~A. Bird, \href{https://doi.org/10.1093/oso/9780198561958.001.0001}{Molecular Gas Dynamics and the Direct simulation of Gas Flows}, Oxford {U}niv. {P}ress, 1994.

\bibitem{brey1999direct}
J.~J. Brey, M.~Ruiz-Montero, \href{https://doi.org/10.1016/S0010-4655(99)00331-8}{Direct {M}onte {C}arlo simulation of dilute granular flow}, Comput. Phys. Commun. 121 (1999) 278--283.

\bibitem{lavine2018fundamentals}
A.~S. Lavine, F.~P. Incropera, D.~P. DeWitt, T.~L. Bergman, \href{http://bcs.wiley.com/he-bcs/Books?action=index&itemId=1119320429&bcsId=10769}{Fundamentals of Heat and Mass Transfer}, 8th Edition, Wiley, 2018.

\bibitem{yang2024exploring}
X.~Yang, Y.~Shan, Y.~Hong, Z.~Zhang, S.~Liu, X.~Yan, X.~Gong, G.~Zhang, Z.~Yang, \href{https://doi.org/10.1039/D3MH01869E}{Exploring the {M}pemba effect: A universal ice pressing enables porous ceramics}, Mater. Horiz. 11 (2024) 1899--1907.

\bibitem{bechhoefer2021control}
J.~Bechhoefer, \href{https://doi.org/10.1017/9780511734809}{Control Theory for Physicists}, Cambridge Univ. Press, 2021.

\bibitem{nocedal2006numerical}
J.~Nocedal, S.~J. Wright, \href{https://doi.org/10.1007/978-0-387-40065-5}{Numerical Optimization}, 2nd Edition, Springer Series in Operations Research, Springer, 2006.

\bibitem{pottier2023accelerating}
B.~Pottier, C.~A. Plata, E.~Trizac, D.~Gu{\'e}ry-Odelin, L.~Bellon, \href{https://link.aps.org/doi/10.1103/PhysRevApplied.19.034072}{Accelerating the Heat Diffusion: {F}ast Thermal Relaxation of a Microcantilever}, Phys. Rev. App. 19 (2023) 034072.

\bibitem{walker2023optimal}
M.~R. Walker, S.~Bera, M.~Vucelja, \href{https://doi.org/10.48550/arXiv.2307.16103}{Optimal transport and anomalous thermal relaxations}, arXiv preprint arXiv:2307.16103 (2023).

\bibitem{bao2023designing}
R.~Bao, Z.~Cao, J.~Zheng, Z.~Hou, \href{https://doi.org/10.1103/PhysRevResearch.5.043066}{Designing autonomous {M}axwell's demon via stochastic resetting}, Phys. Rev. Research 5~(4) (2023) 043066.

\bibitem{maxwell1871theory}
J.~Clerk~Maxwell, Theory of Heat, Longmans, Green, and Co., 1871.

\bibitem{landauer1961irreversibility}
R.~Landauer, \href{https://ieeexplore.ieee.org/document/5392446}{Irreversibility and heat generation in the computing process}, IBM J. Res. Dev. 5~(3) (1961) 183--191.

\bibitem{bennett1982the}
C.~H. Bennett, \href{https://doi.org/10.1007/BF02084158}{The thermodynamics of computation---a review}, Int. J. Theor. Phys. 21 (1982) 905--940.

\bibitem{penrose1970foundations}
O.~Penrose, Foundations of Statistical Mechanics: A Deductive Treatment, Pergamon Press, 1970.

\bibitem{parrondo2015thermodynamics}
J.~M.~R. Parrondo, J.~M. Horowitz, T.~Sagawa, \href{https://www.nature.com/articles/nphys3230}{Thermodynamics of information}, Nat. Phys. 11 (2015) 131--139.

\bibitem{mandal2012work}
D.~Mandal, C.~Jarzynski, \href{https://www.pnas.org/doi/full/10.1073/pnas.1204263109}{Work and information processing in a solvable model of {M}axwell's demon}, Proc. Natl. Acad. Sci. USA 109~(29) (2012) 11641--11645.

\bibitem{lin2022power}
J.~Lin, K.~Li, J.~He, J.~Ren, J.~Wang, \href{https://doi.org/10.1103/PhysRevE.105.014104}{Power statistics of {O}tto heat engines with the {M}pemba effect}, Phys.~Rev. E 105~(1) (2022) 014104.

\bibitem{liu2024speeding}
D.~Liu, J.~Yuan, H.~Ruan, Y.~Xu, S.~Luo, J.~He, X.~He, Y.~Ma, J.~Wang, \href{https://link.aps.org/doi/10.1103/PhysRevA.110.042218}{Speeding up quantum heat engines by the {Mpemba} effect}, Phys. Rev. A 110 (2024) 042218.

\bibitem{chen2024boosting}
J.-F. Chen, K.~S. Rai, P.~Emonts, D.~Farina, M.~P{\l}odzie{\'n}, P.~Grzybowski, M.~Lewenstein, J.~Tura, \href{https://doi.org/10.48550/arXiv.2411.03420}{Boosting thermalization of classical and quantum many-body systems}, arXiv preprint arXiv:2411.03420 (2024).

\bibitem{verstraete2009quantum}
F.~Verstraete, M.~M. Wolf, J.~Ignacio~Cirac, \href{https://doi.org/10.1038/nphys1342}{Quantum computation and quantum-state engineering driven by dissipation}, Nat. Phys. 5~(9) (2009) 633--636.

\bibitem{harrington2022engineered}
P.~M. Harrington, E.~J. Mueller, K.~W. Murch, \href{https://doi.org/10.1038/s42254-022-00494-8}{Engineered dissipation for quantum information science}, Nat. Rev. Phys. 4~(10) (2022) 660--671.

\bibitem{pagare2024mpemba}
A.~Pagare, Z.~Lu, \href{https://link.aps.org/doi/10.1103/PRXLife.2.043019}{Mpemba-Like Sensory Withdrawal Effect}, PRX Life 2~(4) (2024) 043019.

\bibitem{hatakeyama2024enzymatic}
T.~S. Hatakeyama, \href{https://arxiv.org/abs/2411.08058}{Enzymatic {M}pemba Effect: Slowing of biochemical reactions by increasing enzyme concentration}, arXiv preprint arXiv:2411.08058 (2024).

\bibitem{michaelis1913kinetik}
L.~Michaelis, M.~L. Menten, Die kinetik der invertinwirkung, Biochemische Zeitschrift 49 (1913) 333.

\bibitem{kolomeisky2007molecular}
A.~B. Kolomeisky, M.~E. Fisher, \href{https://iopscience.iop.org/article/10.1088/0953-8984/25/46/463101}{Molecular motors: {A} theorist's perspective}, Annu. Rev. Phys. Chem. 58~(1) (2007) 675--695.

\bibitem{kolomeisky2013motor}
A.~B. Kolomeisky, \href{https://dx.doi.org/10.1088/0953-8984/25/46/463101}{Motor proteins and molecular motors: {H}ow to operate machines at the nanoscale}, J. Condens. Matter Phys. 25~(46) (2013) 463101.

\bibitem{teza2020rate}
G.~Teza, S.~Iubini, M.~Baiesi, A.~L. Stella, C.~Vanderzande, \href{https://doi.org/10.1016/j.physa.2019.123176}{Rate dependence of current and fluctuations in jump models with negative differential mobility}, Physica A 552 (2020) 123176, tributes of Non-equilibrium Statistical Physics.

\bibitem{remlein2021optimality}
B.~Remlein, U.~Seifert, \href{https://doi.org/10.1103/PhysRevE.103.L050105}{Optimality of nonconservative driving for finite-time processes with discrete states}, Phys. Rev. E 103~(5) (2021) L050105.

\bibitem{katz1983phase}
S.~Katz, J.~L. Lebowitz, H.~Spohn, \href{https://link.aps.org/doi/10.1103/PhysRevB.28.1655}{Phase transitions in stationary nonequilibrium states of model lattice systems}, Phys. Rev. B 28 (1983) 1655--1658.

\bibitem{degunther2022anomalous}
J.~Deg{\"u}nther, U.~Seifert, \href{https://iopscience.iop.org/article/10.1209/0295-5075/ac8573}{Anomalous relaxation from a non-equilibrium steady state: An isothermal analog of the {M}pemba effect}, Europhys. Lett. 139~(4) (2022) 41002.

\bibitem{bechinger2016active}
C.~Bechinger, R.~D. Leonardo, H.~L{\"o}wen, C.~Reichhardt, G.~Volpe, G.~Volpe, \href{https://doi.org/10.1103/RevModPhys.88.045006}{Active particles in complex and crowded environments}, Rev. Mod. Phys. 88 (2016) 045006.

\bibitem{schwarzendahl2022anomalous}
F.~J. Schwarzendahl, H.~L{\"o}wen, \title{https://journals.aps.org/prl/abstract/10.1103/PhysRevLett.129.138002}{Anomalous cooling and overcooling of active colloids}, Phys. Rev. Lett. 129~(13) (2022) 138002.

\bibitem{biswas2024mpemba}
A.~Biswas, A.~Pal, \href{https://doi.org/10.48550/arXiv.2403.17547}{Mpemba effect on non-equilibrium active {M}arkov chains}, arXiv preprint arXiv:2403.17547 (2024).

\bibitem{biswas2024mpemba1}
A.~Biswas, A.~Pal, \href{https://doi.org/10.48550/arXiv.2411.02652}{Mpemba effect in the relaxation of an active {B}rownian particle in a trap without metastable states}, arXiv preprint arXiv:2411.02652 (2024).

\bibitem{evans2011diffusion}
M.~R. Evans, S.~N. Majumdar, \href{https://link.aps.org/doi/10.1103/PhysRevLett.106.160601}{Diffusion with stochastic resetting}, Phys.~Rev. Lett. 106~(16) (2011) 160601.

\bibitem{evans2020stochastic}
M.~R. Evans, S.~N. Majumdar, G.~Schehr, \href{https://iopscience.iop.org/article/10.1088/1751-8121/ab7cfe}{Stochastic resetting and applications}, J. Phys. A: Math. Theor. 53~(19) (2020) 193001.

\bibitem{kundu2024stochastic}
A.~Kundu, S.~Reuveni, \href{https://iopscience.iop.org/article/10.1088/1751-8121/ad1e1b/meta}{Stochastic Resetting: Theory and Applications}, J. Phys. A: Math. Theor 57 (2024) 060301.

\bibitem{busiello2021inducing}
D.~M. Busiello, D.~Gupta, A.~Maritan, \href{https://iopscience.iop.org/article/10.1088/1367-2630/ac2922}{Inducing and optimizing {M}arkovian {M}pemba effect with stochastic reset}, New J. Phys. 23~(10) (2021) 103012.

\bibitem{bao2022accelerating}
R.~Bao, Z.~Hou, \href{https://arxiv.org/abs/2212.11170}{Accelerating relaxation in {M}arkovian open quantum systems through quantum reset processes}, arXiv preprint arXiv:2212.11170 (2022).

\bibitem{krauth2006statistical}
W.~Krauth, {S}tatistical {M}echanics: {A}lgorithms and {C}omputations, Oxford {U}niv. {P}ress, 2006.

\bibitem{neal2011mcmc}
R.~M. Neal, {MCMC} {U}sing {H}amiltonian {D}ynamics, in: Handbook of Markov Chain Monte Carlo, Chapman and Hall/CRC, 2011, pp. 113--162.

\bibitem{chen1999lifting}
F.~Chen, L.~Lov{\'a}sz, I.~Pak, \href{https://doi.org/10.1145/301250.301315}{Lifting {M}arkov chains to speed up mixing}, in: Proceedings of the thirty-first annual ACM symposium on Theory of computing, 1999, pp. 275--281.

\bibitem{diaconis2000analysis}
P.~Diaconis, S.~Holmes, R.~M. Neal, \href{http://www.jstor.org/stable/2667319}{Analysis of a nonreversible {M}arkov chain sampler}, Ann. Appl. Probab. (2000) 726--752.

\bibitem{zhao2022nonreversible}
H.~Zhao, M.~Vucelja, \href{https://doi.org/10.3389/fphy.2021.782156}{Nonreversible {M}arkov Chain {M}onte {C}arlo algorithm for efficient generation of self-avoiding walks}, Front. Phys. 9 (2022) 782156.

\bibitem{ballard2009replica}
A.~J. Ballard, C.~Jarzynski, \href{https://doi.org/10.1073/pnas.0900406106}{Replica exchange with nonequilibrium switches}, Proc. Natl. Acad. Sci. 106~(30) (2009) 12224--12229.

\bibitem{machta2009strengths}
J.~Machta, \href{https://link.aps.org/doi/10.1103/PhysRevE.80.056706}{Strengths and weaknesses of parallel tempering}, Phys. Rev. E 80~(5) (2009) 056706.

\bibitem{bierkens2019zig}
J.~Bierkens, P.~Fearnhead, G.~Roberts, \href{https://doi.org/10.1214/18-AOS1715}{The zig-zag process and super-efficient sampling for {B}ayesian analysis of big data}, Ann. Stat. 47~(3) (2019) 1288--1320.

\bibitem{krauth2021event}
W.~Krauth, \href{https://doi.org/10.3389/fphy.2021.663457}{Event-chain {M}onte {C}arlo: {F}oundations, applications, and prospects}, Front. Phys. 9 (2021) 663457.

\bibitem{song2017nice}
J.~Song, S.~Zhao, S.~Ermon, \href{https://proceedings.neurips.cc/paper_files/paper/2017/file/2417dc8af8570f274e6775d4d60496da-Paper.pdf}{A-{NICE}-{MC}: Adversarial training for {MCMC}}, Advances in neural information processing systems 30 (2017).

\bibitem{wang2018meta}
T.~Wang, Y.~Wu, D.~Moore, S.~J. Russell, \href{https://proceedings.neurips.cc/paper_files/paper/2018/file/584b98aac2dddf59ee2cf19ca4ccb75e-Paper.pdf}{Meta-learning {MCMC} proposals}, Advances in neural information processing systems 31 (2018).

\bibitem{gueryodelin2023driving}
D.~Gu{\'e}ry-Odelin, C.~Jarzynski, C.~A. Plata, A.~Prados, E.~Trizac, \href{https://iopscience.iop.org/article/10.1088/1361-6633/acacad}{Driving rapidly while remaining in control: {C}lassical shortcuts from {H}amiltonian to stochastic dynamics}, Rep. Prog. Phys. 86 (2023) 035902.

\bibitem{patra2017shortcuts}
A.~Patra, C.~Jarzynski, \href{https://iopscience.iop.org/article/10.1088/1367-2630/aa924c}{Shortcuts to adiabaticity using flow fields}, New J. Phys. 19~(12) (2017) 125009.

\bibitem{martinez2016engineered}
I.~A. Mart{\'\i}nez, A.~Petrosyan, D.~Gu{\'e}ry-Odelin, E.~Trizac, S.~Ciliberto, \href{https://doi.org/10.1038/nphys3758}{Engineered swift equilibration of a {B}rownian particle}, Nat. Phys. 12~(9) (2016) 843--846.

\bibitem{patron2022thermal}
A.~Patr{\'o}n, A.~Prados, C.~A. Plata, \href{https://doi.org/10.1140/epjp/s13360-022-03150-3}{Thermal brachistochrone for harmonically confined {B}rownian particles}, Eur. Phys. J. Plus 137 (2022) 1011.

\bibitem{boyer2011history}
C.~B. Boyer, U.~C. Merzbach, A History of Mathematics, 3rd Edition, John Wiley \& Sons, Inc., 2011.

\bibitem{patron2024minimum}
A.~Patr{\'o}n, C.~A. Plata, A.~Prados, \href{https://doi.org/10.48550/arXiv.2407.21110}{Minimum time connection between non-equilibrium steady states: the {B}rownian gyrator}, J. Phys. A: Math. Theor. 57 (2024) 495004.

\bibitem{boubakour2024dynamical}
M.~Boubakour, S.~Endo, T.~Fogarty, T.~Busch, \href{https://doi.org/10.48550/arXiv.2401.11659}{Dynamical invariant based shortcut to equilibration}, arXiv preprint arXiv:2401.11659 (2024).

\bibitem{chittari2023geometric}
S.~S. Chittari, Z.~Lu, \href{https://doi.org/10.1063/5.0157846}{Geometric approach to nonequilibrium hasty shortcuts}, J.~{C}hem.~{P}hys 159~(8) (2023).

\bibitem{kovacs1964transition}
A.~J. Kovacs, \href{https://doi.org/10.1007/BFb0050366}{Transition vitreuse dans les polym{\`e}res amorphes. {E}tude ph{\'e}nom{\'e}nologique}, Fortschr.~Hochpolymeren-Forsch.~(Adv.~Polymer~Sci.) 3 (1964) 394--507.

\bibitem{prados2014kovacs}
A.~Prados, E.~Trizac, \href{https://link.aps.org/doi/10.1103/PhysRevLett.112.198001}{Kovacs-like memory effect in driven granular gases}, Phys. Rev. Lett. 112~(19) (2014) 198001.

\bibitem{kovacs1979isobaric}
A.~J. Kovacs, J.~J. Aklonis, J.~M. Hutchinson, A.~R. Ramos, \href{https://doi.org/10.1002/pol.1979.180170701Cit}{Isobaric volume and enthalpy recovery of glasses. {II. A} transparent multiparameter theory}, J. Polym. Sci. B 17 (1979) 1097--1162.

\bibitem{ruiz2014kovacs}
M.~Ruiz-Garc{\'\i}a, A.~Prados, \href{https://link.aps.org/doi/10.1103/PhysRevE.89.012140}{Kovacs effect in the one-dimensional {I}sing model: {A} linear response analysis}, Phys. Rev. E 89~(1) (2014) 012140.

\bibitem{militaru2021kovacs}
A.~Militaru, A.~Lasanta, M.~Frimmer, L.~L. Bonilla, L.~Novotny, R.~A. Rica, \href{https://link.aps.org/doi/10.1103/PhysRevLett.127.130603}{Kovacs memory effect with an optically levitated nanoparticle}, Phys. Rev. Lett. 127 (2021) 130603.

\bibitem{mossa2004crossover}
S.~Mossa, F.~Sciotino, \href{https://link.aps.org/doi/10.1103/PhysRevLett.92.045504}{Crossover (or {K}ovacs) Effect in an Aging Molecular Liquid}, Phys. Rev. Lett. 92 (2004) 045504.

\bibitem{berthier2002geometrical}
L.~Berthier, J.-P. Bouchaud, \href{https://link.aps.org/doi/10.1103/PhysRevB.66.054404}{Geometrical aspects of aging and rejuvenation in the {I}sing spin glass: {A} numerical study}, Phys. Rev. B 66~(5) (2002) 054404.

\bibitem{lahini2017nonmonotonic}
Y.~Lahini, O.~Gottesman, A.~Amir, S.~M. Rubinstein, \href{https://link.aps.org/doi/10.1103/PhysRevLett.118.085501}{Nonmonotonic Aging and Memory Retention in Disordered Mechanical Systems}, Phys. Rev. Lett. 118 (2017) 085501.

\bibitem{dillavou2018nonmonotonic}
S.~Dillavou, S.~M. Rubinstein, \href{https://link.aps.org/doi/10.1103/PhysRevLett.120.224101}{Nonmonotonic Aging and Memory in a Frictional Interface}, Phys. Rev. Lett. 120 (2018) 224101.

\bibitem{he2019nonmonotonic}
Y.~He, D.~Jiang, J.~Chen, R.~Liu, J.~Fan, X.~Jiang, \href{https://doi.org/10.1007/s00603-018-1718-4}{Non-monotonic relaxation and memory effect of rock salt}, Rock Mech. and Rock Eng. 52 (2019) 2471--2479.
\newblock \href {https://doi.org/10.1007/s00603-018-1718-4} {\path{doi:10.1007/s00603-018-1718-4}}.
\newline\urlprefix\url{https://doi.org/10.1007/s00603-018-1718-4}

\bibitem{mandal2021memory}
R.~Mandal, D.~Tapias, P.~Sollich, \href{https://link.aps.org/doi/10.1103/PhysRevResearch.3.043153}{Memory in non-monotonic stress response of an athermal disordered solid}, Phys. Rev. Res. 3 (2021) 043153.

\bibitem{morgan2020glassy}
I.~L. Morgan, R.~Avinery, G.~Rahamim, R.~Beck, O.~A. Saleh, \href{https://link.aps.org/doi/10.1103/PhysRevLett.125.058001}{Glassy dynamics and memory effects in an intrinsically disordered protein construct}, Phys. Rev. Lett. 125 (2020) 058001.

\bibitem{josserand2000memory}
C.~Josserand, A.~V. Tkachenko, D.~M. Mueth, H.~M. Jaeger, \href{https://link.aps.org/doi/10.1103/PhysRevLett.85.3632}{Memory Effects in Granular Materials}, Phys. Rev. Lett. 85 (2000) 3632--3635.

\bibitem{trizac2014memory}
E.~Trizac, A.~Prados, \href{https://link.aps.org/doi/10.1103/PhysRevE.90.012204}{Memory effect in uniformly heated granular gases}, Phys. Rev. E 90~(1) (2014) 012204.

\bibitem{lasanta2019emergence}
A.~Lasanta, F.~Vega~Reyes, A.~Prados, A.~Santos, \href{https://dx.doi.org/10.1088/1367-2630/ab0a7b}{On the emergence of large and complex memory effects in nonequilibrium fluids}, New J. Phys. 21 (2019) 033042.

\bibitem{kursten2017giant}
R.~K{\"u}rsten, V.~Sushkov, T.~Ihle, \href{https://link.aps.org/doi/10.1103/PhysRevLett.119.188001}{Giant {K}ovacs-Like Memory Effect for Active Particles}, Phys. Rev. Lett. 119 (2017) 188001.

\bibitem{peyrard2020memory}
M.~Peyrard, J.-L. Garden, \href{https://link.aps.org/doi/10.1103/PhysRevE.102.052122}{Memory effects in glasses: {I}nsights into the thermodynamics of out-of-equilibrium systems revealed by a simple model of the {K}ovacs effect}, Phys. Rev. E 102 (2020) 052122.

\bibitem{santos2024mpemba}
A.~Santos, \href{https://link.aps.org/doi/10.1103/PhysRevE.109.044149}{{Mpemba} meets {Newton}: Exploring the {Mpemba} and {Kovacs} effects in the time-delayed cooling law}, Phys. Rev. E 109 (2024) 044149.

\bibitem{martinez2013effective}
I.~A. Mart{\'\i}nez, E.~Rold{\'a}n, J.~M.~R. Parrondo, D.~Petrov, \href{https://link.aps.org/doi/10.1103/PhysRevE.87.032159}{Effective heating to several thousand kelvins of an optically trapped sphere in a liquid}, Phys. Rev. E 87 (2013) 032159.

\bibitem{bouchbinder2010nonequilibrium}
E.~Bouchbinder, J.~S. Langer, \href{http://dx.doi.org/10.1039/C001388A}{Nonequilibrium thermodynamics of the {K}ovacs effect}, Soft Matter 6 (2010) 3065--3073.

\bibitem{godreche2022the}
C.~Godr{\`e}che, J.-M. Luck, \href{https://dx.doi.org/10.1088/1751-8121/aca84c}{The {G}lauber-{I}sing chain under low-temperature protocols}, J. Phys. A: Math. Theor. 55 (2022) 495001.

\bibitem{keim2019memory}
N.~C. Keim, J.~D. Paulsen, Z.~\v{Z}erav\v{c}i\'c, S.~Sastry, S.~R. Nagel, \href{https://link.aps.org/doi/10.1103/RevModPhys.91.035002}{Memory formation in matter}, Rev. Mod. Phys. 91 (2019) 035002.

\bibitem{parravicini2012equalizing}
J.~Parravicini, A.~J. Agranat, C.~Conti, E.~DelRe, \href{https://doi.org/10.1063/1.4751847}{Equalizing disordered ferroelectrics for diffraction cancellation}, Appl. Phys. Lett. 101 (2012) 111104.

\bibitem{lapolla2020faster}
A.~Lapolla, A.~Godec, \href{https://link.aps.org/doi/10.1103/PhysRevLett.125.110602}{Faster uphill relaxation in thermodynamically equidistant temperature quenches}, Phys.~Rev. Lett. 125~(11) (2020) 110602.

\bibitem{degroot1984nonequilibrium}
S.~R. de~Groot, P.~Mazur, Non-equilibrium {T}hermodynamics, Dover, 1984.

\bibitem{peliti2021stochastic}
L.~Peliti, S.~Pigolotti, Stochastic {T}hermodynamics: {A}n {I}ntroduction, Princeton {U}niv. {P}ress, 2021.

\bibitem{jordan1997free}
R.~Jordan, D.~Kinderlehrer, F.~Otto, \href{https://www.sciencedirect.com/science/article/pii/S0167278997000936}{Free energy and the {F}okker-{P}lanck equation}, Physica D 107~(2-4) (1997) 265--271.

\bibitem{lapolla2019manifestations}
A.~Lapolla, A.~Godec, \href{https://www.frontiersin.org/journals/physics/articles/10.3389/fphy.2019.00182}{Manifestations of projection-induced memory: {G}eneral theory and the tilted single file}, Front. Phys. 7 (2019) 182.

\bibitem{ibanez2024heating}
M.~Ib{\'a}{\~n}ez, C.~Dieball, A.~Lasanta, A.~Godec, R.~A. Rica, \href{https://doi.org/10.1038/s41567-023-02269-z}{Heating and Cooling are Fundamentally Asymmetric and Evolve Along Distinct Pathways}, Nat. Phys. 20 (2024) 135--141.

\bibitem{ito2020stochastic}
S.~Ito, A.~Dechant, \href{https://link.aps.org/doi/10.1103/PhysRevX.10.021056}{Stochastic time evolution, information geometry, and the {C}ram{\'e}r-{R}ao bound}, Phys.~Rev. X 10~(2) (2020) 021056.

\bibitem{meibohm2021relaxation}
J.~Meibohm, D.~Forastiere, T.~Adeleke-Larodo, K.~Proesmans, \href{https://link.aps.org/doi/10.1103/PhysRevE.104.L032105}{Relaxation-speed crossover in anharmonic potentials}, Phys. Rev. E 104 (2021) L032105.

\bibitem{van2021toward}
T.~Van~Vu, Y.~Hasegawa, \href{https://link.aps.org/doi/10.1103/PhysRevResearch.3.043160}{Toward relaxation asymmetry: {H}eating is faster than cooling}, Phys.~Rev. Res. 3~(4) (2021) 043160.

\bibitem{manikandan2021equidistant}
S.~Manikandan, \href{https://link.aps.org/doi/10.1103/PhysRevResearch.3.043108}{Equidistant quenches in few-level quantum systems}, Phys. Rev. Res. 3 (2021) 043108.

\bibitem{dieball2023asymmetric}
C.~Dieball, G.~Wellecke, A.~Godec, \href{https://link.aps.org/doi/10.1103/PhysRevResearch.5.L042030}{Asymmetric thermal relaxation in driven systems: {R}otations go opposite ways}, Phys. Rev. Res. 5 (2023) L042030.

\bibitem{tejero2024asymmetries}
{\'A}.~Tejero, R.~S{\'a}nchez, L.~E. Kaoutit, D.~Manzano, A.~Lasanta, \href{https://arxiv.org/abs/2406.19829}{Asymmetries of thermal processes in open quantum systems}, arXiv preprint arXiv:2406.19829 (2024).

\bibitem{nielsen2010quantum}
M.~A. Nielsen, I.~L. Chuang, Quantum {C}omputation and {Q}uantum {I}nformation, Cambridge {U}niv. {P}ress, 2010.

\bibitem{bravetti2024asymmetric}
A.~Bravetti, M.~A.~G. Ariza, P.~Padilla, \href{https://arxiv.org/abs/2402.14267}{Asymmetric relaxations through the lens of information geometry}, arXiv preprint arXiv:2402.14267 (2024).

\bibitem{hoffmann2008introduction}
K.~H. Hoffmann, \href{https://cab.unime.it/journals/index.php/AAPP/article/view/C1S0801011}{An introduction to endoreversible thermodynamics}, Atti della Accademia Peloritana dei Pericolanti-Classe di Scienze Fisiche, Matematiche e Naturali 86~(S1) (2008).

\bibitem{pemartin2024shortcuts}
I.~G.-A. Pemart\'{\i}n, E.~Momp\'o, A.~Lasanta, V.~Mart\'{\i}n-Mayor, J.~Salas, \href{https://link.aps.org/doi/10.1103/PhysRevLett.132.117102}{Shortcuts of freely relaxing systems Using equilibrium physical observables}, Phys. Rev. Lett. 132 (2024) 117102.

\bibitem{sakurai2020modern}
J.~J. Sakurai, J.~Napolitano, Modern {Q}uantum {M}echanics, Cambridge {U}niv. {P}ress, 2020.

\bibitem{pemartin2021slow}
I.~G.-A. Pemart{\'\i}n, E.~Momp{\'o}, A.~Lasanta, V.~Mart{\'\i}n-Mayor, J.~Salas, \href{https://link.aps.org/doi/10.1103/PhysRevE.104.044114}{Slow growth of magnetic domains helps fast evolution routes for out-of-equilibrium dynamics}, Phys.~Rev. E 104~(4) (2021) 044114.

\bibitem{hohenberg1977theory}
P.~C. Hohenberg, B.~I. Halperin, \href{https://link.aps.org/doi/10.1103/RevModPhys.49.435}{Theory of dynamic critical phenomena}, Rev. Mod. Phys. 49 (1977) 435--479.

\end{thebibliography}
\end{document}